\documentclass[aps,preprint,floats,epsf,epsfig,nofootinbib,letter]{revtex4}
\usepackage{graphicx}% Include figure files
\usepackage{dcolumn}% Align table columns on decimal point
\usepackage{bm}% bold math
\usepackage{subfigure}
\usepackage{mathrsfs}% Special tex fonts: \mathscr{ABC..}
\usepackage{amssymb}
\usepackage{color}
\usepackage{caption}
\usepackage{ragged2e}

\def\be{\begin{eqnarray}}
\def\en{\end{eqnarray}}
\def\non{\nonumber}
\def\la{\langle}
\def\ra{\rangle}
\def\pslash{\!\not{\!\partial}}

\def\pslash{\!\not{\! p}}
\def\eslash{\!\not{\! \epsilon}}

\begin{document}

\renewcommand{\baselinestretch}{1.10}
\font\el=cmbx10 scaled \magstep2{\obeylines\hfill June, 2016}

\vskip 1.5 cm

\centerline{\Large\bf Study of Majorana Fermionic Dark Matter}

\bigskip
\centerline{\bf Chun-Khiang Chua, Gwo-Guang Wong}
\medskip
\centerline{Department of Physics and Chung Yuan Center for High Energy Physics,}
\centerline{Chung Yuan Christian University,}
\centerline{Taoyuan, Taiwan 32023, Republic of China}
\medskip

%\date{\today}
%\bigskip
\centerline{\bf Abstract}
\bigskip
We construct a generic model of Majorana fermionic dark matter (DM). Starting with two Weyl spinor multiplets $\eta_{1,2}\sim (I,\mp Y)$ coupled to the Standard Model (SM) Higgs, six additional Weyl spinor multiplets with $(I\pm 1/2, \pm(Y\pm 1/2))$ are needed in general. It has 13 parameters in total, five mass parameters and eight Yukawa couplings. The DM sector of the minimal supersymmetric standard model (MSSM) is a special case of the model with $(I,Y)=(1/2,1/2)$. Therefore, this model can be viewed as an extension of the neutralino DM sector. We consider three typical cases: the neutralino-like, the reduced and the extended cases. For each case, we survey the DM mass $m_\chi$ in the range of $(1,2500)$ GeV by random sampling from the model parameter space and study the constraints from the observed DM relic density, the direct search of LUX, XENON100 and PICO experiments, and the indirect search of Fermi-LAT data. We investigate the interplay of these constraints and the differences among these cases. It is found that the direct detection of spin-independent DM scattering off nuclei and the indirect detection of DM annihilation to $W^+W^-$ channel are more sensitive to the DM searches in the near future. The allowed mass for finding $\tilde H$-, $\tilde B$-, $\tilde W$- and non neutralino-like DM particles and the predictions on $\langle\sigma (\chi{\chi}\rightarrow ZZ, ZH, t{\bar t}) v\rangle$ in the indirect search are given.
\bigskip
\small

%\pacs{Valid PACS appear here}

\maketitle

\section{Introduction}
It has been more than eighty years since the first evidence of dark matter (DM) was observed by Fritz Zwicky~\cite{Zwicky}. So for, all the astrophysical and cosmological observations of DM evidence show that DM exists everywhere no matter whether it is from the galactic scale~\cite{RF,BBSnBxxx,PDG1}, the scale of galaxy clusters~\cite{Carroll,Cxxx} or the cosmological scale~\cite{WMAPa,SDSS}.
Even though DM contains about $85\%$ for the total mass in the universe~\cite{WMAPb,Plank}, we still do not know much about its nature.
A leading class of DM candidates is the so-called weakly interacting massive particles (WIMPs)~\cite{LW,Ellis} which are non-luminous and non-baryonic cold DM (CDM) matter.
The WIMPs are assumed to be created thermally during the big bang, and froze out of thermal equilibrium escaping the Boltzmann suppression in the early universe.
The DM relic density is approximately related to the velocity averaged DM annihilation cross section by a simple relation~\cite{JKG}.
\be
\Omega_\chi h^2\approx {\frac{0.1 {\rm pb} \times c}{<\sigma v>}}.
\en
On the other hand, the recent measured value of CDM relic density is~\cite{PDG}
\be
\Omega^{\rm obs}_{\chi} h^2=0.1198\pm 0.0026.
\en
It suggests the case of DM with mass in the range of 100 GeV to few TeV and an electroweak size interaction.
That is the so-called âWIMP miracle.

The searches of DM particles in experiments have made much progress in recent years. Several complementary searching strategies have been continuously executed including the direct detection of DM-nucleus scattering in underground laboratories, the indirect detection of DM annihilation processes in astrophysical observation (see~\cite{DG} for a brief review) and the DM direct production at colliders~\cite{Mitsou,GIST,FHKT}. The null results of finding the DM from LUX~\cite{LUXSI}, XENON100~\cite{XENON100SD}, PICO~\cite{PICO-2L,PICO-60} and Fermi-LAT~\cite{Fermi-LAT} experiments put the related upper limits on spin-independent (SI)~\cite{SI1,SI2}, spin-dependent (SD)~\cite{SD1,SD2} DM-nucleus scattering cross sections and the velocity averaged DM annihilation cross sections respectively.
Except working on the well-known models such as the minimal supersymmetric models (MSSM) directly~\cite{JKG,Haber,RSW,Fowlie}, analyzing in the model-independent research with the effective operators of dark matter coupled to standard model (SM) particles ~\cite{Agrawal,Zheng,Cheung} is a way to search the properties of DM due to the little-known nature of DM.
Some authors also constructed models that the DM couples to the SM particles via a mediator, see for example, Higgs portal models~\cite{SS,BPV,PW,He1,Dutra}, 2HDM portal models~\cite{BBEG,He2}, fermion portal models~\cite{BB}, dark $Z^{'}$ portal~\cite{Alves:2013tqa},
left-right model~\cite{DS,GWZ} and so on.

In the DM-nucleus elastic scattering the DM is highly nonrelativistic.
Basically only the scalar-scalar (SS), vector-vector (VV), axial vector-axial vector (AA) and tensor-tensor (TT) DM-quark interactions are non-vanishing~\cite{Agrawal}.~\footnote{We will return to this point and take a closer look in Sec. II C.}
In Ref.~\cite{chua}, one of author (CKC) studied pure weak eigenstate Dirac
fermionic dark matter with renormalizable interaction.
It is well known that a Dirac fermionic DM particle, without a special choice of quantum number, usually gives an oversized SI DM-nucleus cross section through VV-interaction
from the $Z$-exchange diagram. 
To accommodate the bounds from direct searches,
the quantum number of DM is determined to be $I_3=Y=0$. There are only
two possible cases: either the DM has non-vanishing weak isospin ($I\neq
0$) but with $I_3=Y=0$ or it is an isosinglet ($I=0$) with $Y=0$.
In the first case, it is possible to have a
sizable $\chi{\bar\chi}\to W^+ W^-$ cross section, which is comparable
to the latest bounds from indirect searches.
There is no tree level diagram in DM-nucleus elastic scattering.
It successfully evades the SI bounds, but it pays the price of detectability in direct search.
In the second case, to couple DM to the SM particles, a SM-singlet vector mediator $X$ is required
from the renormalizability and the SM gauge quantum numbers.
The allowed parameter space and the consequences were studied.
To satisfy
the latest bounds of direct searches and to reproduce the DM relic
density at the same time, resonant enhancement via the $X$-pole in the DM annihilation
diagram is needed. Thus, the masses of DM and the mediator are
related.
It is arguable that the phenomenology of Dirac fermionic DM is not very rich.

The Majorana DM can naturally evade the dangerous $Z$-exchange diagram from the $VV$ interaction and can have rich phenomenology.
A well known example is the lightest neutralino in MSSM~\cite{JKG,Haber}.
In this work, we construct a generic class of Majorana fermionic DM models having arbitrary weak isospin quantum number. As we shall see the MSSM DM sector is a special case in this model, therefore, this model can be viewed as an extension of the neutralino DM sector. We consider three typical cases: the neutralino-like, the reduced and the extended cases. Note that a somewhat related study to the reduced case has been given in Ref.~\cite{CMT}.

This paper is organized as follows. In Sec. II we construct a generic model of Majorana fermionic DM and give the formulas for the DM annihilation to the SM particles as well as DM-nucleus elastic scattering. We give the results of the neutralino-like, the reduced and the extended cases in Sec. III. We discuss the coannihilation and give the conclusions in Sec. IV. 
We present explicitly the relevant Lagrangian of the WIMP mass term in Appendix~A.
The 4-component Majorana and Dirac mass eigenstates for neutral and single charged WIMPs are constructed respectively in Appendix~B. We present the Lagrangian of WIMPs interacting with the SM particles in Appendix~C, give the matrix elements 
%used in the processes of 
of DM annihilation to the SM particles in Appendix~D, and show that the Lagrangian is CP conserved in Appendix~E. The formulas used in DM-nucleus elastic scattering are derived in Appendix~F. The formulation and the corresponding matrix elements for WIMP coannihilation are given in Appendices~G and~H, respectively.

\section{Formalism}

\subsection{A Generic Model of Majorana Fermionic Dark Matter}

Starting with the SM, we add 
% add $Z_2$-odd
two $Z_2$-odd, 
2-component Weyl spinor multiplets  $\eta_{1,2}\sim (2I+1,\mp Y)$ under $SU_L(2)\times U(1)_Y$ and all SM particles are assigned to be $Z_2$ even.
% add The introducing of $Z_2$ symmetry assures the stability of DM.
The introducing of $Z_2$ symmetry assures the stability of DM.
Without loss of generality we take $Y\geq 0$.
A mass term can be constructed as
\be
-{\cal L}_m=\mu \lambda_{ij} \eta^i_2\eta^j_1+\mu \lambda^*_{ij}\bar\eta^i_2\bar\eta^j_1,
\en
with
\be
\lambda_{ij}\equiv\sqrt{2I+1}\la I I; 0 0 |I i, I j\ra
\en
proportional to the Clebsch-Gordan coefficient
and $i,j=-I,\dots,I$.
This is actually a Dirac particle multiplet.
The reason is explained below.
We define
\be
\xi^i\equiv\eta^i_2,
\quad
\bar\eta^i\equiv\lambda_{ij}\bar\eta^j_1,
\en
and
the Dirac field with $ i^{\rm th}$ component of isospin
\be
\psi^i
\equiv
\left(
\begin{array}{c}
\xi^i \\
\bar\eta^i
\end{array}
\right).
\en
Note that the hypercharge of $\psi$ is $Y$.
Since under SU(2) transformation, we have
\be
\xi'^i&=&U_{ij} \eta_2^j=U_{ij}\xi^j,
\non\\
\bar\eta'^i
&=&\lambda_{ik} U^*_{kl}\lambda^{-1}_{lj}\lambda_{jr}\bar\eta^r_1
=U_{ij}\bar\eta^j,
\en
where we have used the similarity transformation of the SU(2) transformation matrix,
\footnote
{This can be seen from
$-(\vec I)^*_{ij}=(-)^{-i} (\vec I)_{-i,-j}(-)^j=[(-)^{I-i}\delta_{-i,k}](\vec I)_{kl}[(-)^{-I+j}\delta_{l,-j}]$ and $\lambda_{ij}=(-)^{-I+i}\delta_{i,-j}$,
i.e. $-(\vec I)^*_{ij}=\lambda^{-1}_{ik}(\vec I)_{kl}\lambda_{lj}$.}
\be
\lambda_{ik} U^*_{kl}\lambda^{-1}_{lj}=U_{ij}.
\en
Hence the transform of ($2I+1$)-multiplet of Dirac fields in $\psi$ under SU(2) is
\be
\psi'^i=U_{ij}\psi^j,
\en
and the above mass term is simply
\be
-{\cal L}_m=\mu\bar\psi\psi.
\en
The component $\psi^{-Y}$ with neutral charge could be a dark matter candidate. But in the
$I\neq 0$ and $Y\neq 0$ case, $\psi^{-Y}$ will induce a sizable SI-scattering cross section via $Z$-boson exchange ($\sim 10^{-39} {\rm cm}^2$)~\cite{chua}, which is ruled out by the present direct search data~\cite{LUXSI}.
To clarify the situation, we switch back to the $\eta_{1,2}$ basis. By diagonalizing the mass matrix, we find that
there are two neutral Majorana degenerate states $\chi_{1,2}\propto(\eta_1\pm\eta_2)/\sqrt2$ with mass $|\mu \lambda_{Y,-Y}|=\mu$.
Both of them can be dark matter, since their masses are degenerate. The dangerous $Z$-boson exchange diagram is from the $\chi_1\to\chi_2$ vector current (the $\chi_i\to\chi_i$ current can only be an axial one).
The above situation can be avoided, if one lift the mass degeneracy of $\chi_{1,2}$.
To do so, we enlarge the mass matrix. The $Z_2$-odd WIMPs, $\eta_{1,2}$, can mix with additional 
% add $Z_2$-odd
$Z_2$-odd 
WIMPs in the presence of the Higgs field $\phi$ [with quantum number ($2$, $1/2$)] and obtain new mass term after spontaneous symmetry breaking (SSB). 
We consider all possible combinations of renormalizable interactions with $\eta_{1,2}$ coupled to the Higgs field:
\be
(i)\,\phi\times\eta_1\times [new],
\quad
(ii)\,\phi\times\eta_2\times [new],
\quad
(iii)\,\tilde\phi\times\eta_1\times [new],
\quad
(iv)\,\tilde\phi\times\eta_2\times [new],
\label{eq:induced}
\en
where $\tilde\phi^i\equiv\epsilon_{ij}\phi^{*j}$ with $\epsilon_{ij}=\lambda_{ij}$ for $I=1/2$ (i.e. $\epsilon_{ij}=-\epsilon_{ji}$ and $\epsilon_{1/2,-1/2}=1$).
The allowed quantum numbers of these new particles are given in Table.~\ref{tab: QN}.

\captionsetup{
justification=raggedright,
}
\begin{table}[t]
\begin{ruledtabular}
\begin{tabular}{l c c c c}
[new]
 &$SU(2)(I_\eta)$
 & $U_Y(1)$
 & type
 & couples with
 \\
 \hline
$\eta_3$
 & $I-1/2$
 & $-(Y-\frac{1}{2})$
 & (iv)
 & $\tilde\phi\times\eta_2$, $\eta_4$
 \\
$\eta_4$
 & $I-1/2$
 & $Y-\frac{1}{2}$
 & (i)
 & $\phi\times\eta_1$, $\eta_3$
  \\
$\eta_5$
 & $I+1/2$
 & $-(Y-\frac{1}{2})$
 & (iv)
 & $\tilde\phi\times\eta_2$, $\eta_6$
 \\
$\eta_6$
 & $I+1/2$
 & $Y-\frac{1}{2}$
 & (i)
 & $\phi\times\eta_1$, $\eta_5$
 \\
$\eta_7$
 & $I-1/2$
 & $-(Y+\frac{1}{2})$
 & (ii)
 & $\phi\times\eta_2$, $\eta_8$
 \\
$\eta_8$
 & $I-1/2$
 & $Y+\frac{1}{2}$
 & (iii)
 & $\tilde\phi\times\eta_1$, $\eta_7$
 \\
$\eta_{9}$
 & $I+1/2$
 & $-(Y+\frac{1}{2})$
 & (ii)
 & $\phi\times\eta_2$, $\eta_{10}$
 \\
$\eta_{10}$
 & $I+1/2$
 & $Y+\frac{1}{2}$
 & (iii)
 & $\tilde\phi\times\eta_1$, $\eta_{9}$
 \\
 \end{tabular}
 \caption{Summary of the eight types of additional multiplets induced by the 4 general types of couplings involving the Higgs field and $\eta_{1,2}$.}
\label{tab: QN}
\end{ruledtabular}
\end{table}

The generic Lagrangian is given by
\be
-{\cal L}_m&=&
\sum_{p=1}^5 \mu_p\lambda^{p}_{ij}\eta^i_{2p}\eta^j_{2p-1}
+\sum_{p=2}^3(g_{2p-1}\lambda^{p}_{ijk}\tilde\phi^i \eta^j_2\eta^k_{2p-1}
+g_{2p}\lambda^{p}_{ijk}\phi^i\eta^j_1\eta^k_{2p})
\non\\
&&+\sum_{p=4}^5(g_{2p-1}\lambda^{p}_{ijk}\phi^i \eta^j_2\eta^k_{2p-1}
+g_{2p}\lambda^{p}_{ijk}\tilde\phi^i\eta^j_1\eta^k_{2p})+h.c.,
\label{eq: generic}
\en
with
\be
\lambda^1_{ij}&=&\sqrt{2I+1}\la I I; 0 0 |I i, I j\ra,
\non\\
\lambda^2_{ij}=\lambda^4_{ij}
&=&\sqrt{2I}\left\la \left(I-\frac{1}{2}\right) \left(I-\frac{1}{2}\right); 0 0 \Bigg|\left(I-\frac{1}{2}\right) i, \left(I-\frac{1}{2}\right) j\right\ra,
\non\\
\lambda^3_{ij}=\lambda^5_{ij}
&=&\sqrt{2I+2}\left\la \left(I+\frac{1}{2}\right) \left(I+\frac{1}{2}\right); 0 0 \Bigg|\left(I+\frac{1}{2}\right) i, \left(I+\frac{1}{2}\right) j\right\ra,
\en
\be
\lambda^2_{ijk}=\lambda^4_{ijk}&=&\epsilon_{ir} \sqrt{2I}\left\la I \left(I-\frac{1}{2}\right); \frac{1}{2} r \Bigg| I j, \left(I-\frac{1}{2}\right) k\right\ra,
\non\\
\lambda^3_{ijk}=\lambda^5_{ijk}&=&\epsilon_{ir} \sqrt{2I+2}\left\la I \left(I+\frac{1}{2}\right); \frac{1}{2} r \Bigg| I j, \left(I+\frac{1}{2}\right) k\right\ra.
\en

%Note that the above Lagrangian satisfies all the SM symmetries, including the lepton number conservation, or we may impose a $Z_2$ symmetry to avoid the mixing with leptons and $\eta_i$. 
Note that the imposed $Z_2$ symmetry can protect the DM against decays. Otherwise, DM can decay through, for example, the lepton number violation term and become unstable.
Eq.~(\ref{eq: generic}) can be used as a building block to built other multiplets.
In principle, one can replace $\eta_{1,2}$ by the induced fields in Eq.~(\ref{eq:induced}) and involve additional fields. For simplicity, we do not do it here. In fact, a more complicated case can be readily generated by using the present case as a module.

These fields can be combined into Dirac fields with definite isospin and hypercharge quantum numbers:
\be
\psi^i_{(p)}\equiv
\left(
\begin{array}{c}
\xi^i_{(p)}\\
\bar\eta^i_{(p)}
\end{array}
\right)
\en
with
\be
\xi^i_{(p)}\equiv \eta^i_{2p},
\quad
\bar\eta^i_{(p)}\equiv\lambda^p_{ij}\bar\eta^j_{2p-1},
\en
for $p=1,\cdots,5$.
Consequently, we have
\be
\mu_p\lambda^p_{ij}\eta^j_{2p-1}\eta^i_{2p}
&=&\mu_p\eta^i_{(p)}\xi^i_{(p)}
\non\\
&=&\mu_p\bar\psi^i_{(p)R}\psi^i_{(p)L},
\en
for $p=1,\cdots,5$,
\be
g_{2p-1}\lambda^{p}_{ijk}\tilde\phi^i \eta^j_2\eta^k_{2p-1}
&=&g_{2p-1}[\lambda^p_{ijk}(\lambda^p)^{-1}_{kl}]\tilde\phi^i \xi^j_{(1)}\eta^l_{(p)}
\non\\
&=&g_{2p-1}[\lambda^p_{ijk}(\lambda^p)^{-1}_{kl}]\tilde\phi^i \bar\psi^l_{(p)R}\psi^j_{(1)L}
\en
and
\be
g_{2p}\lambda^p_{ijk}\phi^i\eta^j_1\eta^k_{2p}
&=&g_{2p}[\lambda^p_{ijk}(\lambda^1)^{-1}_{jl}]\phi^i\eta^l_{(1)}\xi^k_{(2p)}
\non\\
&=&g_{2p}[\lambda^p_{ijk}(\lambda^1)^{-1}_{jl}]\phi^i\bar\psi^l_{(1)R}\psi^k_{(p)L},
\en
for $p=2,\cdots,5$, giving
\be
-{\cal L}_m&=&
\sum_{p=1}^5 \mu_p\bar\psi^i_{(p)R}\psi^i_{(p)L}
+\sum_{p=2}^3\{ g_{2p-1}[\lambda^p_{ijk}(\lambda^p)^{-1}_{kl}]\tilde\phi^i \bar\psi^l_{(p)R}\psi^j_{(1)L}
+g_{2p}[\lambda^p_{ijk}(\lambda^1)^{-1}_{jl}]\phi^i\bar\psi^l_{(1)R}\psi^k_{(p)L}\}
\non\\
&&
+\sum_{p=4}^5\{g_{2p-1}[\lambda^p_{ijk}(\lambda^p)^{-1}_{kl}]\phi^i \bar\psi^l_{(p)R}\psi^j_{(1)L}
+g_{2p}[\lambda^p_{ijk}(\lambda^1)^{-1}_{jl}]\tilde\phi^i\bar\psi^l_{(1)R}\psi^k_{(p)L}\}+h.c..
\en
After SSB, the above Lagrangian will generate the mixing in these Dirac fields.
We still do not have any Majorana particle.

The MSSM case can shed some light on this issue. In fact,
the relevant MSSM multiplet corresponds to
\be
I=Y=\frac{1}{2},
\quad
\eta_{1,2}=\tilde H_{1,2},\quad
\eta_3,\eta_4\propto\tilde B,
\quad
\eta_5,\eta_6\propto\tilde W,
\quad
\mbox {without $\eta_{7,8,9,10}$}.
\label{eq:mssm}
\en
The Majorana particles can only
enter when $Y=1/2$, where the quantum numbers of $\eta_{3(5)}$ and $\eta_{4(6)}$ are identical, and to have neutral particles $I$ can only be half-integers. Consequently, we have
\be
Y=\frac{1}{2},
\quad
I=\frac{2n+1}{2},
\quad
\eta_3={\rm sign}(\mu_2)(-1)^n\eta_4,
\quad
\eta_5={\rm sign}(\mu_3)(-1)^{n+1}\eta_6,
\en
and $\mu_{2,3}$ change to $\mu_{2,3}/2$,
which we will stick to
this
throughout this work. Note that the additional signs in the relations of $\eta_{3,4}$ and $\eta_{5,6}$ are designed to absorb the signs of the corresponding Majorana mass terms ($\mu_{2,3}$, see Eq.~(\ref{eq: L0m}) below).

The Lagrangian for neutral WIMP mass term is
\be
-{\cal L}^0_m&=&
\mu_1\lambda^{1}_{-\frac{1}{2},\frac{1}{2}}\eta^{-\frac{1}{2}}_2\eta^{\frac{1}{2}}_1
+\frac{1}{2}\mu_2\lambda^{2}_{0,0}\eta^{0}_4\eta^{0}_3
+\frac{1}{2}\mu_3\lambda^{3}_{0,0}\eta^{0}_6\eta^{0}_5
\non\\
&&
+\mu_4\lambda^{4}_{-1,1}\eta^{-1}_8\eta^{1}_7
+\mu_5\lambda^{5}_{-1,1}\eta^{-1}_{10}\eta^{1}_{9}
\non\\
&&
+g_3\lambda^2_{\frac{1}{2},-{\frac{1}{2}},0}\la\tilde\phi^{\frac{1}{2}}\ra \eta^{-{\frac{1}{2}}}_2\eta^{0}_3
+g_4\lambda^2_{-\frac{1}{2},{\frac{1}{2}},0}\la\phi^{-\frac{1}{2}}\ra\eta^{\frac{1}{2}}_1\eta^{0}_4
\non\\
&&
+g_5\lambda^3_{\frac{1}{2},-{\frac{1}{2}},0}\la\tilde\phi^{\frac{1}{2}}\ra \eta^{-{\frac{1}{2}}}_2\eta^{0}_5
+g_6\lambda^3_{-\frac{1}{2},{\frac{1}{2}},0}\la\phi^{-\frac{1}{2}}\ra\eta^{\frac{1}{2}}_1\eta^{0}_6
\non\\
&&
+g_7\lambda^4_{-\frac{1}{2},-{\frac{1}{2}},1}\la\phi^{-\frac{1}{2}}\ra\eta^{-{\frac{1}{2}}}_2\eta^{1}_7
+g_8\lambda^4_{\frac{1}{2},{\frac{1}{2}},-1}\la\tilde\phi^{\frac{1}{2}}\ra \eta^{\frac{1}{2}}_1\eta^{-1}_8
\non\\
&&
+g_{9}\lambda^5_{-\frac{1}{2},-{\frac{1}{2}},1}\la\phi^{-\frac{1}{2}}\ra\eta^{-{\frac{1}{2}}}_2\eta^{1}_{9}
+g_{10}\lambda^5_{\frac{1}{2},{\frac{1}{2}},-1}\la\tilde\phi^{\frac{1}{2}}\ra \eta^{\frac{1}{2}}_1\eta^{-1}_{10}
+h.c..
\en
It can be simplified as
\be
-{\cal L}^0_m&=&
\mu_1(-1)^{n+1}\eta^{-\frac{1}{2}}_2\eta^{\frac{1}{2}}_1
+\frac{1}{2}\mu_2(-1)^n\eta^{0}_4\eta^{0}_3
+\frac{1}{2}\mu_3(-1)^{n+1}\eta^{0}_6\eta^{0}_5
\non\\
&&
+\mu_4(-1)^{n+1}\eta^{-1}_8\eta^{1}_7
+\mu_5(-1)^n\eta^{-1}_{10}\eta^{1}_{9}
\non\\
&&
+g_3(-1)^n\la\tilde\phi^{\frac{1}{2}}\ra \eta^{-{\frac{1}{2}}}_2\eta^{0}_3
+g_4(-1)^{n+1}\la\phi^{-\frac{1}{2}}\ra\eta^{\frac{1}{2}}_1\eta^{0}_4
\non\\
&&
+g_5(-1)^{1-n}\la\tilde\phi^{\frac{1}{2}}\ra \eta^{-{\frac{1}{2}}}_2\eta^{0}_5
+g_6(-1)^{1-n}\la\phi^{-\frac{1}{2}}\ra\eta^{\frac{1}{2}}_1\eta^{0}_6
\non\\
&&
+g_7(-1)^n\sqrt{\frac{n}{n+1}}\la\phi^{-\frac{1}{2}}\ra\eta^{-{\frac{1}{2}}}_2\eta^{1}_7
+g_8(-1)^{n+1}\sqrt{\frac{n}{n+1}}\la\tilde\phi^{\frac{1}{2}}\ra \eta^{\frac{1}{2}}_1\eta^{-1}_8
\non\\
&&
+g_{9}(-1)^{-n}\sqrt{\frac{n+2}{n+1}}\la\phi^{-\frac{1}{2}}\ra\eta^{-{\frac{1}{2}}}_2\eta^{1}_{9}
+g_{10}(-1)^{-n}\sqrt{\frac{n+2}{n+1}}\la\tilde\phi^{\frac{1}{2}}\ra \eta^{\frac{1}{2}}_1\eta^{-1}_{10}
+h.c..
\label{eq: L0m}
\en
With the basis $\Psi^{0T}_i=(\eta^{1/2}_1,\eta^{-1/2}_2,\eta^0_3,\eta^0_5,\eta^1_7,\eta^{-1}_8,
\eta^1_9,\eta^{-1}_{10})$, the above 
Lagrangian after SSB can be written as
\be
{\cal L}^0_m=-\frac{1}{2}\Psi^{0T}{Y}\Psi^0+h.c.,
\label{eq: L1m}
\en
where the corresponding mass matrix $Y$ takes the form
\be
\footnotesize
\left(
\begin{array}{cccccccc}
0
 &(-)^{n+1}\mu_1
 &-\frac{g_4 v}{\sqrt2}
 &\frac{g_6 v}{\sqrt2}
 &0
 &\frac{(-1)^{n+1}g_8 v\sqrt n}{\sqrt{2(n+1)}}
 &0
 &\frac{(-1)^{n}g_{10} v\sqrt{n+2}}{\sqrt{2(n+1)}}
 \\
(-)^{n+1}\mu_1
 &0
 &\frac{(-1)^n g_3 v}{\sqrt2}
 &\frac{(-1)^{n+1}g_5 v}{\sqrt2}
 &\frac{(-1)^{n}g_7 v\sqrt n}{\sqrt{2(n+1)}}
 &0
 &\frac{(-1)^{n}g_9 v\sqrt{n+2}}{\sqrt{2(n+1)}}
 &0
 \\
-\frac{g_4 v}{\sqrt2}
 &\frac{ (-1)^n g_3 v}{\sqrt2}
 &\mu_2
 &0
 &0
 &0
 &0
 &0
\\
\frac{g_6 v}{\sqrt2}
 &\frac{(-)^{n+1}g_5 v}{\sqrt2}
 &0
 &\mu_3
 &0
 &0
 &0
 &0
 \\
0
 &\frac{(-1)^{n}g_7 v\sqrt n}{\sqrt{2(n+1)}}
 &0
 &0
 &0
 &(-1)^{n+1}\mu_4
 &0
 &0
 \\
\frac{(-1)^{n+1} g_8 v\sqrt n}{\sqrt{2(n+1)}}
 &0
 &0
 &0
 &(-1)^{n+1}\mu_4
 &0
 &0
 &0
 \\
0
 &\frac{(-1)^{n}g_9 v\sqrt{n+2}}{\sqrt{2(n+1)}}
 &0
 &0
 &0
 &0
 &0
 &(-1)^n\mu_5
 \\
\frac{(-1)^{n} g_{10} v\sqrt{n+2}}{\sqrt{2(n+1)}}
 &0
 &0
 &0
 &0
 &0
 &(-1)^n\mu_5
 &0
\end{array}
\right).
\non \\
\label{eq: MY}
\en

In parallel with the neutralino sector in MSSM, we work the model with $I=Y=1/2$ and the Lagrangian for the neutral WIMP mass term must be modified as in Appendix A.
Note that the sign convention of Clebsch-Gordan coefficient is different from those usually used in quantum field theory. For example we usually use $\pi^{\pm}=(\pi_1\mp i\pi_2)/\sqrt2$, while the Clebsch-Gordan convention is $\pi^{\pm}=\mp(\pi_1\mp i\pi_2)/\sqrt2$.
Comparing to MSSM, we then have the following correspondences:
\be
&\eta_1=\tilde H_1,\eta_2=\tilde H_2,\eta_3=-i\lambda',\eta^{\pm,0}_5=-i(\mp\lambda_\pm,\lambda_3),&
\non\\
&g_3 v=\sqrt2 m_Z\cos\beta \sin\theta_W,
\quad
g_4 v=\sqrt2 m_Z\sin\beta \sin\theta_W,&
\non\\
&
g_5 v=\sqrt2 m_Z\cos\beta \cos\theta_W,
\quad
g_6 v=\sqrt2 m_Z\sin\beta \cos\theta_W,&
\non\\
&
\mu_4=\mu_5=0,
\quad
g_{7,8,9,10}=0,&
\label{eq: MSSM}
\en
where the additional sign in front of $\lambda_+$ is to absorb the sign from the Clebsch-Gordan sign convention.

When diagonalizing the mass matrix in Eq.~(\ref{eq: MY}) and producing nonnegative mass eigenvalues, one sometimes needs to absorb a negative sign resulting in purely imaginary matrix elements in the transition matrix. 
On the other hand,
one should note that all parameters in the Lagrangian are assumed to be real before transforming the gauge eigenstates to mass eigenstates in this model. The whole Lagrangian in this model is then $CP$ conserved. As noted after field redefinition, some couplings become purely imaginary. However, the whole Lagrangian should still be $CP$ conserved (see Appendix E). 

The Lagrangian for single charged WIMP mass term is
\be
-{\cal L}^{\pm}_m&=&
\mu_1(-1)^{n}(\eta^{\frac{1}{2}}_2\eta^{-\frac{1}{2}}_1+\eta^{-\frac{3}{2}}_2\eta^{\frac{3}{2}}_1)
+\frac{1}{2}\mu_2(-1)^{n+1}(\eta^{1}_4\eta^{-1}_3+\eta^{-1}_4\eta^{1}_3)
+\frac{1}{2}\mu_3(-1)^{n}(\eta^{1}_6\eta^{-1}_5+\eta^{-1}_6\eta^{1}_5)
\non\\
&&
+\mu_4(-1)^{n}(\eta^{0}_8\eta^{0}_7+\eta^{-2}_8\eta^{2}_7)
+\mu_5(-1)^{n+1}(\eta^{0}_{10}\eta^{0}_{9}+\eta^{-2}_{10}\eta^{2}_{9})
\non\\
&&
+g_3(-1)^{n+1}\left(
  \sqrt{\frac {n}{n+1}}\la\tilde\phi^{\frac{1}{2}}\ra \eta^{{\frac{1}{2}}}_2\eta^{-1}_3
+\sqrt{\frac {n+2}{n+1}}\la\tilde\phi^{\frac{1}{2}}\ra \eta^{{-\frac{3}{2}}}_2\eta^{1}_3\right)
\non\\
&&
+g_4(-1)^{n}\left(
  \sqrt{\frac {n+2}{n+1}}\la\phi^{-\frac{1}{2}}\ra\eta^{\frac{3}{2}}_1\eta^{-1}_4
+\sqrt{\frac {n}{n+1}}\la\phi^{-\frac{1}{2}}\ra\eta^{-\frac{1}{2}}_1\eta^{1}_4\right)
\non\\
&&
+g_5(-1)^{n}\left(
  \sqrt{\frac {n+2}{n+1}}\la\tilde\phi^{\frac{1}{2}}\ra \eta^{\frac{1}{2}}_2\eta^{-1}_5
+\sqrt{\frac {n}{n+1}}\la\tilde\phi^{\frac{1}{2}}\ra \eta^{-\frac{3}{2}}_2\eta^{1}_5\right)
\non\\
&&
+g_6(-1)^{n}\left(
  \sqrt{\frac {n}{n+1}}\la\phi^{-\frac{1}{2}}\ra\eta^{-\frac{1}{2}}_1\eta^{1}_6
+\sqrt{\frac {n+2}{n+1}}\la\phi^{-\frac{1}{2}}\ra\eta^{\frac{3}{2}}_1\eta^{-1}_6\right)
\non\\
&&
+g_7(-1)^{n+1}\left(
  \la\phi^{-\frac{1}{2}}\ra\eta^{-\frac{1}{2}}_2\eta^{0}_7
+\sqrt{\frac {n-1}{n+1}}\ra\eta^{-\frac{-3}{2}}_2\eta^{2}_7\right)
\non\\
&&
+g_8(-1)^{n}\left(
\sqrt{\frac{n-1}{n+1}}\la\tilde\phi^{\frac{1}{2}}\ra \eta^{\frac{3}{2}}_1\eta^{-2}_8
+\la\tilde\phi^{\frac{1}{2}}\ra \eta^{\frac{-1}{2}}_1\eta^{0}_8\right)
\non\\
&&
+g_{9}(-1)^{n-1}\left(
\la\phi^{-\frac{1}{2}}\ra\eta^{\frac{1}{2}}_2\eta^{0}_{9}
+\sqrt{\frac{n+3}{n+1}}\la\phi^{-\frac{1}{2}}\ra\eta^{-{\frac{3}{2}}}_2\eta^{2}_{9}\right)
\non\\
&&
+g_{10}(-1)^{n-1}\left(
\sqrt{\frac{n+3}{n+1}}\la\tilde\phi^{\frac{1}{2}}\ra \eta^{\frac{3}{2}}_1\eta^{-2}_{10}
+\la\tilde\phi^{\frac{1}{2}}\ra \eta^{-\frac{1}{2}}_1\eta^{0}_{10}\right)
+h.c..
\label{eq: L2m}
\en

As mentioned previously, $<i|T^+|j>$ used in quantum field theory is connected to Clebsch-Gordan coefficient $<m'|J^+|m>$ used in quantum mechanics by a similarity transformation $V$
\be
<I_k,i|T^+|I_k,j>=\sum_{m,m'}<I_k,i|V^\dagger|I_k,m'><I_k,m'|VJ^+V^\dagger|I_k,m><I_k,m|V|I_k,j>.
\en
When dealing with the single charged particles, the similarity transformation only changes the sign of positive charged particles with an integer isospin; namely,  we only need to do the following transform
\be
\eta^{q_k+1}_k\rightarrow\eta'^{q_k+1}_k\equiv V\eta^{q_k+1}_k=(-1)^{{\rm mod}(2I_k,2)+1}\eta^{q_k+1}_k,
\en
where $q_k$ in $\eta^{q_k+1}_k$ is defined as the the third component of isospin corresponding to the neutral particle in the multiplet $\eta_k$ with isospin $I_k$.
With the basis
$\Psi^{+T}_i=(\eta'^{3/2}_1,\eta'^{1/2}_2,\eta'^1_3,\eta'^1_5,\eta'^2_7,\eta'^0_8,\eta'^2_9,\eta'^0_{10})$ and
$\Psi^{-T}_i=(\eta^{-1/2}_1,\eta^{-3/2}_2,\eta^{-1}_3,\eta^{-1}_5,\eta^0_7,\eta^{-2}_8,
\eta^0_9,\eta^{-2}_{10})$,
the Lagrangian in Eq. (\ref{eq: L2m}) becomes
\be
-{\cal L}^{\pm}_m&=&
\mu_1(-1)^{n}(\eta^+_2\eta^-_1+\eta^-_2\eta^+_1)
+\frac{1}{2}\mu_2(\eta'^+_3\eta^-_3+\eta^-_3\eta'^+_3)
+\frac{1}{2}\mu_3(\eta'^+_5\eta^-_5+\eta^-_5\eta'^+_5)
\non\\
&&
-\mu_4(-1)^{n}(\eta'^{+}_8\eta^-_7+\eta^-_8\eta'^+_7)
-\mu_5(-1)^{n+1}(\eta'^{+}_{10}\eta^{-}_{9}+\eta^{-}_{10}\eta'^{+}_{9})
\non\\
&&
+g_3(-1)^{n+1}\left(
  \sqrt{\frac {n}{n+1}}\la\tilde\phi^{0}\ra \eta^{+}_2\eta^{-}_3
-\sqrt{\frac {n+2}{n+1}}\la\tilde\phi^{0}\ra \eta^{-}_2\eta'^{+}_3\right)
\non\\
&&
+g_4\left(
  \sqrt{\frac {n+2}{n+1}}\la\phi^{0}\ra\eta^{+}_1\eta^{-}_3
+\sqrt{\frac {n}{n+1}}\la\phi^{0}\ra\eta^{-}_1\eta'^{+}_3\right)
\non\\
&&
+g_5(-1)^{n}\left(
  \sqrt{\frac {n+2}{n+1}}\la\tilde\phi^{0}\ra \eta^{+}_2\eta^{-}_5
-\sqrt{\frac {n}{n+1}}\la\tilde\phi^{0}\ra \eta^{-}_2\eta'^{+}_5\right)
\non\\
&&
+g_6\left(
-\sqrt{\frac {n}{n+1}}\la\phi^{0}\ra\eta^{+}_1\eta^{-}_5
+\sqrt{\frac {n+2}{n+1}}\la\phi^{0}\ra\eta^{-}_1\eta'^{+}_5\right)
\non\\
&&
+g_7(-1)^{n+1}\left(
  \la\phi^{0}\ra\eta^{+}_2\eta^{-}_7
-\sqrt{\frac {n-1}{n+1}}\la\phi^{0}\ra\eta^{-}_2\eta'^{+}_7\right)
\non\\
&&
+g_8(-1)^{n}\left(
\sqrt{\frac{n-1}{n+1}}\la\tilde\phi^{0}\ra \eta^{+}_1\eta^{-}_8
-\la\tilde\phi^{0}\ra \eta^-_1\eta'^{+}_8\right)
\non\\
&&
+g_{9}(-1)^{n-1}\left(
\la\phi^{0}\ra\eta^{+}_2\eta^{-}_{9}
-\sqrt{\frac{n+3}{n+1}}\la\phi^{0}\ra\eta^{-}_2\eta'^{+}_{9}\right)
\non\\
&&
+g_{10}(-1)^{n-1}\left(
\sqrt{\frac{n+3}{n+1}}\la\tilde\phi^{0}\ra \eta^{+}_1\eta^{-}_{10}
-\la\tilde\phi^{0}\ra \eta^{-}_1\eta'^{+}_{10}\right)
+h.c..
\label{eq: L3m}
\en
After SSB, it can be written as a compact form as follows
\be
{\cal L}^{\pm}_m=-\frac{1}{2}(\Psi^+, \Psi^-)
\left(
\begin{array}{cc}
0&X^T
\\
X&0
\end{array}
\right)
\left(
\begin{array}{c}
\Psi^+
\\
\Psi^-
\end{array}
\right)+h.c..
\label{eq: L4m}
\en
where $X$ takes the form
\be
\footnotesize
\left(
\begin{array}{cccccccc}
0
 &(-)^{n}\mu_1
 &\frac{-g_4 v \sqrt n}{\sqrt{2(n+1)}}
 &\frac{g_6 v \sqrt{n+2}}{\sqrt{2(n+1)}}
 &0
 &\frac{(-1)^{n+1}g_8 v}{\sqrt2}
 &0
 &\frac{(-1)^{n}g_{10} v}{\sqrt2}
 \\
(-)^{n}\mu_1
 &0
 &\frac{(-)^{n} g_3 v \sqrt{n+2}}{\sqrt{2(n+1)}}
 &\frac{(-)^{n+1}g_5 v \sqrt{n}}{\sqrt{2(n+1)}}
 &\frac{(-)^{n}g_7 v \sqrt{n-1}}{\sqrt{2(n+1)}}
 &0
 &\frac{(-)^{n}g_9 v\sqrt{n+3}}{\sqrt{2(n+1)}}
 &0
 \\
\frac{g_4 v \sqrt{n+2}}{\sqrt{2(n+1)}}
 &\frac{ (-)^{n+1} g_3 v \sqrt{n}}{\sqrt{2(n+1)}}
 &\mu_2
 &0
 &0
 &0
 &0
 &0
\\
\frac{-g_6 v \sqrt{n}}{\sqrt{2(n+1)}}
 &\frac{(-)^{n}g_5 v  \sqrt{n+2}}{\sqrt{2(n+1)}}
 &0
 &\mu_3
 &0
 &0
 &0
 &0
 \\
0
 &\frac{(-1)^{n+1}g_7 v}{\sqrt{2}}
 &0
 &0
 &0
 &(-)^{n+1}\mu_4
 &0
 &0
 \\
\frac{(-1)^{n} g_8 v\sqrt {n-1}}{\sqrt{2(n+1)}}
 &0
 &0
 &0
 &(-)^{n+1}\mu_4
 &0
 &0
 &0
 \\
0
 &\frac{(-)^{n-1}g_9 v}{\sqrt{2}}
 &0
 &0
 &0
 &0
 &0
 &(-)^{n+1}\mu_5
 \\
\frac{(-)^{n-1} g_{10} v\sqrt{n+3}}{\sqrt{2(n+1)}}
 &0
 &0
 &0
 &0
 &0
 &(-)^{n+1}\mu_5
 &0
\end{array}
\right).
\non \\
\label{eq: MX}
\en

Comparing to the chargino sector in MSSM with $\psi^{+T}_i=(-i\lambda^+,\psi^1_{H_2})$ and
$\psi^{-T}_j=(-i\lambda^-,\psi^2_{H_1})$, we have the following correspondences:
\be
&\eta^-_1=\psi^2_{H_1},\eta^+_2=\psi^1_{H_2} ,\eta'^+_5=-i\lambda^+,\eta^-_5=-i\lambda^-,&
\non\\
&
g_5 v=\sqrt2 m_Z\cos\beta \cos\theta_W,
\quad
g_6 v=\sqrt2 m_Z\sin\beta \cos\theta_W,&
\non\\
&
\mu_4=\mu_5=0,
\quad
g_{7,8,9,10}=0.&
\en
Note that the Lagrangian for single charged WIMP mass term with $I=Y=1/2$ also need to be modified as in Appendix A and the mass eigenstates of the neutral as well as single charged particles in the 4-component notation are constructed in the Appendix B.

\subsection{Dark Matter Annihilation}

 The DM particles are thought to be created thermally during the big bang, and froze out of thermal equilibrium in the early universe with a relic density. The evolution of DM abundance is described by the Boltzmann eqution:
 \be
 \frac{dn_\chi}{dt}+3Hn_\chi=-\la\sigma_{\rm ann} v_{\rm M\phi l}\ra[n_\chi n_{\bar\chi}-n_\chi^{\rm eq} n_{\bar\chi}^{\rm eq}],
 \label{eq: Boltzmann}
 \en
where $H\equiv \dot a/a=\sqrt{4\pi^3g_*(T)T^4/(45M_{PL}^2})$ is the Hubble parameter, $M_{PL}$ is the Plank mass, $g_*$ is the total effective number of relativistic degrees of freedom~\cite{Kolb,Coleman}. $n_\chi (n_{\bar\chi})$ is the number density of DM (antiDM) particles, and $n_{\bar\chi}=n_\chi$ for Majorana fermions (that is, $\chi = \bar\chi$) as in this model. Eq.~(\ref{eq: Boltzmann}) is measured in the cosmic comoving frame~\cite{GG} and  $\la\sigma_{\rm ann} v_{\rm M\phi l}\ra$ is the thermal averaged annihilation cross section times M$\phi$ller velocity which is defined by $v_{\rm M\phi l}\equiv \sqrt{(p_1\cdot p_2)^2-m_1^2m_2^2}/(E_1E_2)=\sqrt{|{\bf v}_1-{\bf  v}_2|^2-|{\bf v}_1\times {\bf v}_2|^2}$ with subscripts 1 and 2 labeling the two initial DM particles and velocities ${\bf v}_i\equiv {\bf p}_i/E_i (i=1,2)$.~\footnote{ In general, the collision is not collinear in the comoving frame. Hence the M$\phi$ller velocity is not equal to the relative velocity $v_{\rm rel}\equiv |{\bf v}_1-{\bf v_2}|$. Nevertheless, it has been shown~\cite{GG} that  $\la\sigma_{\rm ann} v_{\rm M\phi l}\ra=\la\sigma_{\rm ann}v_{\rm lab}\ra^{\rm lab}$ where $v_{\rm lab}\equiv |{\bf v}_{1,{\rm lab}}- {\bf v}_{2,{\rm lab}}|$ is calculated in the lab frame with one of two initial particles being at rest.}

The DM particles became non-relativistic when they froze out of thermal equilibrium in the early universe. In this non-relativistic (NR) limit, $\sigma_{\rm ann}(\chi{\chi}\rightarrow {\rm all})v= a+bv^2+O(v^4)$ where $v\equiv v_{\rm lab}=\sqrt{s(s-4m_\chi^2)}/(s-2m_\chi^2)$ and the Mandelstam variable $s=2m_\chi^2(1+1/\sqrt{1-v^2)}$ in the lab frame.
The velocity averaged DM annihilation cross section via Maxwell velocity distribution can be calculated~\cite{chua} to be $\la\sigma_{\rm ann}v\ra= a+6b/x+O(1/x^2)$ with the freeze-out temperature parameter $x\equiv m_\chi/T$. At the freeze-out temperature, the interaction rate of DM particles is equal to the expansion rate of universe, namely $\Gamma_f\equiv n_\chi^{eq}\la\sigma_{\rm ann}v\ra=H(T_f)$. From this freeze-out condition,
$x_f$ can be solved numerically by the following equation~\cite{Kolb,JKG}
\be
x_f={\rm ln}\left [c(c+2)\sqrt{\frac{45}{8}}\frac{g_\chi m_\chi M_{\rm PL}(a+6b/x_f)}{2\pi^3\sqrt{g_*(m_\chi/ x_f)}x_f^{1/2}}\right ],
\label{eq: xf}
\en
where $c$ is an order of unity parameter determined by matching the late-time and early-time in the freeze-out criterion. We take the usual value  $c=1/2$ since the exact value of $c$ is not so significant to solve the numerical solution for $x_f$ due to the logarithmic dependence in Eq.~(\ref{eq: xf}). Following the standard procedure~\cite{Kolb} to solve Eq.~(\ref{eq: Boltzmann}), the relic CDM density $\Omega_{\rm DM}\equiv \rho_\chi / \rho_{\rm crit}$  can be approximately related to the velocity averaged annihilation cross section $\la\sigma_{\rm ann}v\ra$ as

\begin{figure}[t]
  \centering
  \includegraphics[width=14cm]{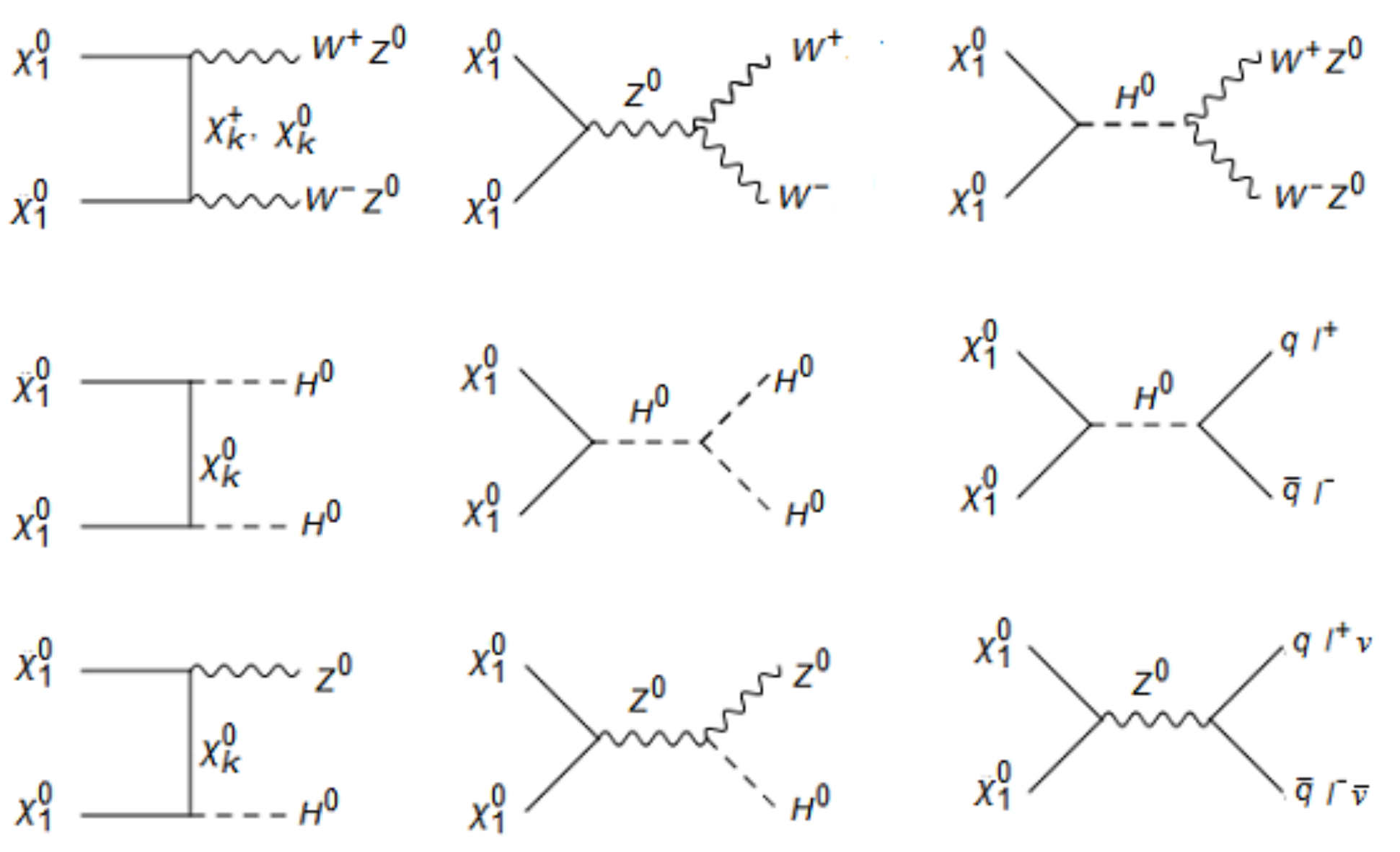}\\
  \caption{The annihilation processes $(\chi{\chi}\rightarrow W^+W^-, ZZ, ZH, HH, f{\bar f})$}
  \label{fey1}
\end{figure}

\be
\Omega_{\rm DM}h^2\approx 1.04\times 10^9\frac{\rm GeV^{-1}}{M_{\rm PL}\sqrt{g_*(m_\chi)}J(x_f)},
\label{eq:Omegah2}
\en
where
\be
J(x_f)\equiv \int_{x_f}^\infty\frac{\la\sigma_{\rm ann}v\ra}{x^2}dx= ax_f^{-1}+3bx_f^{-2}+O(x_f^{-3}).
\label{J-factor}
\en

When doing the calculation of DM relic density, we need to consider three exceptions~\cite{GS}: coannihilation, forbidden channel annihilation, and annihilation near the pole. In this article, we focus on the model building and mainly consider the annihilation processes. The leading effect on coannihilation in this model will be discussed in Sec. IV. To solve the last two exceptions, we do not take the Taylor series expansion on $v^2$ in s-channel, and for each annihilation channel we put a step function for the allowed threshold energy  in the thermal average cross section as follows:
\be
\la\sigma_{\rm ann}v\ra&=&\frac{x^{3/2}}{2\pi}\sum_{A,B}\int_0^{\infty}dv v^2 e^{-x v^2/4}
\non\\
& &\times [\sigma_{\rm ann} (\chi{\chi}\rightarrow A+B)v] 
\theta[2m_\chi^2(1+\frac{1}{\sqrt{1-v^2}})-(m_A+m_B)^2].
\en 
In stead of $a+6 b/x_f$, we replace with the above thermal averaged cross section with $x=x_f$ in Eq.~(\ref{eq: xf}) and solve the value of  $x_f$ numerically. Then we can get the DM relic density by modifying $J(x_f)$ in Eq.~(\ref{J-factor}) as follows:
\be
J(x_f)&\equiv& \int_{x_f}^\infty\frac{\la\sigma_{\rm ann}v\ra}{x^2}dx
\non\\
&=& \sum_{A,B}\int_0^{\infty}dv  [\sigma_{\rm ann} (\chi{\chi}\rightarrow A+B)v] [1-{\rm erf}(v\sqrt {x_f}/2)]
\non\\
& &\times \theta[2m_\chi^2(1+\frac{1}{\sqrt{1-v^2}})-(m_A+m_B)^2].
\label{J-factor2}
\en

We will calculate the relic density in the early universe
through the DM annihilation processes $(\chi{\chi}\rightarrow W^+W^-, ZZ$, ZH, $HH, f{\bar f}$). Fig.~\ref{fey1} shows the corresponding Feynman diagrams. The corresponding Lagrangian and the matrix elements are shown in Appendices C and D, respectively, and it is straightforward to obtain $\la \sigma_{\rm ann} v\ra$.
Although the present DM relic density is determined by the velocity averaged cross section $\langle \sigma_{\rm ann} v\rangle$ of DM annihilation processes which have been ceased after the freeze-out stage in the cosmological scale, the DM annihilation to the SM particles would still occur today in regions of high DM density and result in the indirect search for end products as excesses relative to products from SM astrophysical processes.
The results on $\la \sigma_{\rm ann} v\ra$ can be readily applied to the indirect search processes by using a typical velocity $v\simeq 300 $km/s (explained in Sec. III).

As we know that in the nonrelativistic limit $\sigma_{\rm ann}v=a+b v^2+O(v^4)$ where $a$ is the $s$-wave contribution at zero relative velocity and $b$ contains contributions from both the $s$ and $p$ waves. 
$\sigma_{\rm ann}v$ is dominated by the $s$-wave term in indirect-detection calculations, while both $s$ and $p$ wave terms becomes important when dealing with the calculation of DM relic density. 

It will be useful to recall some qualitative properties of the DM annihilation amplitudes in the channels of 
$\chi{\chi}\rightarrow W^+W^-, ZZ$, $ZH$, $HH, f{\bar f}$~\cite{Drees,JKG}. Fermi statistics forces the two identical Majorana fermions with orbital angular momentum $L$ and total spin $S$ to satisfy $(-)^S=(-)^L$. The total angular momentum of the $s$-wave state is $J=0$ and the $CP$ is given by $CP=(-1)^{L+1}=-1$, while the $p$-wave state has $CP=+1$ [see Eqs.~(\ref{eq: Fermi}) and (\ref{eq: DM CP})]. 

The final state $W^+W^-$ can be produced via $t$-channel exchange of a single charged WIMP and $s$-channel exchange of a Higgs scalar or a $Z$ boson (see Fig.~\ref{fey1}). The final state $ZZ$
can be produced via $t$-channel exchange of a neutral WIMP and $s$-channel exchange of a Higgs scalar (see Fig.~\ref{fey1}). Note that in the $s$-wave DM amplitude both gauge bosons in final state are transversely polarized and governed via the $t$-channel exchange diagrams~\cite{JKG,Drees}. Also note that a bino-like DM pair does not contribute to the $s$-wave amplitude~\cite{Drees}.

The DM particles can annihilate into $ZH$  via $t$-channel exchange of a neutral WIMP and $s$-channel exchange of a $Z$ boson (see Fig.~\ref{fey1}). %For the final state $ZH$, the total spin is $S=1$. 
The final state $ZH$ in a $L=1$ configuration
can match the angular momentum and the $CP$ of the $s$-wave DM pair.
Hence the $s$-wave amplitude is allowed in this channel~\cite{JKG,Drees}.

The DM particles can annihilate into two Higgs bosons via $t$-channel exchange of a neutral WIMP and $s$-channel exchange of a scalar Higgs (see Fig.~\ref{fey1}). The $s$-wave scattering amplitude is vanishing since two scalar can not be in a state with $J=0$ and $CP=-1$~\cite{JKG,Drees}.

The final state fermion-antifermion pair $f \bar f$ can be produced
via the $s$-channel exchange of a  Higgs scalar or a $Z$ boson (see Fig.~\ref{fey1}). The $Z$-exchange contributes to both the $s$ and $p$ wave matrix elements with chiral conserving interactions~\cite{Drees}. The final state $f\bar f$ has $CP=(-)^{S+1}$. The $s$-wave DM pair requires the total spin $S=0$ in final state to conserve $CP$ so that both fermion and antifermion should have the same helicity. 
The $Z$-$f$-$\bar f$ couplings implies the fermion and the antifermion in opposite chirality and hence results in the helicity suppression of the $s$-wave amplitude.
The Higgs scalar exchange only contributes to $p$-wave matrix elements (since the $CP$ of Higgs boson is $+1$) with fermion mass factor.  
Hence the process $\chi{\chi}\rightarrow f{\bar f}$ favors a heavy fermion pair~\cite{JKG,Drees}.

\subsection{DM-nucleus elastic scattering cross section}\label{Direct}

To compare with the results of LUX, XENON100, and PICO-60 experiments, we calculate the
SI and SD cross sections of DM scattering off $^{129, 131}{\rm Xe}$ nuclei and the SD cross section of DM scattering off ${\rm CF_3I}$ nuclei.
We shall obtain $\overline{\sum} |M_{fi}|^2$ at $q^2=0$ first.
In this model, the DM is composed of Majorana fermions so that the DM vector current matrix elements are vanishing. Hence
the Lagrangian in this model is given by
\be
{\cal L}=\bar\chi\gamma_\mu\gamma_5\chi j_{Ah}^\mu
+\bar\chi\gamma_\mu\gamma_5\chi j_{Vh}^\mu
+\bar\chi\chi s_h+\bar\chi\gamma_5\chi s'_h.
\label{eq:Lchi}%??
\en
where
\be
s_h=a^q\bar q q,
\quad
s'_h=a'^q \bar q q.
\quad
j_{Vh}^\mu=b^q j_{Vq}^\mu=b^q\bar q\gamma_\mu q,
\quad
j_{Ah}^\mu=d^q j_{Aq}^\mu=d^q\bar q\gamma\gamma_5 q,
\en
and $a^q$, $a'^q$, $b^q$ and $d^q$ are given in Appendix F.
The corresponding scattering amplitude is
\be
iM_{fi}&=&\la \chi(p'_\chi, s'_\chi), {\cal N}(p',s')|i{\cal L}(0)|\chi(p_\chi,s_\chi),{\cal N}(p,s)\ra
\non\\
&=&i\kappa_\chi\bar u(p'_\chi,s'_\chi)\gamma_\mu\gamma_5 u(p_\chi,s_\chi)
\la {\cal N}(p',s')| j_{Ah}^\mu+j_{Vh}^\mu|{\cal N}(p,s)\ra
\non\\
&&
+i\kappa_\chi\bar u(p'_\chi,s'_\chi)u(p_\chi,s_\chi)
\la {\cal N}(p',s')| s_{h}|{\cal N}(p,s)\ra
\non\\
&&+i\kappa_\chi\bar u(p'_\chi,s'_\chi)\gamma_5u(p_\chi,s_\chi)
\la {\cal N}(p',s')| s'_{h}|{\cal N}(p,s)\ra.
\en
In the above, $\kappa_\chi =2$ for the Majorana fermions in this model and $\kappa_\chi =1$ for the Dirac fermions.

It is useful to define
\be
\chi^{XY}&\equiv&\frac{1}{2J_\chi+1}
\sum_{\rm spins}
\la \chi (p_\chi,s_\chi) |(\bar\chi\chi)_X|\chi(p'_\chi,s'_\chi)\ra
\la \chi (p'_\chi,s'_\chi) |(\bar\chi\chi)_Y|\chi(p_\chi,s_\chi)\ra
\non\\
W^{XY}&\equiv&\frac{1}{2 J_{\cal N}+1}
\sum_{\rm spins}\la {\cal N}(p',s')|O_{hX}|{\cal N}(p',s')\ra \la {\cal N}(p',s')|O_{hY}|{\cal N}(p,s)\ra,
\en
where $X,Y=A,V,S,P$ and $O_{hX}$ is the corresponding operator. 
For example, we have
\be
\chi^{AA}_{\mu\nu}&\equiv&\frac{1}{2}
\sum_{\rm spins} \la \chi (p'_\chi,s'_\chi) |\bar\chi\gamma_\mu\gamma_5\chi(0)|\chi(p_\chi,s_\chi)\ra
\la \chi (p_\chi,s_\chi) |\bar\chi\gamma_\nu\gamma_5\chi(0)|\chi(p'_\chi,s'_\chi)\ra,
\en
or explicitly,
\be
\chi^{AA}_{\mu\nu}
=
\left(
(p_\chi+p'_\chi)_\mu (p_\chi+p'_\chi)_\nu-g_{\mu\nu} 4m_\chi^2+g_{\mu\nu}q^2
-q_\mu q_\nu
\right)\kappa^2_\chi.
\label{eq: Wchi}
\en
Similarly, for $X,Y=A,V$, we have
\be
W^{XY}_{\mu\nu}
&\equiv&\frac{1}{2 J_{\cal N}+1}\sum_{s ,s'}
\la {\cal N}(p',s')|j_{Xh,\mu}(0)|{\cal N}(p,s)\ra\la {\cal N}(p,s)|j_{Yh,\nu}(0)|{\cal N}(p',s')\ra,
\label{eq: WXY}
\en
\be
W^{SS}_{\mu\nu}
&\equiv&\frac{1}{2 J_{\cal N}+1}\sum_{s ,s'}
\la {\cal N}(p',s')|s_h(0)|{\cal N}(p,s)\ra\la {\cal N}(p,s)|s_h(0)|{\cal N}(p',s')\ra,
\label{eq: WSS}
\en
and so on.

Note that $q^2=0$ means $q=0$ in all frames (see Appendix F).
It is simpler to work in the lab frame (the rest frame of ${\cal N}$). The matrix elements of scalar, vector and axial vector current operators with initial and final state nucleus at rest are given by
\be
\la {\cal N}(m_{\cal N},s')| s_{h}(0)|{\cal N}(m_{\cal N},s)\ra
&=&2m_{\cal N}f_{s\cal N} \delta_{ss'},
\non\\
\la {\cal N}(m_{\cal N},s')| s'_{h}(0)|{\cal N}(m_{\cal N},s)\ra
&=&2m_{\cal N}f'_{s\cal N} \delta_{ss'},
\non\\
\la {\cal N}(m_{\cal N},s')|j_{Vh,\mu}(0)|{\cal N}(m_{\cal N},s)\ra
&=&2 g_\mu^ 0m_{\cal N}\delta_{ss'} Q_{V\cal N},%,
\non\\
\la {\cal N}(m_{\cal N}, s')|j_{Ah}^\mu(0)|{\cal N}(m_{\cal N},s)\ra
&=&4 g^\mu_i m_{\cal N} Q_{A\cal N}\la J_{\cal N}, s'|(\vec S_{\cal N})_i|J_{\cal N},s\ra,
\en
with
\be
Q_{V\cal N}&=&Z (2b^u+b^d)+(A-Z) (2b^d+b^u),
\non\\
Q_{A\cal N}&=&d^q (\Delta_q^p \lambda_p+\Delta_q^n \lambda_n),
\non\\
f^{(')}_{s\cal N}&=&a^{('q)} (Z f_{sp}+(A-Z) f_{sn}),
\non\\
f_{s p(n)}&=&\sum_{q=u,d,s} \frac{m_{p(n)}}{m_q} f^{(p(n))}_{Tq}+
\sum_{q=c,b,t} \frac{2}{27} \frac{m_{p(n)}}{m_q} \left(1-\sum_{q'=u,d,s} f^{(p(n))}_{Tq'}\right),
\non\\
\lambda_{p,n}
&=&
\frac{\la S_{p,n,z}\ra_{\rm eff}}{J_{\cal N}}.
\label{eq:MAq=0}
\en
The derivation of the above formulae are given in Appendix F.
Using
\be
\chi^{AA,\mu\nu}(q=0)
=\kappa^2_\chi 4(p_\chi^\mu p_\chi^\nu-g^{\mu\nu} m_\chi^2),
\quad
\chi^{SS}(q=0)=4 m_\chi^2,
\non\\
\chi^{AS,\mu}=\chi^{AP,\mu}=\chi^{SP}=0,
\quad\qquad\ \ \ \ \ \ \ \ \,
\chi^{PP}(q=0)=0,\ \ \
\en
with
$p_\chi=p'_\chi=(E_\chi,0,0,p^3_\chi)$,
\be
(p_\chi^3)^2= \frac{m_\chi^2 v^2}{1-v^2},
\en
in the nucleus rest frame
and 
\be
\sum_{s,s'} \la J_{\cal N}, s'|(\vec S_{\cal N})_z|J_{\cal N},s\ra\delta_{ss'}
&=&
0,
\non\\
\sum_{s,s'} \la J_{\cal N}, s|(\vec S_{\cal N})_z|J_{\cal N},s'\ra\la J_{\cal N}, s'|(\vec S_{\cal N})_z|J_{\cal N},s\ra
&=&
\frac{1}{3}J_{\cal N}(J_{\cal N}+1)(2J_{\cal N}+1),
\non\\
\sum_{s,s'} \la J_{\cal N}, s|(\vec S_{\cal N})_i|J_{\cal N},s'\ra\la J_{\cal N}, s'|(\vec S_{\cal N})_i|J_{\cal N},s\ra
&=&
J_{\cal N}(J_{\cal N}+1)(2J_{\cal N}+1),
\en
we obtain,
\be
\overline{\sum}|M_{fi}|^2&=&\chi^{AA,\mu\nu} W^{AA}_{\mu\nu}
+\chi^{AA,\mu\nu} W^{VV}_{\mu\nu}
+\chi^{SS} W^{SS},
\en
where
\be
\chi^{AA,\mu\nu} W^{AA}_{\mu\nu}
&=&64 \kappa^2_\chi m_{\cal N}^2 m^2_\chi \left(1+\frac{v^2}{3(1-v^2)}\right)Q_{A\cal N}^2
J_{\cal N}(J_{\cal N}+1),
\label{eq: AAAA}
\\
\chi^{AA,\mu\nu} W^{VV}_{\mu\nu}
&=&16 \kappa^2_\chi m_{\cal N}^2m_\chi^2 \frac{v^2}{1-v^2} Q_{V\cal N}^2,
\label{eq: AAVV}
\\
\chi^{SS} W^{SS}&=&16 \kappa_\chi^2 m_\chi^2 m^2_{\cal N}f^2_{\cal N}.
\en
Consequently, we have
\be
\overline{\sum}|M_{fi}|^2(q^2=0)
&=&
16 m_{\cal N}^2 m^2_\chi \kappa^2_\chi
\left[\left(4+\frac{4v^2}{3(1-v^2)}\right)  Q_{A\cal N}^2 J_{\cal N}(J_{\cal N}+1)
+\frac{v^2}{1-v^2} Q_{V\cal N}^2
+f^2_{s\cal N}\right].
\non\\
\label{eq: iM^2}
\en

Several comments are in order:
(i) Note that there is no interference between various
interaction
terms in $iM_{fi}$.
(ii) In the
nucleus rest frame and at $q=0$,
the matrix element of the space component of the vector current is vanishing,
while the one of the time component of the axial vector current is also vanishing,
see Eq.~(\ref{eq: q=0}).
It seems that the matrix elements of $j_{A\chi\mu}$ and $j_{V h,\mu}$ is orthogonal and hence the decay amplitude from the $j_{A\chi\mu} j^\mu_{Vh}$ contribution, i.e.  $\chi^{AA,\mu\nu} W^{AA}_{\mu\nu}$, is vanishing.
This is however untrue, since the rest frame of $\chi$ is not the rest frame of ${\cal N}$.
Although the decay amplitude, see Eq. (\ref{eq: AAVV}), is indeed suppressed by $v$ [$v={\cal O}(10^{-3})$], it is enhanced by $Q_{V\cal N}$, which contains large factors such as $Z$ and $A$. The contribution from this term needs to be kept.

Usually the direct search experiments report the cross section normalized to the interaction with a single nucleon (neutron/proton) since the target materials used in different direct search experiments are not the same. The normalization procedure is shown in Appendix F, we summarize the formulas in below.
The differential cross section is given by [see Eq. (\ref{eq: dsigma})]
\be
\frac{d\sigma_{A_i}}{d |{\bf q}|^2}
=\frac{1}{4\mu_{A_i}^2 v^2} (\sigma^{SI}_0 F^2_{SI}(|{\bf q}|^2)
+\sigma^{SD}_{0,pp} F^2_{pp}(|{\bf q}|^2)+\sigma^{SD}_{0,nn} F^2_{nn}(|{\bf q}|^2)+\sigma^{SD}_{0,pn} F^2_{pn}(|{\bf q}|^2)),
\en
where
\be
\sigma_0^{SI}&=&\frac{\mu_{A_i}^2}{\pi}
\kappa^2_\chi
\left[
\frac{v^2}{1-v^2} Q_{V A_i}^2
+f^2_{s A_i}\right],
\non\\
\sigma^{SD}_{0,pp(nn)}&=&\frac{\mu_{A_i}^2}{\pi} \kappa^2_\chi
\left[\left(4+\frac{4v^2}{3(1-v^2)}\right)  (\sum d^q \Delta_q^{p(n)})^2 \lambda^2_{p(n)}
J_{A_i}(J_{A_i}+1)
\right],
\non\\
\sigma^{SD}_{0,pn}&=&\frac{\mu_{A_i}^2}{\pi} \kappa^2_\chi
\left[\left(4+\frac{4v^2}{3(1-v^2)}\right)  2(\sum d^q d^{q'} \Delta_q^p \Delta_{q'}^n) \lambda_p\lambda_n
J_{A_i}(J_{A_i}+1)
\right].
\en
Note that in the above formulas the form factors do not depends on $a^q$, $a'^q$, $b^q$ and $d^q$ in Eq. (\ref{eq:Lchi}).
It is better than those usually used in literature, where $d^q$s are involved in the form factors.
The DM-nucleus scattering cross section is
\be
\sigma_{A_i}=\int d|{\bf q}|^2 \frac{d\sigma}{d|{\bf q}|^2}=
 (\sigma^{SI}_0 r_{SI} +\sigma^{SD}_{0,pp} r_{pp}
+\sigma^{SD}_{0,nn} r_{nn}
+\sigma^{SD}_{0,pn} r_{pn}) ,
\en
where
\be
r_j\equiv \int_0^{4\mu_{A_i}^2 v^2} \frac{d|{\bf q}|^2}{4\mu_{A_i}^2 v^2} F^2_{j}(|{\bf q}|),
\en
with $j=SI, pp, nn, pn$ and
\be
F^2_{pp(nn)}(|{\bf q}|)&\equiv&\frac{S_{00}(|{\bf q}|)+S_{11}(|{\bf q}|)\pm S_{01}(|{\bf q}|)}{S_{00}(0)+S_{11}(0)\pm S_{01}(0)},
\quad
F^2_{pn}(|{\bf q}|)\equiv\frac{S_{00}(|{\bf q}|)-S_{11}(|{\bf q}|)}{S_{00}(0)-S_{11}(0)}.
\en
Finally, the spin-independent and spin-dependent scaled cross sections are defined as
\be
\sigma^Z_N&\equiv&\frac{\sum_i \eta_i\sigma_{A_i}}
{\sum_j \eta_j  A^2_j\frac{\mu^2_{A_j}}{\mu^2_p}}
\label{eq: sigma Z N}
\en
and
\be
\sigma^{SD}_{p,n}\equiv
(\sum_i\eta_i \sigma_{A_i})
\left(
\sum_j \eta_j\frac{4\mu_{A_j}^2 \la S_{p,n}\ra_{\rm eff}^2  (J_{A_j}+1)}{3\mu_{p,n}^2J_{A_j}}\right)^{-1},
\label{eq: SD p n}
\en
respectively.
In this way, the data obtained from different experiments can be compared using $\sigma^Z_N$ and $\sigma^{SD}_{p,n}$.~\footnote{The terminology of spin-(in)dependent cross section is somewhat misleading. There are, in fact, two different normalizations, where 
both spin-dependent and spin-independent interactions are involved in $\sigma^{SD}_{p,n}$ and $\sigma^Z_N$.}

\section{Results}

In parallel with the DM sector of MSSM~\cite{JKG,Haber}, we analyze the model with $I=1/2$ and $Y=1/2$.  In this model, there are 13 parameters in total, five mass parameters $\mu_i (i=1\sim 5)$ and eight Yukawa couplings $g_i (i=3\sim 10)$, as shown in the mass matrices of
neutral as well as single charged WIMPs in Eq.~(\ref{eq: MY2}) and Eq.~(\ref{eq: MX2}), respectively. 
In principle the 13 parameters can be reduced to fewer parameters under different considerations.
First of all, let us see what is the minimal particle content which can make up the DM. In this model the Majorana fermion can be generated purely by the singlet $\eta_3$, namely, only the mass parameter $\mu_2$ being nonzero. Due to its quantum number $(2I,-(Y-1/2))=(1,0)$, it doe not couple to the SM gauge bosons. It also does not couple to the SM Higgs boson since all Yakawa couplings are set to be zeros. Hence it is inert and impossible to be a WIMP, unless some exotic Higgs boson is introduced~\cite{GIV}.
Next we consider the Majorana fermion generated by the two doublets $\eta_1$ and $\eta_2$, namely, only the parameter $\mu_1$ being nonzero. Due to their quantum numbers $(2I+1,\mp Y)=(2,\mp 1/2)$, they couple to the SM gauge bosons, but still do not couple to the SM Higgs boson. As mentioned previously, they are two degenerate Majorana states $\chi_{1,2}\propto(\eta_1\pm\eta_2)/\sqrt2$ with the same mass $\mu_1$.
It results in 
an oversized DM-nucleus scattering cross section via $Z$ boson exchange from $\chi_{1(2)}\rightarrow\chi_{2(1)}$ vector current. Nevertheless the problem can be solved if one can lift the mass degeneracy of $\chi_{1,2}$. Hence the minimal particle content to make up the DM is to combine these fermion doublets $\eta_1$, $\eta_2$ and the singlet $\eta_3$.

\begin{table}[t]
\begin{tabular}{|cccc|c|c|}
  \hline
 &
 & \hspace{-2.7cm} Case A
 &
 & Case B
 & Case C
 \\
 neutralino-like I
 & neutralino-like II
 & neutralino-like III
 & neutralino-like IV
 & Reduced
 & Extended
 \\
  \hline
  GUT 
  & GUT 
  & No GUT 
  & No GUT 
  & 
  & 
 \\
 %MSSM with
 %&MSSM with
% &MSSM with
 %&No MSSM
 %&
% &
 %\\
     $\tan\beta=2$ 
  &$\tan\beta=20$ 
  &$\tan\beta=2$ 
  & 
  & 
  & 
  \\
  $\eta_{1\sim3,5}$
  &  $\eta_{1\sim3,5}$
  &  $\eta_{1\sim3,5}$
  &  $\eta_{1\sim3,5}$
  & $\eta_{1\sim3}$
  & $\eta_{1\sim 3,5,7\sim10}$
  \\
  \hline
\end{tabular}
\caption{Summary of three typical cases.}
\label{tab:cases}
\end{table}

To have an overall understanding of the model, we will consider the following three typical
cases: the neutralino-like, the reduced and the extended cases (see Table~\ref{tab:cases}).
For the neutralino-like case, only the parameters $\mu_{1\sim 3}$ and $g_{3\sim 6}$ are nonzero and the Majorana DM is generated by $\eta_{1,2,3}$ and the triplet $\eta_5$. It contains 4 neutral Majorana fermions and 2 single charged fermions. Furthermore, depending on whether the grand unified theory (GUT) relation ($\mu_2=\frac{5}{3}\mu_3\tan^2\theta_W$)~\cite{IKKT} or the $\tan\beta$ relation (note that $g_3 v=\sqrt2 m_Z\cos\beta \sin\theta_W$,
$g_4 v=\sqrt2 m_Z\sin\beta \sin\theta_W$,
$g_5 v=\sqrt2 m_Z\cos\beta \cos\theta_W$ and
$g_6 v=\sqrt2 m_Z\sin\beta \cos\theta_W$) is imposed or not, we classify the neutralino-like case into four subcases:
the neutralino-like I case with the GUT relation and $\tan\beta=2$,
%with $\tan\beta=2$,
the neutralino-like II case with the GUT relation and $\tan\beta=20$, the neutralino-like III without the GUT relation but with $\tan\beta=2$, and the neutralino-like IV case without the GUT and the $\tan\beta$ relations.

For the reduced case, only the parameters $\mu_1, \mu_2, g_3$ and $g_4$ are free
%nonzero 
with the minimal particle content (i.e., $\eta_{1,2,3}$). It contains 3 neutral Majorana fermions and 1 single charged fermions. For the extended case, all of 13 model parameters are free with the maximal particle content (i.e., all $\eta$ fields) and it contains 6 neutral Majorana fermions and 4 single charged fermions.
In each case, we generate 10,000 random samples and survey the DM mass $m_\chi$ in the range of $1\sim2500$ GeV by random sampling the mass couplings $\mu_i (i=1\sim 5)$ linearly in the range of $0\sim8000$ GeV and the Yukawa coupling $g_i (i=3\sim 10)$ linearly in the range of $0\sim1$ if these parameters are active.

For each sample, we numerically solve the mass eigenstates and eigenvalues, find the freeze-out temperature parameter $x_f$ [see Eq.~(\ref{eq: xf})] and obtain the DM thermal relic density $\Omega_\chi h^2$ via the calculations of DM annihilation processes $\chi{\chi}\rightarrow W^+W^-, ZZ, ZH, HH, f{\bar f}$ to compare with the observed relic density.
We calculate the normalized SI, SD elastic cross sections ($\sigma^{SI}_N$, $\sigma_n^{SD}$ and $\sigma_p^{SD}$) of DM scattering off $^{129, 131}{\rm Xe}$ nuclei to compare with the results of direct search experiments of LUX SI and XENON100 SD elastic cross sections of DM scattering off $^{129, 131}{\rm Xe}$ nuclei, respectively. We also calculate $\sigma_p^{SD}$ for DM scattering off $\rm{CF_3I}$ nuclei to compare with the result of PICO-60 experiment using $\rm{CF_3I}$ as material target.

In calculation of $\sigma^{SI}_N$, we adopt the exponential form factor~\cite{JKG,SI1,SI2} for $F_{SI}(|{\bf q}|)$ and we use the data in Ref.~\cite{EFO} for the nucleon parameters $f^{(p,n)}_{Tq}$ in Eq.~(\ref{eq:MAq=0}). In calculation of $\sigma_{n,p}^{SD}$, we adopt the structure factors $S_{00,01,11}(|{\bf q}|)$ for $^{129, 131}{\rm Xe}$ nucleus in Ref.~\cite{Menendez}, and ${^{19}\rm F}$ and ${^{127}\rm I}$ (by Bonn A calculation) nuclei in Ref.~\cite{Bednyakov}, and use
the experimental data in Refs.~\cite{EFO,Mallot} for the quark spin component in a nucleon$\Delta_q^{p,n}$.
For $^{129, 131}{\rm Xe}$ nuclei, we use the nuclear total angular momentum $J$ and the predicted spin expectation values $\la S_{p,n}\ra$ by Menendes ${\it et\ al.}$ calculation in Refs.~\cite{Menendez,XENON100SD} for $\la S_{p,n,z}\ra_{\rm eff}$ and the isotope abundance of $^{129,131}$Xe in Refs.~\cite{XENON100SD} for $\eta_i$.
For $^{19}{\rm F}$ and $^{127}{\rm I }$ nuclei, we use the nuclear total angular momentum and the predicted spin expectation values in Refs.~\cite{nSD2}. For simplicity, we only consider the case that the second lightest neutral particle $\chi_2$ is dynamically forbidden to be produced from 
%$\chi_1\chi_1\to \chi_1\chi_2, \chi_2\chi_2$ annihilation processes.
$\chi_1+{^{129}\rm Xe}\to \chi_2+{^{129}\rm Xe}$ inelastic scattering process.

For indirect search, 
we calculate the present velocity averaged cross section $\langle \sigma (\chi{\chi}\rightarrow W^+W^-, ZZ, ZH, HH, f{\bar f}) v\rangle$ 
%with the typical DM velocity $v\simeq 300$ km/s 
to compare with the Fermi-LAT results which provide six upper limits on $\langle \sigma(\chi{\chi}\rightarrow W^+W^-, b{\bar b}, u{\bar u}, \tau^+\tau^-, \mu^+\mu^-, e^+e^-) v\rangle$ from a combined analysis of 15 dSphs in indirect search \cite{Fermi-LAT}.  
As we know that the DM halo is immersed in the Galaxy. The speed of the sun moving around the Galactic center is about 220 km/s at the local distance $r\approx$ 8.5 kpc and the Galactic circular rotation speed is about 230 km/s at radii $\approx$ 100 kpc~\cite{JKG,Kochanek:1995xv}. On the other hand, the shortest and longest distance of these 15 dSphs from the sun are $\approx$ 23 and 233 kpc, 
respectively~\cite{Fermi-LAT}. Hence we will use a typical DM velocity $v\simeq 300$ km/s in the indirect-detection calculation.

Finally we collect all allowed samples which satisfy all these eleven constraints; namely, one from the observed value of DM relic density, four from the direct detection of LUX, XENON100 and PICO-60 experiments and six from the indirect detection of Fermi-LAT observations such that we can find the lower bound of DM mass with different particle attribute, the allowed range of the model parameters as well as the coupling strengths in this model.

\begin{table}[t]
%\footnotesize
\scriptsize{
\begin{tabular}{|c|cccc|c|c|}
  \hline
  &
  &
  & {\hspace{-2.7cm}Case A}
  &
  & Case B
  & Case C
  \\
   Percentage $(\%)$ & neutralino-like I  & neutralino-like II & neutralino-like III & neutralino-like IV & Reduced & Extended \\
  \hline
  higgsino-like ($\sim\eta_{1,2}$)
  %& 28.51 & 27.59 & 32.76 & 31.26 & 49.73 & 28.62 \\
  & 29 & 28 & 33 & 31 & 50 & 29 \\
  bino-like  ($\sim\eta_3$)
  %& 71.07 & 72.07 & 33.38 & 33.53 & 49.43 & 33.94 \\
  & 71 & 72 & 33 & 34 & 49 & 34 \\
  wino-like ($\sim\eta_5$)
  %& 0 & 0 & 33.31 & 33.97 & 0 & 30.93 \\
  & 0 & 0 & 33 & 34 & 0 & 31 \\
  non neutralino-like ($\sim\eta_{9,10}$)
  %& 0 & 0 & 0 & 0 & 0 & 5.25 \\
  & 0 & 0 & 0 & 0 & 0 & 5 \\
  mixed
  %& 0.42 & 0.33 & 0.55 & 1.24 & 0.84 & 1.26 \\
  & 0.4 & 0.3 & 0.6 & 1.2 & 0.8 & 1.3 \\
  \hline
\end{tabular}
\caption{Particle attribute distribution of sample sets.}
\label{tab:pdf}}
\end{table}

\begin{figure}[t]
  \centering
  \includegraphics[width=8cm]{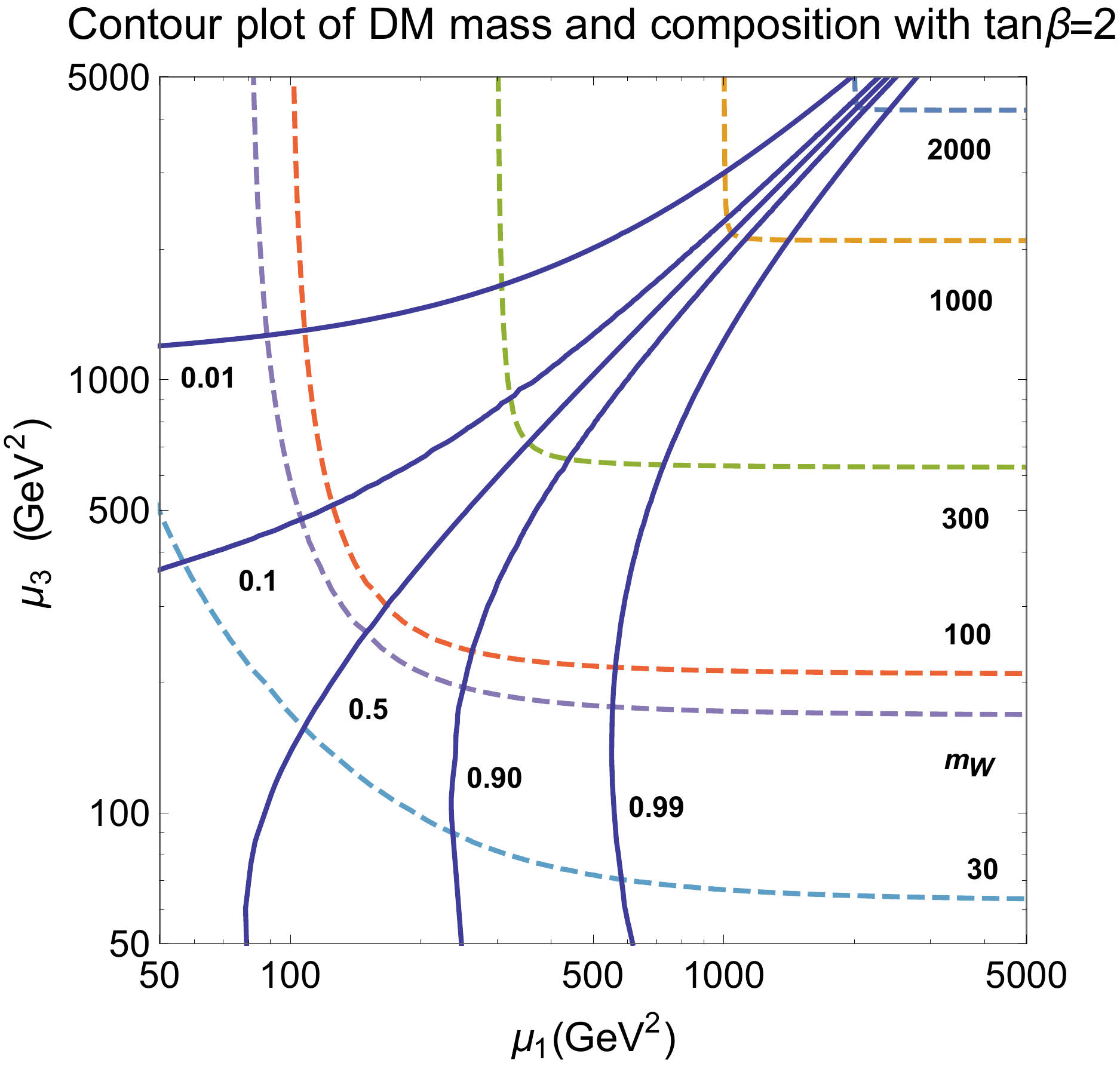}\\
  \caption{Contour plot of the DM mass and composition in the $\mu_1$-$\mu_3$ plane for the neutralino-like I case. The broken curves are contours of DM mass $m_\chi$, and the solid curves are contours of gaugino-like ($\eta_3^0$ or $\eta_5^0$) fraction. Here, the GUT relation $\mu_2=\frac{5}{3} \mu_3 \tan^2\theta_W$ has been used.}
 %\label{fey1}
\label{fig:MassContour}
\end{figure}

Before showing our results, we first define the different particle attribute, namely, higgsino-, bino-, wino-, non neutralino-like particle
if the main ingredient (composition fraction) $\geq 60\%$ of a sample is in the state of
$\eta_{1, 2}$, $\eta_3$, $\eta_5$ and $\eta_{9 , 10}$ and is denoted by $\tilde H$-,
$\tilde B$-, $\tilde W$-like and non neutralino-like $\tilde X$ particle, respectively; otherwise, we call it a mixed particle.
Let us first show the sample structures from six sample sets in Table~\ref{tab:pdf}.
We see that less than $1.3\%$ of the samples are the mixed particles which can be ignored in each case.
For the cases of neutralino-like I and~II, 
the population ratio of $\tilde H$-like to $\tilde B$-like particles is roughly about 3 to 7.
%It is roughly one fourth of samples being $\tilde H$-like and three fourth of samples being $\tilde B$-like.
Due to the GUT relation, the $\tilde W$-like particles do not appear in these two cases.
For the cases of neutralino-like III and IV, now without GUT relation, plenty of $\tilde W$-like particles come out.
In these two cases, $\tilde H$-, $\tilde B$-, $\tilde W$-like particles are roughly equally distributed.
For the reduced case, it is about fifty-fifty equally distributed for $\tilde H$- and
$\tilde B$-like particles.
For the extended case, it contains about $5\%$ non neutralino-like $\tilde X$ particles and is roughly equally distributed for $\tilde H$-, $\tilde B$- and $\tilde W$-like particles.
In the subsequent descriptions, we will use `{\color{red} $\circ$}', `{\color{blue} $\times$}', `{\color{green} $\triangle$}',
`{\color{magenta} $\blacksquare$}' and  `{\color{yellow} $\bullet$}' to denote the higgsino-, bino-, wino-, non neutralino-like and the mixed particles, respectively.
The contour plot of the DM mass and composition in the $\mu_1$-$\mu_3$ plane for the neutralino-like case I 
%and II are 
is shown in Fig.~\ref{fig:MassContour}.
Note that the contour plot of the neutralino mass and composition in MSSM~\cite{JKG} is successfully reproduced in Fig~\ref{fig:MassContour}.
%(a).
Hence the fermion multiplets  $\eta_1$, $\eta_2$, $\eta_3$, and $\eta_5$ correspond to two doublets of higgsinos, a singlet of bino and a triplet of winos in MSSM, respectively [recall Eq.~(\ref{eq:mssm})]. 
Nevertheless, the model does not contain particles corresponding to the sfermions and the second higgs doublet in MSSM so that there does not exist the annihilation channels into the extra scalar states and scattering
diagrams mediated by the extra scalars.
On the other hand, the model do contain more $Z_2$-odd fermion particles with multiplets $\eta_7$, $\eta_8$, $\eta_9$, and $\eta_{10}$. Hence this generic Majorana DM model is still quite different from the MSSM. %Next, we will see later that the slightly different contour plots of DM mass and composition between neutralino-like I and II cases [see Fig.~\ref{fig:MassContour} (b)] does affect the spread in the scatter plots of thermal relic density, SI and SD DM-nucleus scattering cross sections, and the velocity averaged cross section of DM annihilation processes versus DM mass, respectively.

%\begin{figure}[t]
%\centering
%\captionsetup{justification=raggedright}
% \subfigure[\ \ neutralino-like I]{
  %\includegraphics[width=0.45\textwidth]{02aDmassContour.pdf}
%}\subfigure[\ \ neutralino-like II]{
  %\includegraphics[width=0.45\textwidth]{02bEmassContour.pdf}
%}
%\caption{Contour plots of the DM mass and composition in the $\mu_1$-$\mu_3$ plane. The broken curves are contours of DM mass $m_\chi$, and the solid curves are contours of gaugino-like ($\eta_3^0$ or $\eta_5^0$) fraction. Here, the GUT relation $\mu_2=\frac{5}{3} \mu_3 \tan^2\theta_W$ has been used in both of (a) $\tan\beta=2$, and (b) $\tan\beta =20$.}
%\label{fig:MassContour}
%\end{figure}

\subsection{Case A: neutralino-like cases}

\begin{figure}[t!]
\centering
\captionsetup{justification=raggedright}
 \subfigure[\ Constraint on $\Omega^{\rm{obs}}_\chi$]{
  \includegraphics[width=0.45\textwidth,height=0.135\textheight]{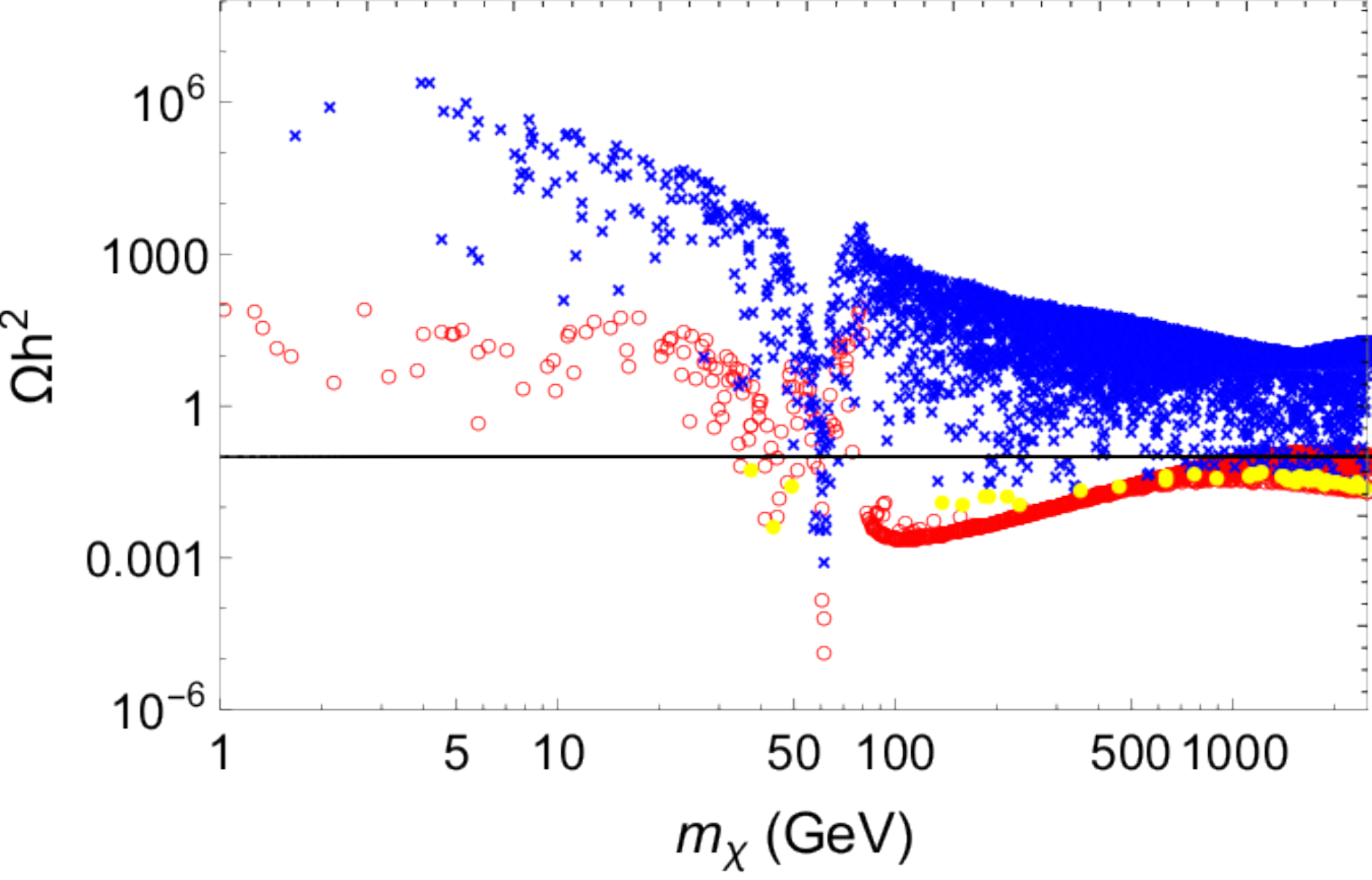}
}\subfigure[\ LUX constraint on $\sigma^{SI}$ with NB limit]{
  \includegraphics[width=0.45\textwidth,height=0.135\textheight]{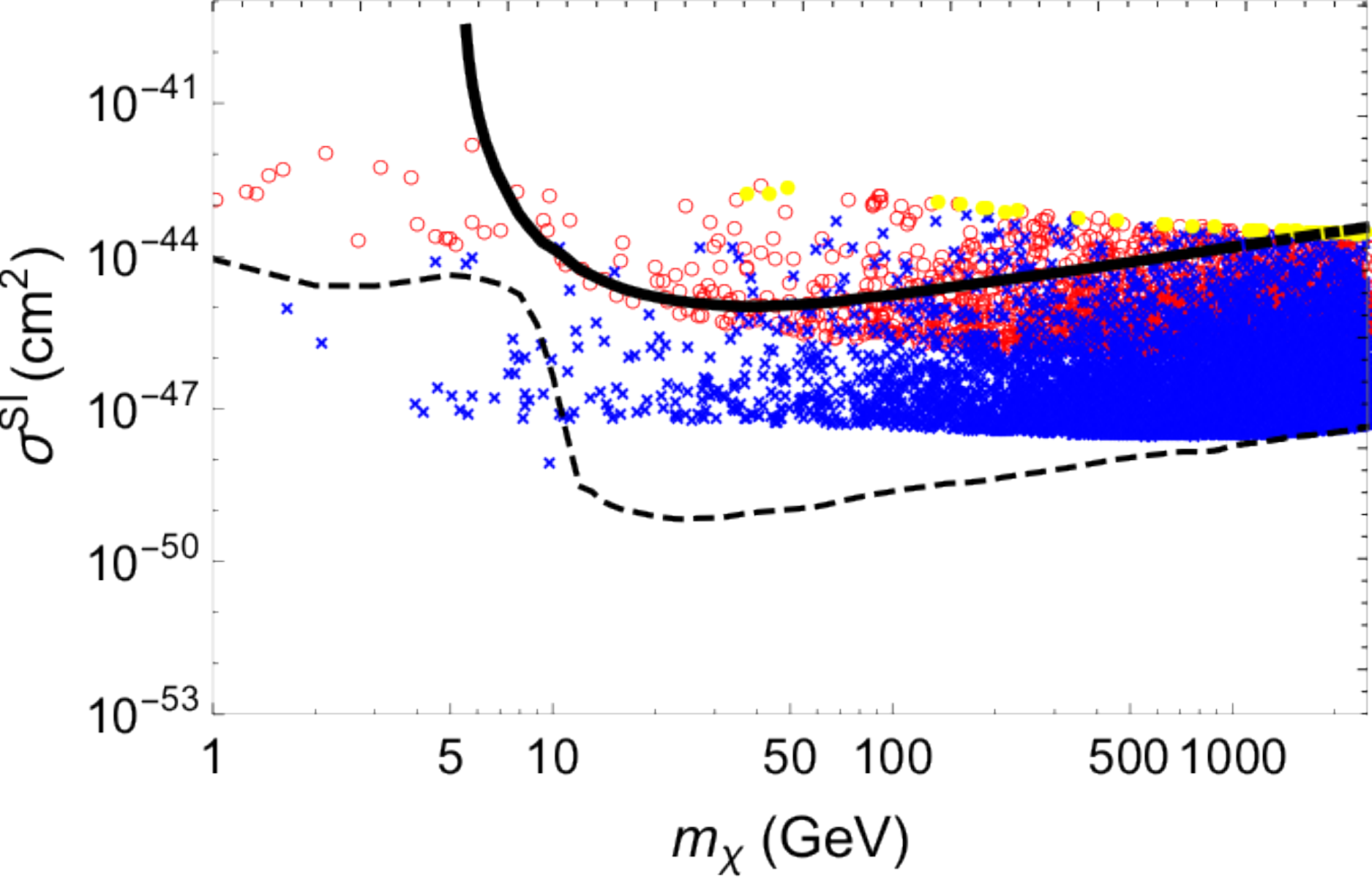}
}\\\subfigure[\ XENON100 constraint on $\sigma^{SD}_n$]{
  \includegraphics[width=0.45\textwidth,height=0.135\textheight]{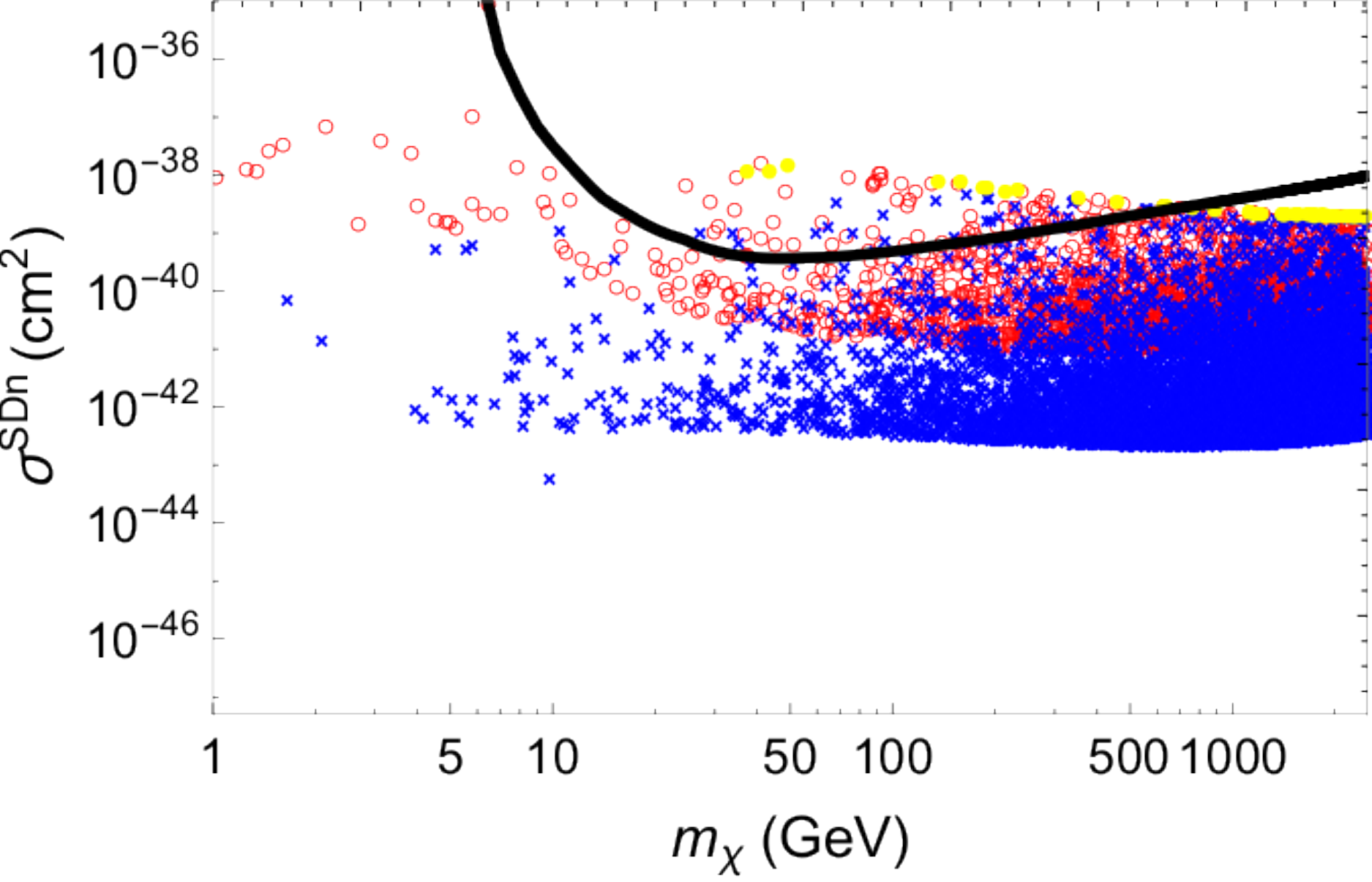}
}\subfigure[\ XENON100 constraint on $\sigma^{SD}_p$]{
  \includegraphics[width=0.45\textwidth,height=0.135\textheight]{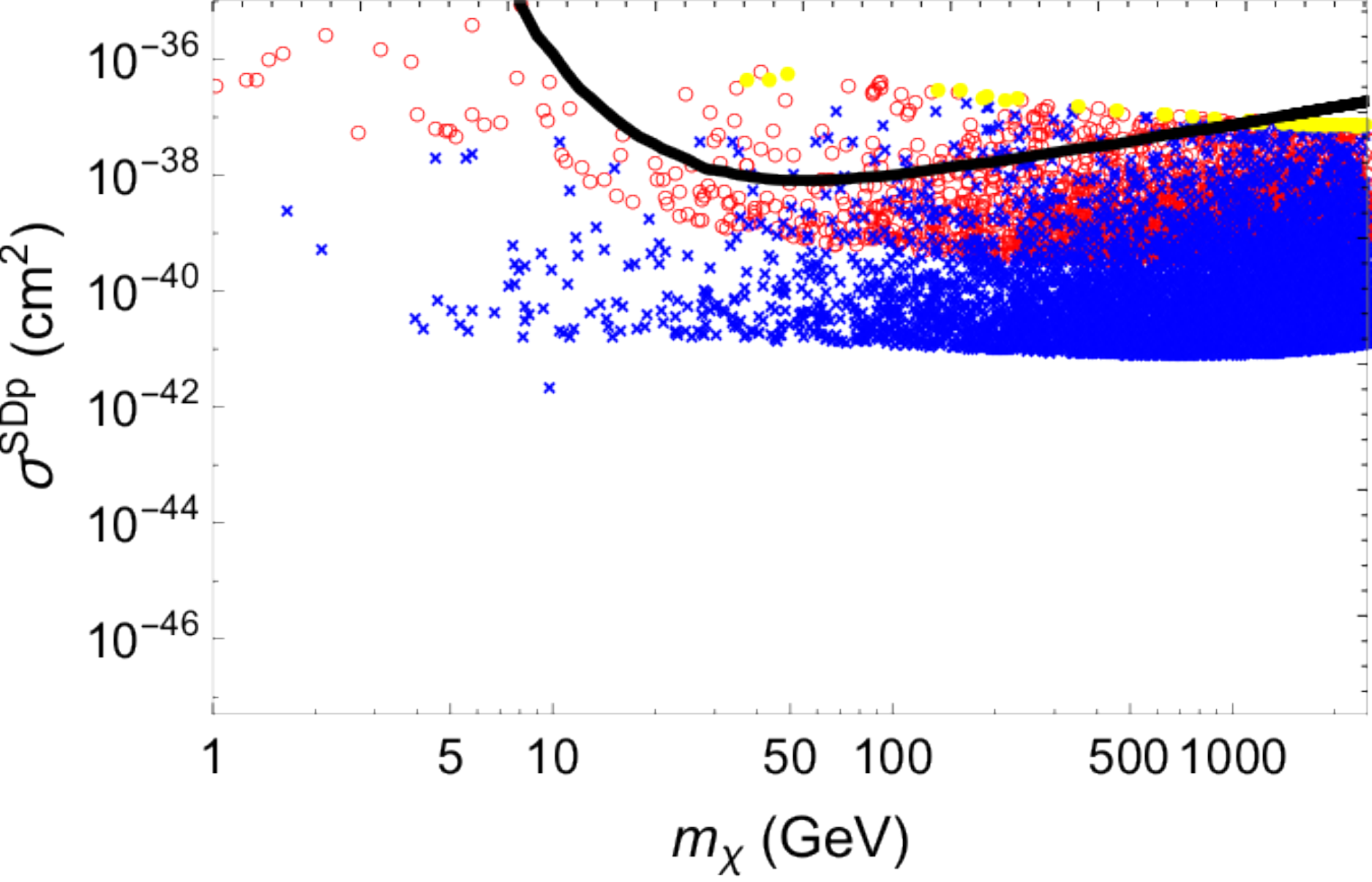}
}\\\subfigure[\ PICO-60 constraint on $\sigma^{SD}_p$]{
  \includegraphics[width=0.45\textwidth,height=0.135\textheight]{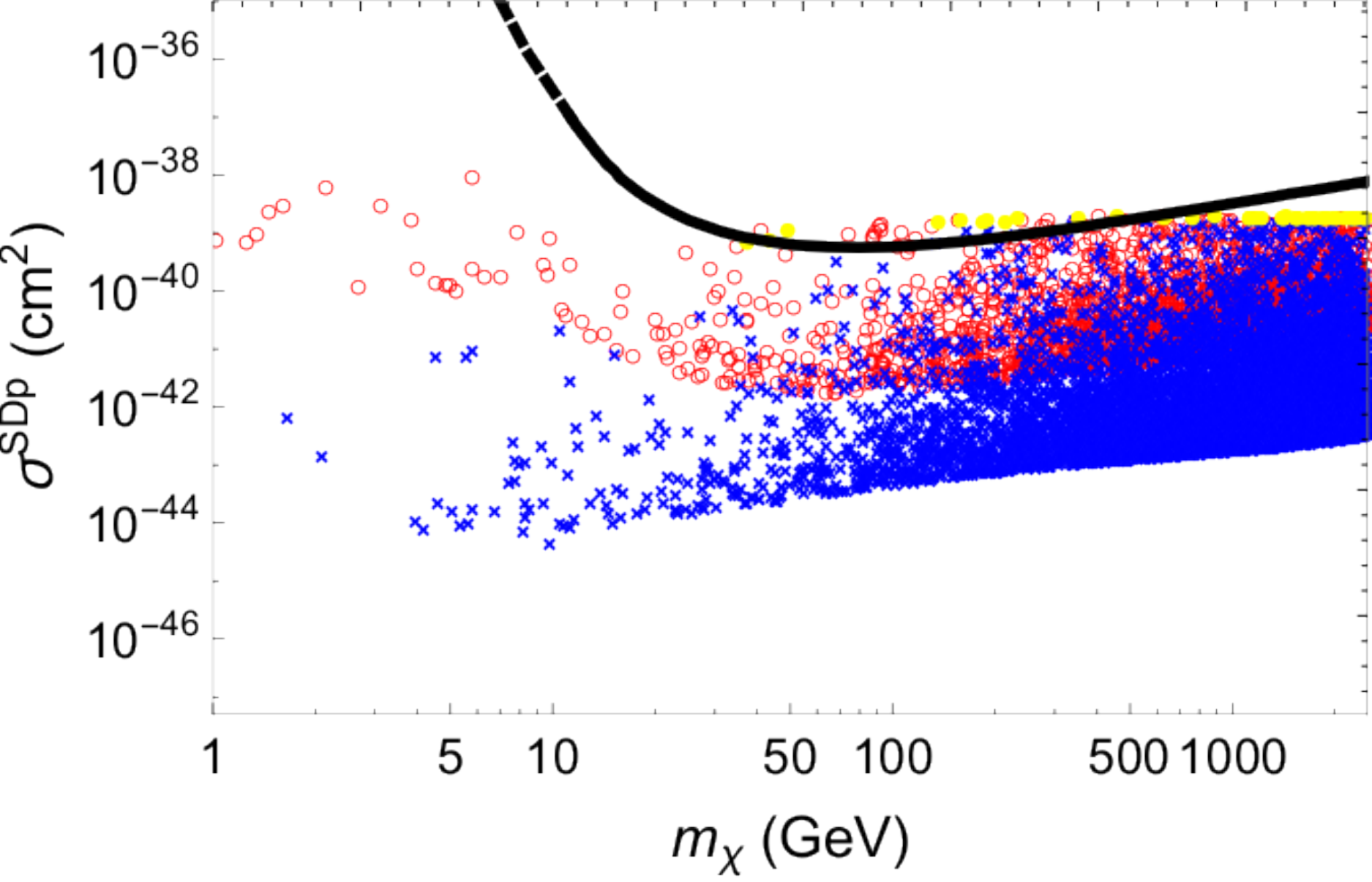}
}\subfigure[\ Fermi-LAT constraint on $\chi^0 {\chi}^0\rightarrow W^+W^-$]{
  \includegraphics[width=0.45\textwidth,height=0.135\textheight]{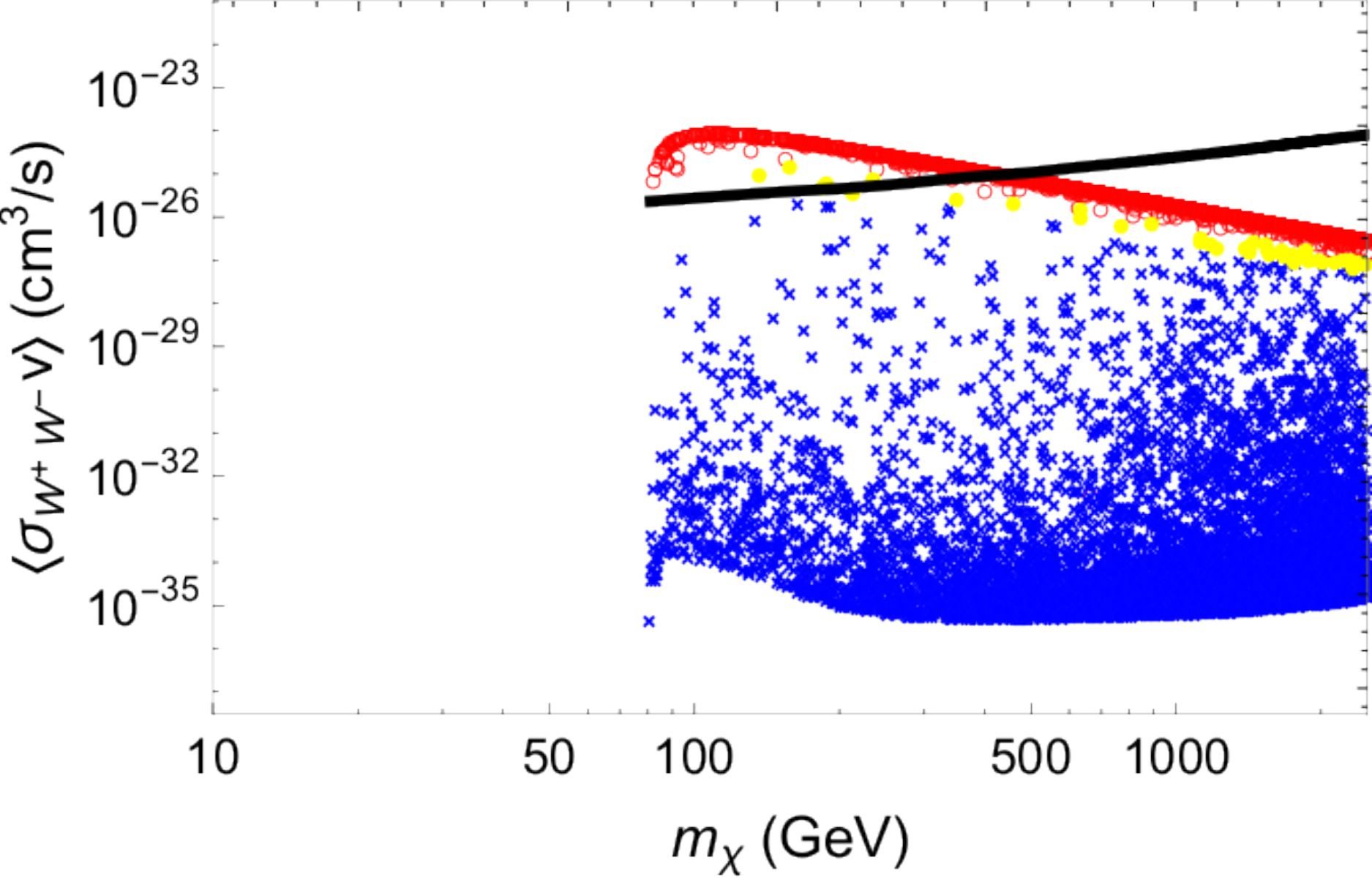}
}\\\subfigure[\ Fermi-LAT constraint on $\chi^0 {\chi}^0\rightarrow b\bar{b}$]{
  \includegraphics[width=0.45\textwidth,height=0.135\textheight]{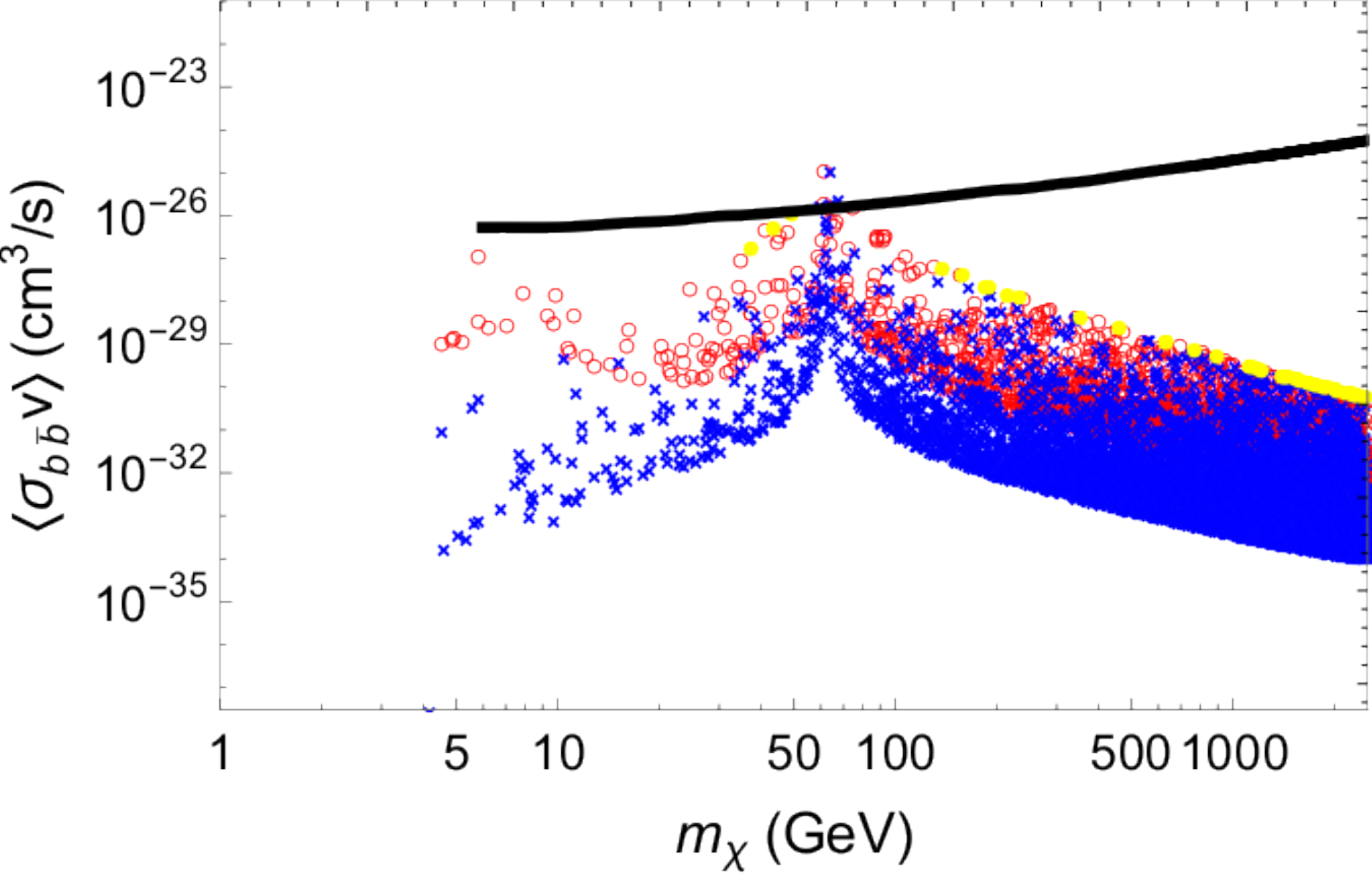}
}\subfigure[\ Fermi-LAT constraint on $\chi^0 {\chi}^0\rightarrow u\bar{u}$]{
  \includegraphics[width=0.45\textwidth,height=0.135\textheight]{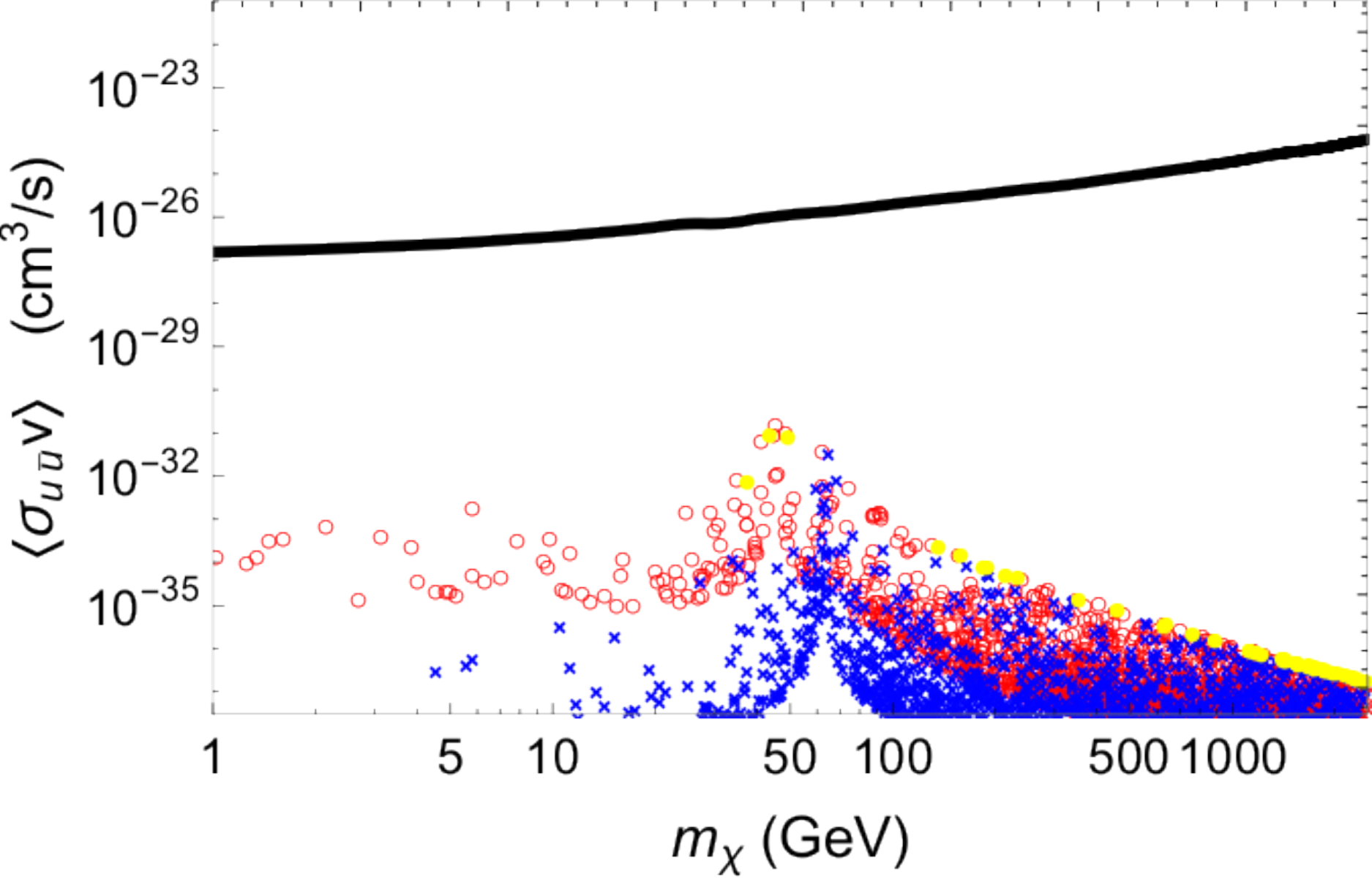}
}\\\subfigure[\ Fermi-LAT constraint on $\chi^0 {\chi}^0\rightarrow \tau^+\tau^-$]{
  \includegraphics[width=0.45\textwidth,height=0.135\textheight]{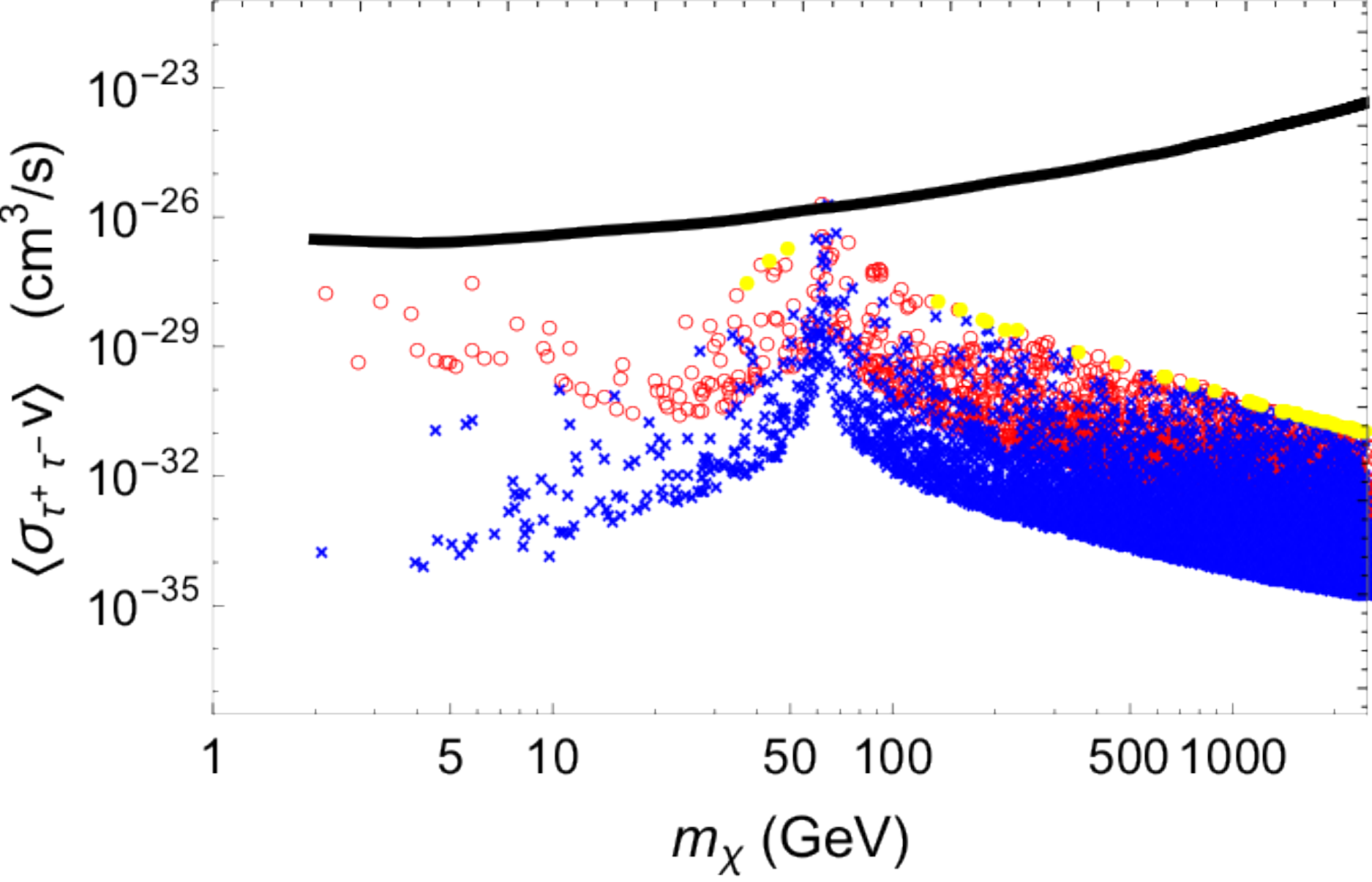}
}\subfigure[\ Fermi-LAT constraint on $\chi^0 {\chi}^0\rightarrow \mu^+\mu^-$]{
  \includegraphics[width=0.45\textwidth,height=0.135\textheight]{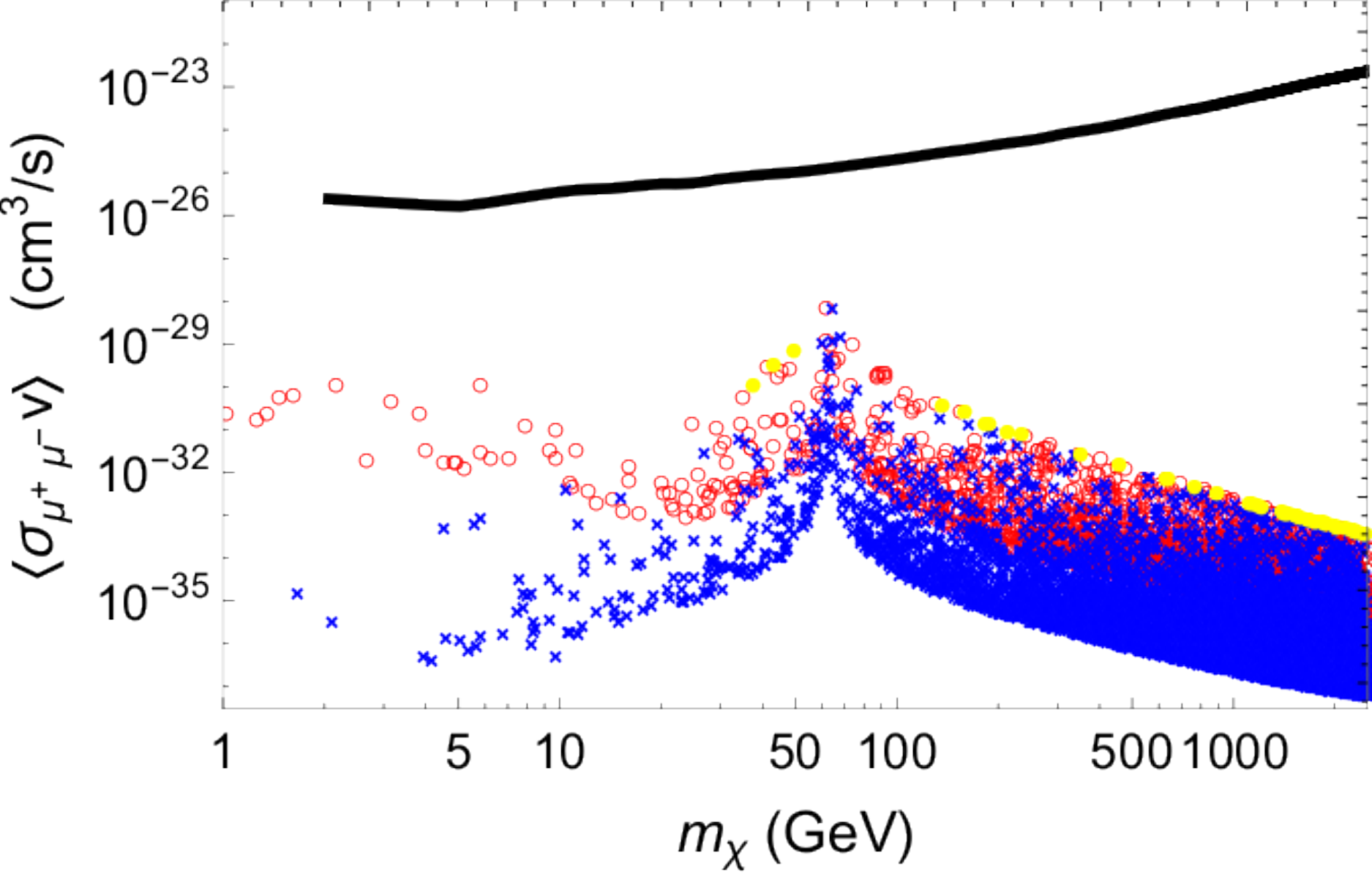}
%}\subfigure[\ Fermi-LAT constraint on $\chi^0 \bar{\chi}^0\rightarrow e^+e^-$]{
  %\includegraphics[width=0.45\textwidth,height=0.135\textheight]{03jDindee.pdf}
}
\caption{Results for all samples with constraints in the case of neutralino-like I
[{\color{red} $\circ$}:~higgsino-like,
{\color{blue} $\times$}: bino-like,
{\color{yellow} $\bullet$}: mixed].}
\label{fig:neutralino-like I}
\end{figure}

\begin{figure}[t!]
\centering
\captionsetup{justification=raggedright}
  \subfigure[$\chi^0 {\chi}^0\rightarrow ZZ$]{
  \includegraphics[width=0.45\textwidth,height=0.15\textheight]{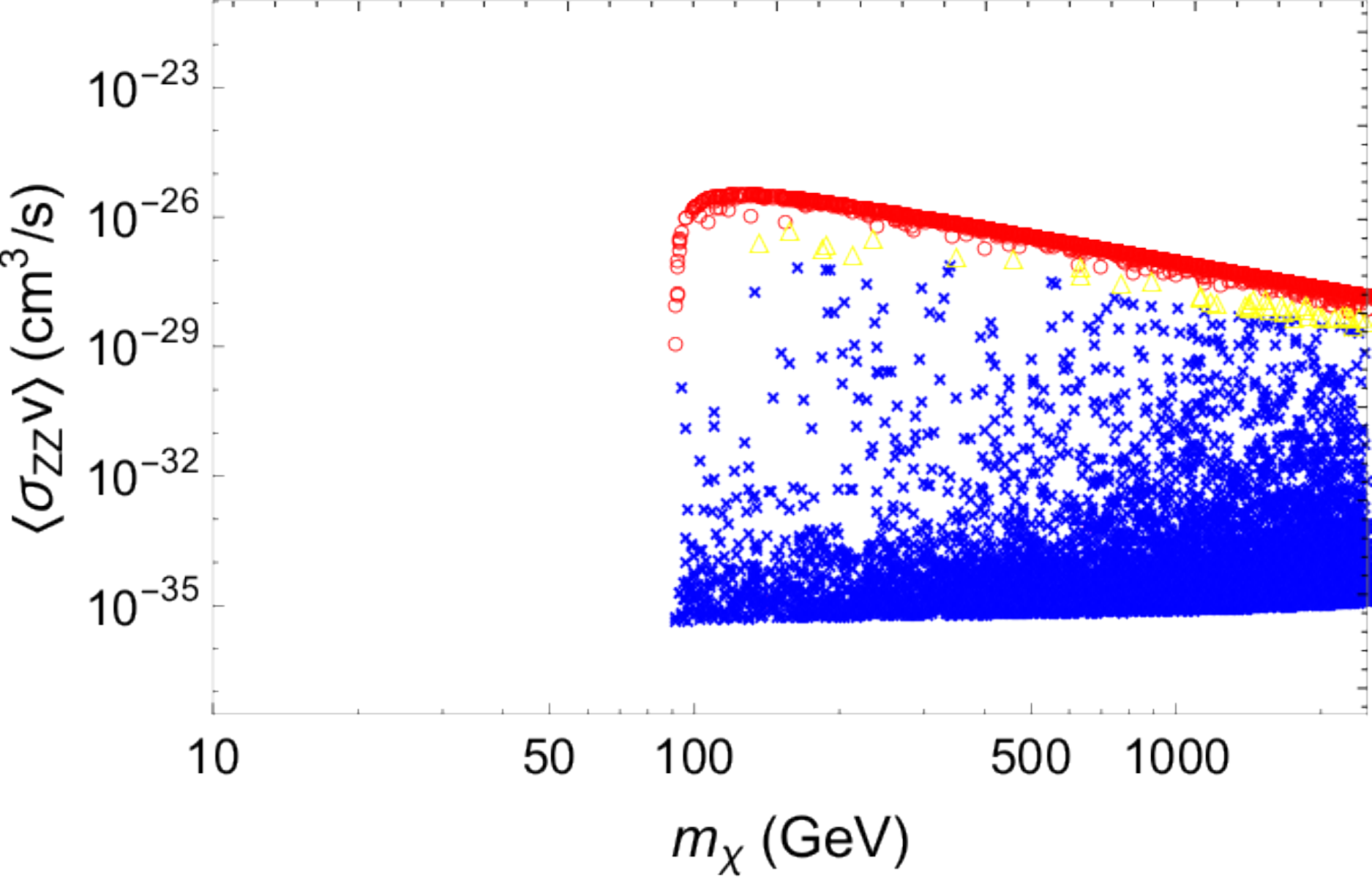}
}\subfigure[$\chi^0 {\chi}^0\rightarrow ZH$]{
  \includegraphics[width=0.45\textwidth,height=0.15\textheight]{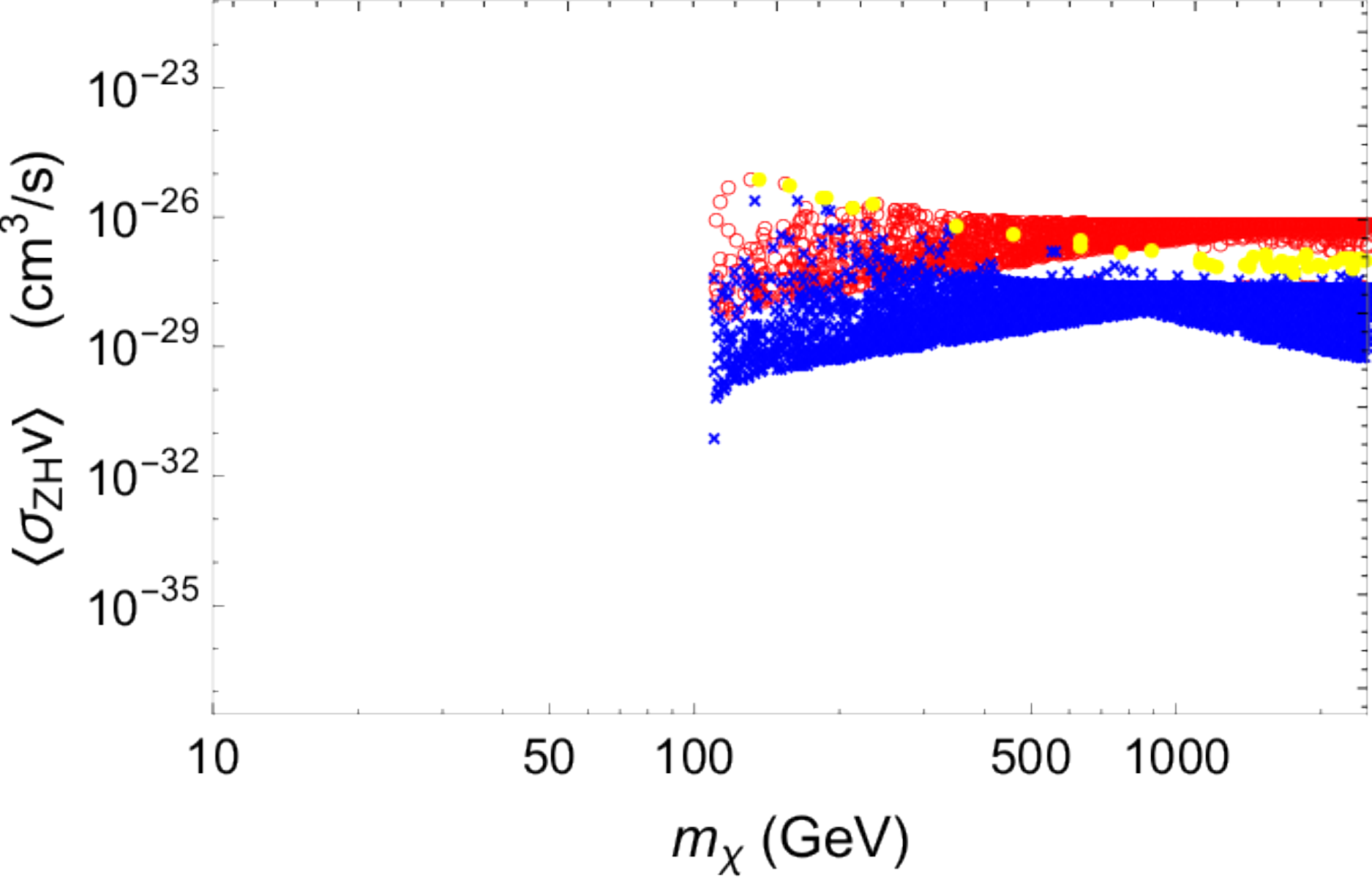}
}\\\subfigure[$\chi^0 {\chi}^0\rightarrow t {\bar t}$]{
  \includegraphics[width=0.45\textwidth,height=0.15\textheight]{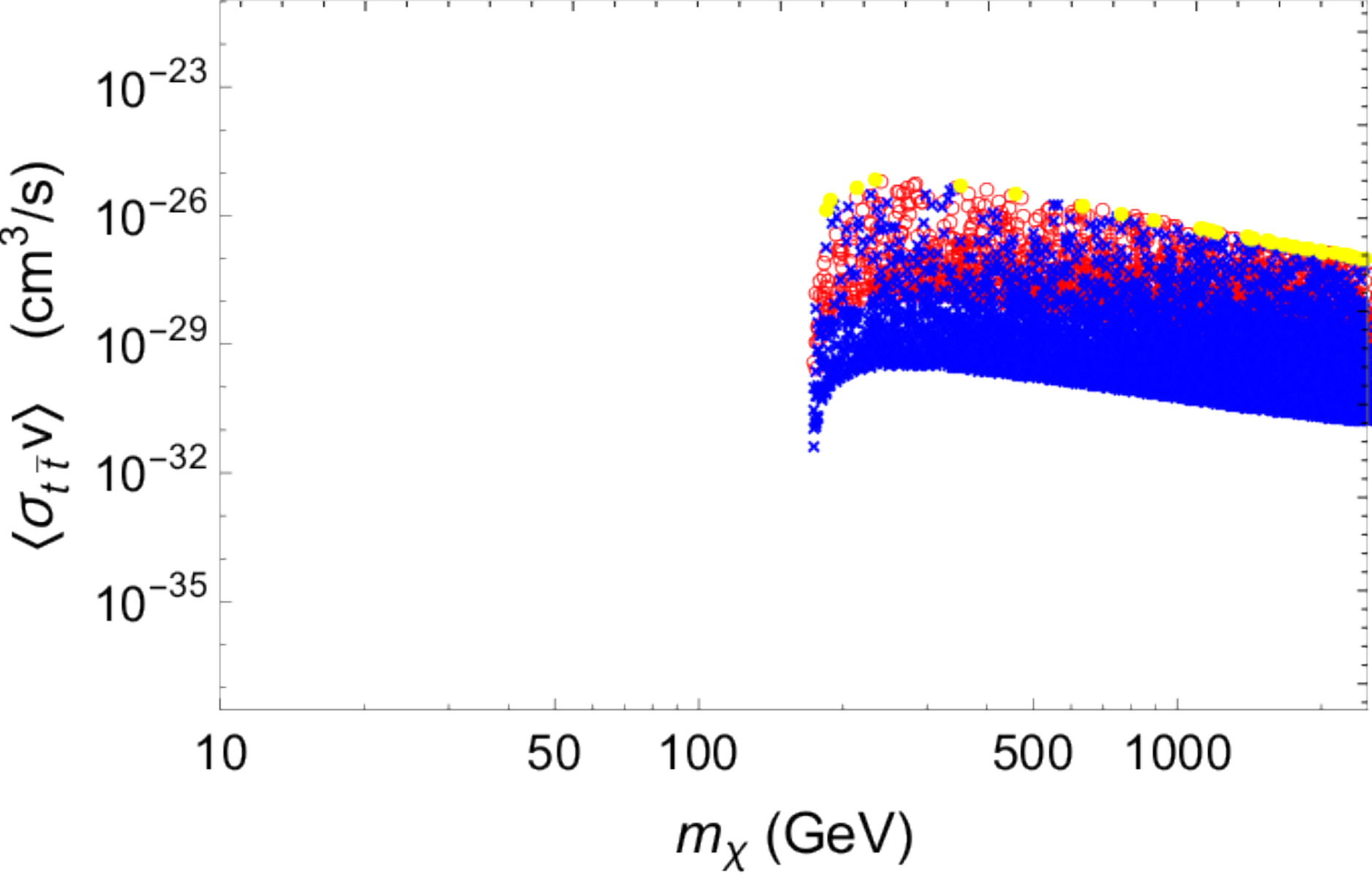}
}\subfigure[$\chi^0 {\chi}^0\rightarrow HH$]{
  \includegraphics[width=0.45\textwidth,height=0.15\textheight]{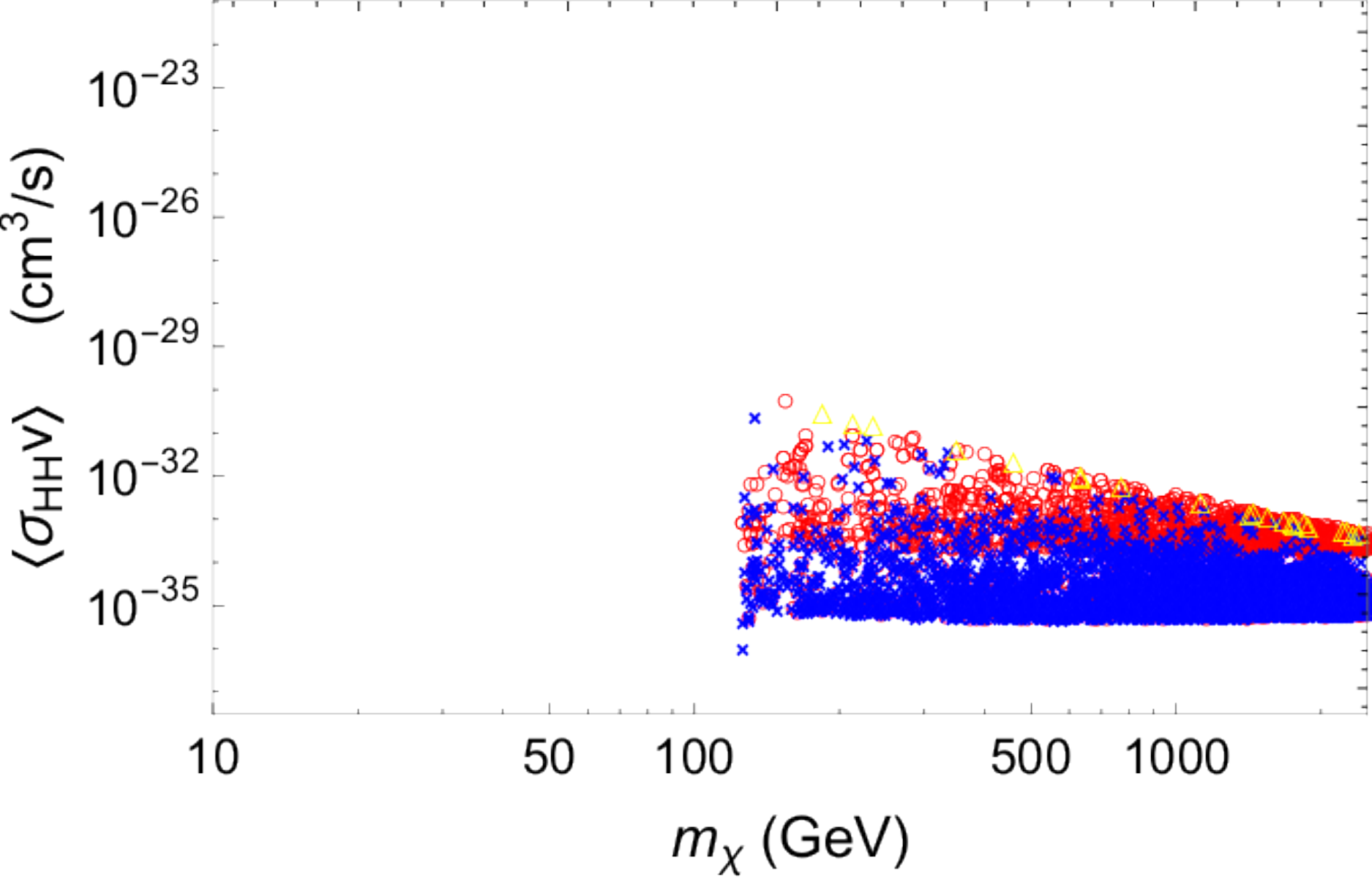}
}
\caption{Scatter plots of  $\la \sigma_{ZZ, ZH, t{\bar t}, HH}\ v \ra$ versus $m_\chi$ in the case of neutralino-like I
[{\color{red} $\circ$}:~higgsino-like~,
{\color{blue} $\times$}:~bino-like,
{\color{yellow} $\bullet$}:~mixed].}
\label{fig:ZZZHttHH}
\end{figure}

\begin{figure}[t!]
\centering
\captionsetup{justification=raggedright}
 \subfigure[\ Constraint on $\Omega^{\rm{obs}}_\chi$]{
  \includegraphics[width=0.45\textwidth,height=0.135\textheight]{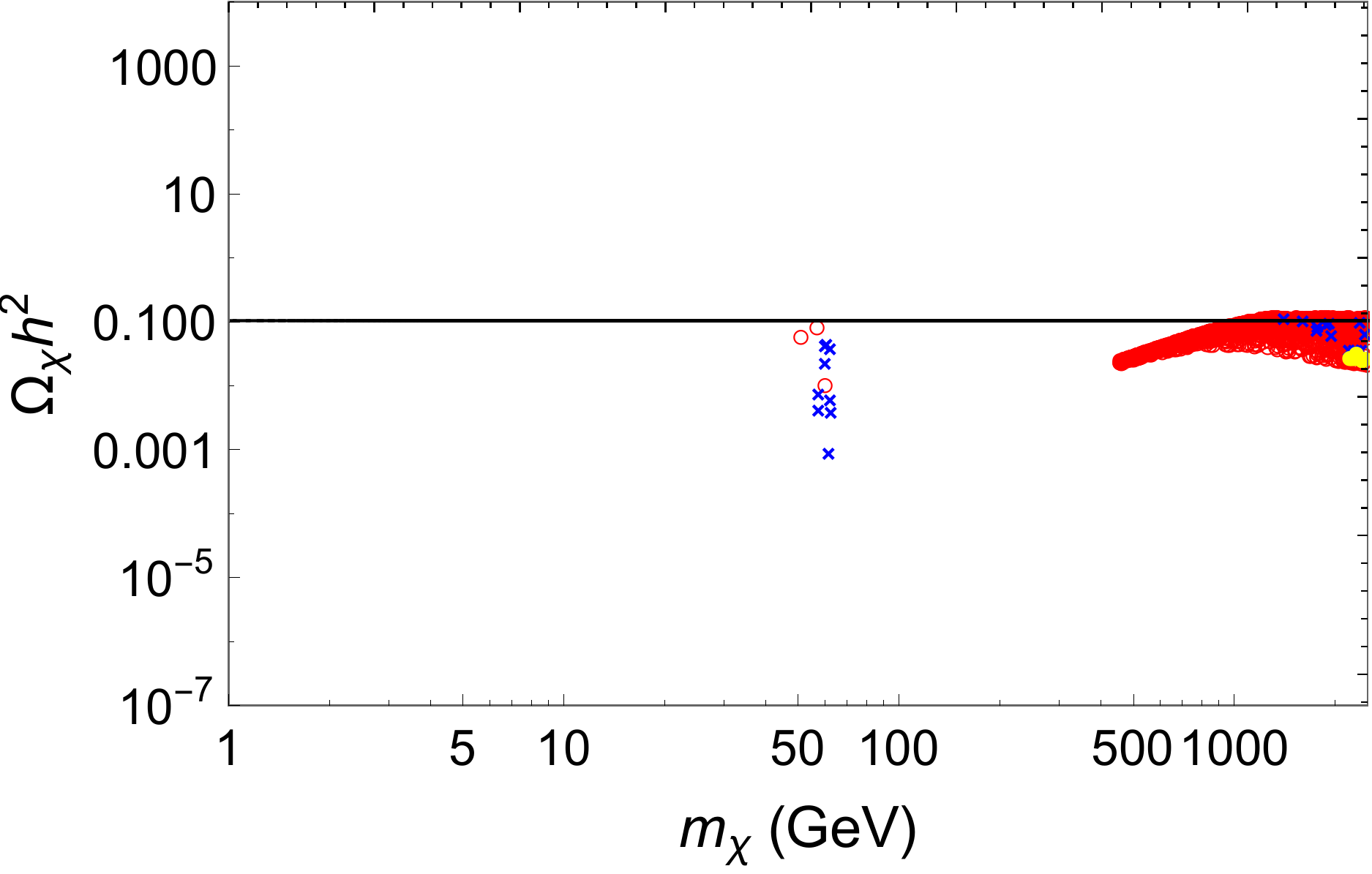}
}\subfigure[\ LUX constraint on $\sigma^{SI}$ with NB limit]{
  \includegraphics[width=0.45\textwidth,height=0.135\textheight]{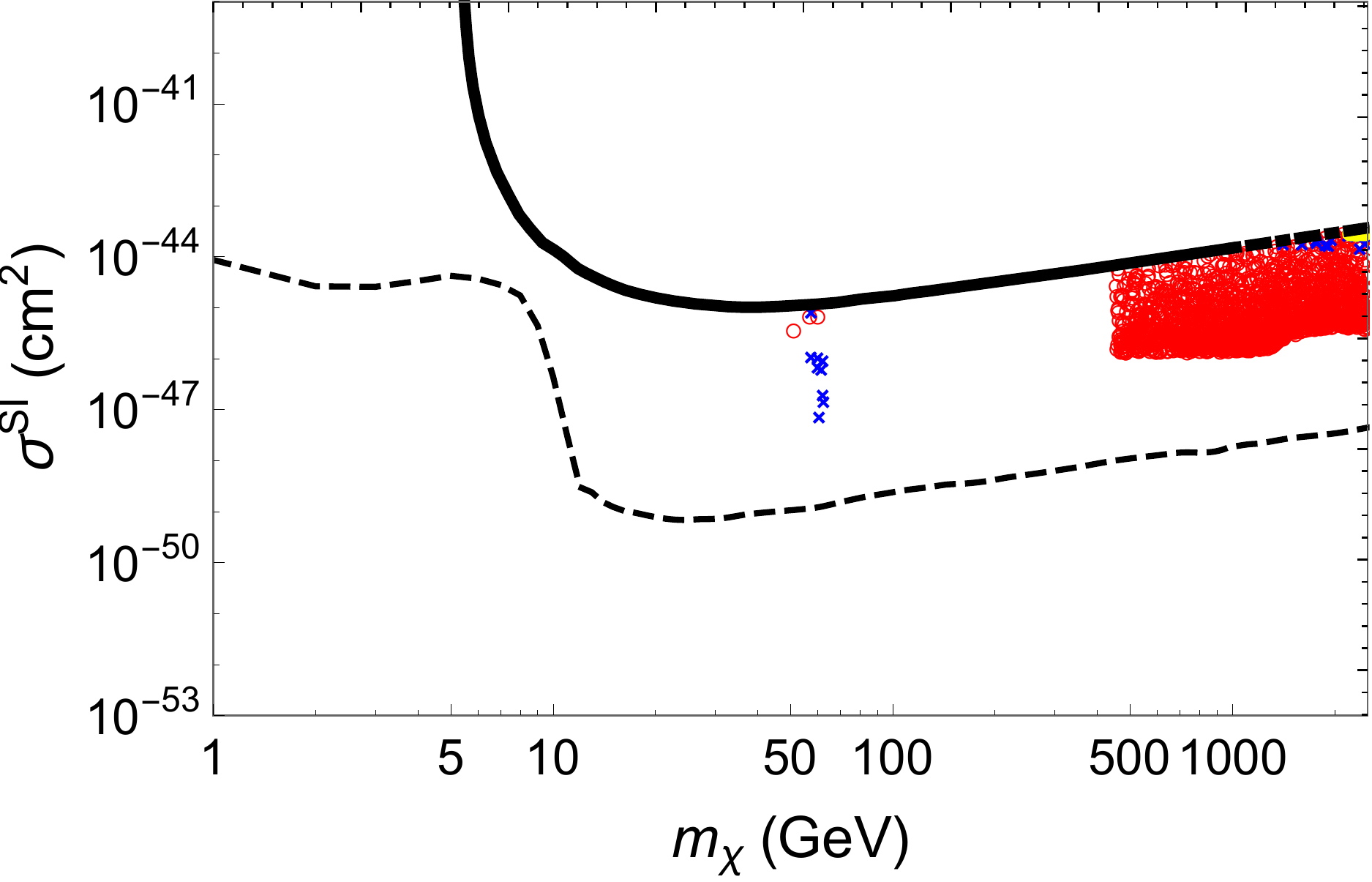}
}\\\subfigure[\ XENON100 constraint on $\sigma^{SD}_n$]{
  \includegraphics[width=0.45\textwidth,height=0.135\textheight]{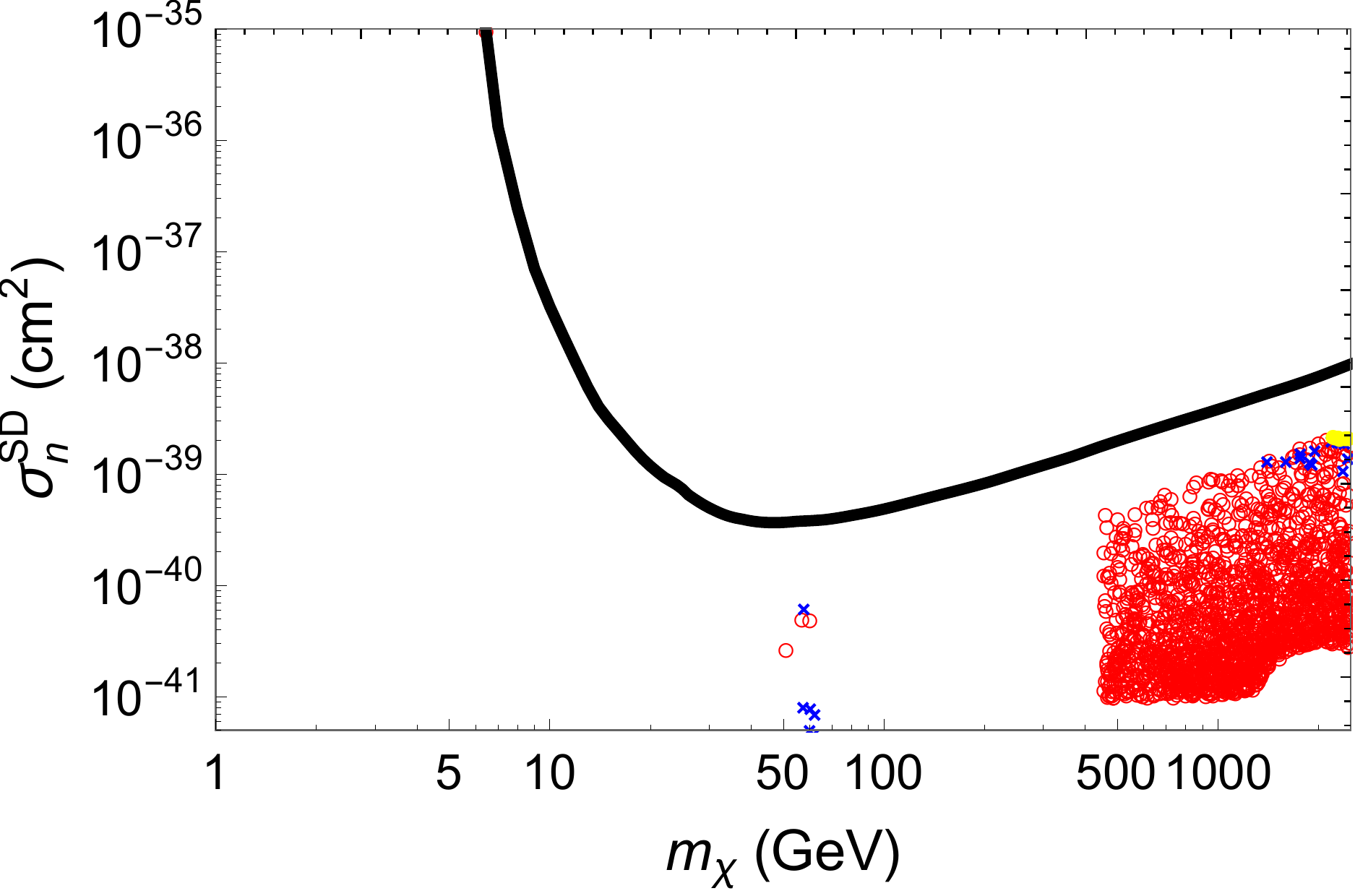}
}\subfigure[\ XENON100 constraint on $\sigma^{SD}_p$]{
  \includegraphics[width=0.45\textwidth,height=0.135\textheight]{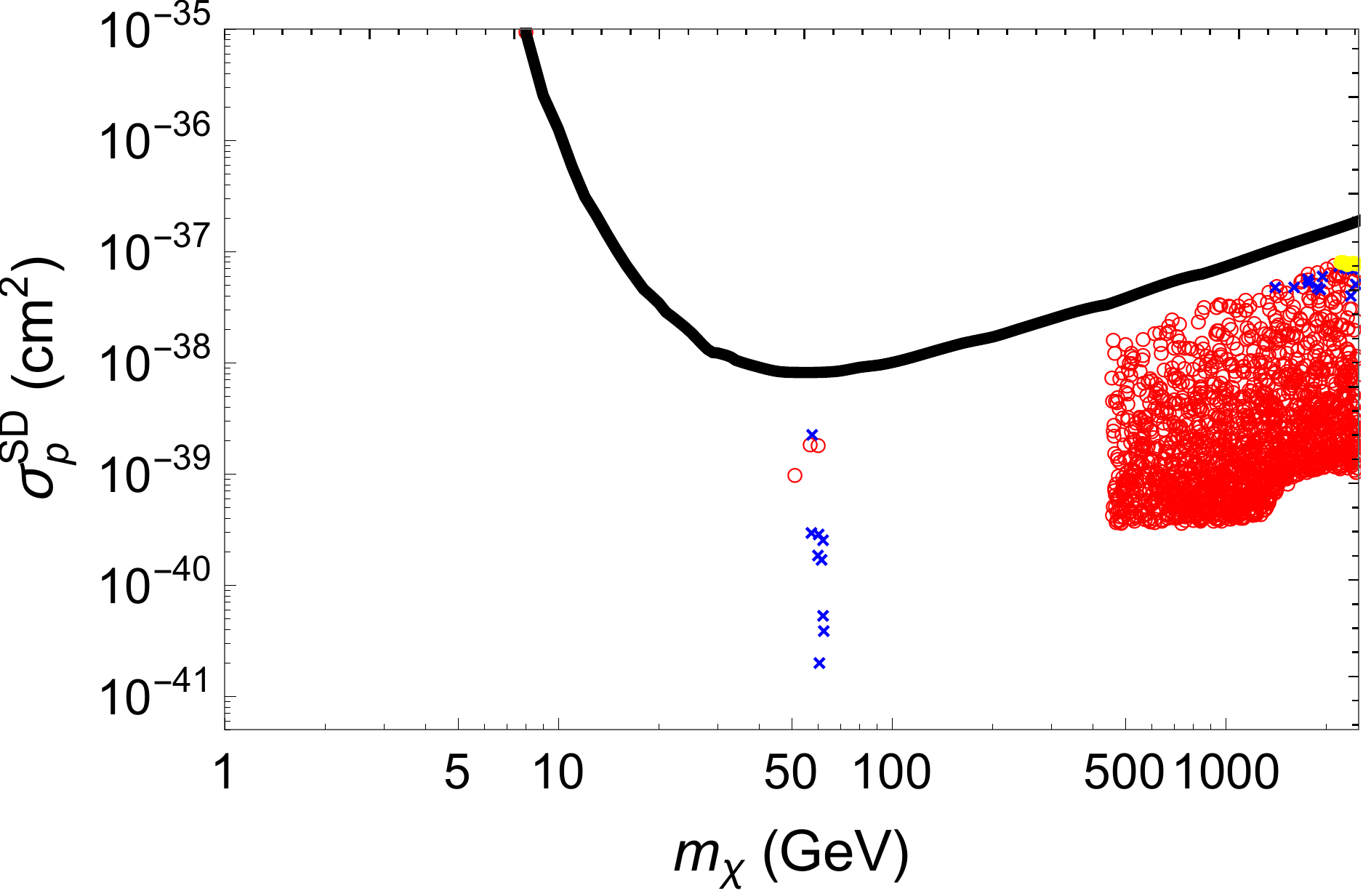}
}\\\subfigure[\ PICO-60 constraint on $\sigma^{SD}_p$]{
  \includegraphics[width=0.45\textwidth,height=0.135\textheight]{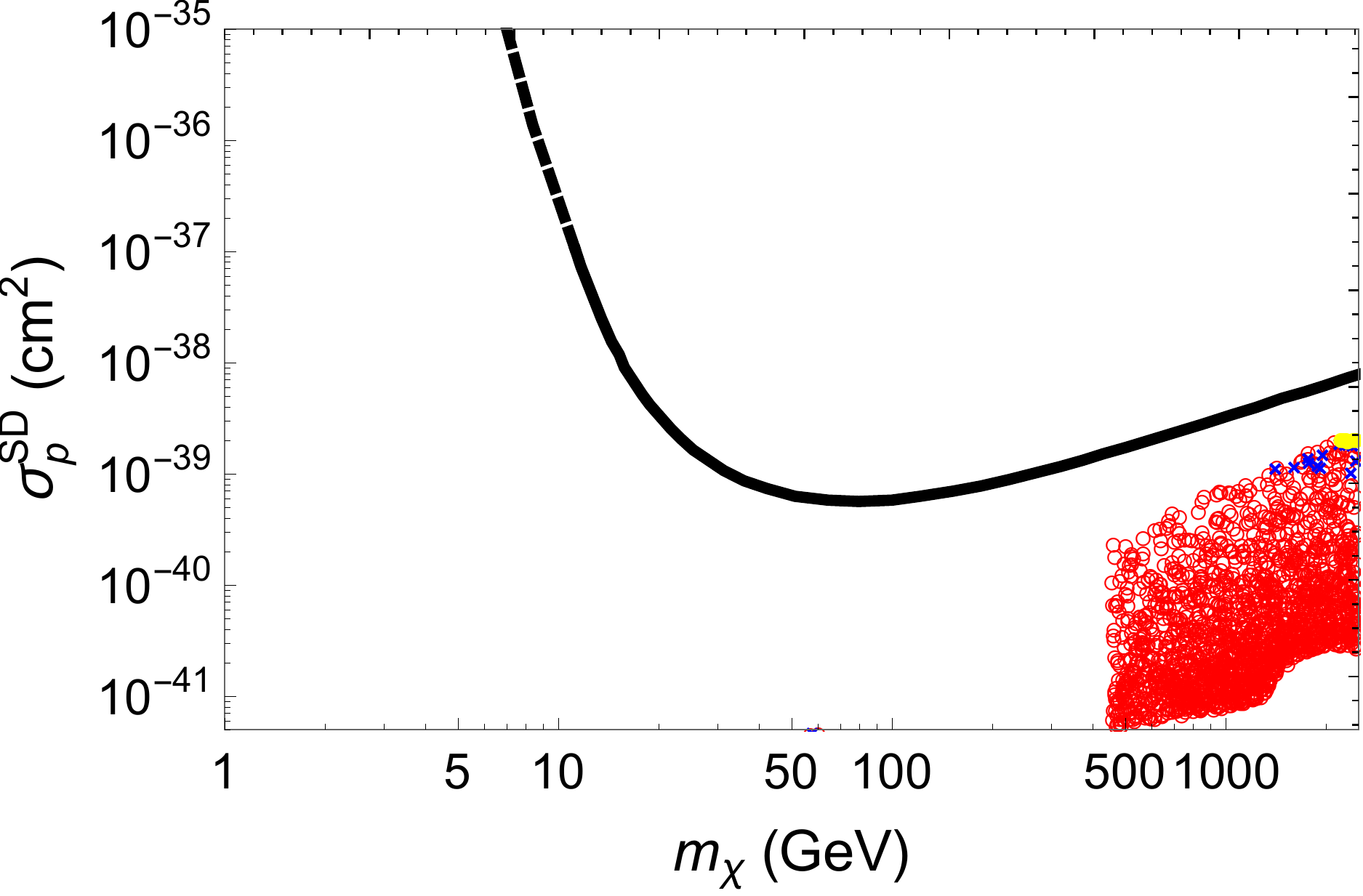} 
}\subfigure[\ Fermi-LAT constraint on $\chi^0 {\chi}^0\rightarrow W^+W^-$]{
  \includegraphics[width=0.45\textwidth,height=0.135\textheight]{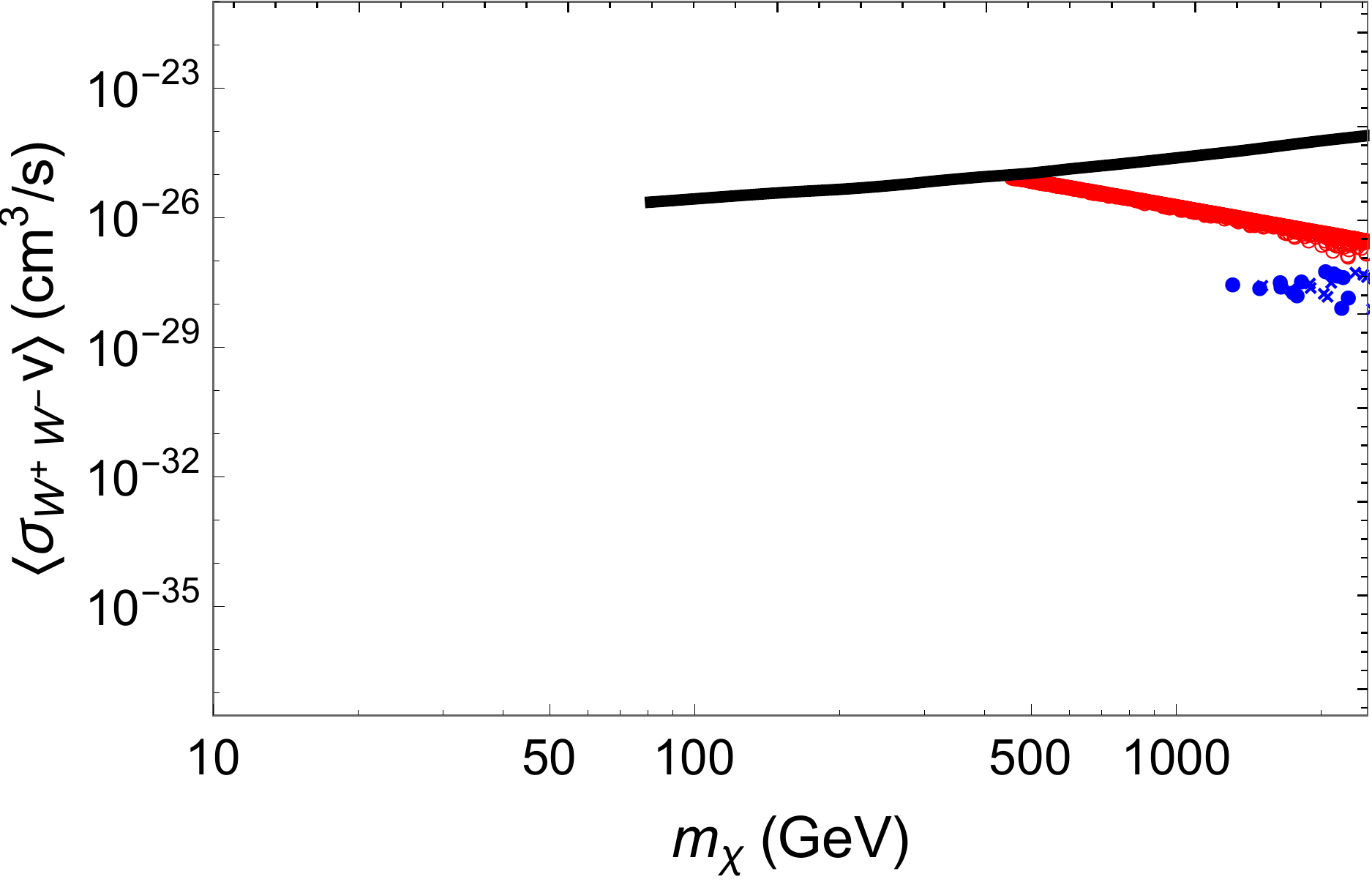}
}\\\subfigure[\ Fermi-LAT constraint on $\chi^0 {\chi}^0\rightarrow b\bar{b}$]{
  \includegraphics[width=0.45\textwidth,height=0.135\textheight]{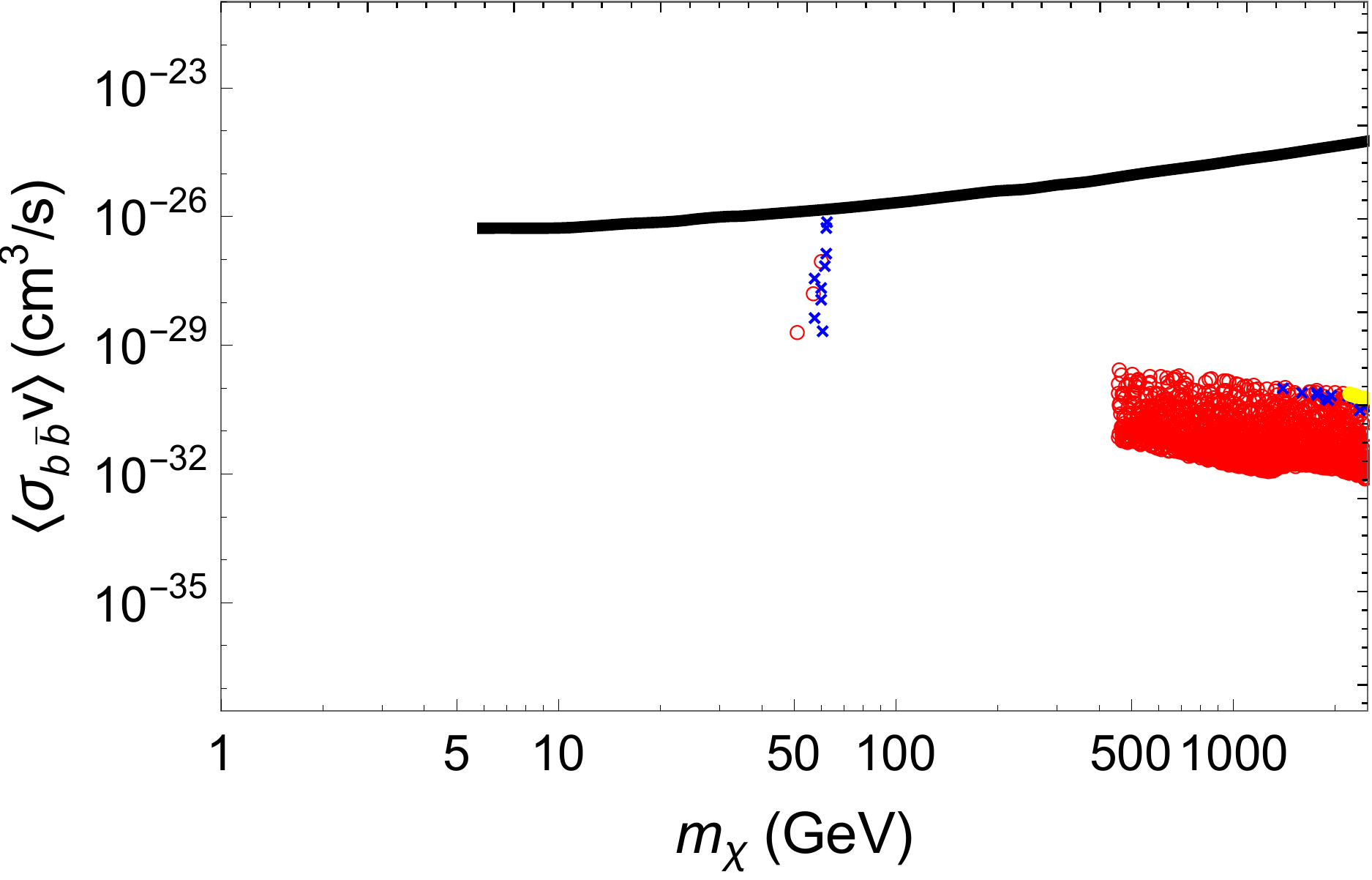}
}\subfigure[\ Fermi-LAT constraint on $\chi^0 {\chi}^0\rightarrow u\bar{u}$]{
  \includegraphics[width=0.45\textwidth,height=0.135\textheight]{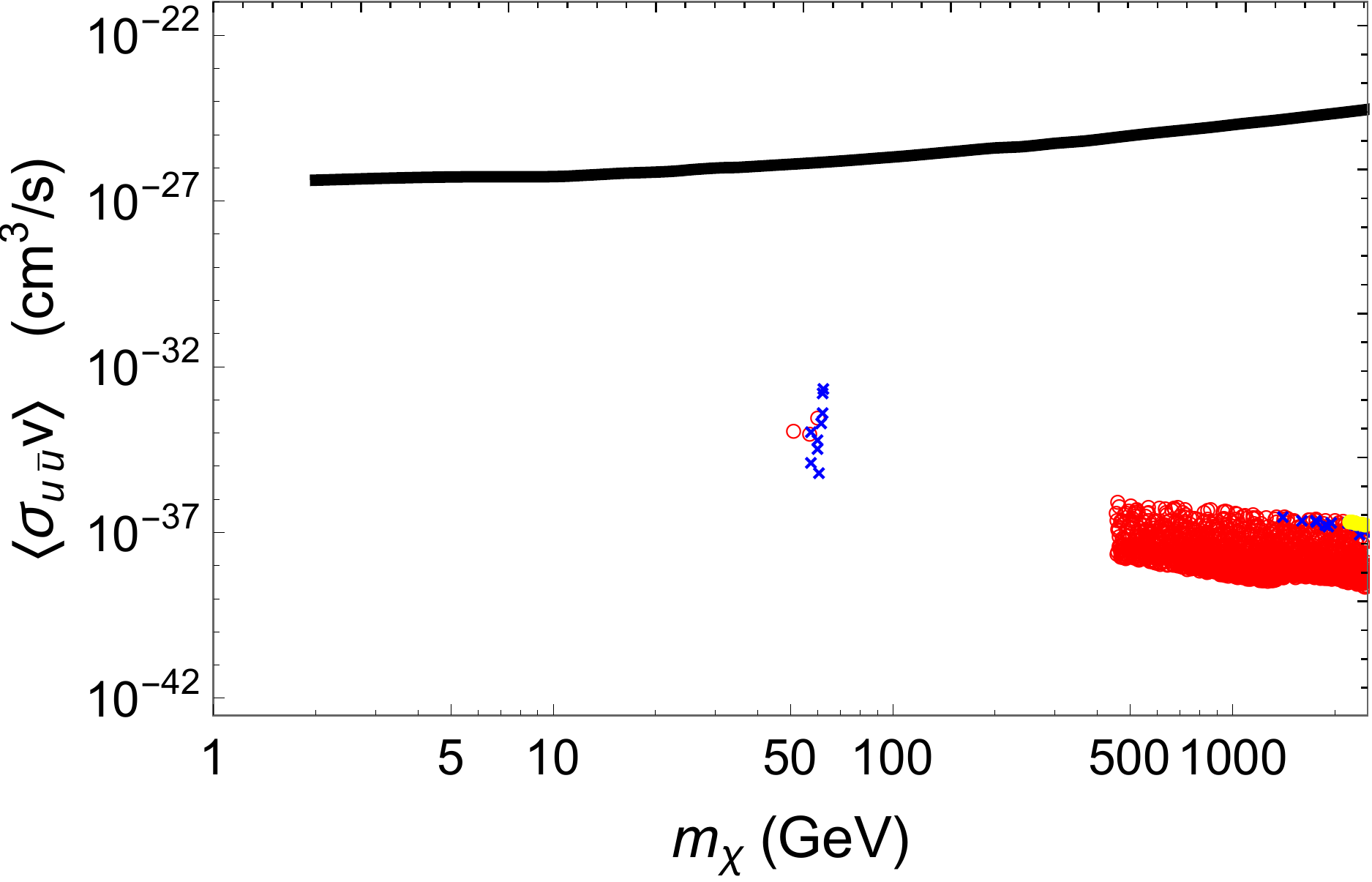}
}\\\subfigure[\ Fermi-LAT constraint on $\chi^0 {\chi}^0\rightarrow \tau^+\tau^-$]{
  \includegraphics[width=0.45\textwidth,height=0.135\textheight]{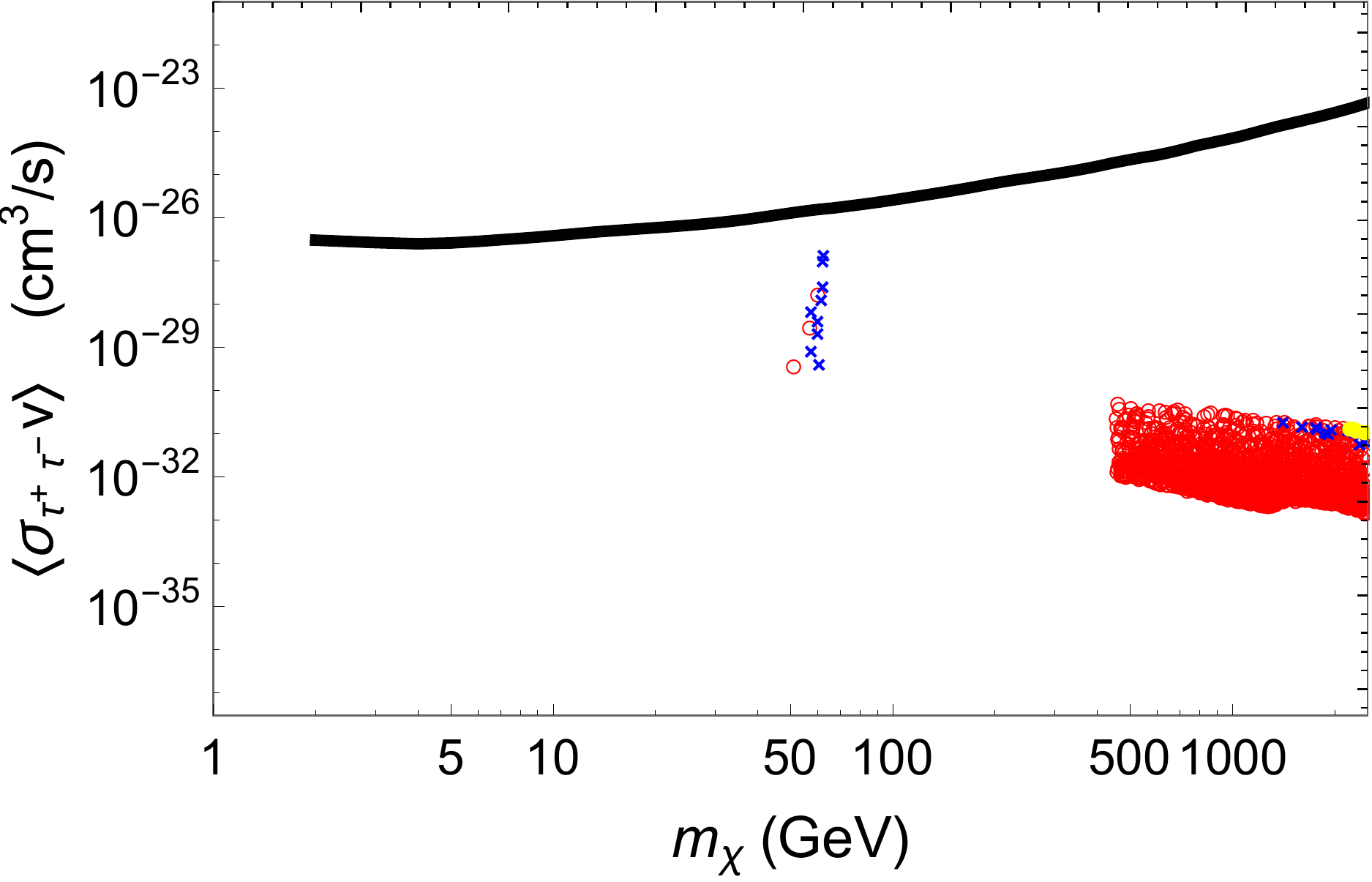}
}\subfigure[\ Fermi-LAT constraint on $\chi^0 {\chi}^0\rightarrow \mu^+\mu^-$]{
  \includegraphics[width=0.45\textwidth,height=0.135\textheight]{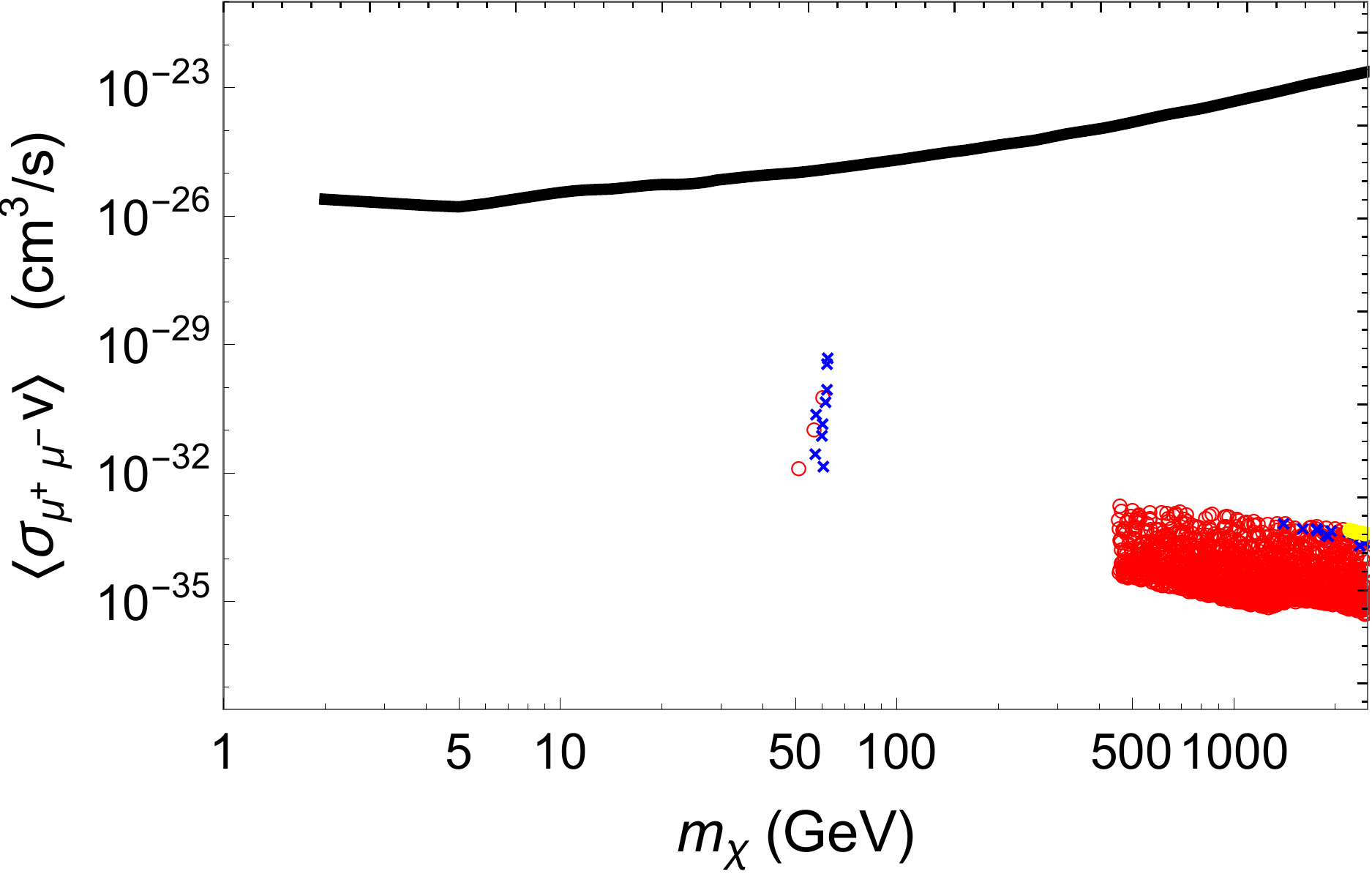}
%}\subfigure[\ Fermi-LAT constraint on $\chi^0 \bar{\chi}^0\rightarrow e^+e^-$]{
%  \includegraphics[width=0.45\textwidth,height=0.135\textheight]{05jDallowee.pdf}
}
\caption{Results for allowed samples satisfying all constraints in the neutralino-like I case
[{\color{red} $\circ$}:~higgsino-like,
{\color{blue} $\times$}:~bino-like,   {\color{yellow} $\bullet$}:~mixed].}
\label{fig:allow neutralino-like I}
\end{figure}

Both the neutralino-like I and II cases contain 7 parameters, $\mu_{1\sim 3}, g_{3\sim 6}$, which are subjected to the GUT and the $\tan\beta$ relations resulting in only two free parameters $\mu_1$ and $\mu_2$ (or $\mu_3$). The neutralino-like III case is only subjected to the $\tan\beta$ relation resulting in three free parameters $\mu_{1\sim 3}$. Without the GUT and the $\tan\beta$ relations, all of these 7 parameters in the neutralino-like IV case are free. 
We first emphasize on the description of the interplay among these constraints with the case of neutralino-like I using Figs.~\ref{fig:neutralino-like I}-\ref{fig:allow neutralino-like I}, and then tell the differences among these neutralino-like cases in this subsection. The reduced case and the extended case are discussed in the next two subsections.
For neutralino-like I case, we show the scatter plot of $\Omega_\chi h^2$ versus $m_\chi$ in Fig.~\ref{fig:neutralino-like I}(a). The horizontal line denote the upper limit using the upper $3\sigma$ value of the
observed relic density $\Omega_{\chi} h^2=0.1198\pm 0.0026$.
The samples sitting above the horizontal line are ruled out. We see that most of the $\tilde B$-like particles are ruled out, while the $\tilde H$-like particles tending to have smaller values in relic density with $m_\chi > M_W$ are safe. The $\Omega_\chi^{\rm obs} h^2$ constraint is the most stringent constraint since about $74 \%$ of samples are ruled out by this constraint.
The results of DM-nucleon elastic scattering cross sections comparing to the LUX $\sigma^{SI}_N$, the XENON100 $\sigma_{n, p}^{SD}$ and the PICO-60 $\sigma_p^{SD}$ constraints are shown in Fig.~\ref{fig:neutralino-like I}(b)-(e), respectively.
Since the LUX constraint on $\sigma^{SI}$ is the most stringent one among these four constraints, we should concentrate on Fig.~\ref{fig:neutralino-like I}(b).
We find that the mixed and the $\tilde H$-like particles tend to have larger values  in the DM-nucleon elastic scattering cross section,
while the $\tilde B$-like particles tend to have smaller values.
The samples sitting below the upper limit of the LUX SI-experiment~\cite{LUXSI} (solid curve) and above the line of  neutrino background (dashed curve) are allowed. We see that most of mixed particles, part of the $\tilde H$-like and a few of $\tilde B$-like particles are ruled out by the LUX constraint so that about $96 \%$ of the samples are safe. However, most $\tilde B$-like particles sitting between these two lines [see Fig~\ref{fig:neutralino-like I}(b)] have been ruled out by the $\Omega_\chi^{\rm obs} h^2$ constraint [see Fig~\ref{fig:neutralino-like I}(a)], and hence only $23 \%$ of the samples are survived. Furthermore near $99 \%$ of the survived samples are $\tilde H$-like.
It shows that the DM relic density and the direct search constraints are complementary to each other.

To compare with the Fermi-LAT constraints, we show the scatter plots of $\la \sigma (\chi {\chi}\rightarrow W^+W^-, b{\bar b}, u{\bar u}, \tau^+ \tau^-, \mu^+ \mu^-) v\ra$ versus $m_{\chi}$ in Fig.~\ref{fig:neutralino-like I}(f)-(j), respectively. We do not show the plot of $\la \sigma (\chi {\chi}\rightarrow e^+e^-) v\ra$ since it is highly helicity suppressed as mentioned in Sec. II-B. The samples sitting above the Fermi-LAT constraints are ruled out. 
For $W^+W^-$ channel [see Fig.~\ref{fig:neutralino-like I}(f)], a $\tilde B$-like DM pair do not contribute to the $s$-wave amplitude (also mentioned in Sec. II-B) so that all values of $\la\sigma_{\rm ann}v\ra$ for the $\tilde B$-like particles are less than those values for the $\tilde H$-like and the mixed particles. We also see that part of the $\tilde H$-like and the mixed particles are ruled out by this constraint so that about $94 \%$ of samples are safe under this constraint. However, most $\tilde B$-like particles sitting below the limit are ruled out by the $\Omega_\chi^{\rm obs} h^2$ constraint and hence only about $20 \%$ of the samples are survived.
In Fig.~\ref{fig:neutralino-like I}(f)-(j), 
we see that, in general, the $\tilde B$-like particles tend to have smaller $\la\sigma_{\rm ann}v\ra$, while the $\tilde H$-like and the mixed particles tend to have larger $\la\sigma_{\rm ann}v\ra$.
Note that all the DM particles annihilating into $f{\bar f}$ with the final fermion mass less than $M_W$ have the similar resonance shapes with peaks at $m_\chi =m_Z/2$, and $m_H/2$.
For $b {\bar b}$ and $\tau^+ \tau^-$ channels, only a few DM candidates are ruled out by these two constraints, and for other channels the constraints become less important
when the final fermion mass is less than $m_\tau$.
Besides, we also give the scatter plots of velocity averaged cross sections  $\la\sigma (\chi{\chi}\rightarrow ZZ, HZ, t{\bar t},HH) v\ra$ versus $m_{\chi}$ in Fig.~\ref{fig:ZZZHttHH}. 
Similar to the case of $W^+W^-$ channel, the $\tilde B$-like particles do not contribute to the $s$-wave amplitude in $ZZ$ channel (mentioned in Sec. II-B) so that all the values of $\la\sigma_{\rm ann} v\ra$ for the $\tilde B$-like particles are less than those values for the $\tilde H$-like particles in $ZZ$ channel [see Fig.~\ref{fig:ZZZHttHH}(a)]. In addition, 
the process $\chi{\chi}\rightarrow HH$ can only proceed from the $p$-wave. It results in that almost all values of $\la\sigma_{\rm ann} v\ra$ in $HH$ channel are less than those values in $ZZ, ZH$ and $t\bar t$ channels [see Fig.~\ref{fig:ZZZHttHH}(a-d)].
%Since $\la \sigma_{\rm ann}v\ra$ is dominated by $s$-wave contribution in this model, the $\la\sigma_{\rm ann} v\ra$ appearing in the relic density and in the indirect search processes with all possible final states summed are of the same magnitude.
Recall that the relic density is proportional to the inverse of $\la\sigma_{\rm ann} v\ra$, while $\la\sigma_{\rm ann} v\ra$ is dominated by the $W^+W^-$ channel for $m_\chi>M_W$ and the $b {\bar b}$ channel for $m_\chi<M_W$.
Therefore, the shape of the relic density in Fig.~\ref{fig:neutralino-like I}(a) can be easily understood from Fig.~\ref{fig:neutralino-like I}(f) and (g).
The interplay of different observables are useful and instructive.

In Fig.~\ref{fig:allow neutralino-like I}, we redraw Fig.~\ref{fig:neutralino-like I} only with the allowed samples which satisfy all the constraints.
These plots are the predictions of the neutralino-like I case.
We will also redraw the plots of Fig.~\ref{fig:ZZZHttHH} only with allowed samples later.
We find that the direct detection of SI cross section from DM scattering off nuclei and the indirect detection of velocity averaged
cross section from DM annihilating to $W^+W^-$ are two more sensitive constraints as the allowed regions touch the corresponding upper limits.
It means that they are more accessible for DM searches in the near future.
Now it is interesting to see how these constraints shape the allowed range of DM mass for a given particle attribute. In the following discussion, we will ignore the
outlier samples with DM mass near the peaks, namely, $m_{\chi}\simeq M_Z/2$ and $M_H/2$ in Fig.~\ref{fig:allow neutralino-like I}.
For the $\tilde B$-like particles, about $99\%$ of them are ruled out by the DM relic density constraint. The 
%SI cross section 
LUX $\sigma^{SI}_N$ constraint is complementary to the relic density constraint such that only the
$\tilde B$-like particles with $m_{\chi}\gtrsim 1411$ GeV could be DM candidates [see Fig.~\ref{fig:neutralino-like I}(b)].
All of the $\tilde H$-like particles with mass $m_{\chi}\lesssim M_W$ GeV are ruled out by the DM relic density constraint, followed by Fermi-LAT $\la (\sigma (\chi\chi\rightarrow b\bar b) v\ra $ constraint around $m_\chi\sim M_W$. 
All the $\tilde H$-like particles with $m_{\chi}> M_W$ are not ruled out by the observed relic density [see Figs.~\ref{fig:neutralino-like I}(a)], and all the $\tilde H$-like particles with $M_W < m_{\chi} \lesssim 456$ GeV are ruled out by Fermi-LAT $\la (\sigma (\chi\chi\rightarrow W^+W^-) v\ra $ constraint [see Figs.~\ref{fig:neutralino-like I}(f)],
while $\tilde H$-like particles with $m_{\chi}\gtrsim 456$ GeV are still subject to the LUX $\sigma^{SI}_N$ constraint.
Therefore without considering the outliers, the allowed mass regions for the $\tilde B$-like and the $\tilde H$-like particles in Fig.~\ref{fig:allow neutralino-like I} can be understood.

\begin{figure}[t!]
\centering
\captionsetup{justification=raggedright}
 \subfigure[\ Constraint on $\Omega^{\rm{obs}}_\chi$]{
  \includegraphics[width=0.45\textwidth,height=0.135\textheight]{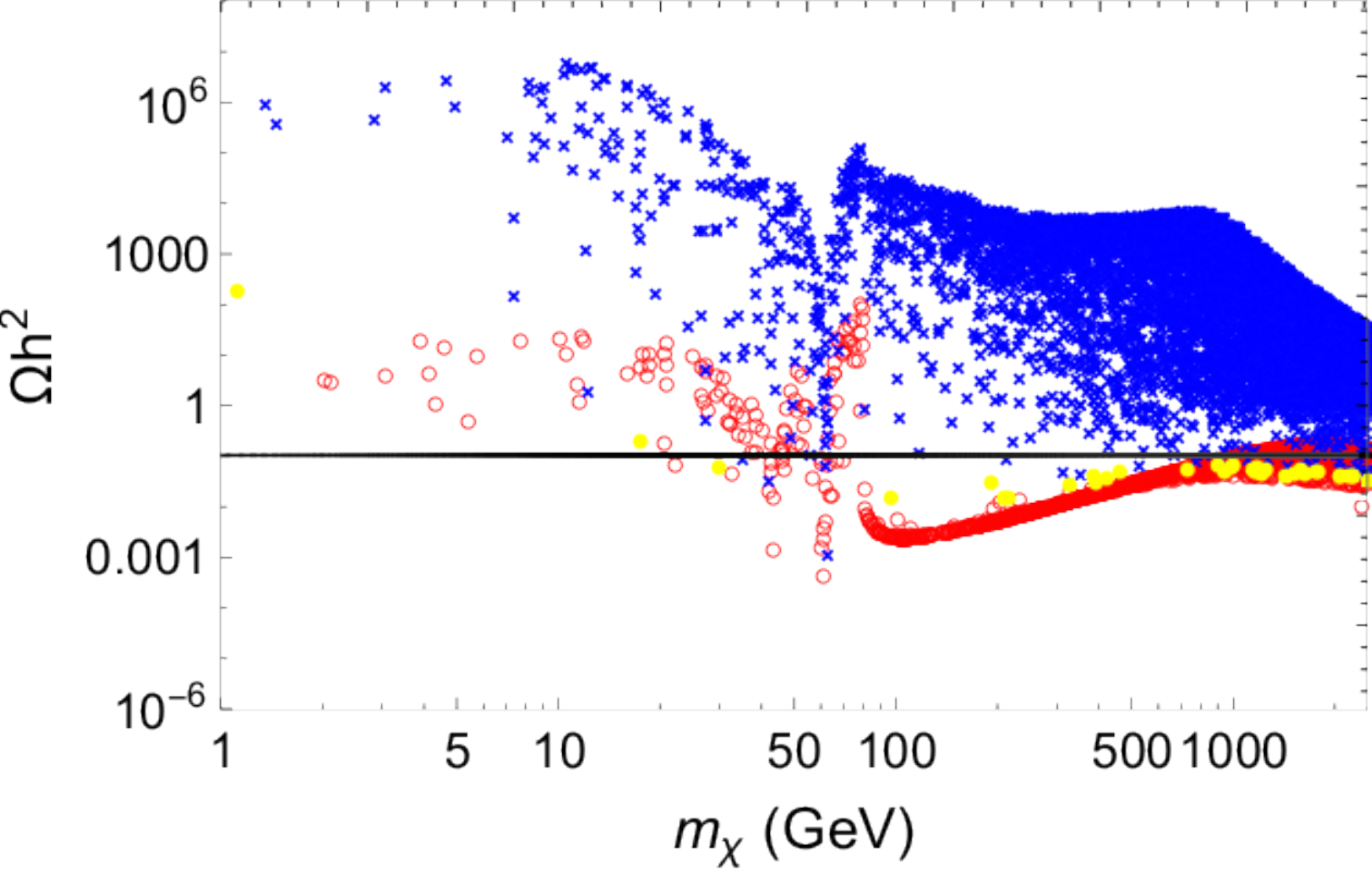}
}\subfigure[\ LUX constraint on $\sigma^{SI}$ with NB limit]{
  \includegraphics[width=0.45\textwidth,height=0.135\textheight]{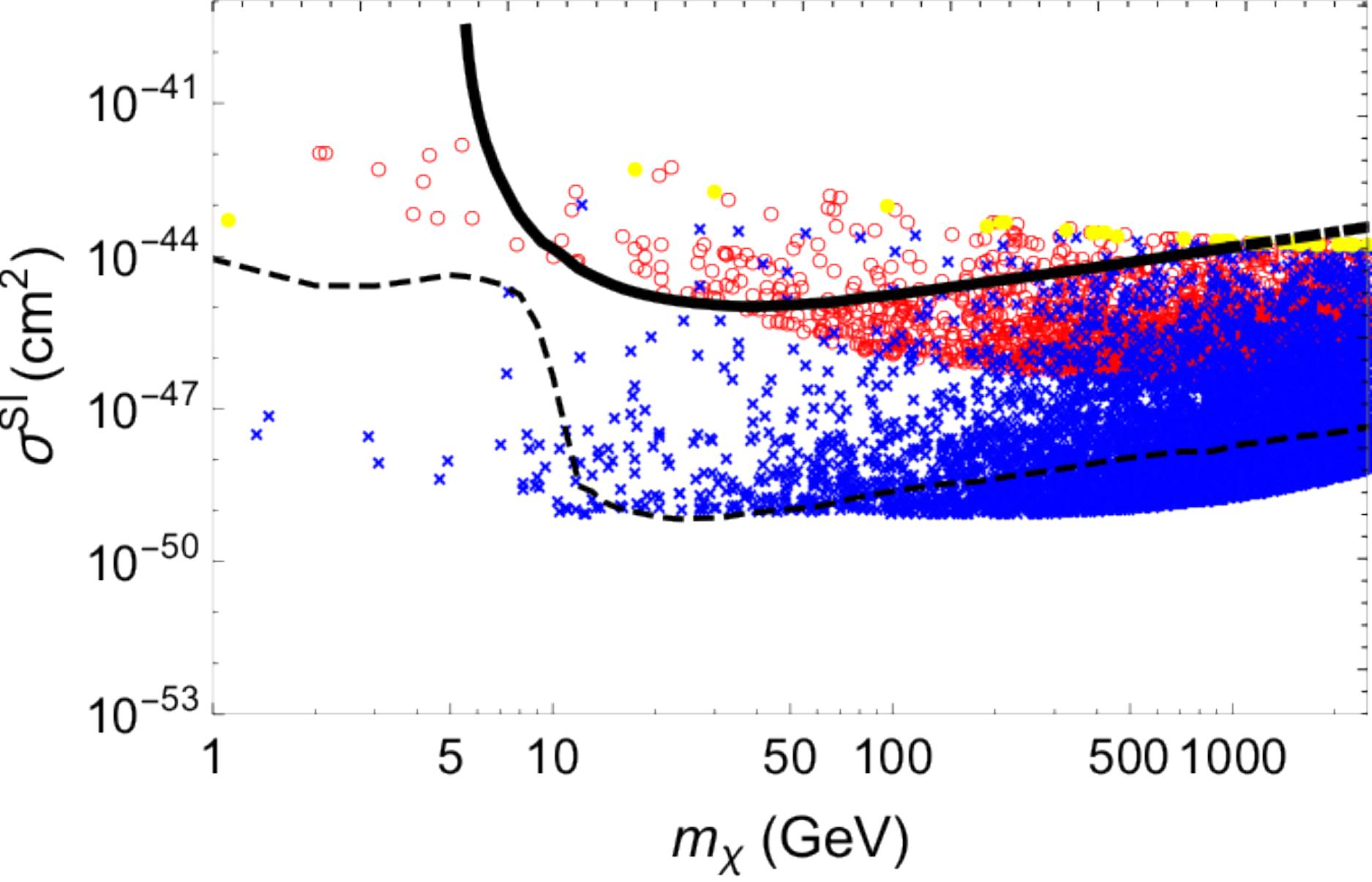}
}\\\subfigure[\ XENON100 constraint on $\sigma^{SD}_n$]{
  \includegraphics[width=0.45\textwidth,height=0.135\textheight]{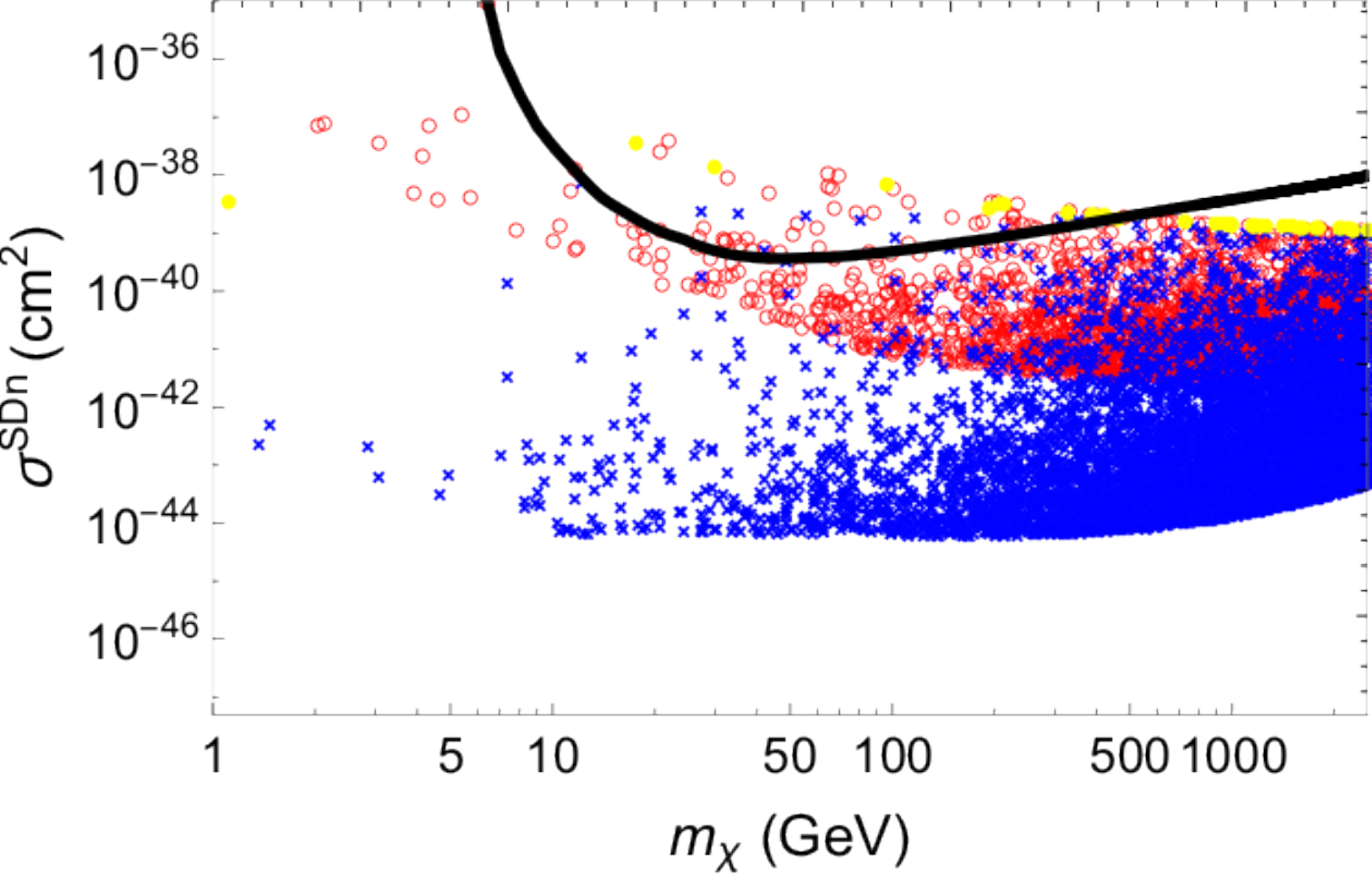}
}\subfigure[\ XENON100 constraint on $\sigma^{SD}_p$]{
  \includegraphics[width=0.45\textwidth,height=0.135\textheight]{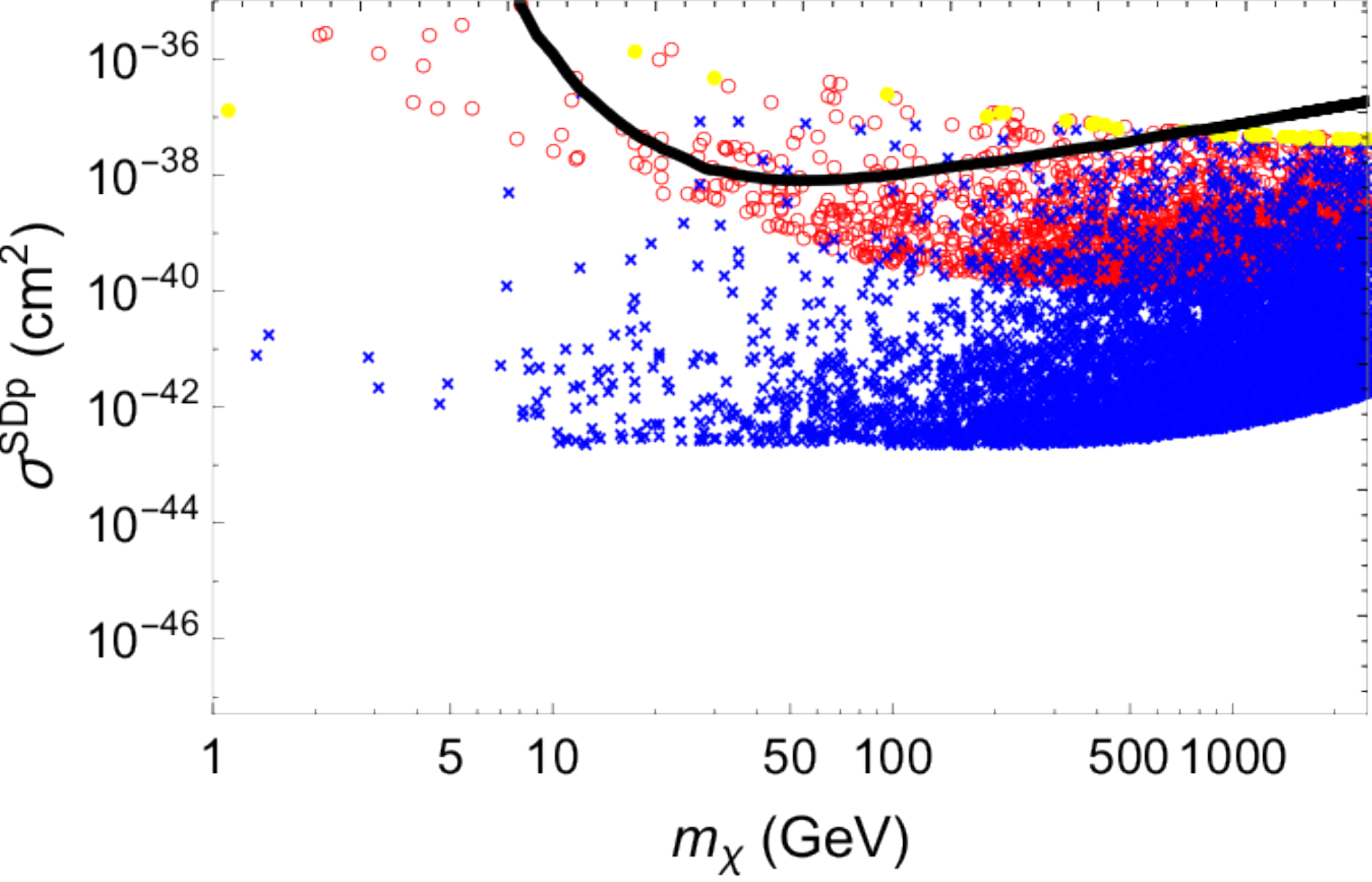}
}\\\subfigure[\ PICO-60 constraint on $\sigma^{SD}_p$]{
  \includegraphics[width=0.45\textwidth,height=0.135\textheight]{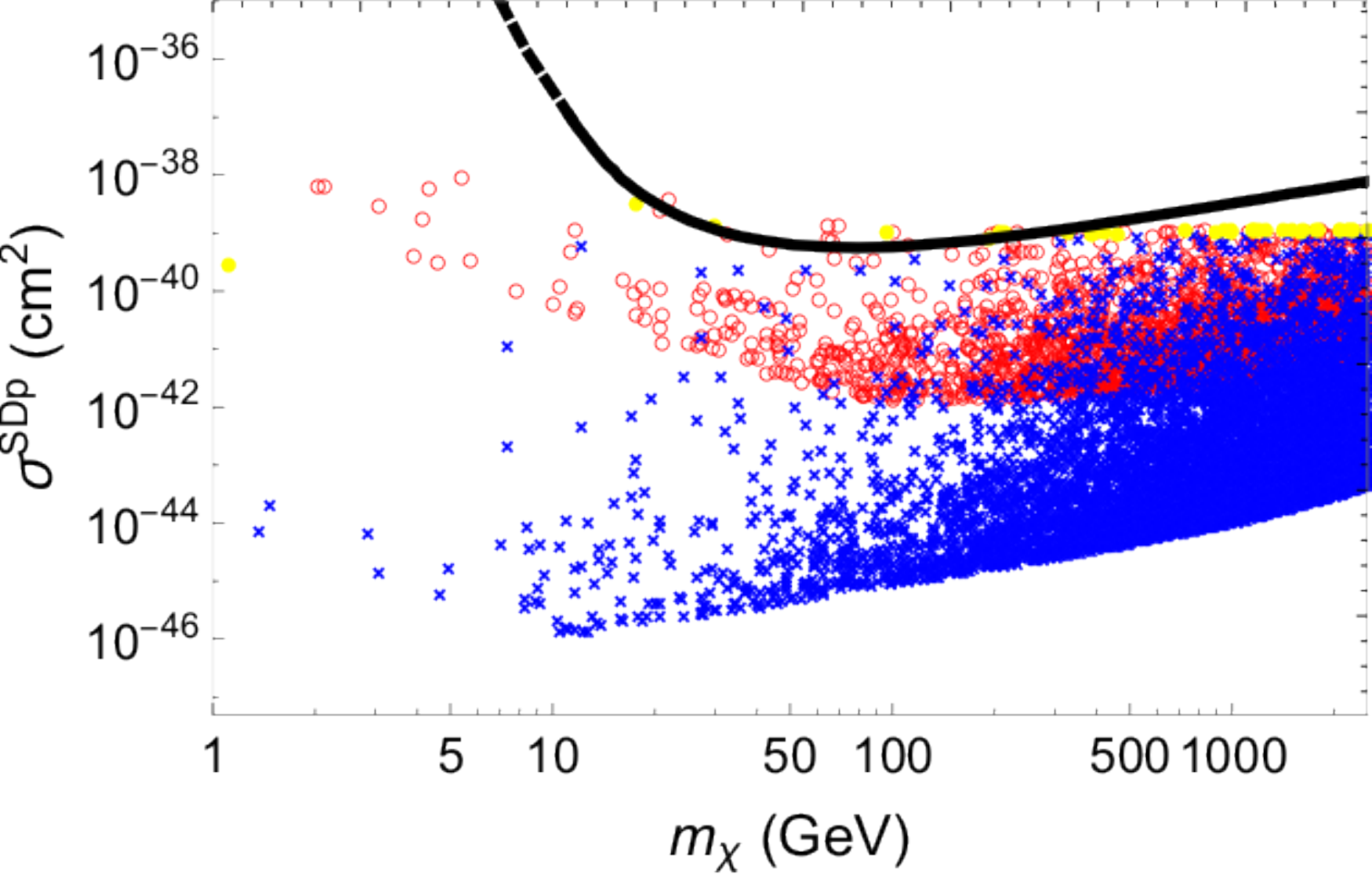}
}\subfigure[\ Fermi-LAT constraint on $\chi^0 {\chi}^0\rightarrow W^+W^-$]{
  \includegraphics[width=0.45\textwidth,height=0.135\textheight]{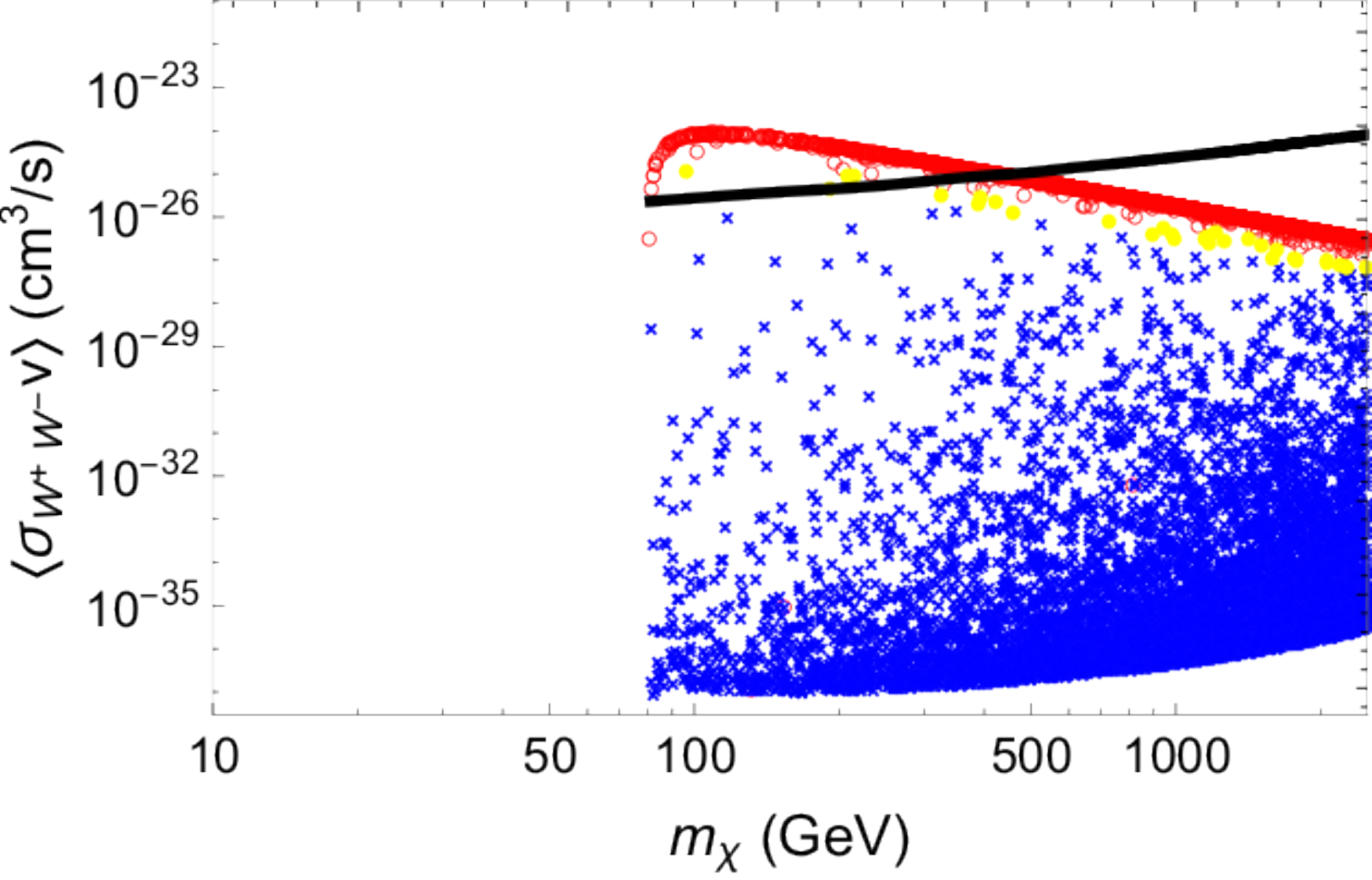}
}\\\subfigure[\ Fermi-LAT constraint on $\chi^0 {\chi}^0\rightarrow b\bar{b}$]{
  \includegraphics[width=0.45\textwidth,height=0.135\textheight]{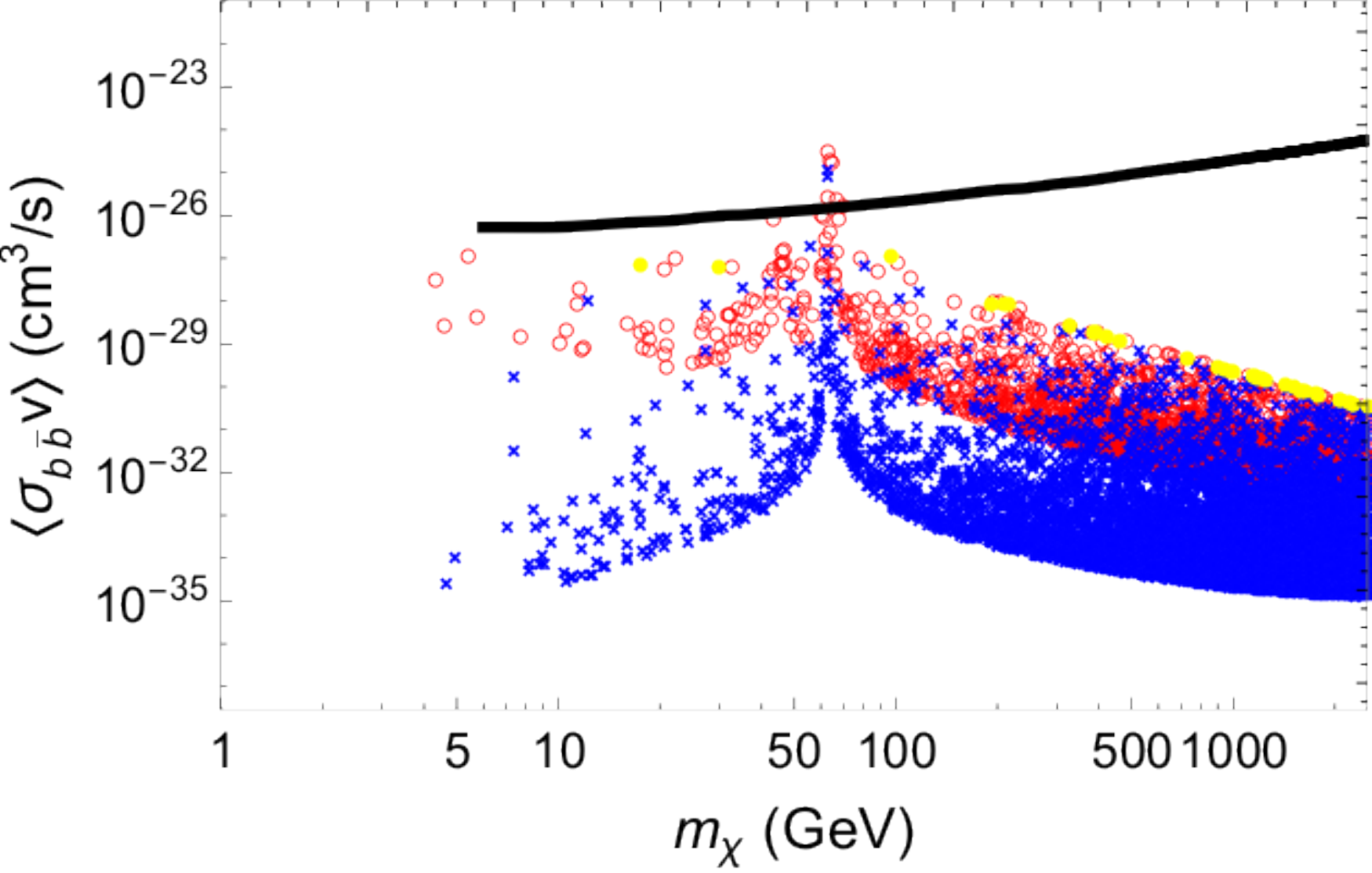}
}\
%\subfigure[\ Fermi-LAT constraint on $\chi^0 \bar{\chi}^0\rightarrow u\bar{u}$]{
  %\includegraphics[width=0.45\textwidth,height=0.135\textheight]{06gEinduu2.pdf}
%}
\subfigure[\ Fermi-LAT constraint on $\chi^0 {\chi}^0\rightarrow \tau^+\tau^-$]{
  \includegraphics[width=0.45\textwidth,height=0.135\textheight]{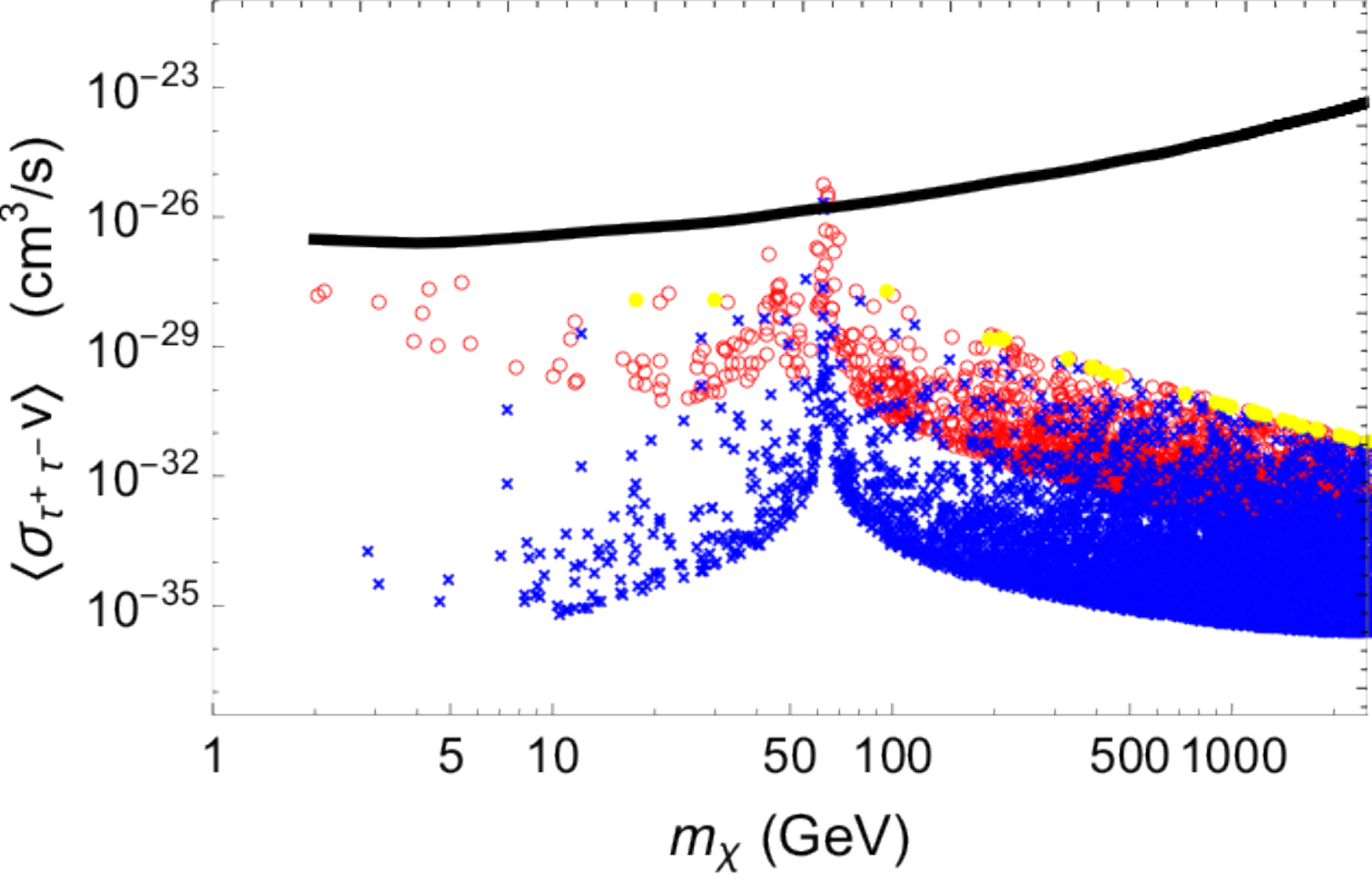}
}%\\\subfigure[\ Fermi-LAT constraint on $\chi^0 \bar{\chi}^0\rightarrow \mu^+\mu^-$]{
  %\includegraphics[width=0.45\textwidth,height=0.135\textheight]{06iEindmumu2.pdf}
%}\subfigure[\ Fermi-LAT constraint on $\chi^0 \bar{\chi}^0\rightarrow e^+e^-$]{
  %\includegraphics[width=0.45\textwidth,height=0.135\textheight]{06jEindee2.pdf}
%}
\caption{Results for all samples with constraints in the neutralino-like II case
 [{\color{red} $\circ$}:~higgsino-like,
 {\color{blue} $\times$}:~bino-like,   {\color{yellow} $\bullet$}:~mixed].}
\label{fig:neutralino-like II}
\end{figure}

\begin{figure}[t!]
\centering
\captionsetup{justification=raggedright}
 \subfigure[\ Constraint on $\Omega^{\rm{obs}}_\chi$]{
  \includegraphics[width=0.45\textwidth,height=0.13\textheight]{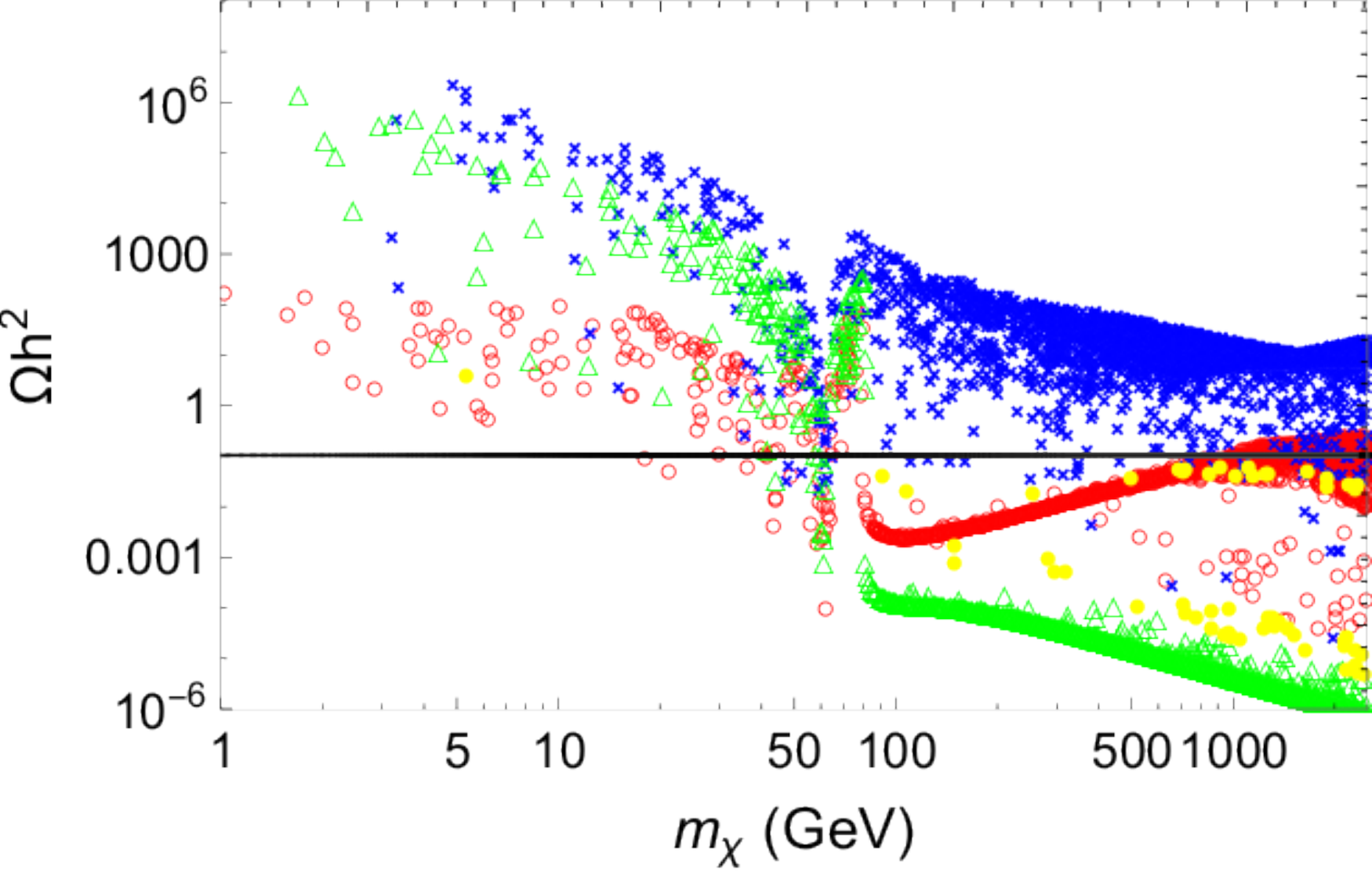}
}\subfigure[\ LUX constraint on $\sigma^{SI}$ with NB limit]{
  \includegraphics[width=0.45\textwidth,height=0.13\textheight]{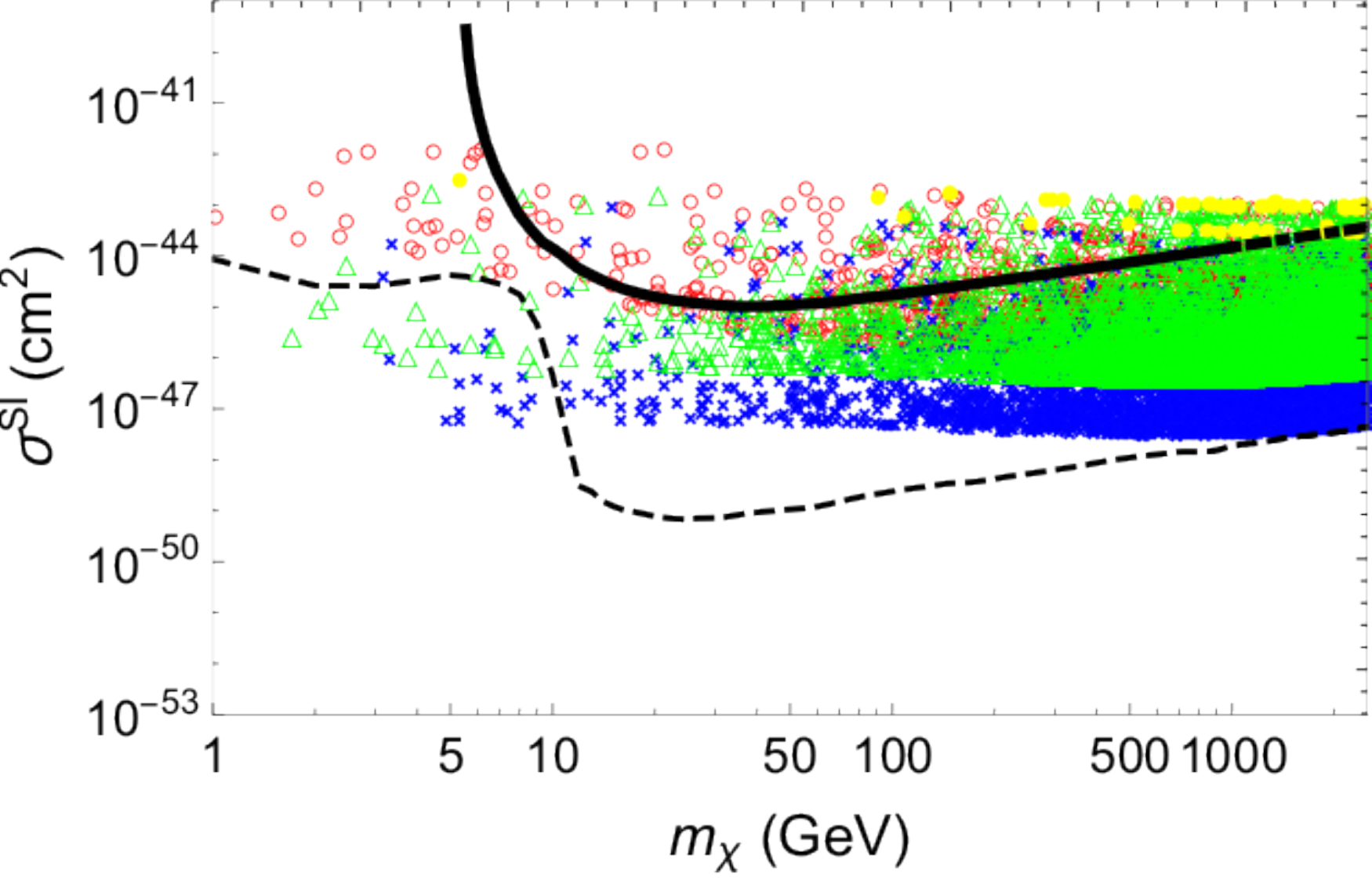}
}\\\subfigure[\ XENON100 constraint on $\sigma^{SD}_n$]{
  \includegraphics[width=0.45\textwidth,height=0.13\textheight]{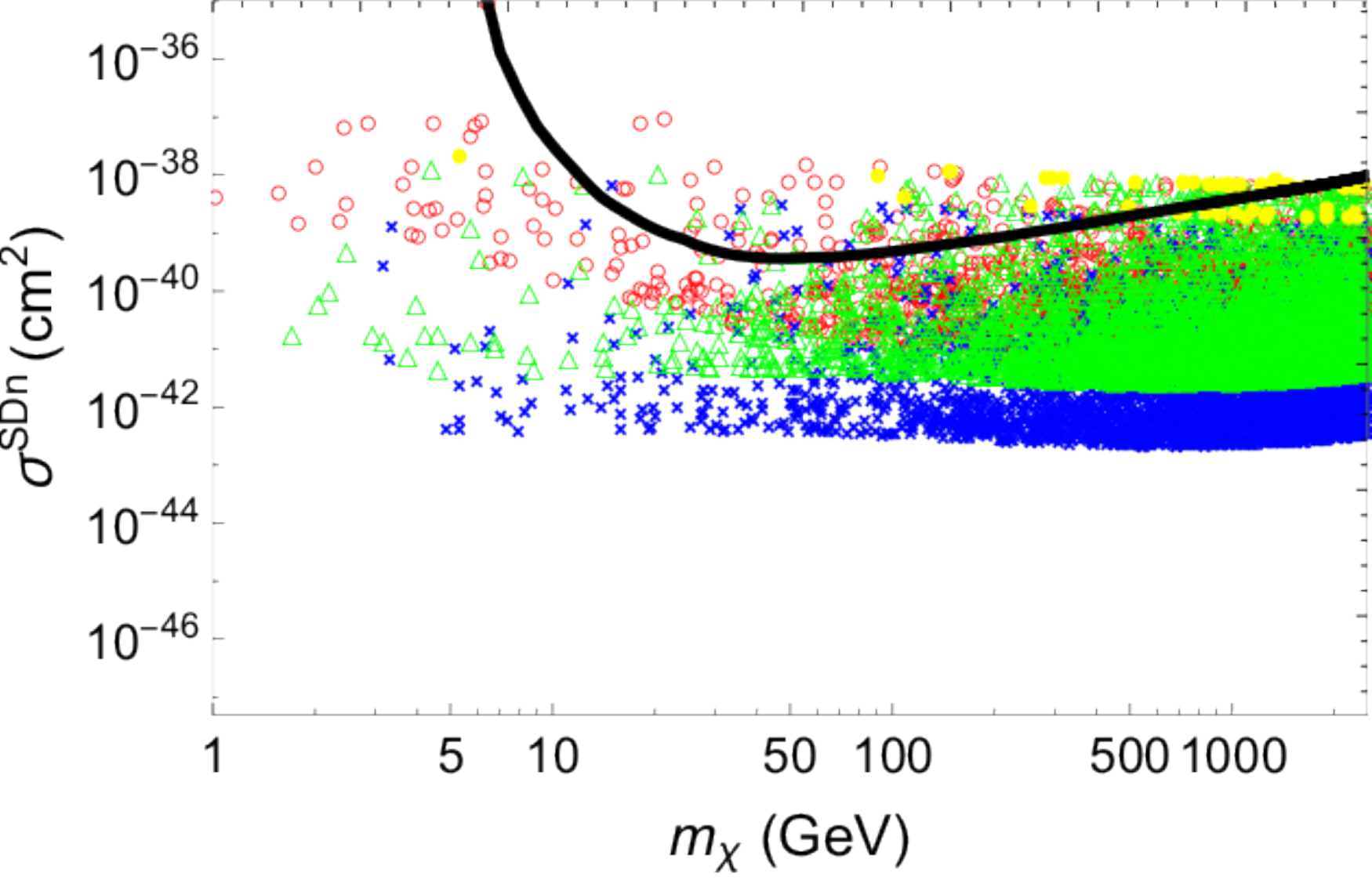}
}\subfigure[\ XENON100 constraint on $\sigma^{SD}_p$]{
  \includegraphics[width=0.45\textwidth,height=0.13\textheight]{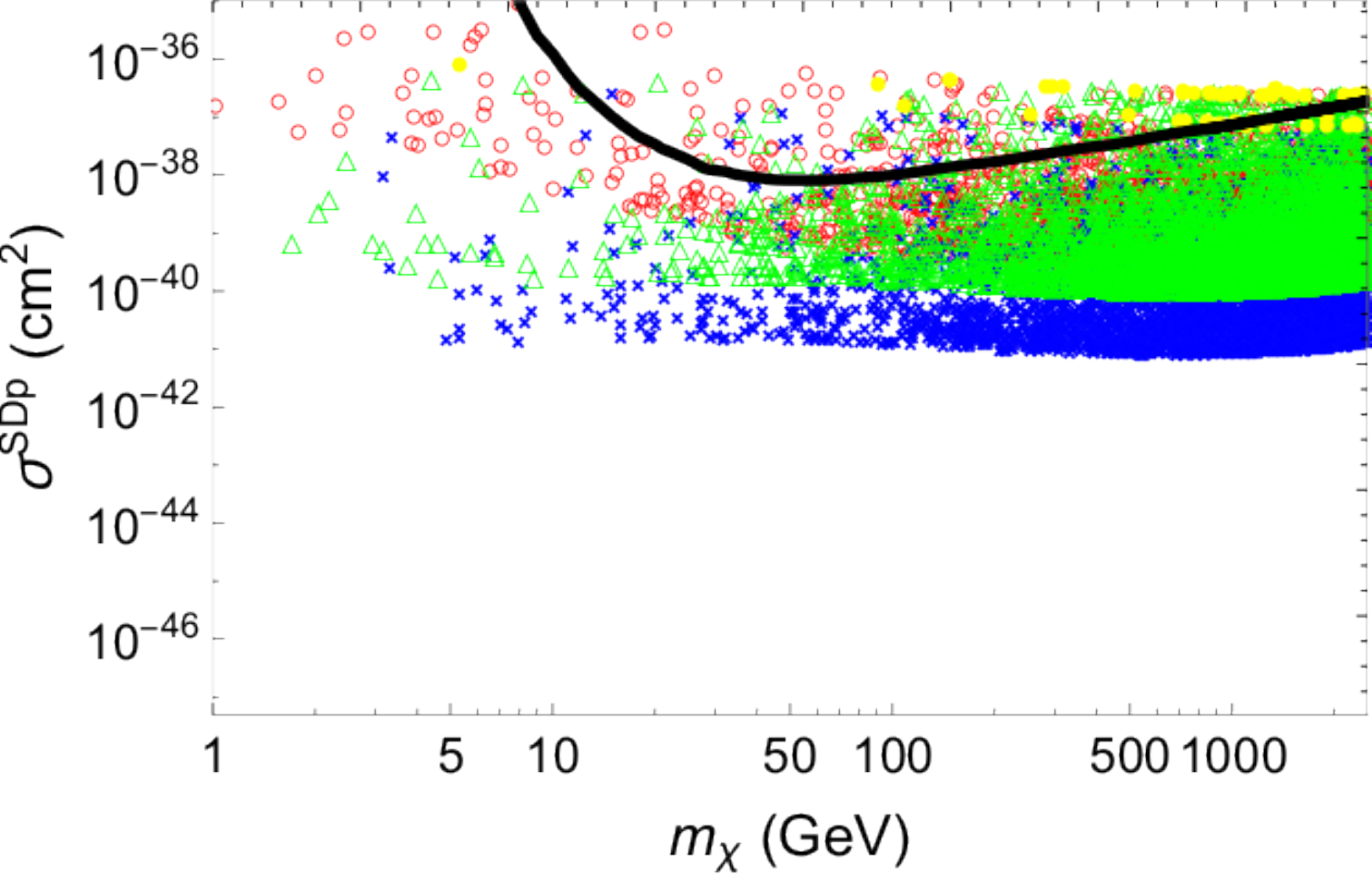}
}\\\subfigure[\ PICO-60 constraint on $\sigma^{SD}_p$]{
  \includegraphics[width=0.45\textwidth,height=0.13\textheight]{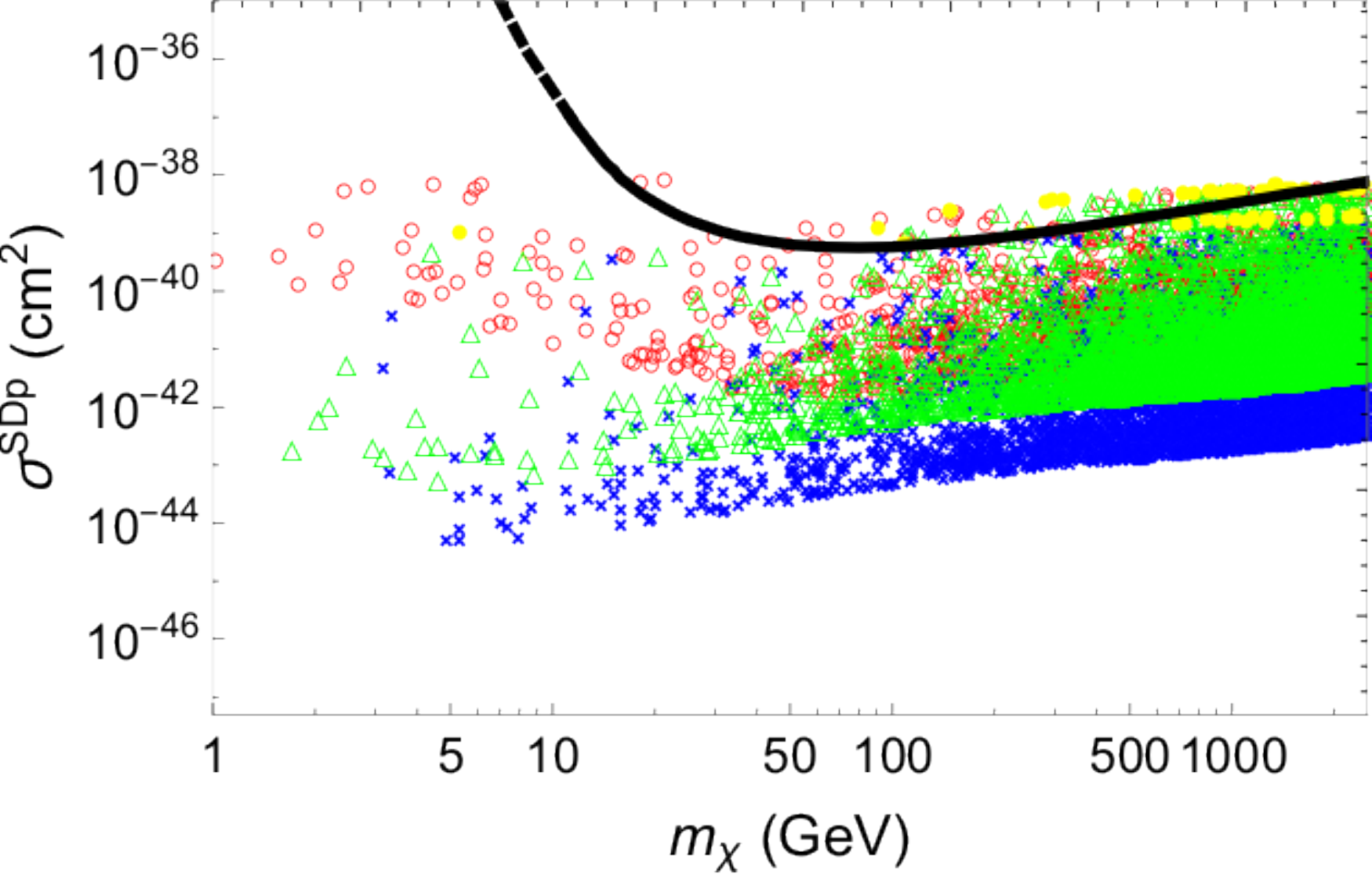}
}\subfigure[\ Fermi-LAT conststraint on $\chi^0 {\chi}^0\rightarrow W^+W^-$]{
  \includegraphics[width=0.45\textwidth,height=0.13\textheight]{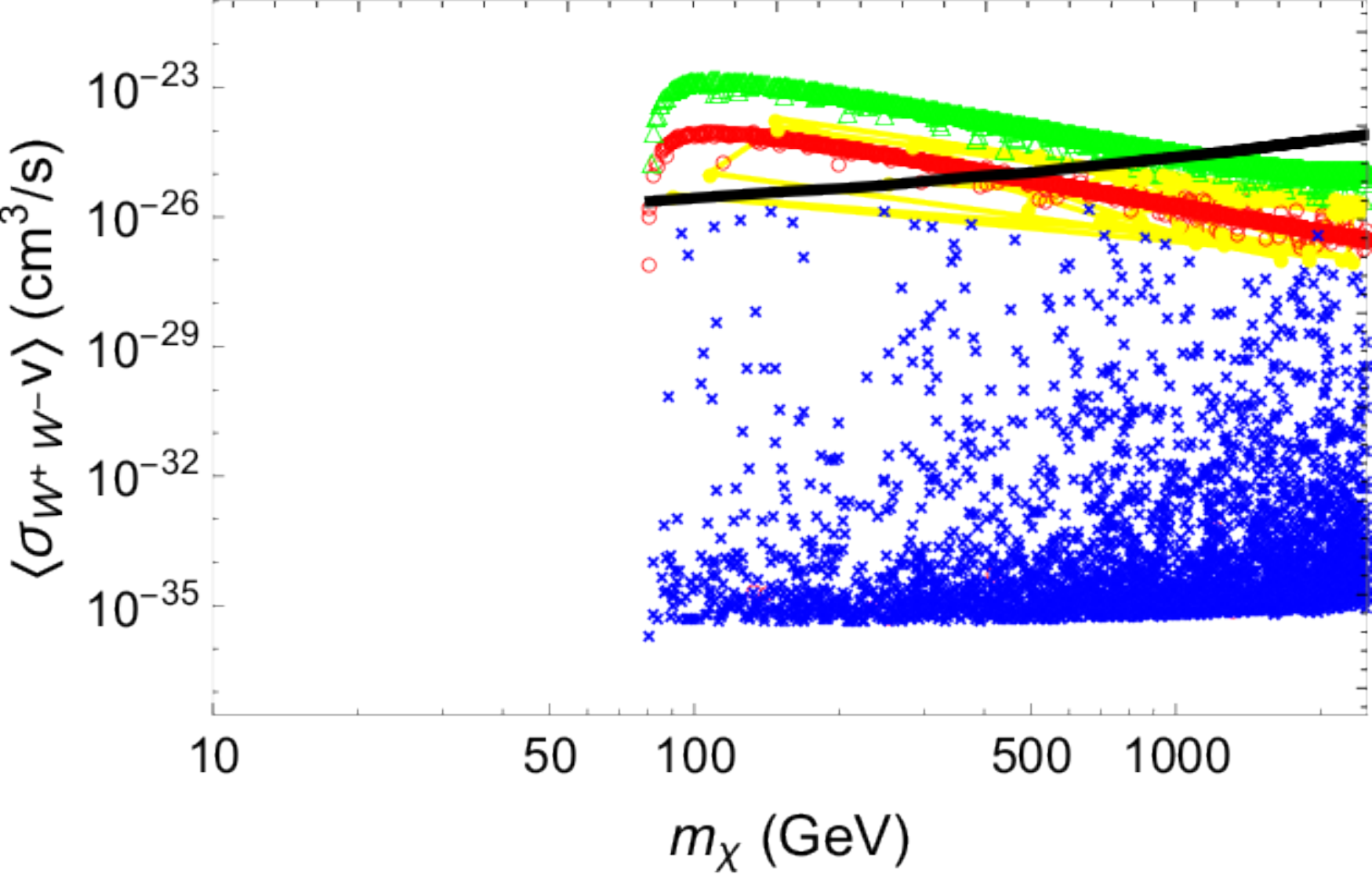}
}\\\subfigure[\ Fermi-LAT conststraint on $\chi^0 {\chi}^0\rightarrow b\bar{b}$]{
  \includegraphics[width=0.45\textwidth,height=0.13\textheight]{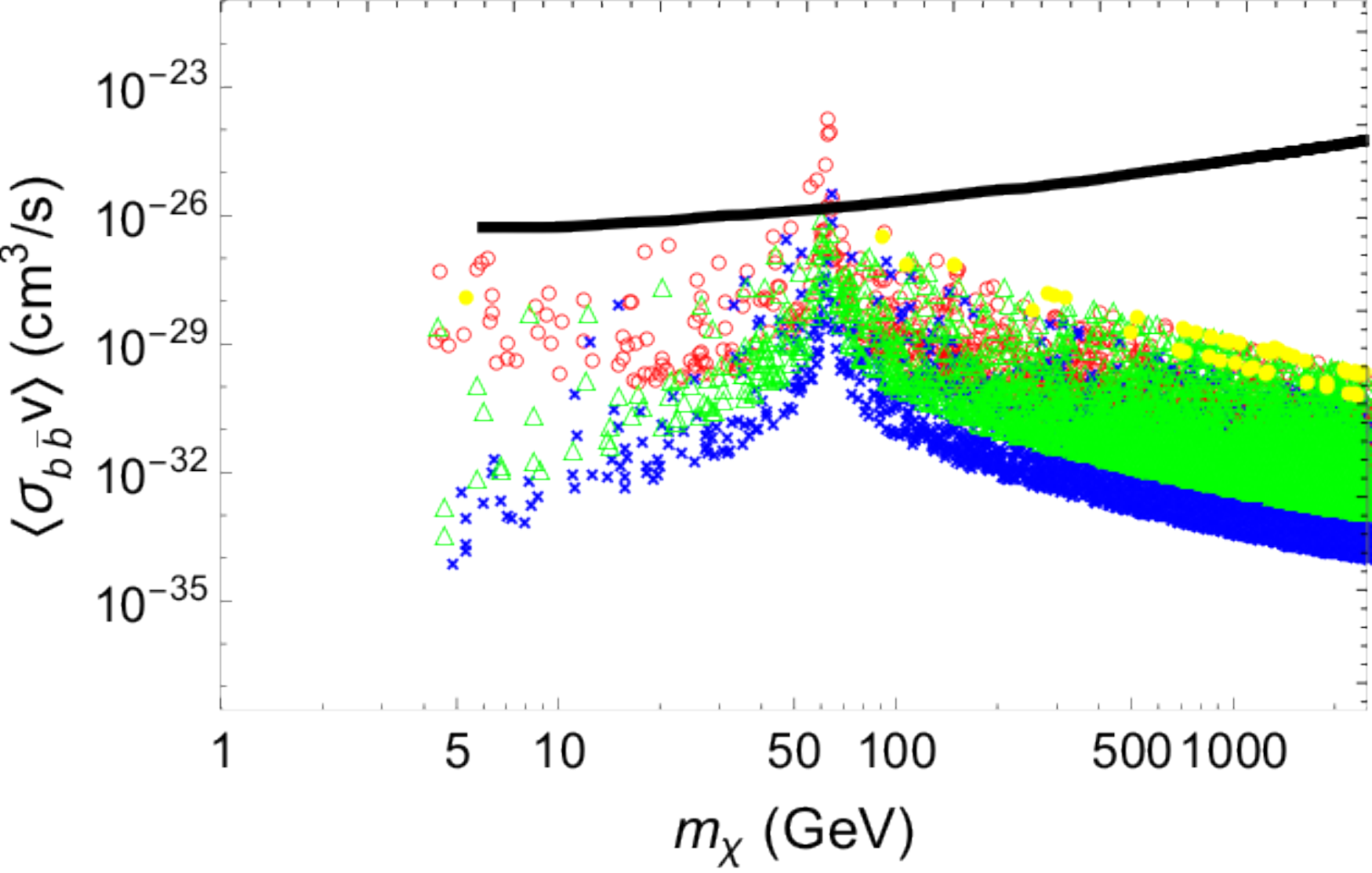}
}\
%\subfigure[\ Fermi-LAT conststraint on $\chi^0 \bar{\chi}^0\rightarrow u\bar{u}$]{
 % \includegraphics[width=0.45\textwidth,height=0.13\textheight]{07gFinduu2.pdf}
%}
\subfigure[\ Fermi-LAT conststraint on $\chi^0 {\chi}^0\rightarrow \tau^+\tau^-$]{
  \includegraphics[width=0.45\textwidth,height=0.13\textheight]{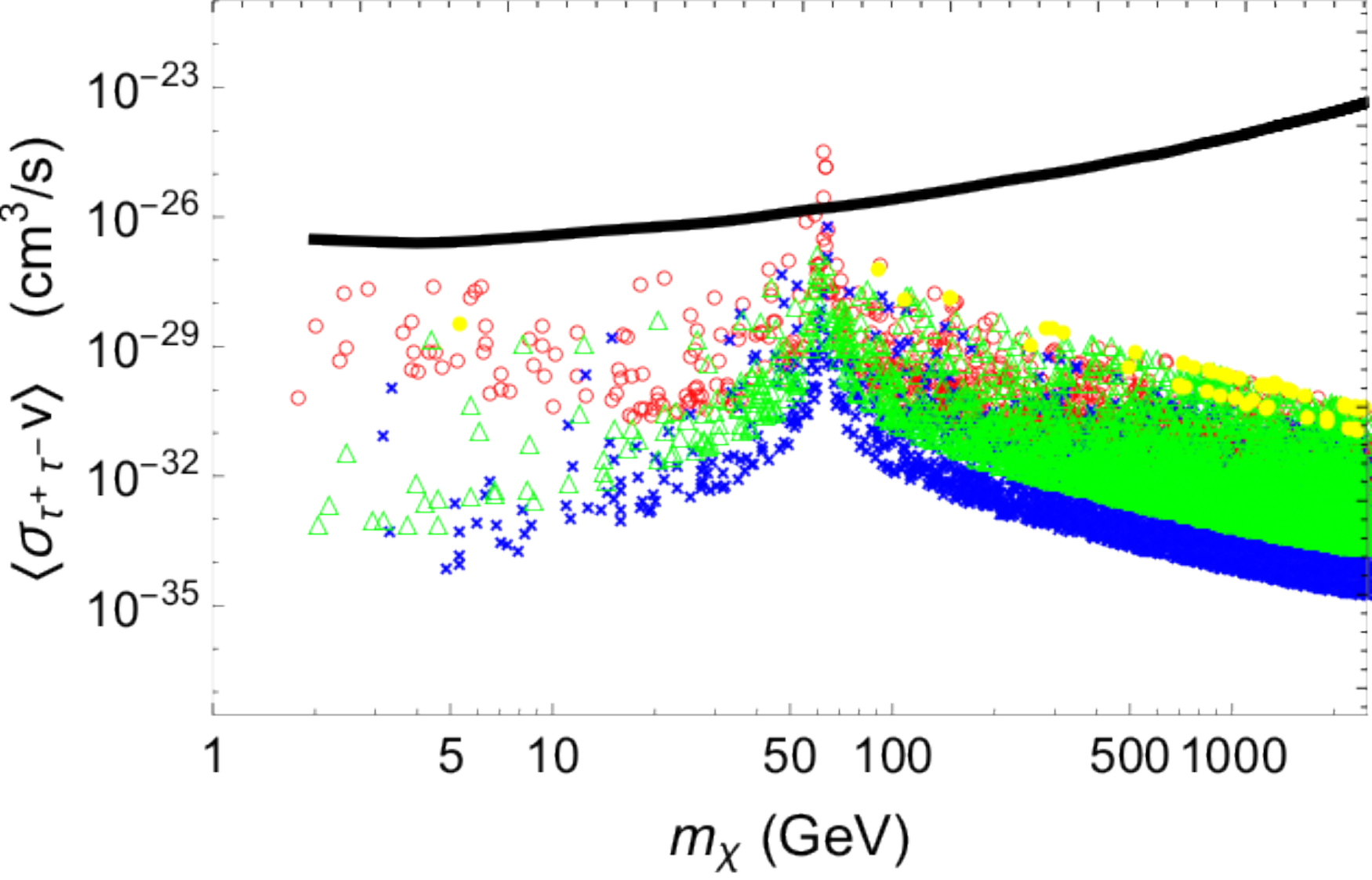}
}%\\\subfigure[\ Fermi-LAT conststraint on $\chi^0 \bar{\chi}^0\rightarrow \mu^+\mu^-$]{
 %\includegraphics[width=0.45\textwidth,height=0.13\textheight]{07iFindmumu2.pdf}
%}\subfigure[\ Fermi-LAT conststraint $\chi^0 \bar{\chi}^0\rightarrow e^+e^-$]{
 % \includegraphics[width=0.45\textwidth,height=0.13\textheight]{07jFindee2.pdf}
%}
\caption{Results for all samples with constraints in the case of neutralino-like III
[{\color{red} $\circ$}:~higgsino-like,
{\color{blue} $\times$}:~bino-like,  {\color{green} $\triangle$}:~wino-like,
{\color{yellow} $\bullet$}:~mixed].}
\label{fig:neutralino-like III}
\end{figure}

\begin{figure}[t!]
\centering
\captionsetup{justification=raggedright}
 \subfigure[\ Constraint on $\Omega^{\rm{obs}}_\chi$]{
  \includegraphics[width=0.45\textwidth,height=0.135\textheight]{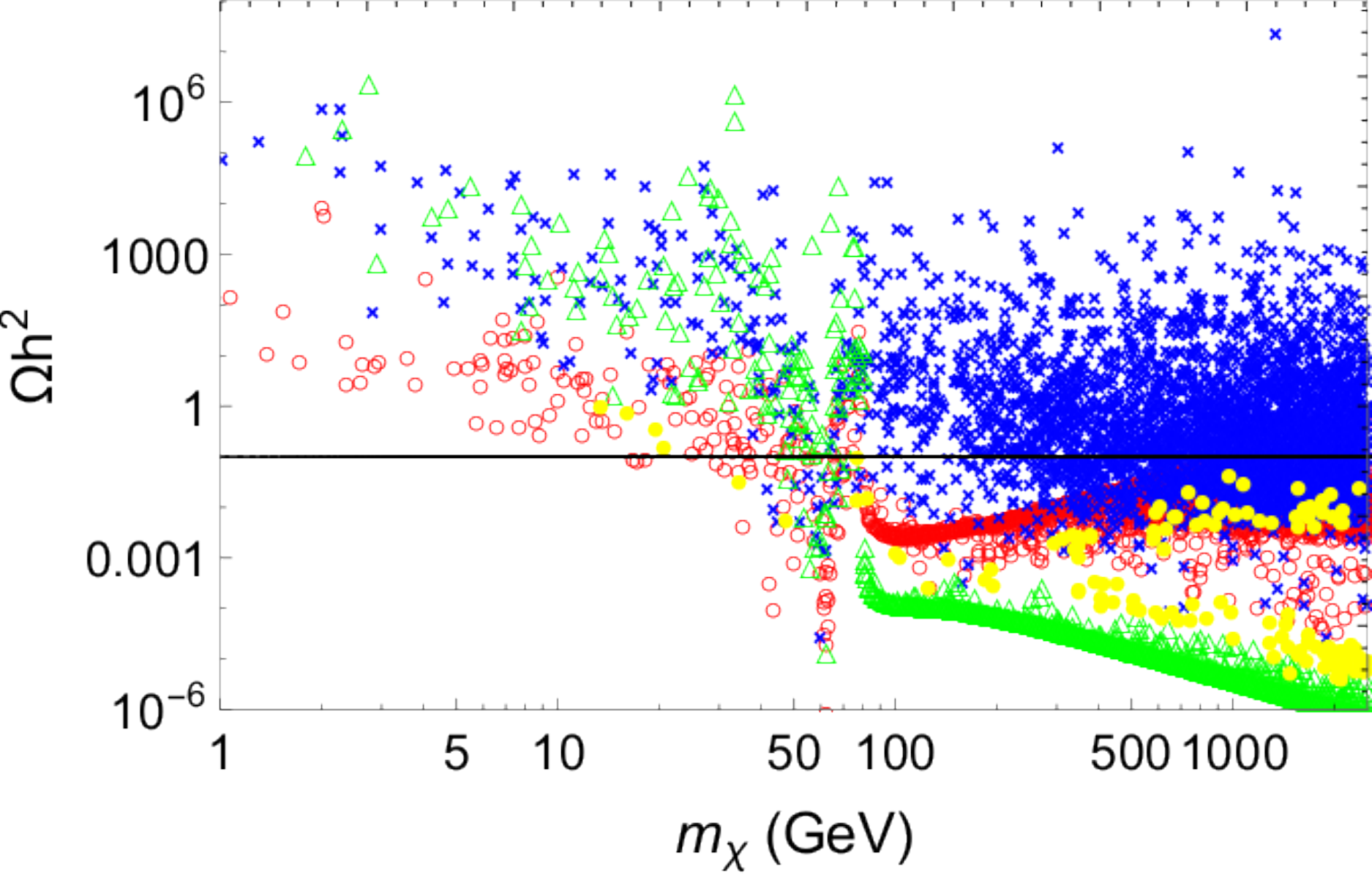}
}\subfigure[\ LUX constraint on $\sigma^{SI}$ with NB limit]{
  \includegraphics[width=0.45\textwidth,height=0.135\textheight]{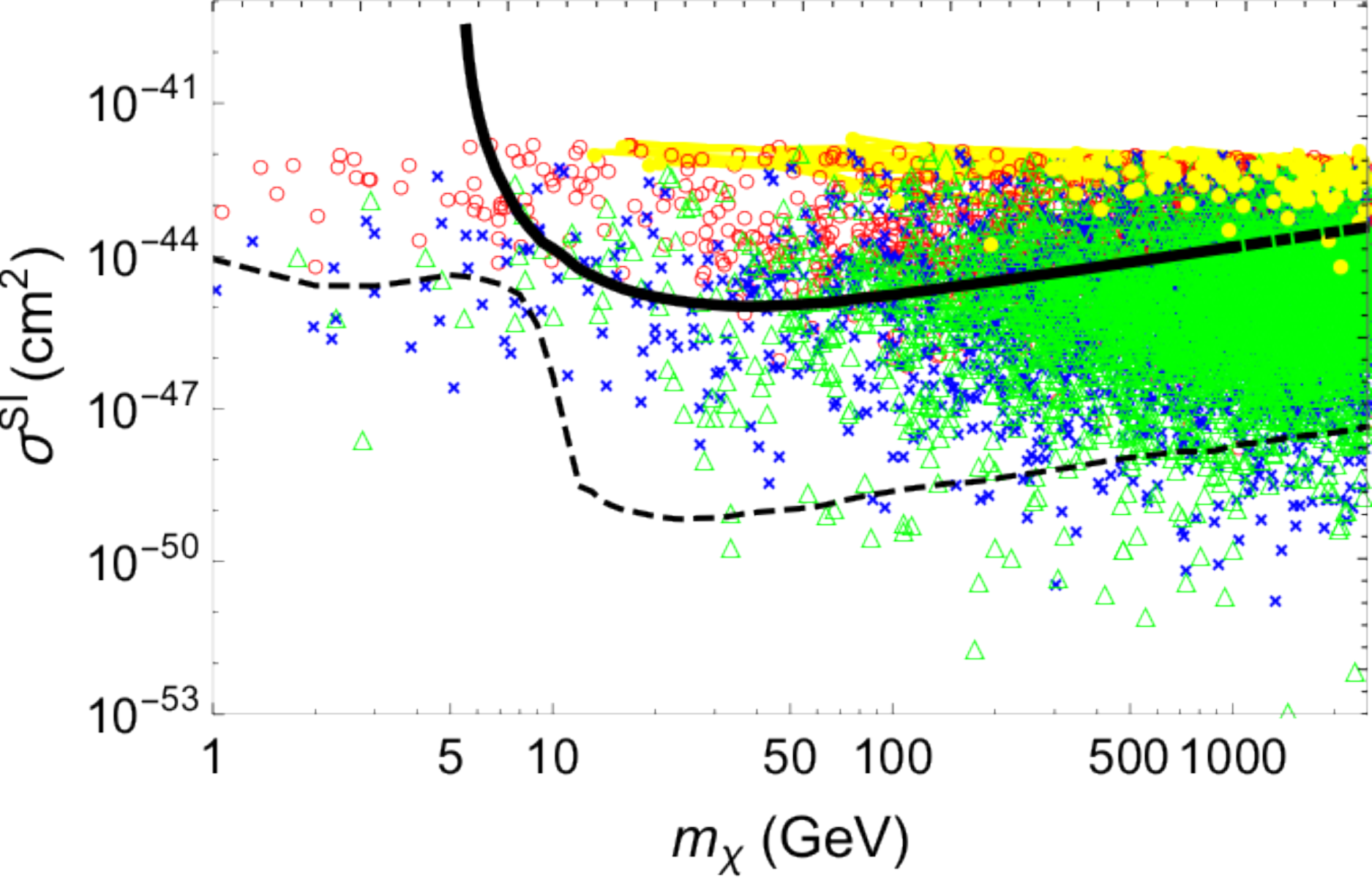}
}\\\subfigure[\ XENON100 constraint on $\sigma^{SD}_n$]{
  \includegraphics[width=0.45\textwidth,height=0.135\textheight]{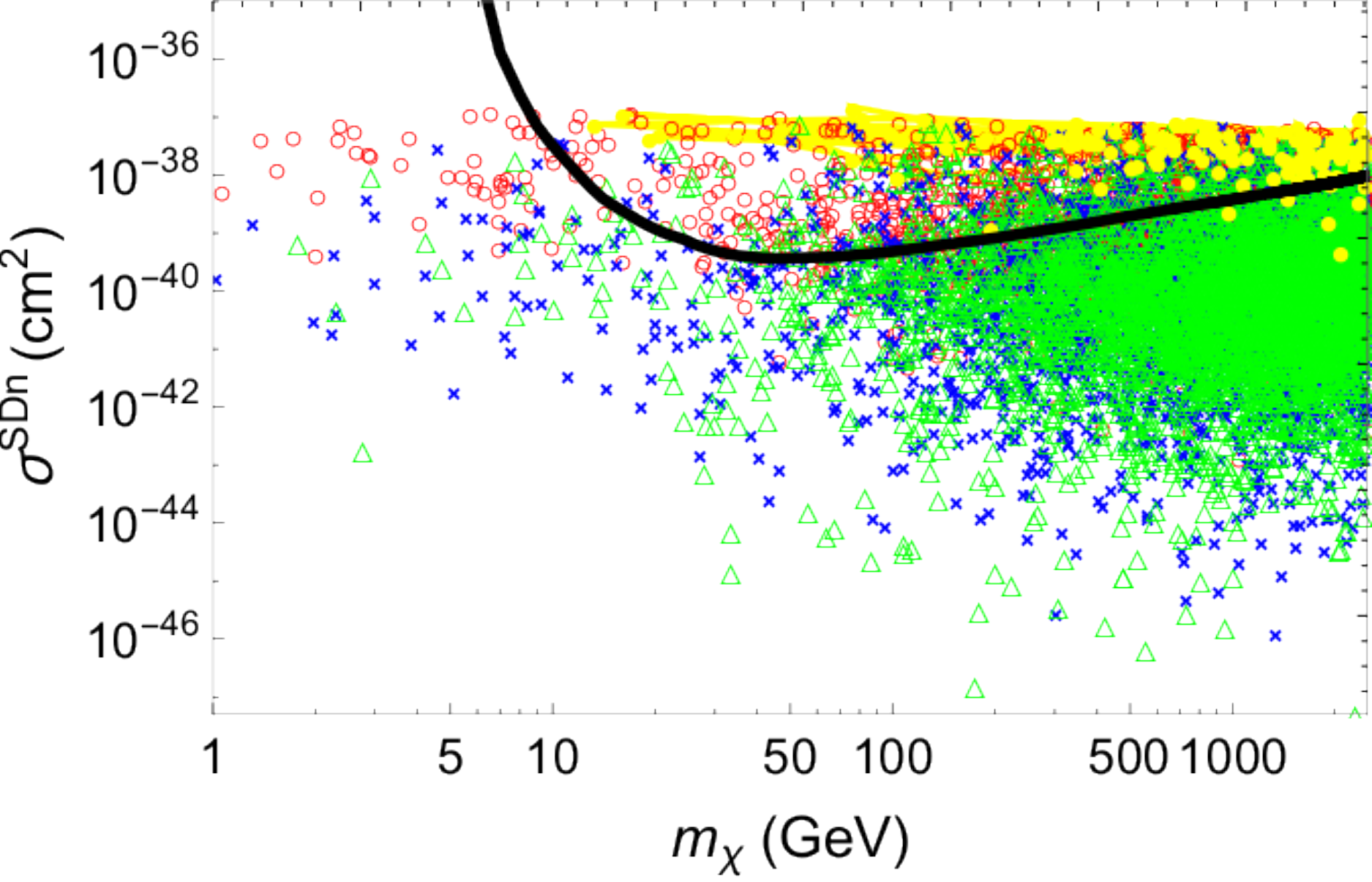}
}\subfigure[\ XENON100 constraint on $\sigma^{SD}_p$]{
  \includegraphics[width=0.45\textwidth,height=0.135\textheight]{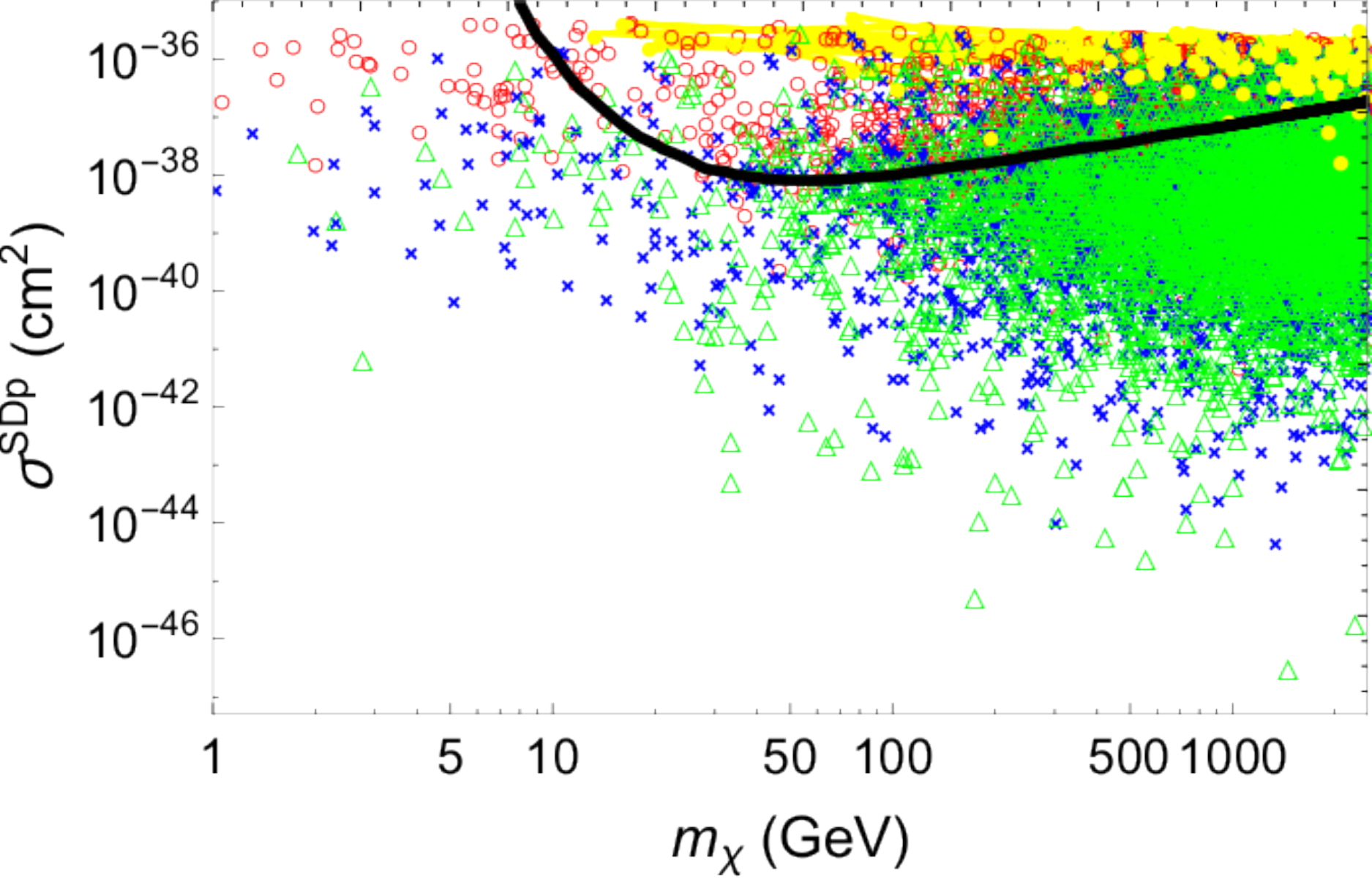}
}\\\subfigure[\ PICO-60 constraint on $\sigma^{SD}_p$]{
  \includegraphics[width=0.45\textwidth,height=0.135\textheight]{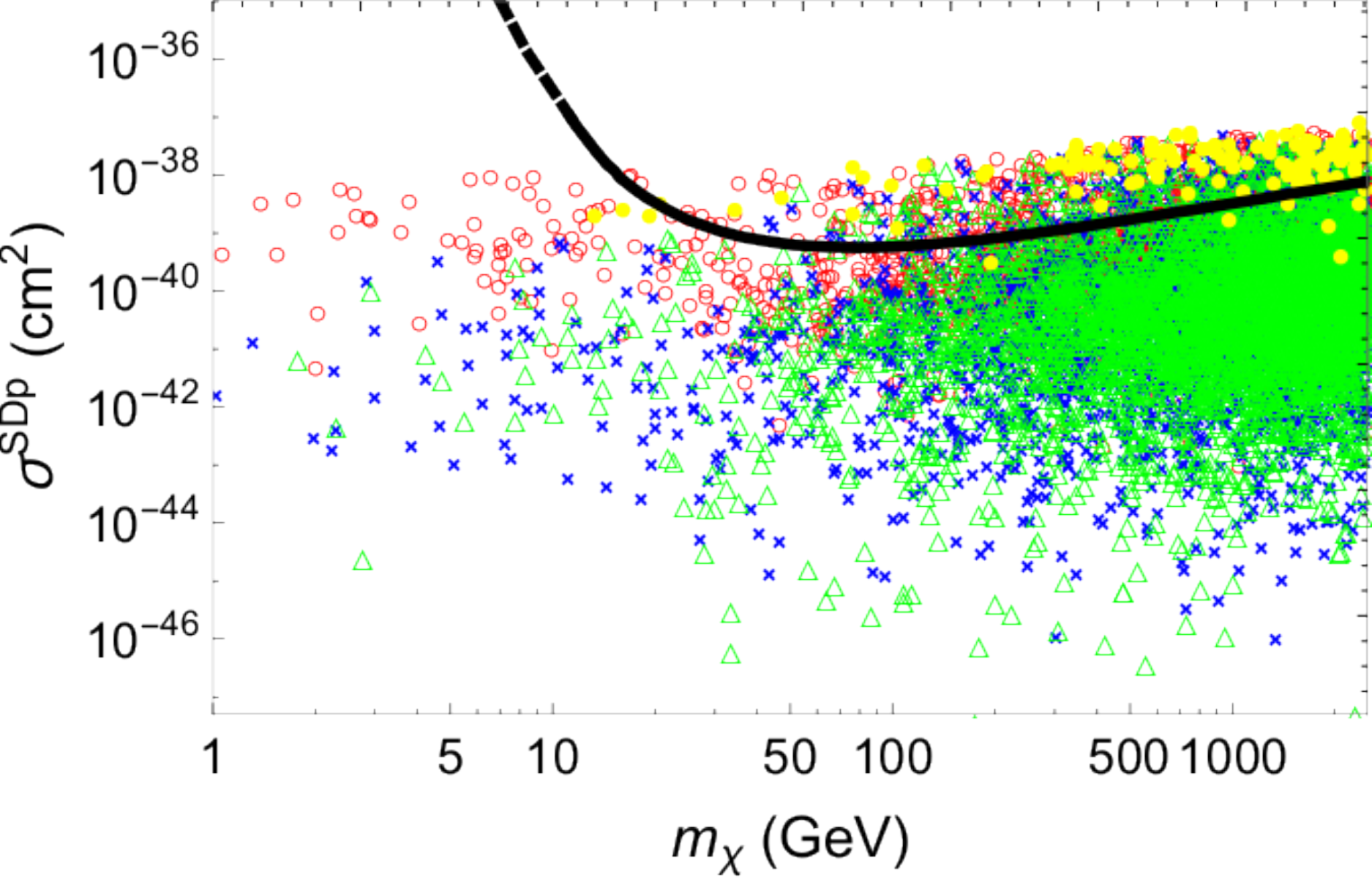}
}\subfigure[\ Fermi-LAT conststraint on $\chi^0 {\chi}^0\rightarrow W^+W^-$]{
  \includegraphics[width=0.45\textwidth,height=0.135\textheight]{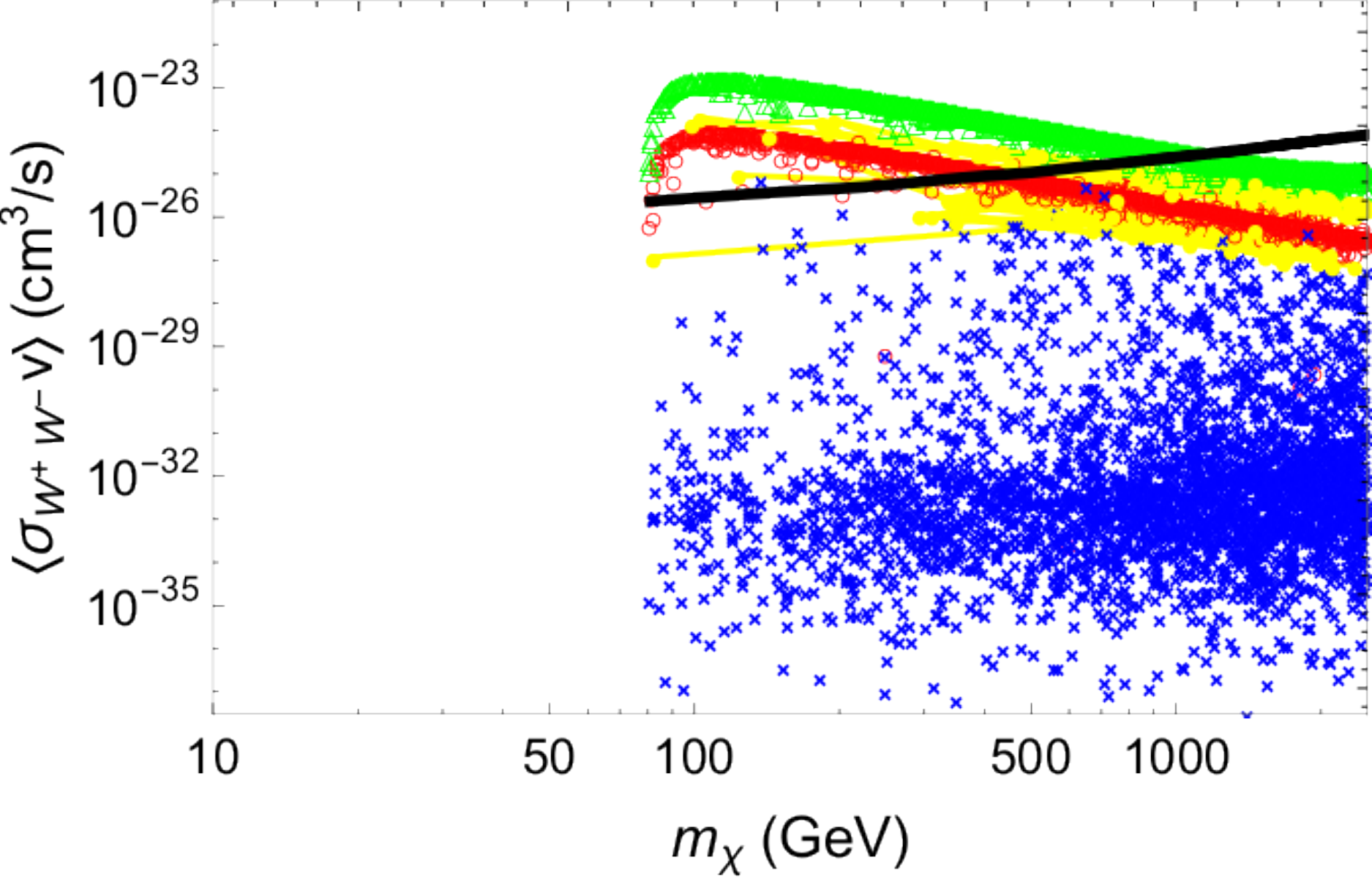}
}\\\subfigure[\ Fermi-LAT conststraint on $\chi^0 {\chi}^0\rightarrow b\bar{b}$]{
  \includegraphics[width=0.45\textwidth,height=0.135\textheight]{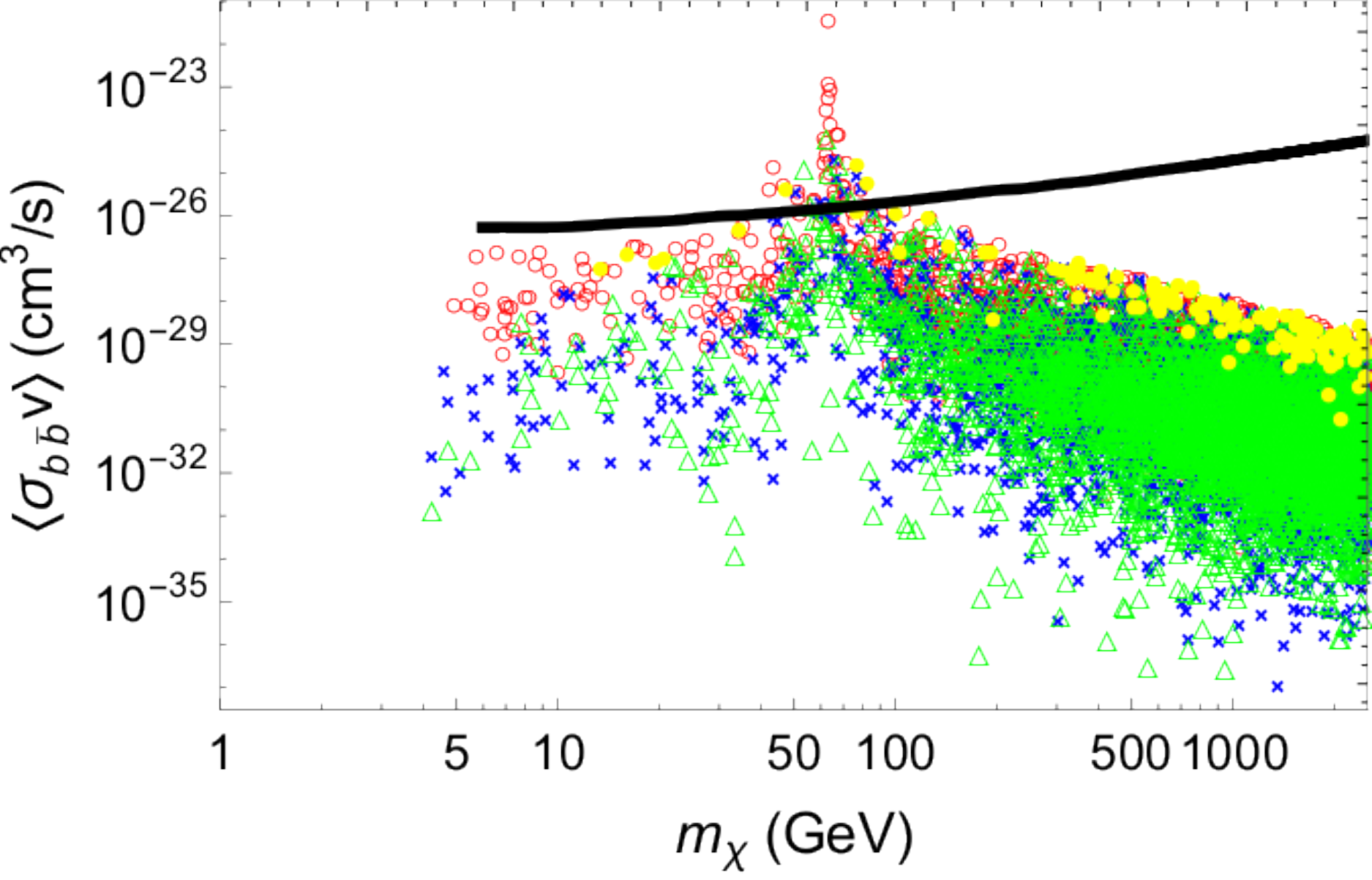}
}\
%\subfigure[\ Fermi-LAT conststraint on $\chi^0 \bar{\chi}^0\rightarrow u\bar{u}$]{
  %\includegraphics[width=0.45\textwidth,height=0.135\textheight]{08gGinduu2.pdf}
%}
\subfigure[\ Fermi-LAT conststraint on $\chi^0 {\chi}^0\rightarrow \tau^+\tau^-$]{
  \includegraphics[width=0.45\textwidth,height=0.135\textheight]{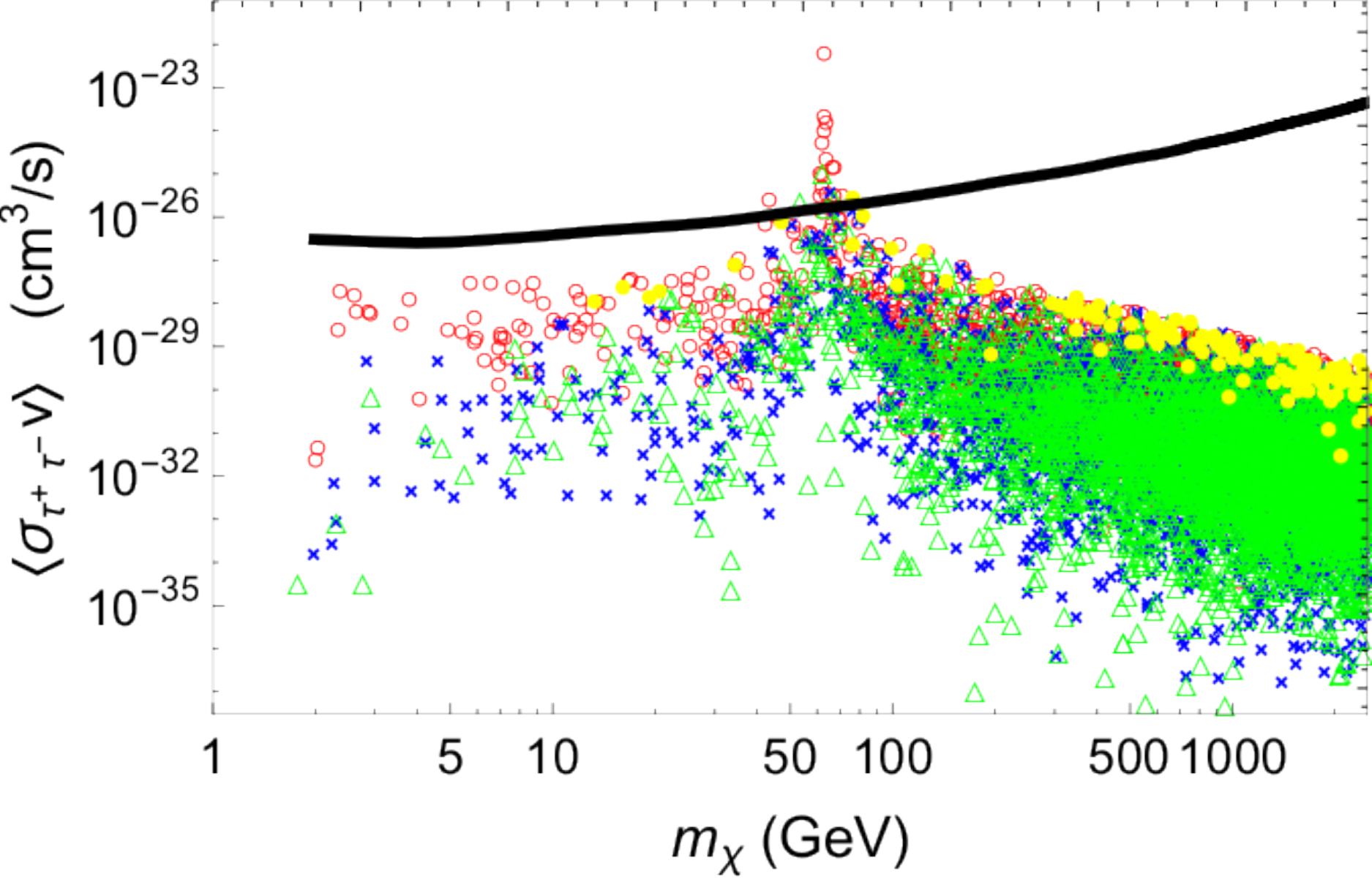}
}%\\\subfigure[\ Fermi-LAT conststraint on $\chi^0 \bar{\chi}^0\rightarrow \mu^+\mu^-$]{
  %\includegraphics[width=0.45\textwidth,height=0.135\textheight]{08iGindmumu2.pdf}
%}\subfigure[\ Fermi-LAT conststraint on $\chi^0 \bar{\chi}^0\rightarrow e^+e^-$]{
  %\includegraphics[width=0.45\textwidth,height=0.135\textheight]{08jGindee2.pdf}
%}
\caption{Results for all samples with constraints in the case of neutralino-like IV
[{\color{red} $\circ$}:~higgsino-like,
{\color{blue} $\times$}:~bino-like,  {\color{green} $\triangle$}:~wino-like,
{\color{yellow} $\bullet$}:~mixed].}
\label{fig:neutralino-like IV}
\end{figure}

After explaining the interplay among these constraints in the case of neutralino-like~I. Now we turn to see the
differences among these neutralino-like cases. The results of other three cases with
all samples are shown in Figs.~\ref{fig:neutralino-like II}-\ref{fig:neutralino-like IV}.
In these figures, we do not show the highly helicity suppressed plots of $\la\sigma_{u{\bar u}} v\ra$, $\la\sigma_{\mu^+\mu^-} v\ra$ and $\la\sigma_{e^+e^-} v\ra$.
First of all, the $\tilde W$-like particles do not appear in the cases of neutralino-like I and II with different $\tan\beta$ values (see Figs.~\ref{fig:neutralino-like I} and \ref{fig:neutralino-like II}).
It is highly unlikely to generate the $\tilde W$-like particles with the GUT relation.~\footnote{It does not mean that the $\tilde W$ component is vanishing, but it is not the dominant composition of DM particles in these cases.}
In contrast, without the GUT relation, plenty of $\tilde W$-like particles can be generated as in the cases of neutralino-like III, IV (see Figs.~\ref{fig:neutralino-like III} and \ref{fig:neutralino-like IV}). 
For neutralino-like III case with a fixed $\tan\beta$, the $\tilde W$-like particles tend to have smaller values in $\Omega_{\chi} h^2$ and larger values in the cross section of DM scattering off nuclei and in the velocity averaged cross section of DM annihilation to the SM particles than the $\tilde B$-like particles (see Fig.~\ref{fig:neutralino-like III}). 
For neutralino-like IV case without fixing $\tan\beta$, only the $\tilde W$-like particles with $m_\chi \gtrsim M_W$ have smaller values in $\Omega_{\chi} h^2$ and greater values in $\la\sigma _{W^+W^-}v\ra$ than the $\tilde B$-like particles (see Fig.~\ref{fig:neutralino-like IV}).
It is originated from the fact that a $\tilde B$-like DM pair does not contribute to the $s$-wave amplitude. 

Among the neutralino-like cases, we see that either ``a higher $\tan\beta$ value" (neutralino-like II, Fig.~\ref{fig:neutralino-like II}) or ``without the GUT relation" (MSSM like-III, IV, Figs.~\ref{fig:neutralino-like III}-\ref{fig:neutralino-like IV}) gives a wider spread in each scatter plot as comparing to Fig.~\ref{fig:neutralino-like I}.
With the DM relic constraint, $99\%$, $99\%$, $98\%$ and $60\%$ of $\tilde B$-like particles are ruled out in the 
neutralino-like I-IV cases, respectively.
%while with the LUX $\sigma^{SI}_N$ constraint, $79\%$, $67\%$, $61\%$ and $51\%$
%$87\%$, $71\%$, $67\%$ and $95\%$ 
%of $\tilde H$-like particles are survived in neutralino-like I-IV cases, respectively.
After considering all constraints,
%more than $44\%$ of $\tilde H$-like particles and 
less than $1\%$ of $\tilde B$-like particles could be DM candidates for the cases of
neutralino-like I - III.
However, for the neutralino-like IV case, without the GUT and the $\tan\beta$ relations,
it has the widest spread in each scatter plot among the neutralino-like cases
%Note that 
%only about $46\%$ of $\tilde H$-like particles and 
so that up to $23\%$ of $\tilde B$-like particles could be DM candidates.
A closer look reveals that in the latter case, more
$\tilde B$-like particles have lower values in DM relic density
% and more $\tilde H$-like particles have larger values 
%in the DM-nucleus scattering cross sections 
[see Fig.~\ref{fig:neutralino-like IV}(a)].
% and (b)]. 
Therefore, more $\tilde B$-like particles are allowed
%, while less $\tilde H$-like particles can survive 
in the neutralino-like IV case.
On the other hand, with the LUX $\sigma^{SI}_N$ constraint, $79\%$, $67\%$, $61\%$ and $51\%$ 
of $\tilde H$-like particles are survived in neutralino-like I-IV cases, respectively [see Figs.~\ref{fig:neutralino-like I},\ref{fig:neutralino-like II}-\ref{fig:neutralino-like IV}(a)].
It means that in the case of either `` a higher $\tan\beta$" or ``without the GUT relation", more $\tilde H$-like particles spread toward larger values in $\sigma^{SI}_N$, namely, less $\tilde H$-like particles (relative to neutralino-like I) can be allowed .
After considering all constraints, $63\%$, $49\%$, $45\%$ and $46\%$ of $\tilde H$-like particles are allowed in neutralino-like I-IV cases, respectively.
As for the mixed particles, it can be ignored since less than $0.1\%$ of samples are allowed as the DM candidates in the neutralino-like cases.

The $\tilde W$-like particles can only appear in the cases without the GUT relation (neutralino-like III, IV, see Figs.~\ref{fig:neutralino-like III} and \ref{fig:neutralino-like IV}).
All the $\tilde W$-like particles with $m_{\chi} < M_W$ are ruled out mainly by the DM relic density constraint [see Figs.~\ref{fig:neutralino-like III}(a), \ref{fig:neutralino-like IV}(a)], followed by the Fermi-LAT constraint via the DM annihilation to $b{\bar b}$ channel around $m_\chi\sim M_W$ [see Figs.~\ref{fig:neutralino-like III}(g), \ref{fig:neutralino-like IV}(g)]. All the $\tilde W$-like particles with $m_{\chi}> M_W$ are not ruled out by the observed relic density [see Figs.~\ref{fig:neutralino-like III}-\ref{fig:neutralino-like IV}(a)], and all the $\tilde W$-like particles with $M_W < m_{\chi} \lesssim 1$ TeV are ruled out by the Fermi-LAT constraint via the DM annihilation to $W^+W^-$ channel [see Figs.~\ref{fig:neutralino-like III}(f), \ref{fig:neutralino-like IV}(f)].
The remaining $\tilde W$-like particles with $m_{\chi} \gtrsim 1$ TeV
are still subjected to the LUX, XENON100 and PICO-60 constraints [see Figs.~\ref{fig:neutralino-like III}-\ref{fig:neutralino-like IV}(b-e)].
It results in about $45\%$ and $39\%$ of $\tilde W$-like particles allowed to be DM candidates in neutralino-like III and IV cases, respectively,
and the allowed $\tilde W$-like particles are heavy ($m_\chi\gtrsim 1$~TeV).

\vfill
\eject

\vfill
\eject

\vfill
\eject

\begin{figure}[t!]
\centering
\captionsetup{justification=raggedright}
 \subfigure[\ Constraint on $\Omega^{\rm{obs}}_\chi$]{
  \includegraphics[width=0.45\textwidth,height=0.13\textheight]{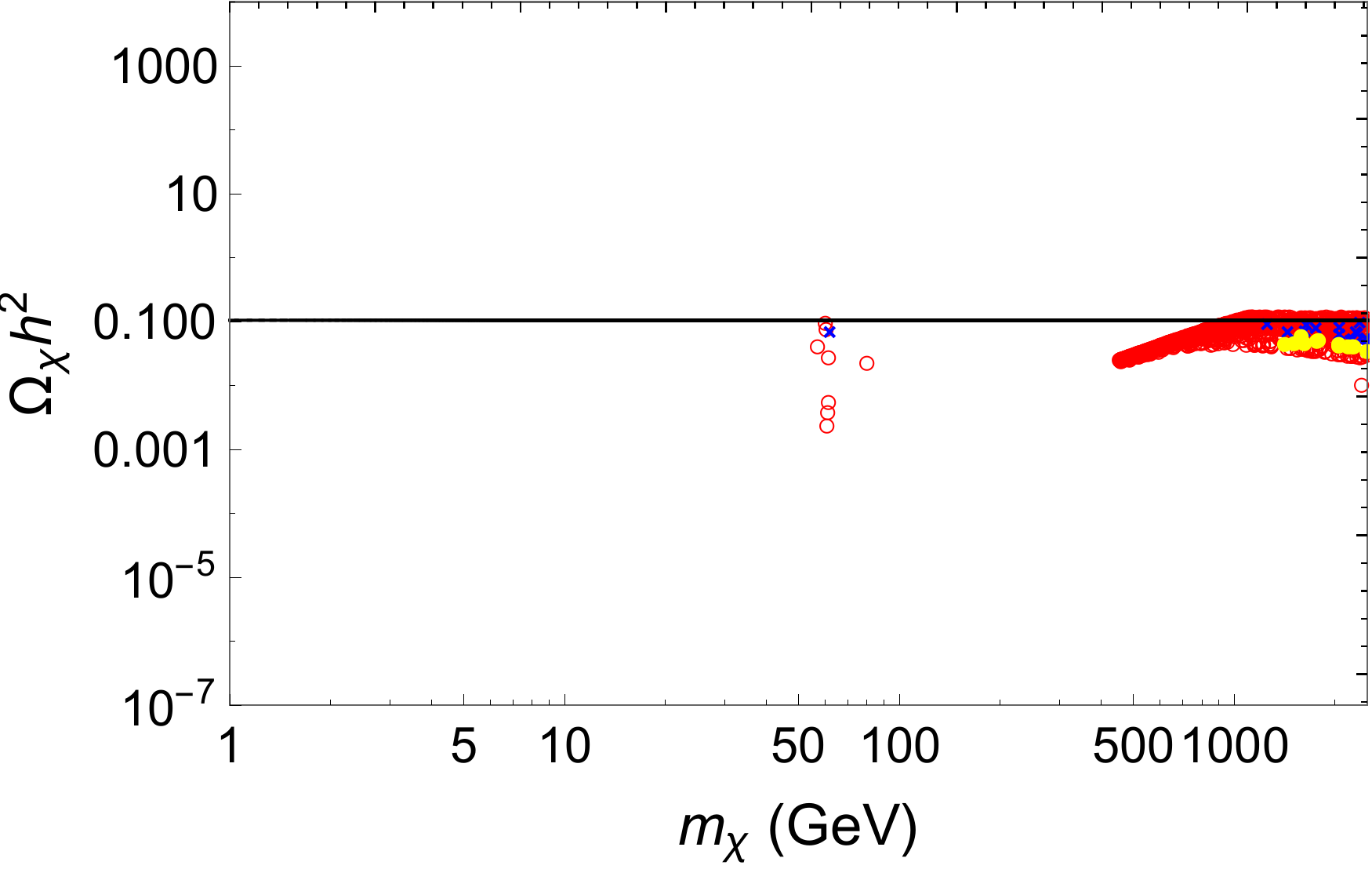}
}\subfigure[\ LUX constraint on $\sigma^{SI}$ with NB limit]{
  \includegraphics[width=0.45\textwidth,height=0.13\textheight]{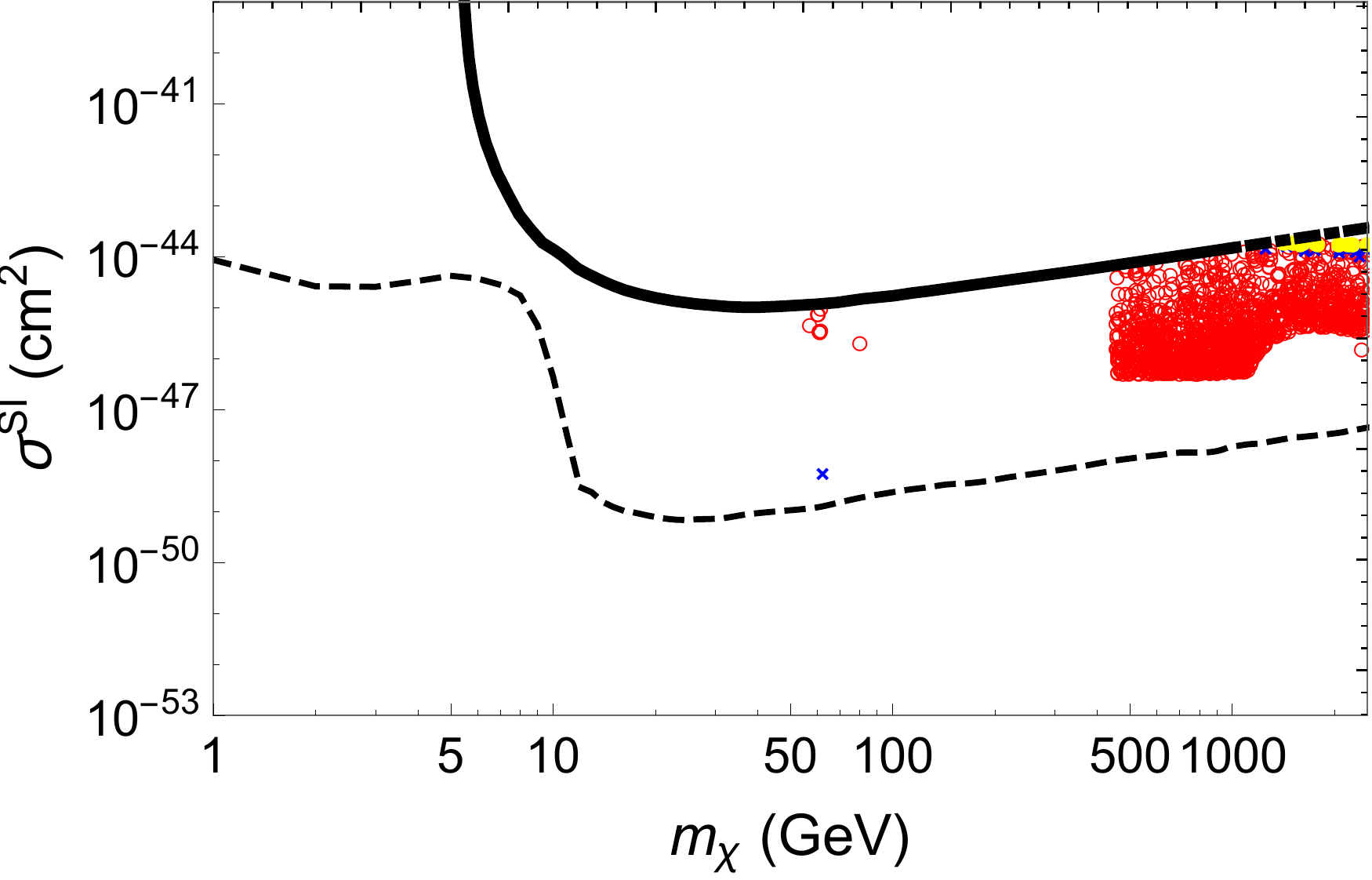}
}\\\subfigure[\ XENON100 constraint on $\sigma^{SD}_n$]{
  \includegraphics[width=0.45\textwidth,height=0.13\textheight]{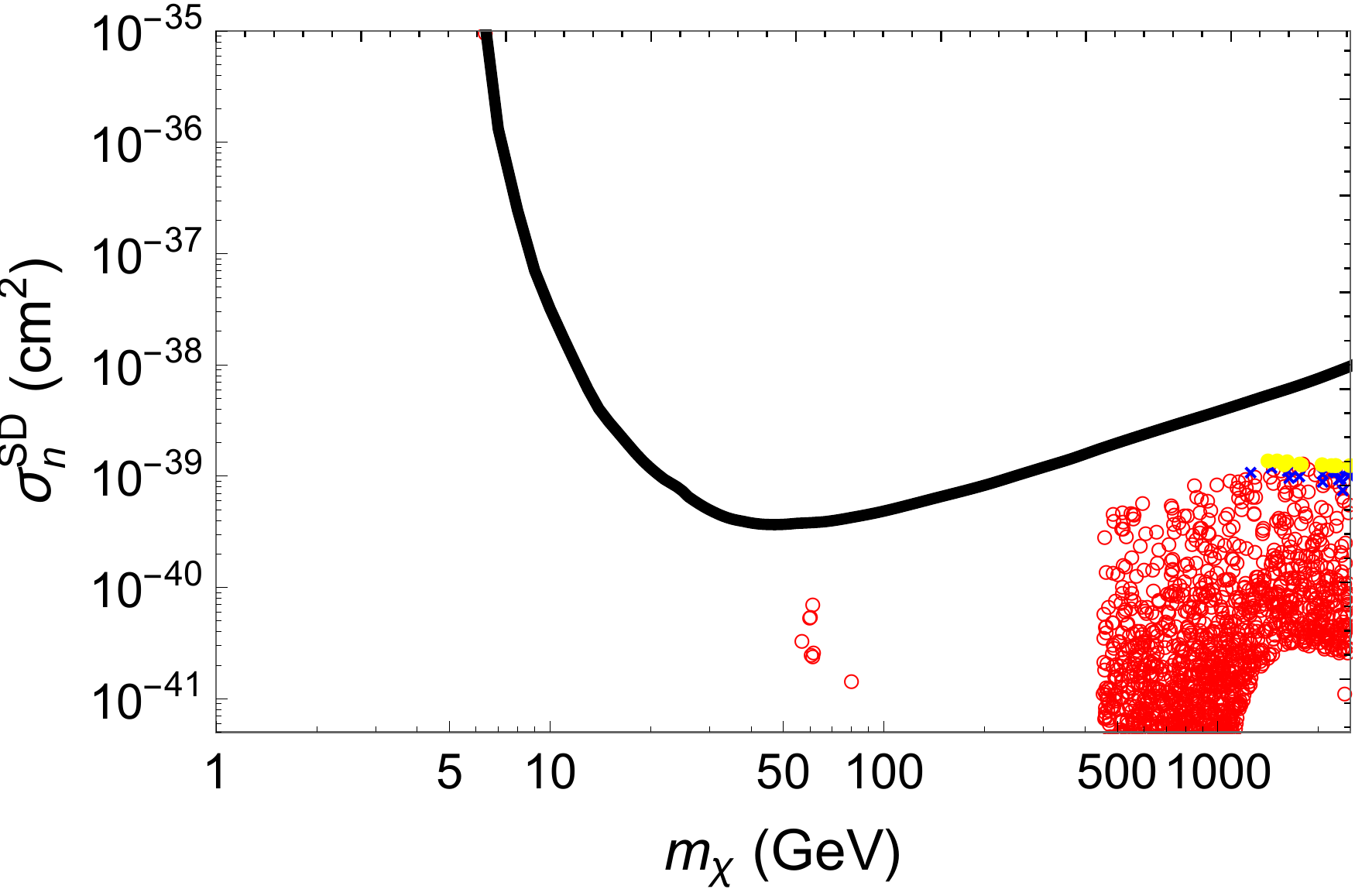}
}\subfigure[\ XENON100 constraint on $\sigma^{SD}_p$]{
  \includegraphics[width=0.45\textwidth,height=0.13\textheight]{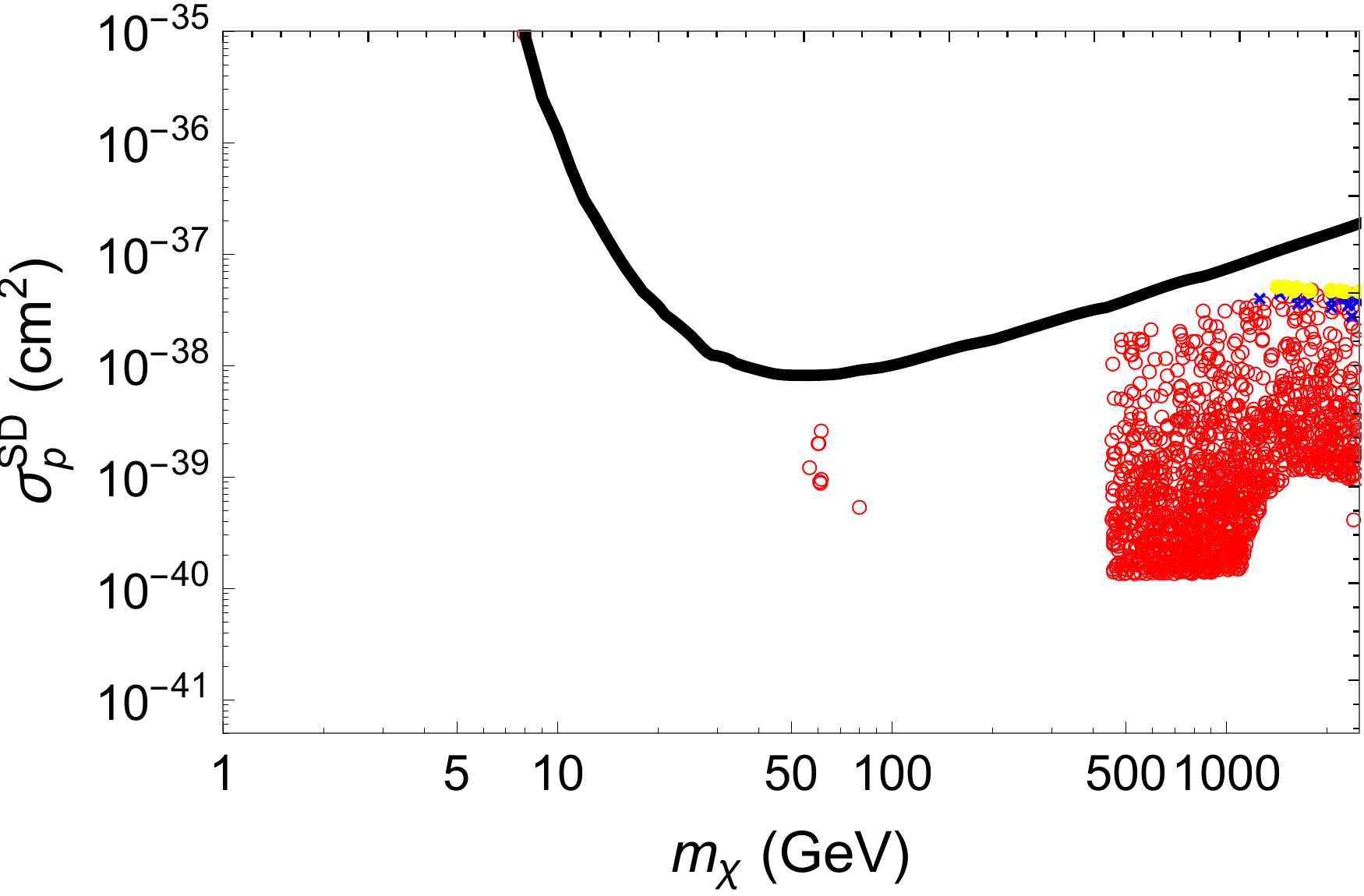}
}\\\subfigure[\ PICO-60 constraint on $\sigma^{SD}_p$]{
  \includegraphics[width=0.45\textwidth,height=0.13\textheight]{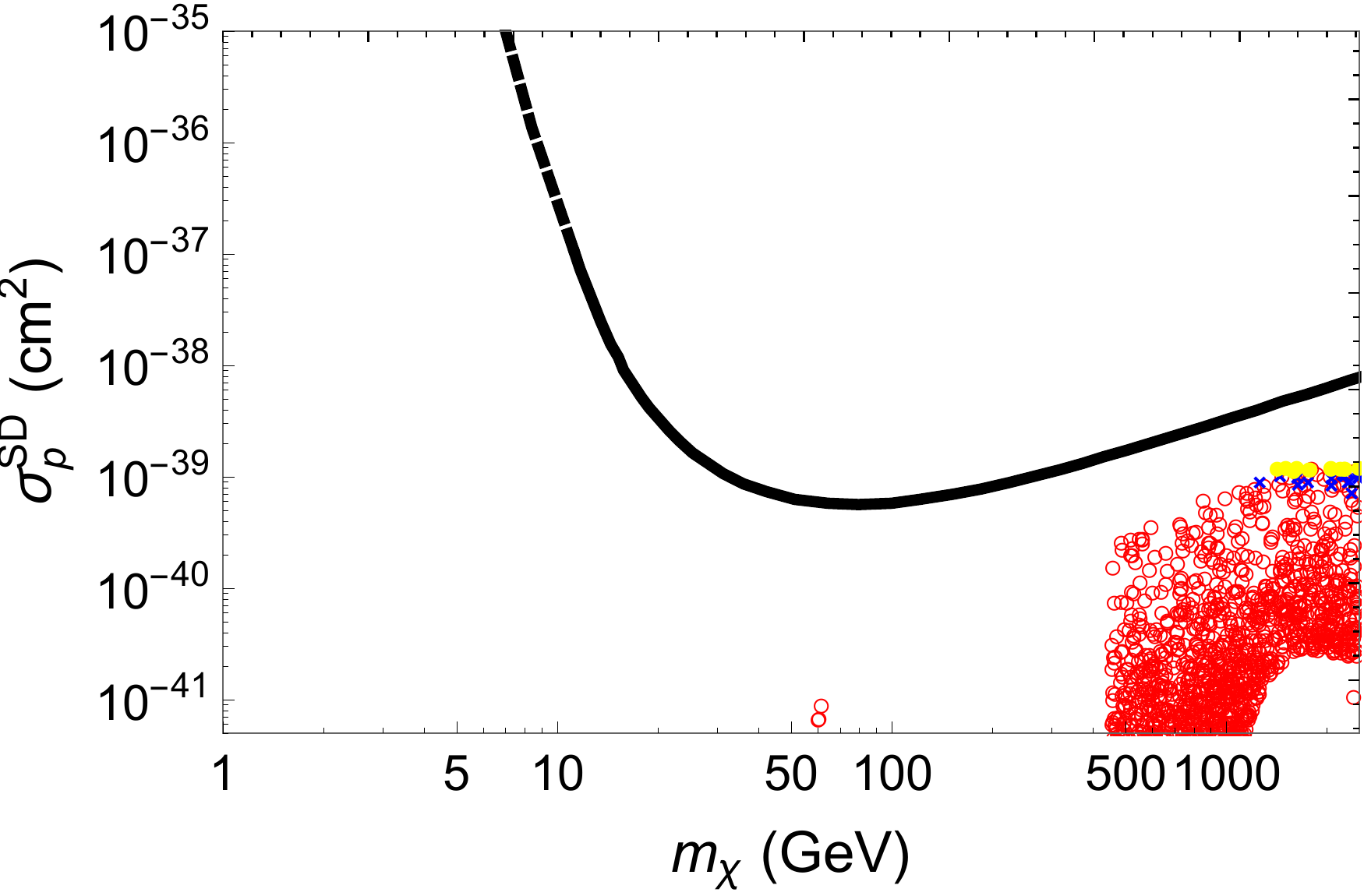}
}\subfigure[\ Fermi-LAT constraint on $\chi^0 {\chi}^0\rightarrow W^+W^-$]{
  \includegraphics[width=0.45\textwidth,height=0.13\textheight]{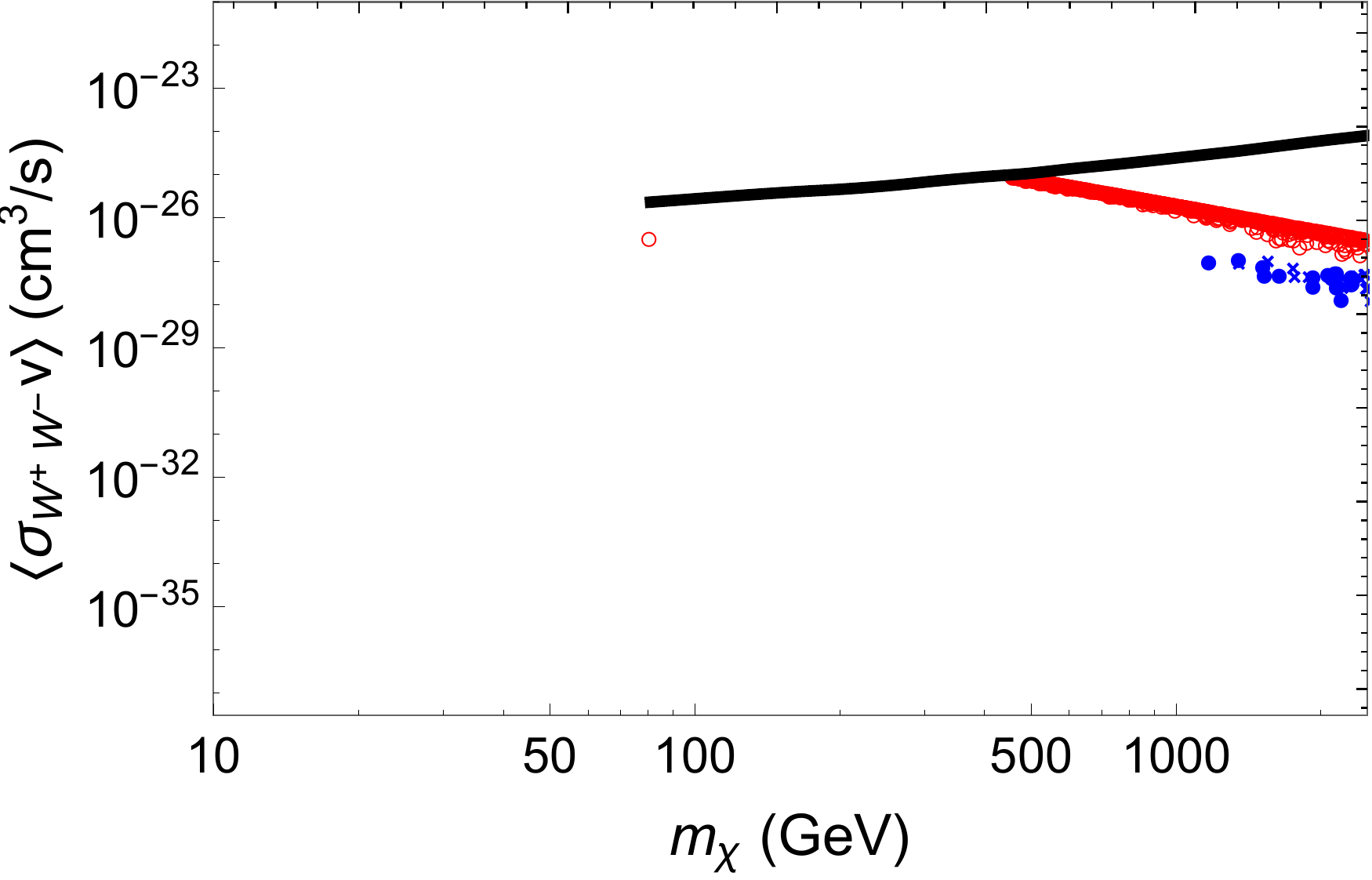}
}\\\subfigure[\ Fermi-LAT constraint on $\chi^0 {\chi}^0\rightarrow b\bar{b}$]{
  \includegraphics[width=0.45\textwidth,height=0.13\textheight]{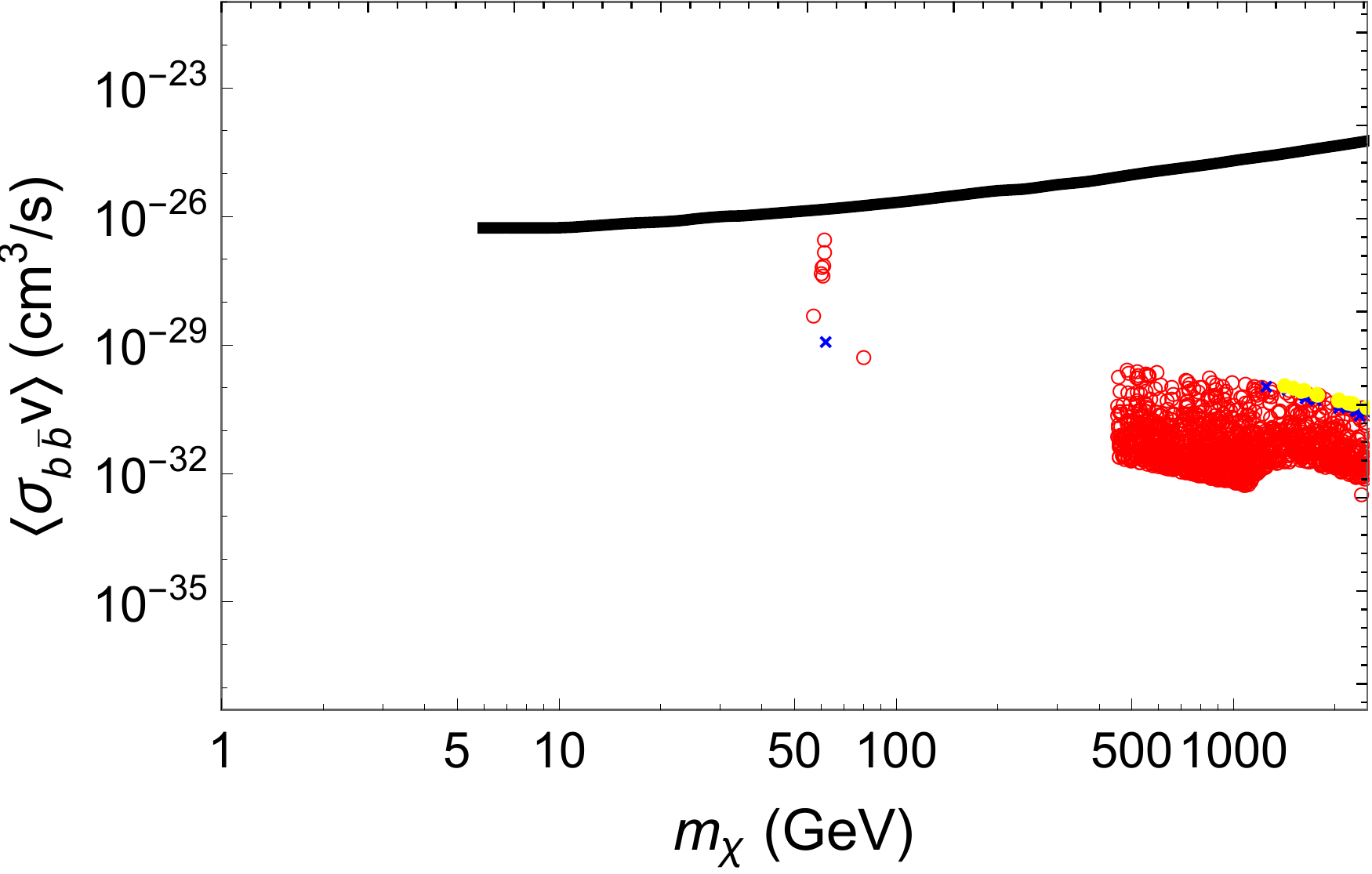}
}\
%\subfigure[\ Fermi-LAT constraint on $\chi^0 \bar{\chi}^0\rightarrow u\bar{u}$]{
  %\includegraphics[width=0.45\textwidth,height=0.13\textheight]{09gEallowuu.pdf}
%}
\subfigure[\ Fermi-LAT constraint on $\chi^0 {\chi}^0\rightarrow \tau^+\tau^-$]{
  \includegraphics[width=0.45\textwidth,height=0.13\textheight]{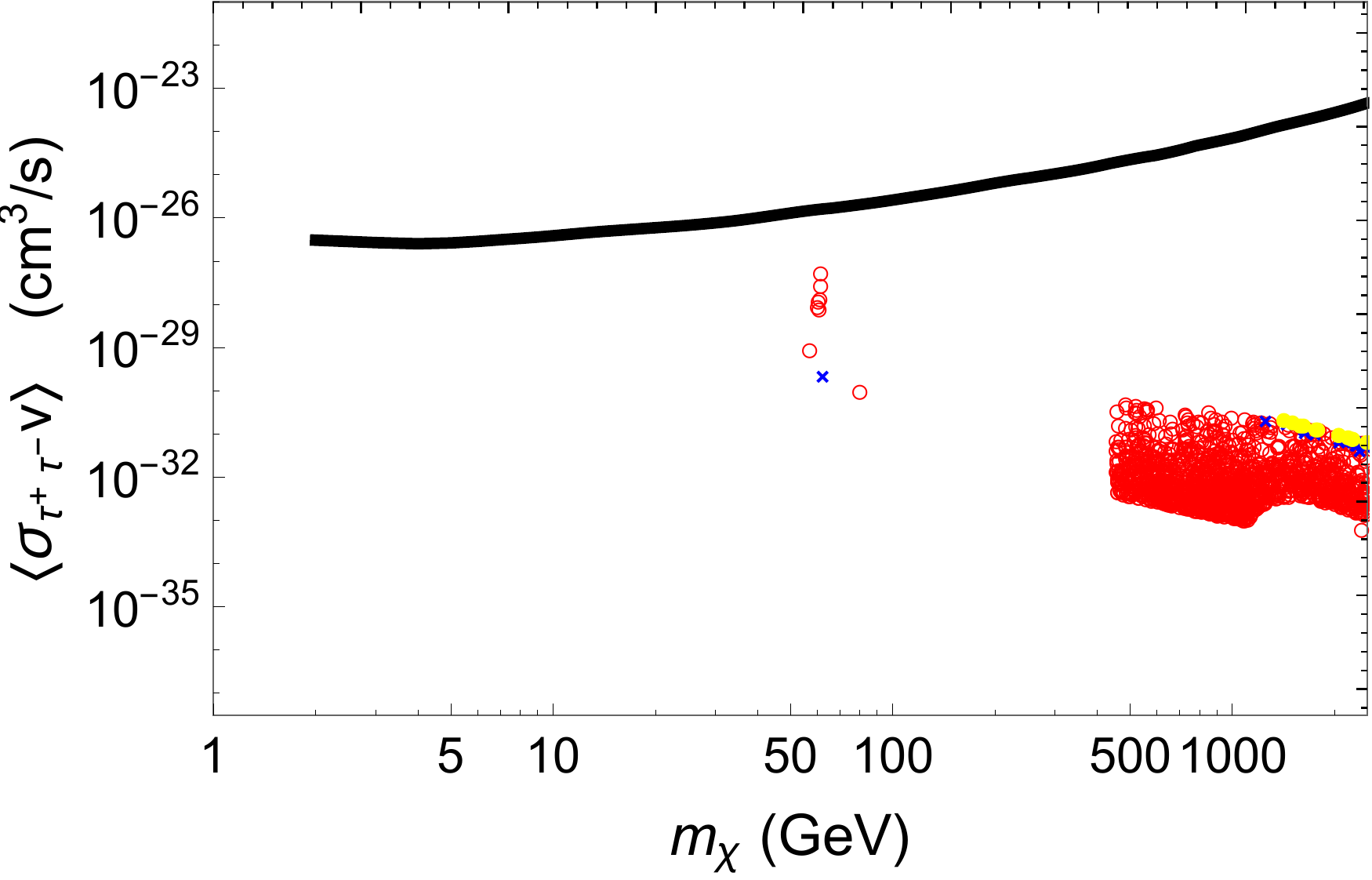}
}%\\\subfigure[\ Fermi-LAT constraint on $\chi^0 \bar{\chi}^0\rightarrow \mu^+\mu^-$]{
 % \includegraphics[width=0.45\textwidth,height=0.13\textheight]{09iEallowmumu.pdf}
%}\subfigure[\ Fermi-LAT constraint on $\chi^0 \bar{\chi}^0\rightarrow e^+e^-$]{
  %\includegraphics[width=0.45\textwidth,height=0.13\textheight]{09jEallowee.pdf}
%}
\caption{Results for allowed samples satisfying all constraints in the neutralino-like II case
[{\color{red} $\circ$}:~higgsino-like,
{\color{blue} $\times$}:~bino-like,   {\color{yellow} $\bullet$}:~mixed].}
\label{fig:allow neutralino-like II}
\end{figure}
\vfill
\eject

\begin{figure}[t!]
\centering
\captionsetup{justification=raggedright}
 \subfigure[\ Constraint on $\Omega^{\rm{obs}}_\chi$]{
  \includegraphics[width=0.45\textwidth,height=0.13\textheight]{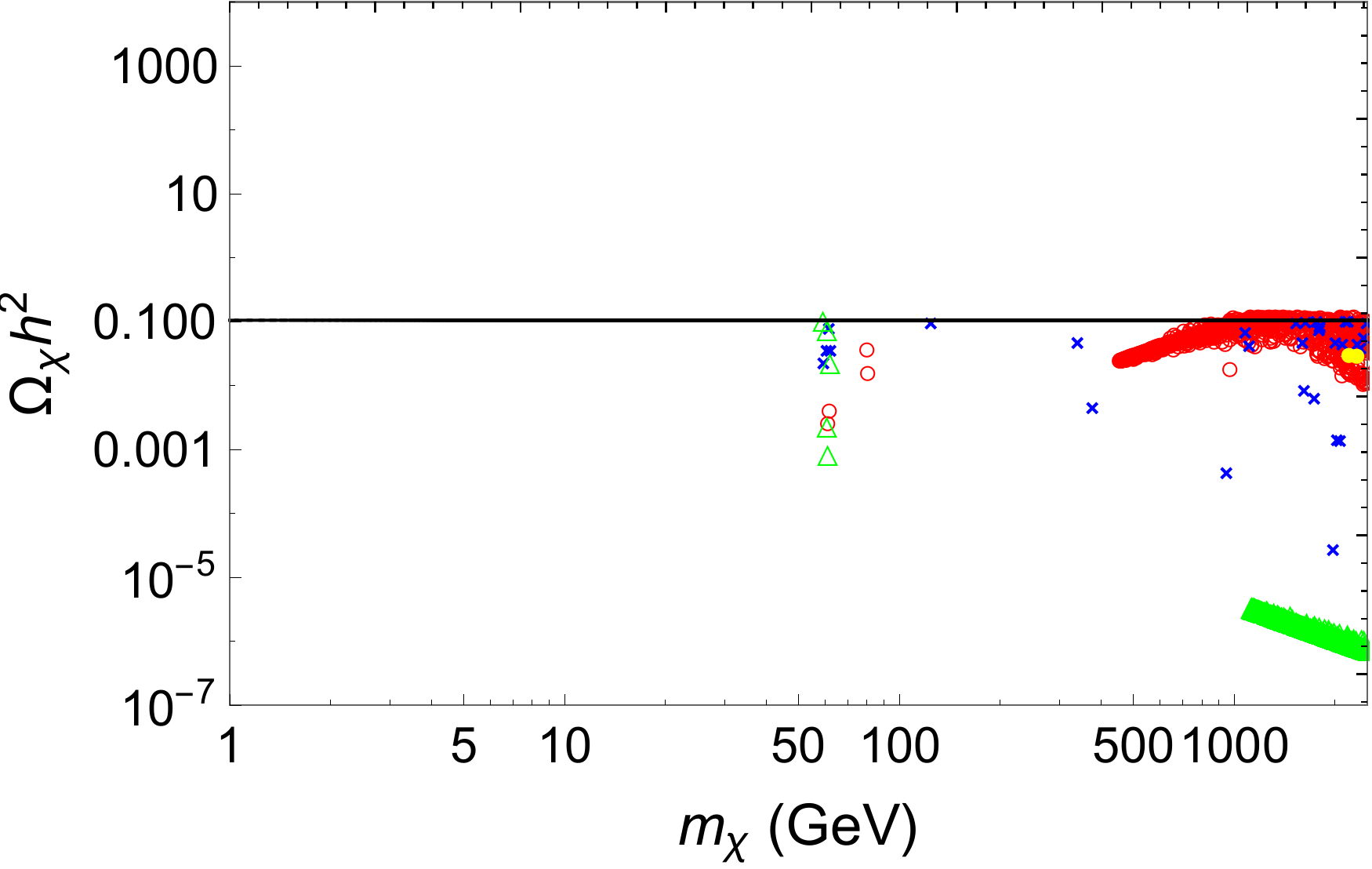}
}\subfigure[\ LUX constraint on $\sigma^{SI}$ with NB limit]{
  \includegraphics[width=0.45\textwidth,height=0.13\textheight]{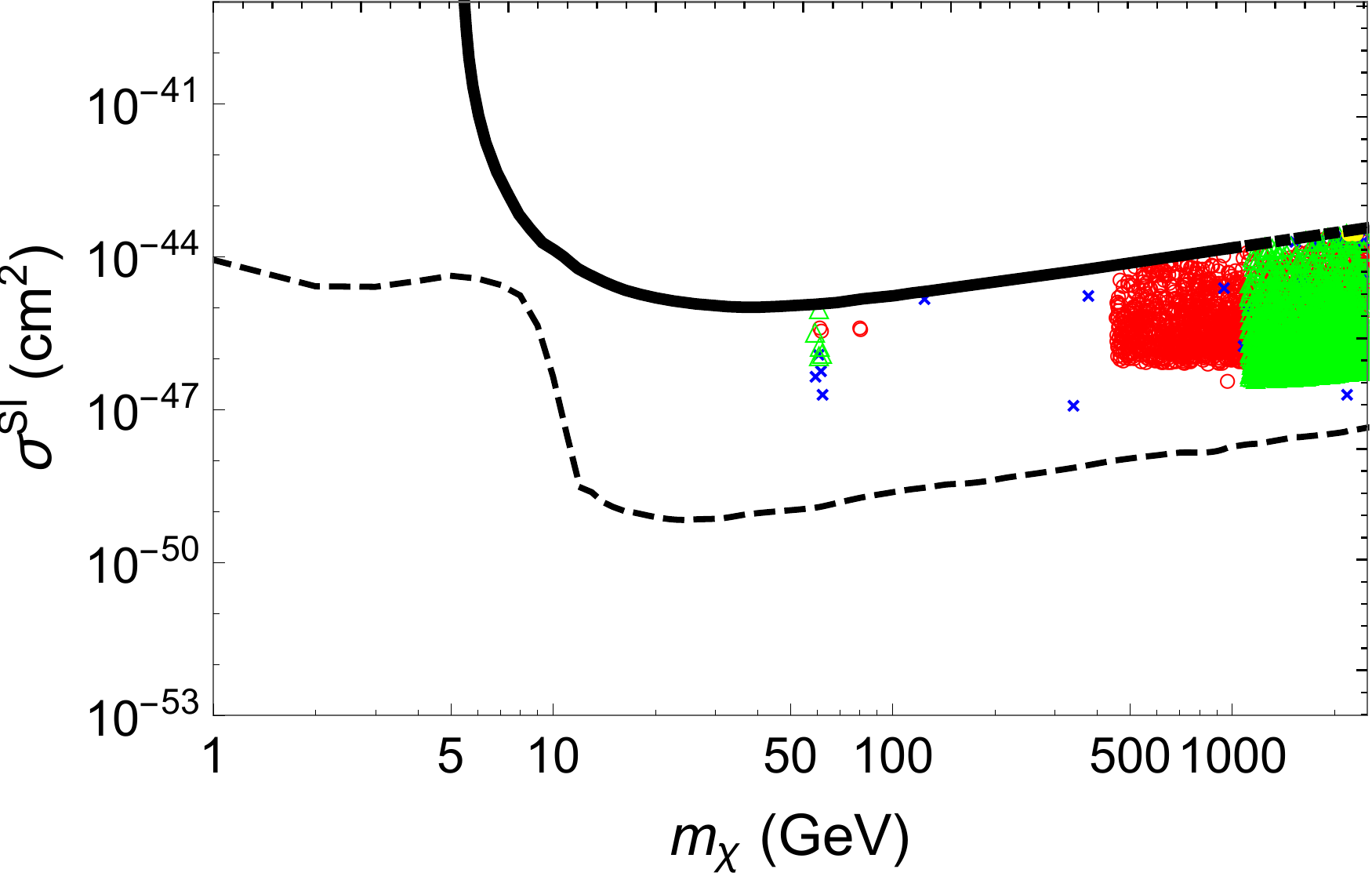}
}\\\subfigure[\ XENON100 constraint on $\sigma^{SD}_n$]{
  \includegraphics[width=0.45\textwidth,height=0.13\textheight]{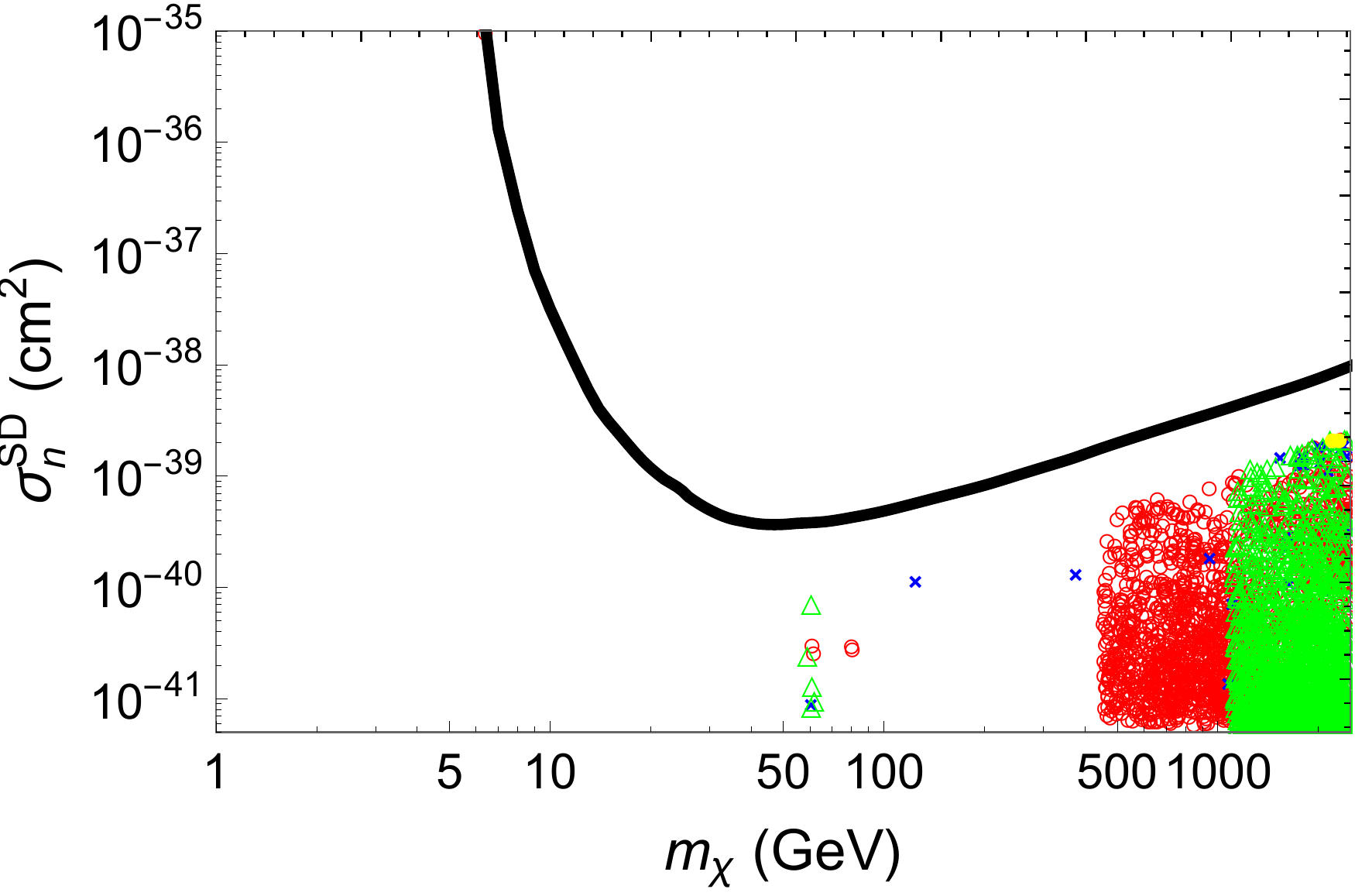}
}\subfigure[\ XENON100 constraint on $\sigma^{SD}_p$]{
  \includegraphics[width=0.45\textwidth,height=0.13\textheight]{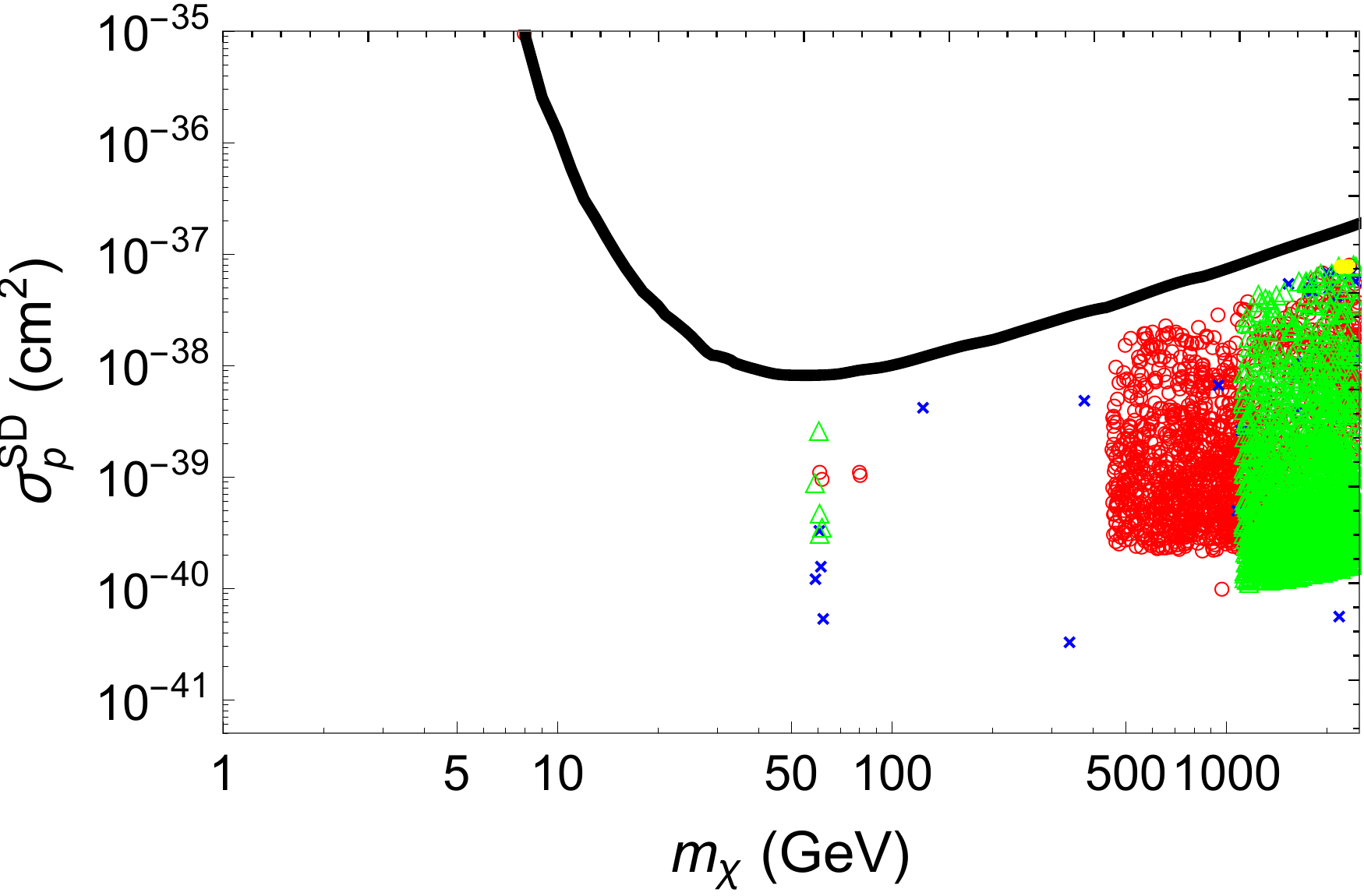}
}\\\subfigure[\ PICO-60 constraint on $\sigma^{SD}_p$]{
  \includegraphics[width=0.45\textwidth,height=0.13\textheight]{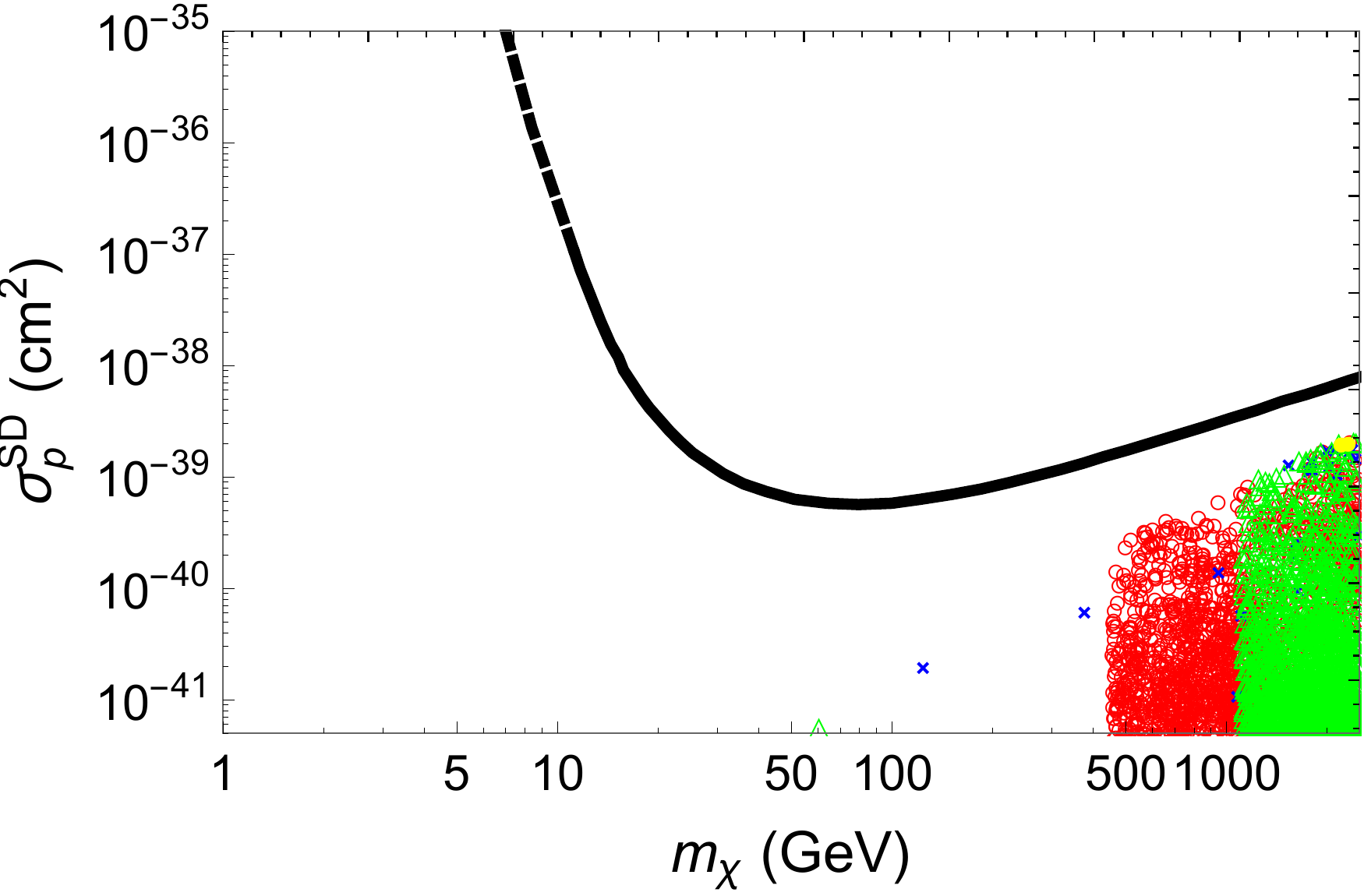}
}\subfigure[\ Fermi-LAT conststraint on $\chi^0 {\chi}^0\rightarrow W^+W^-$]{
  \includegraphics[width=0.45\textwidth,height=0.13\textheight]{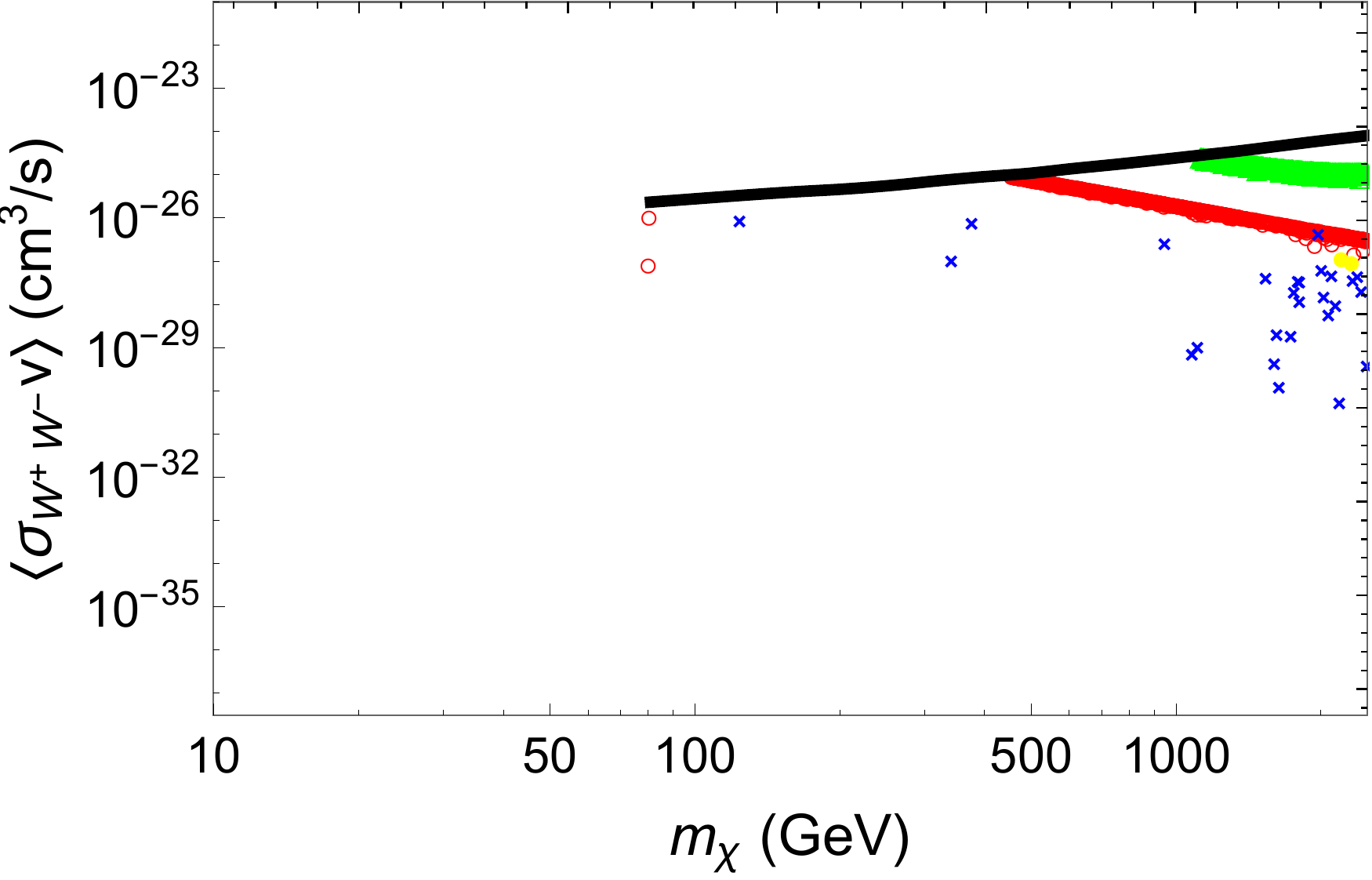}
}\\\subfigure[\ Fermi-LAT conststraint on $\chi^0 {\chi}^0\rightarrow b\bar{b}$]{
  \includegraphics[width=0.45\textwidth,height=0.13\textheight]{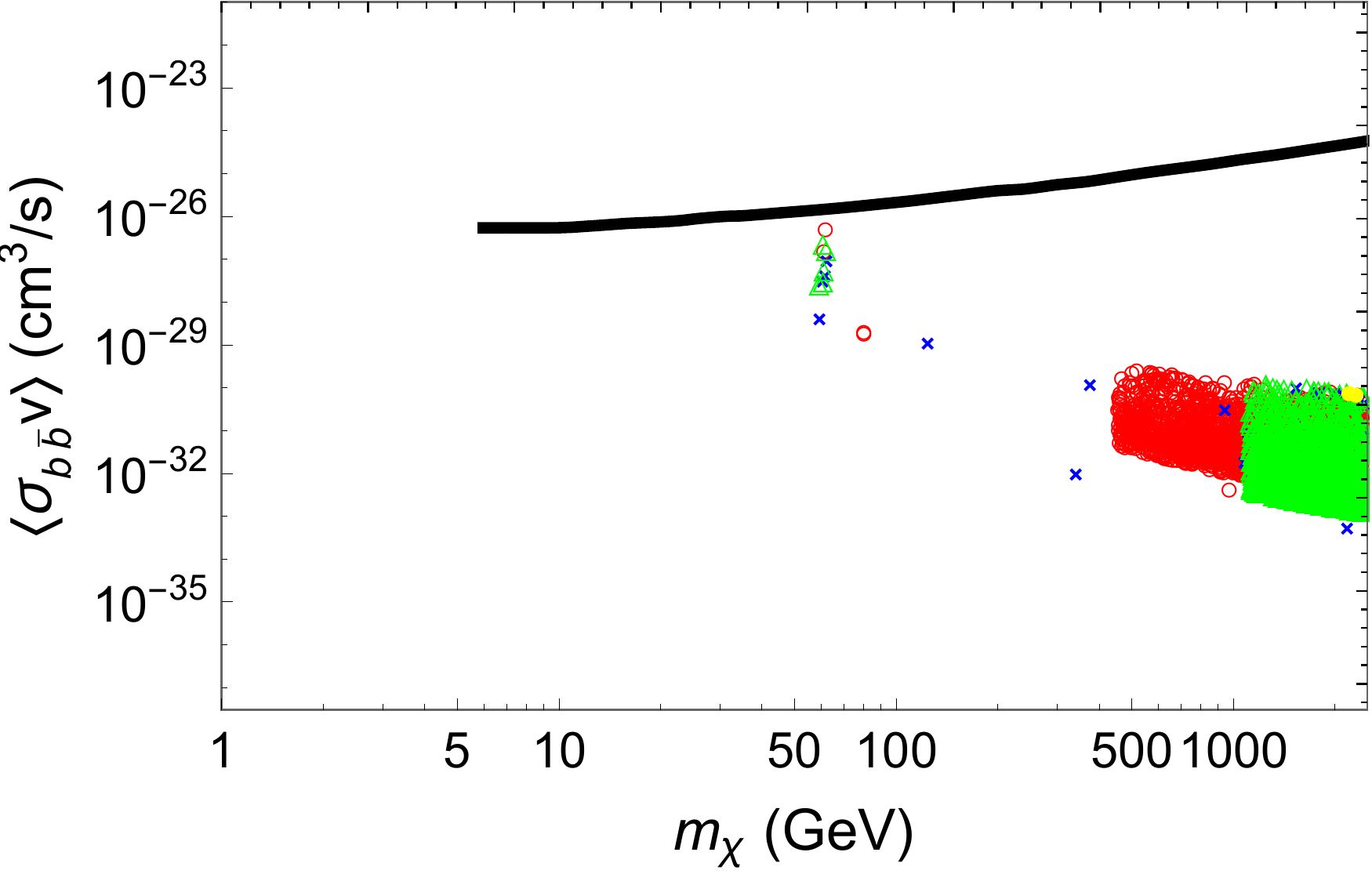}
}\
%\subfigure[\ Fermi-LAT conststraint on $\chi^0 \bar{\chi}^0\rightarrow u\bar{u}$]{
  %\includegraphics[width=0.45\textwidth,height=0.13\textheight]{10gFallowuu.pdf}
%}
\subfigure[\ Fermi-LAT conststraint on $\chi^0 {\chi}^0\rightarrow \tau^+\tau^-$]{
  \includegraphics[width=0.45\textwidth,height=0.13\textheight]{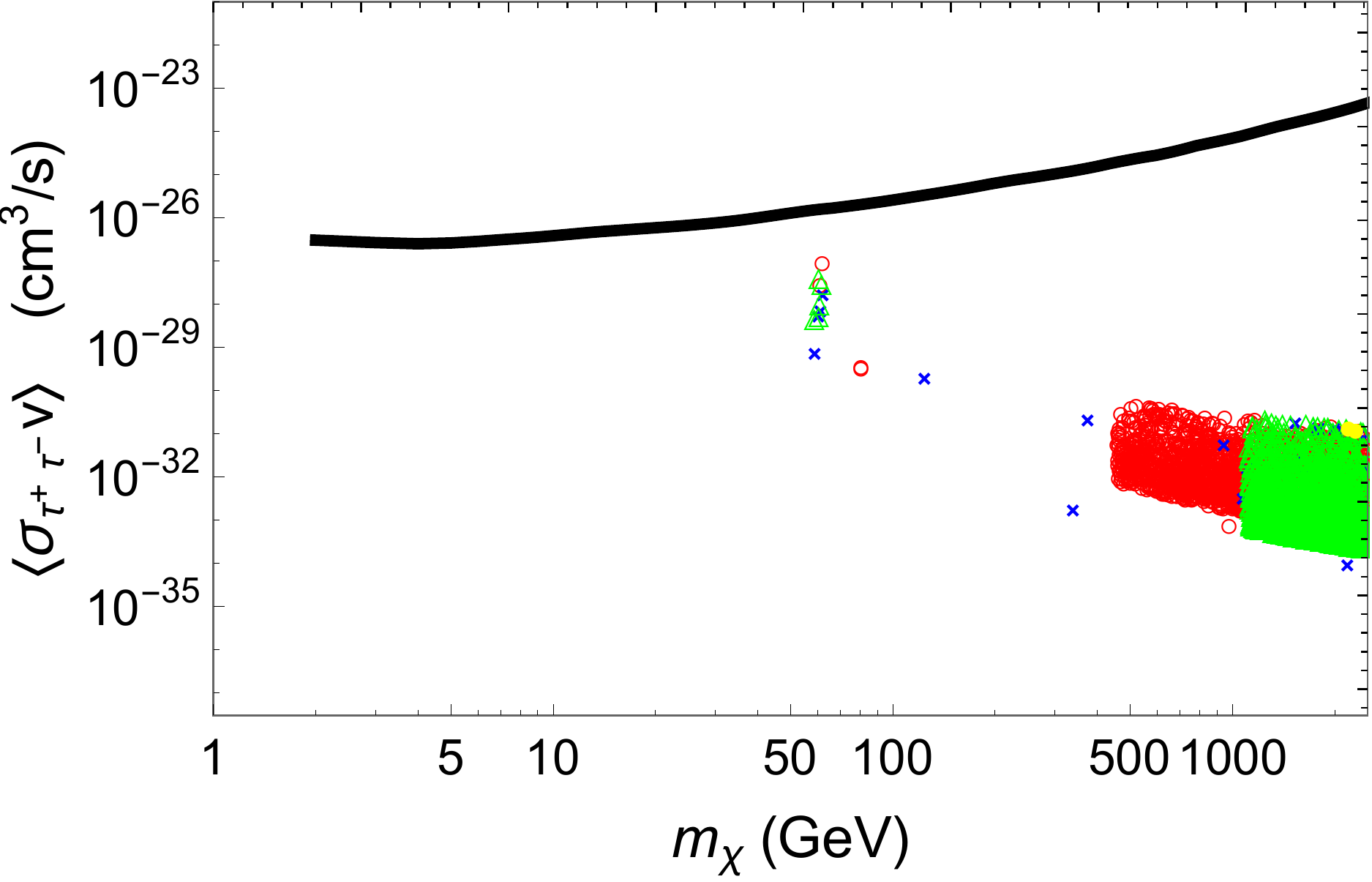}
}%\\\subfigure[\ Fermi-LAT conststraint on $\chi^0 \bar{\chi}^0\rightarrow \mu^+\mu^-$]{
  %\includegraphics[width=0.45\textwidth,height=0.13\textheight]{10iFallowmumu.pdf}
%}\subfigure[\ Fermi-LAT conststraint $\chi^0 \bar{\chi}^0\rightarrow e^+e^-$]{
  %\includegraphics[width=0.45\textwidth,height=0.13\textheight]{10jFallowee.pdf}
%}
\caption{Results allowed samples satisfying all constraints in the neutralino-like III case
[{\color{red} $\circ$}:~higgsino-like,
{\color{blue} $\times$}:~bino-like,  {\color{green} $\triangle$}:~wino-like,
{\color{yellow} $\bullet$}:~mixed].}
\label{fig:allow neutralino-like III}
\end{figure}
\vfill
\eject

\begin{figure}[t!]
\centering
\captionsetup{justification=raggedright}
 \subfigure[\ Constraint on $\Omega^{\rm{obs}}_\chi$]{
  \includegraphics[width=0.45\textwidth,height=0.13\textheight]{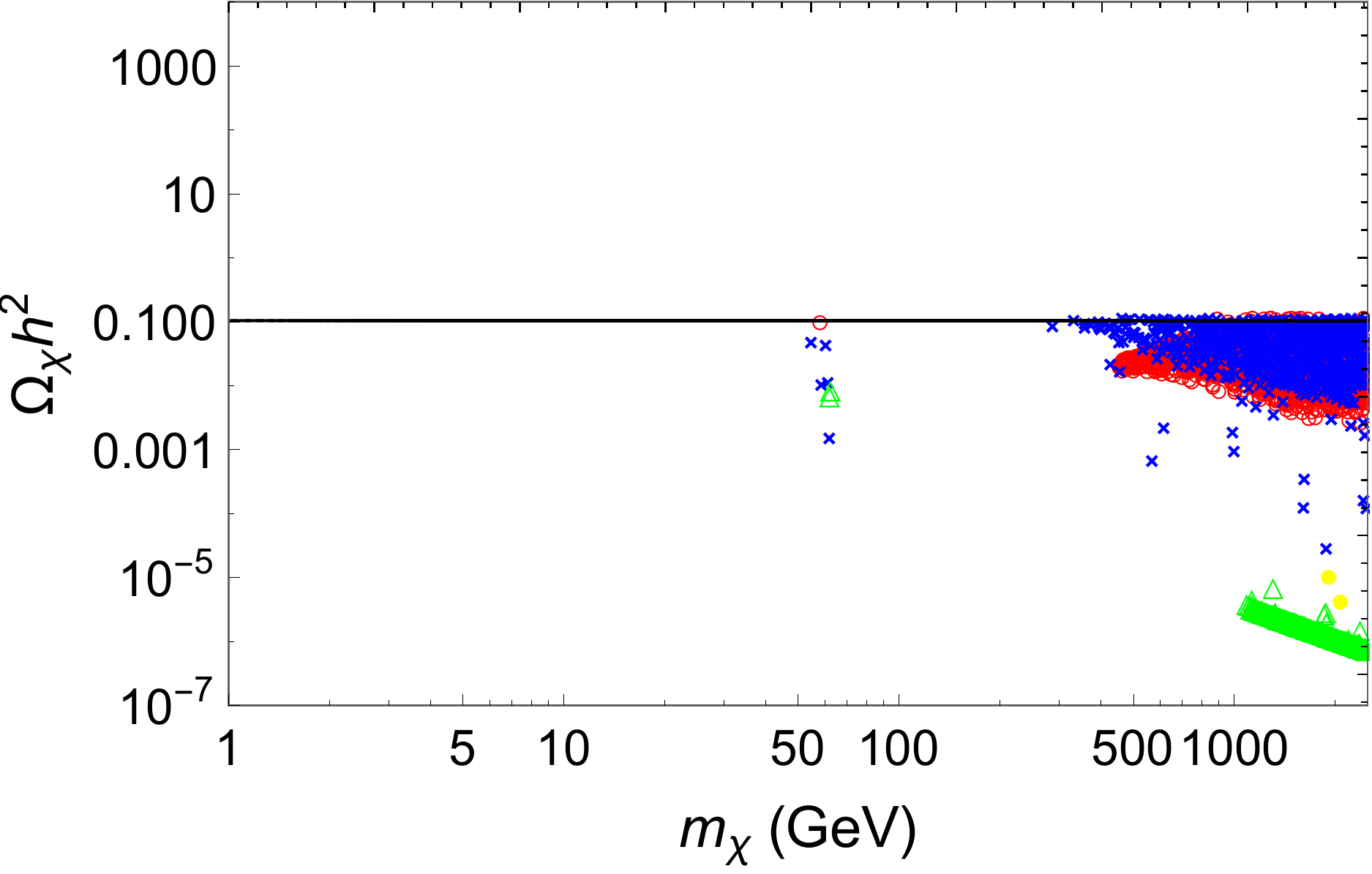}
}\subfigure[\ LUX constraint on $\sigma^{SI}$ with NB limit]{
  \includegraphics[width=0.45\textwidth,height=0.13\textheight]{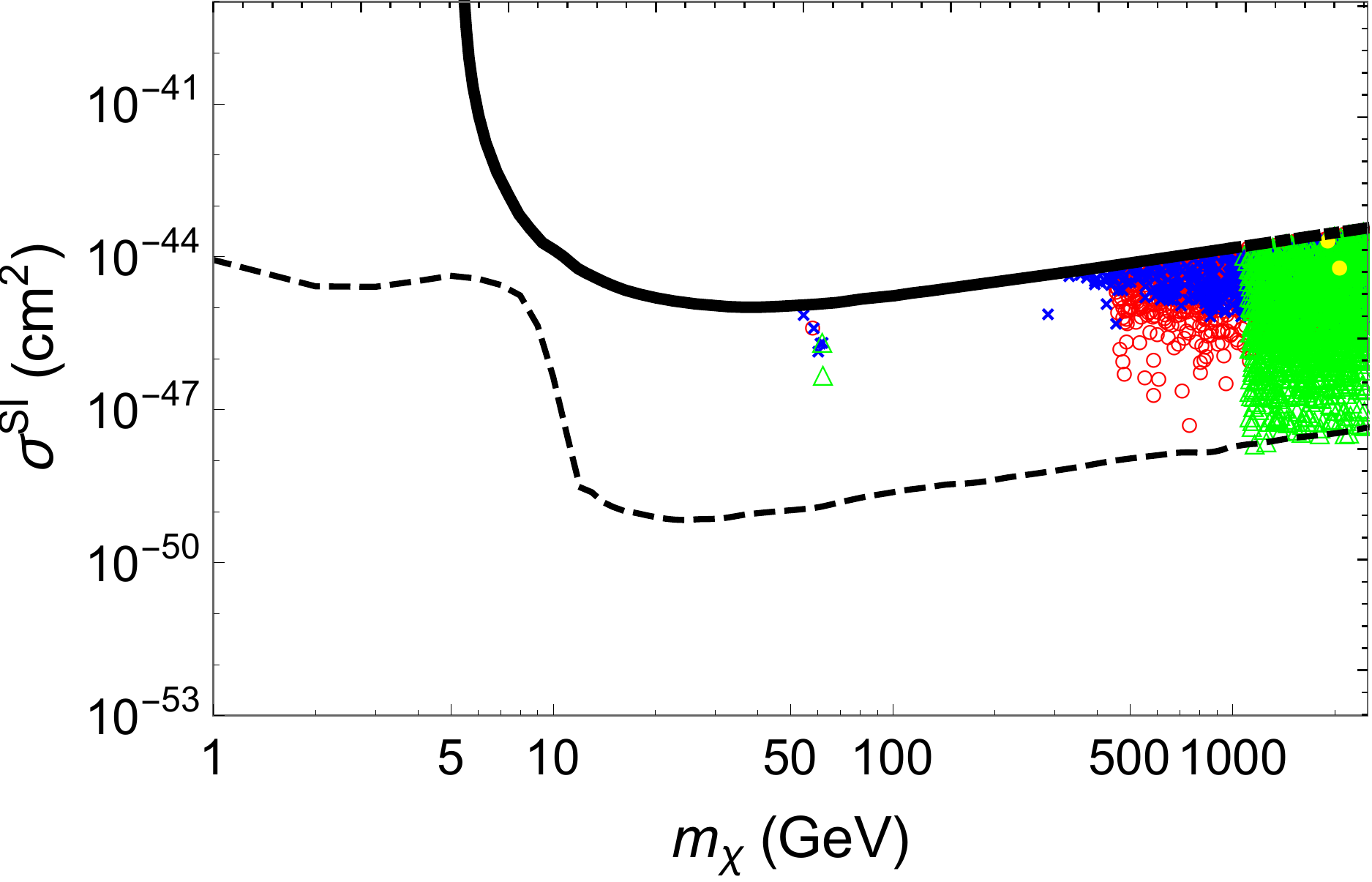}
}\\\subfigure[\ XENON100 constraint on $\sigma^{SD}_n$]{
  \includegraphics[width=0.45\textwidth,height=0.13\textheight]{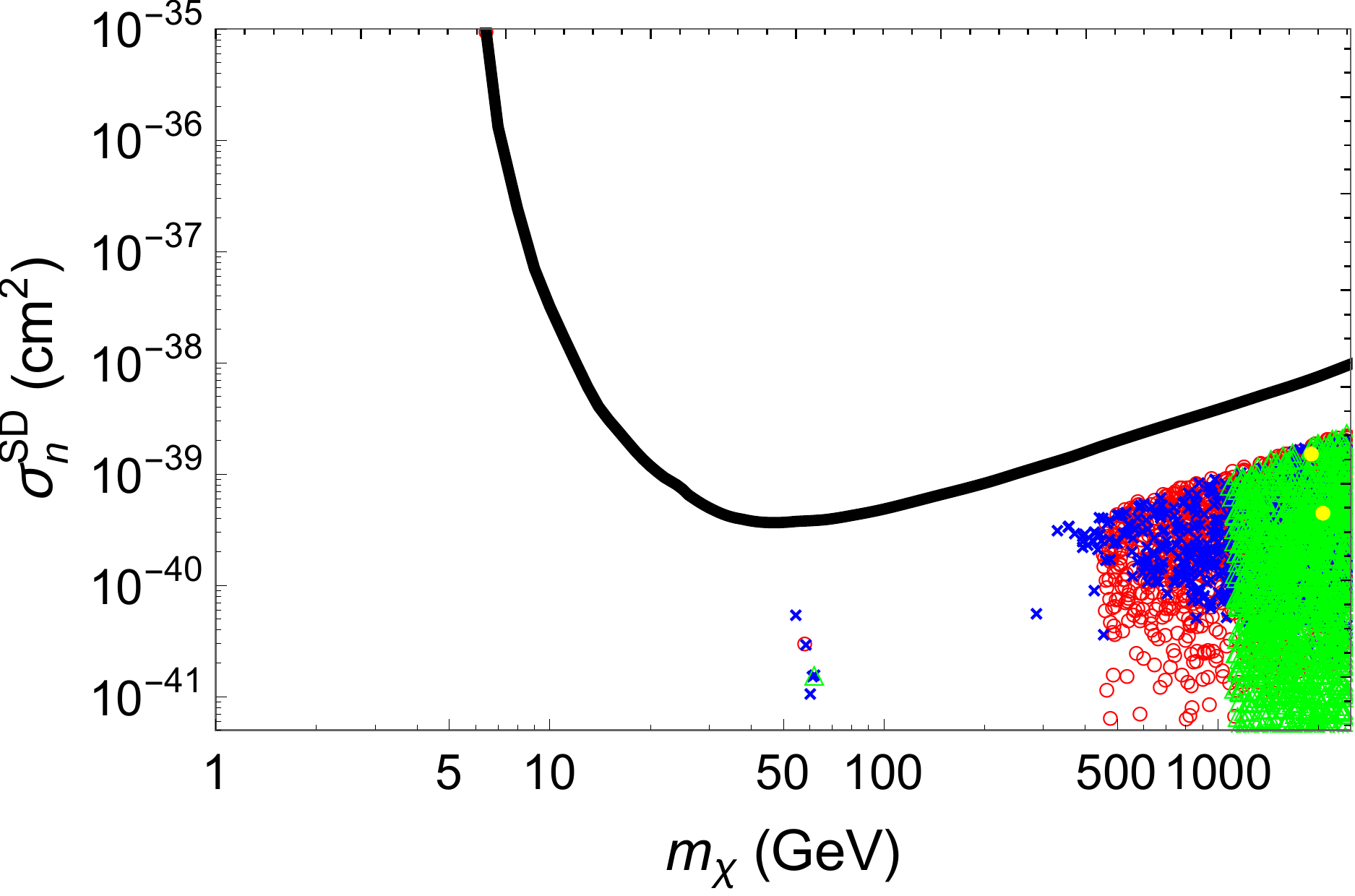}
}\subfigure[\ XENON100 constraint on $\sigma^{SD}_p$]{
  \includegraphics[width=0.45\textwidth,height=0.13\textheight]{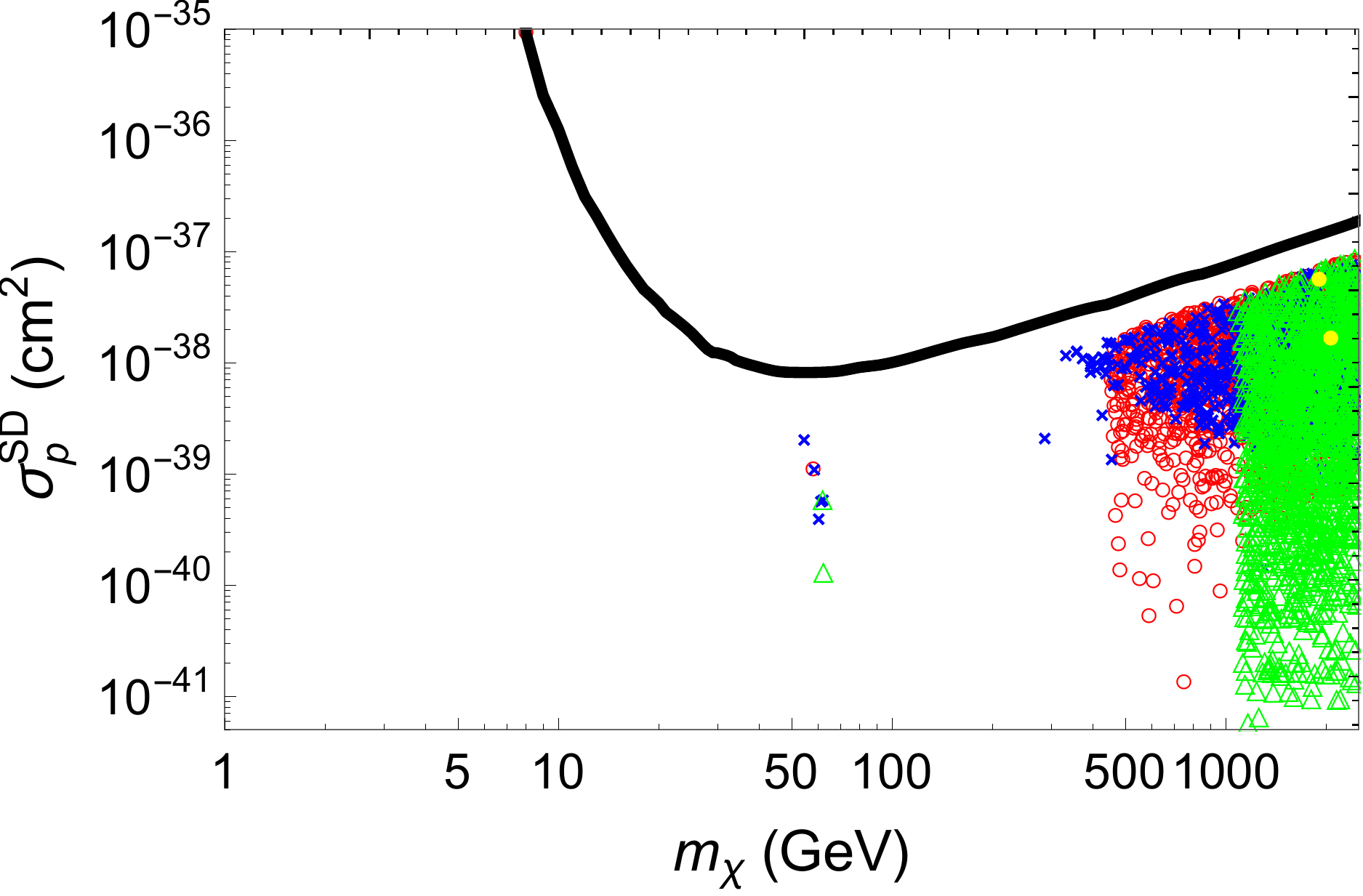}
}\\\subfigure[\ PICO-60 constraint on $\sigma^{SD}_p$]{
  \includegraphics[width=0.45\textwidth,height=0.13\textheight]{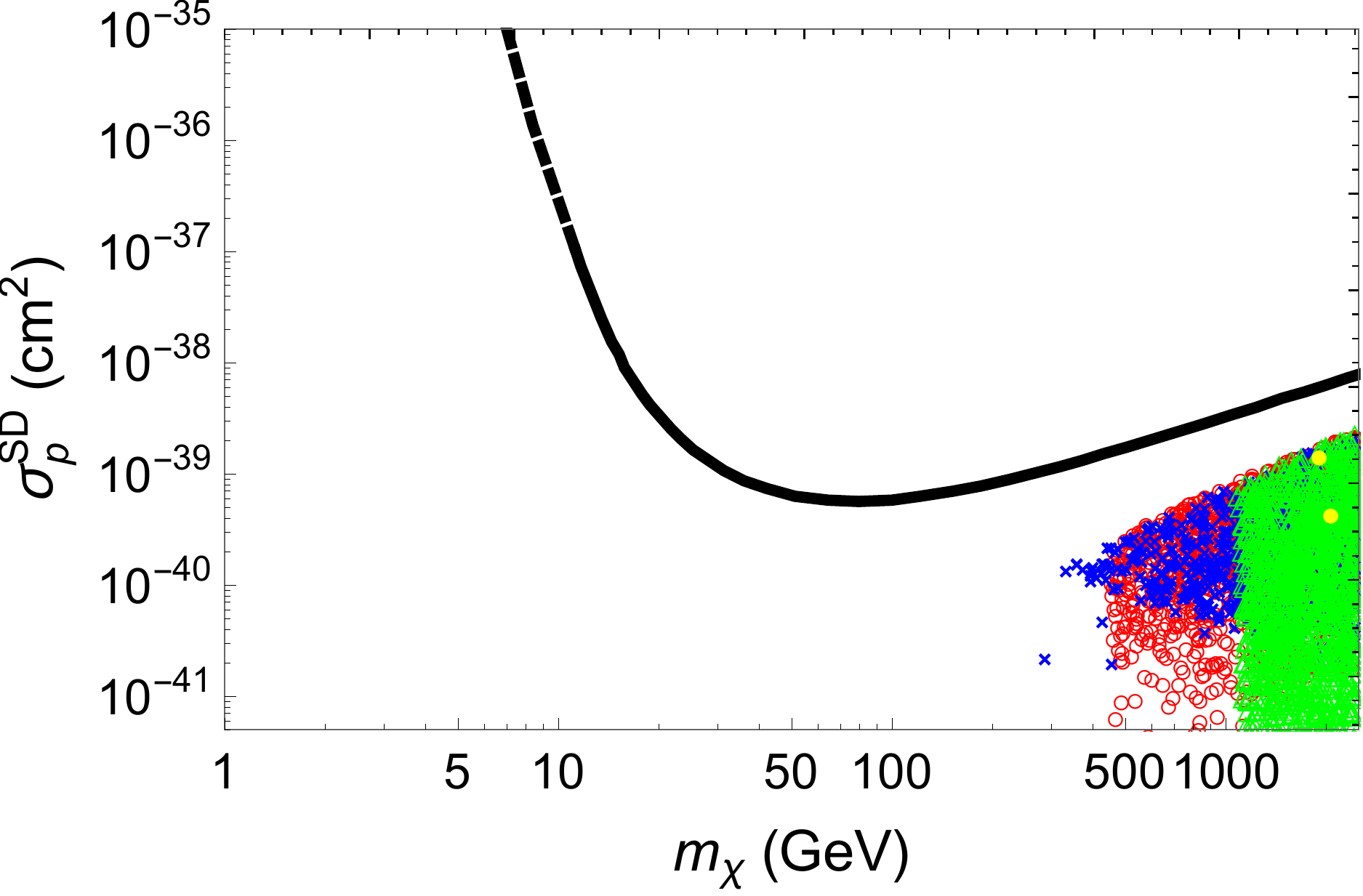}
}\subfigure[\ Fermi-LAT conststraint on $\chi^0 {\chi}^0\rightarrow W^+W^-$]{
  \includegraphics[width=0.45\textwidth,height=0.13\textheight]{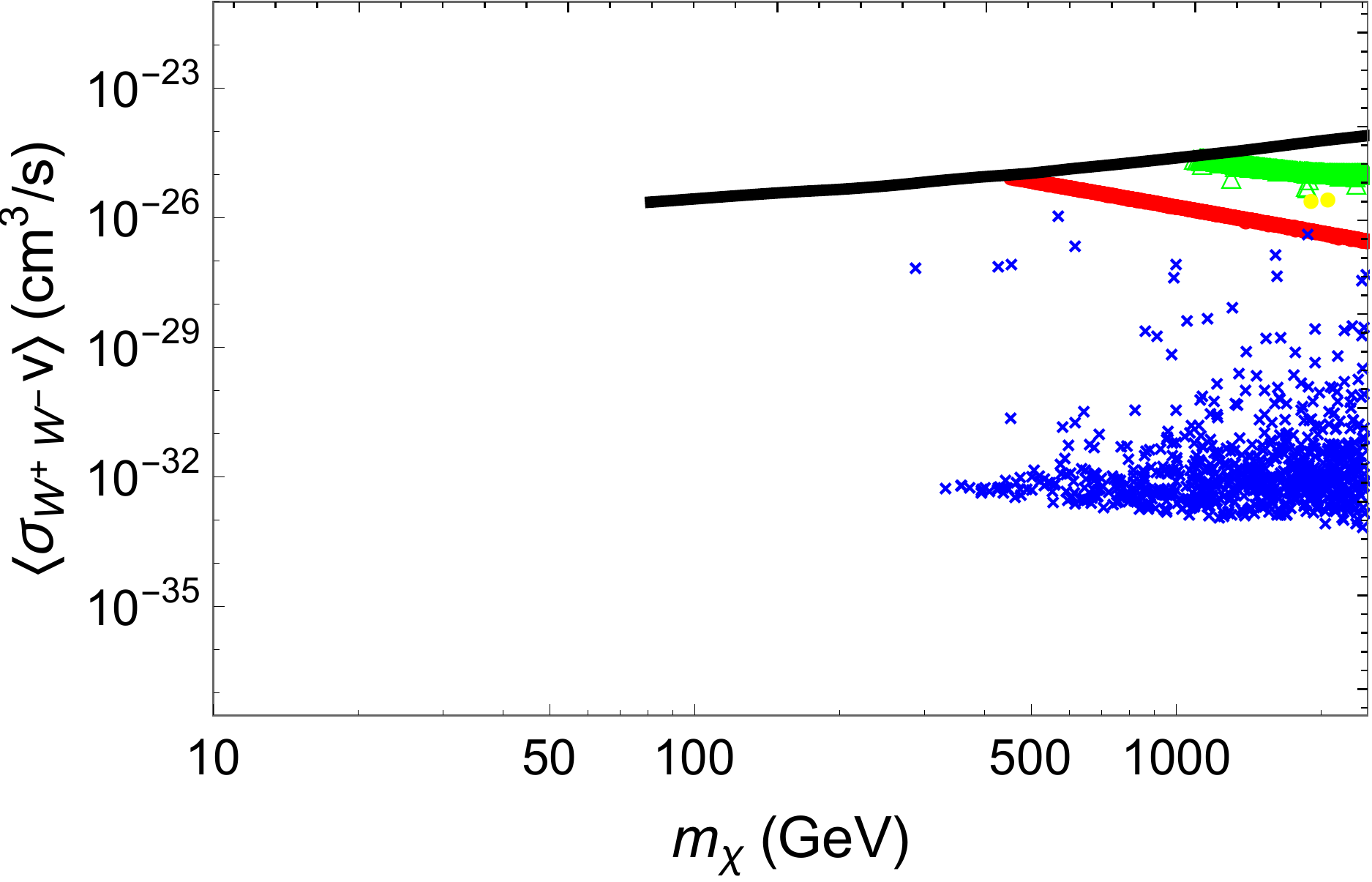}
}\\\subfigure[\ Fermi-LAT conststraint on $\chi^0 {\chi}^0\rightarrow b\bar{b}$]{
  \includegraphics[width=0.45\textwidth,height=0.13\textheight]{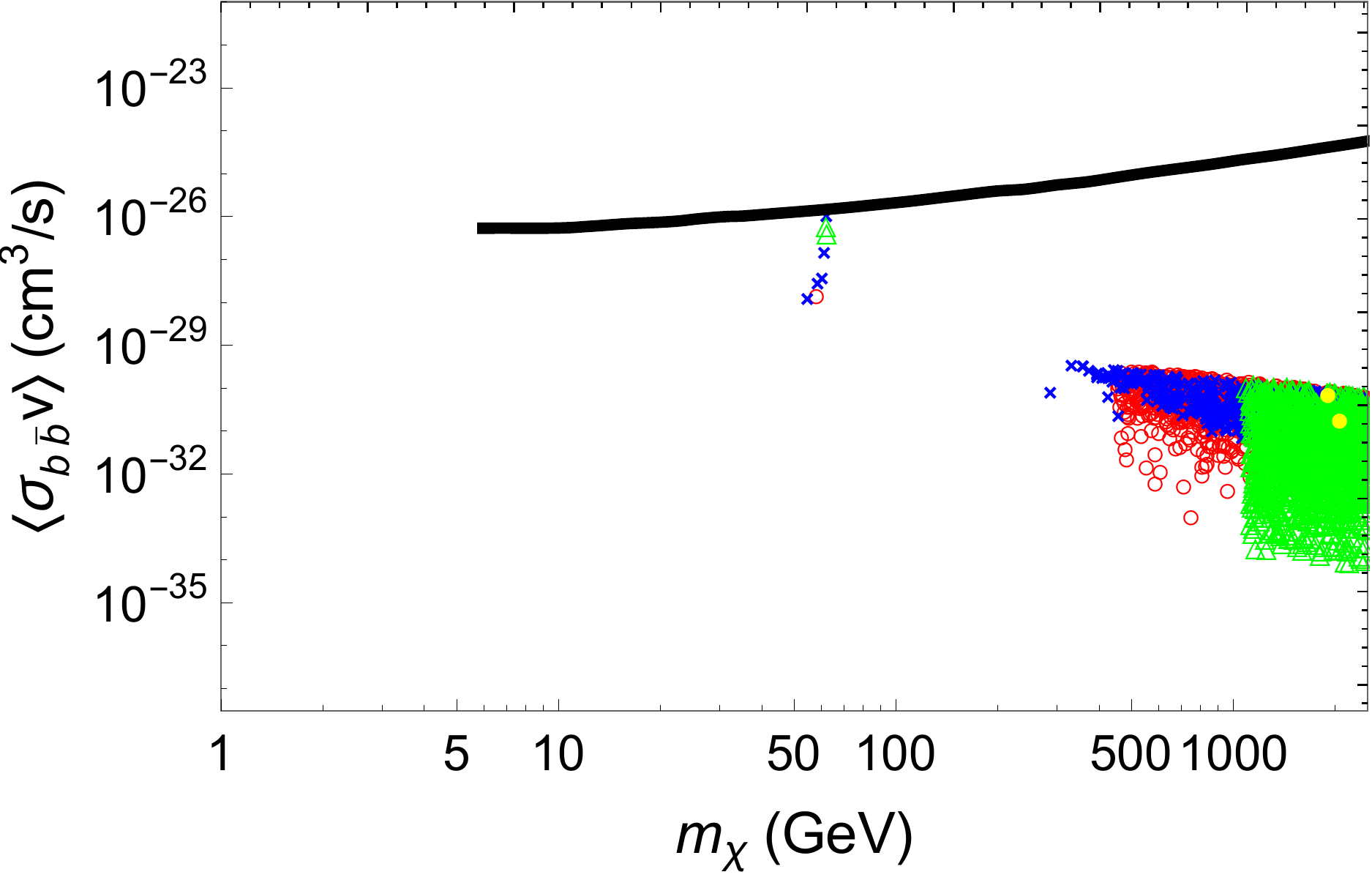}
}\
%\subfigure[\ Fermi-LAT conststraint on $\chi^0 \bar{\chi}^0\rightarrow u\bar{u}$]{
 % \includegraphics[width=0.45\textwidth,height=0.13\textheight]{11gGallowuu.pdf}
%}
\subfigure[\ Fermi-LAT conststraint on $\chi^0 {\chi}^0\rightarrow \tau^+\tau^-$]{
  \includegraphics[width=0.45\textwidth,height=0.13\textheight]{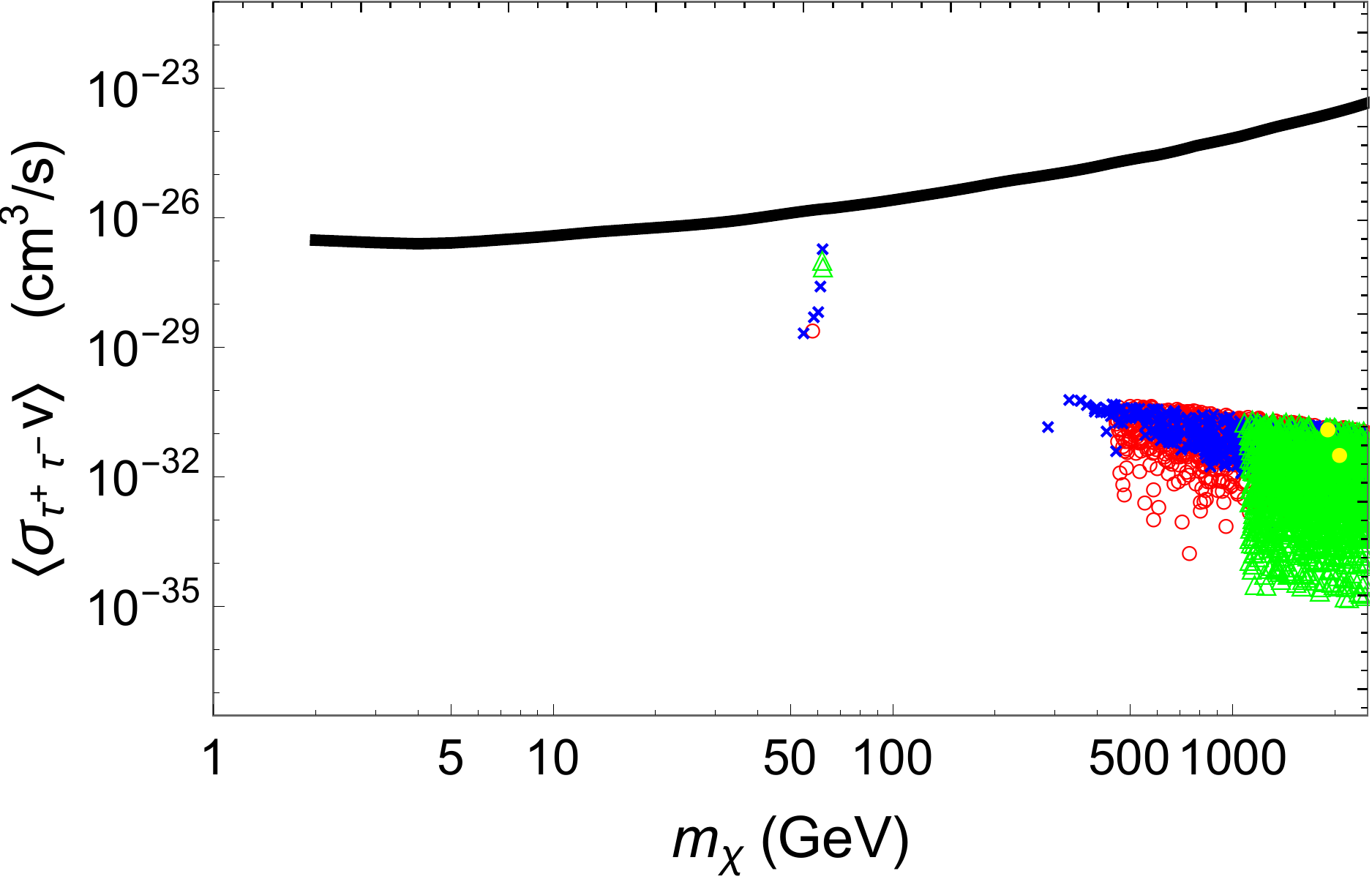}
}%\\\subfigure[\ Fermi-LAT conststraint on $\chi^0 \bar{\chi}^0\rightarrow \mu^+\mu^-$]{
  %\includegraphics[width=0.45\textwidth,height=0.13\textheight]{11iGallowmumu.pdf}
%}\subfigure[\ Fermi-LAT conststraint on $\chi^0 \bar{\chi}^0\rightarrow e^+e^-$]{
  %\includegraphics[width=0.45\textwidth,height=0.13\textheight]{11jGallowee.pdf}
%}
\caption{Results for allowed samples satisfying all constraints in the neutralino-like IV case
[{\color{red} $\circ$}:~higgsino-like,
{\color{blue} $\times$}:~bino-like,  {\color{green} $\triangle$}:~wino-like,
{\color{yellow} $\bullet$}:~mixed].}
\label{fig:allow neutralino-like IV}
\end{figure}
\vfill
\eject

In Figs.~\ref{fig:allow neutralino-like II}-\ref{fig:allow neutralino-like IV}, we redraw the Figs.~\ref{fig:neutralino-like II}-\ref{fig:neutralino-like IV} with the allowed samples, respectively.
As in the case of neutralino-like I, we still see that the direct detection of $\sigma^{SI}_N$ and the indirect detection of $\la\sigma(\chi{\chi}\rightarrow W^+W^-) v\ra$ are more accessible for DM searches in the near future. Hence we focus on these two and the relic density plots in these figures. Note that in the following discussion, we jump over the allowed outlier samples.

We see that most of $\tilde B$-like particles are ruled out by the $\Omega_{\chi} h^2$ constraint [see Figs.~\ref{fig:neutralino-like II}-\ref{fig:neutralino-like IV} (a)], followed by its complementary constraint of $\sigma^{SI}_N$ [see Figs.~\ref{fig:neutralino-like II}-\ref{fig:neutralino-like IV} (b)].
With the GUT relation, the cases of neutralino-like I ($\tan\beta=2$) and neutralino-like II ($\tan\beta=20$) have similar results which only the $\tilde B$-like particle with $m_{\chi}\gtrsim 1411, 1258$ GeV could be DM candidates, respectively (see Fig.~\ref{fig:allow neutralino-like I} and \ref{fig:allow neutralino-like II}). Without the GUT relation, the mass of the allowed $\tilde B$-like particle can lower down with $m_{\chi}\gtrsim 341, 288$ GeV in the cases of neutralino-like III and IV, respectively (see Figs.~\ref{fig:allow neutralino-like III} and \ref{fig:allow neutralino-like IV}).
Less than $0.3\%, 0.3\%$ and $0.9\%$ of $\tilde B$-like samples are allowed in the cases of neutralino-like I, II and III, respectively. Without GUT relation, the allowed $\tilde B$-like samples become sparse in the neutralino-like III case.
Note that the allowed $\tilde B$-like particles only attach to the LUX limit, in other words, the LUX limit is an active constraint and consequently only the experiments of SI DM-nucleus scattering are accessible to the DM searches in the near future.

The $\tilde H$- and $\tilde W$-like particles with $m_{\chi}\lesssim M_W$ are ruled out by the $\Omega_{\chi} h^2$ constraint [see Figs.~\ref{fig:neutralino-like II}-\ref{fig:neutralino-like IV} (a)], followed by the $\la\sigma(\chi{\chi}\rightarrow b{\bar b}) v\ra$ constraint [see Figs.~\ref{fig:neutralino-like II}-\ref{fig:neutralino-like IV} (g)], while the $\tilde H$- and $\tilde W$-like particles with $m_{\chi}\gtrsim M_W$ are mainly ruled out by the
$\la\sigma(\chi{\chi}\rightarrow W^+W^-) v\ra$ [see Figs.~\ref{fig:neutralino-like II}-\ref{fig:neutralino-like IV} (f)] and the $\sigma^{SI}_N$ constraints [see Figs.~\ref{fig:neutralino-like II}-\ref{fig:neutralino-like IV} (b)].
We see that the allowed lower mass bound of $\tilde H$-like DM candidates 
is about
%should be greater than 
$455$ GeV for all the neutralino-like cases (see Figs.~\ref{fig:allow neutralino-like II}-\ref{fig:allow neutralino-like IV}), namely, independent of the GUT and the $\tan\beta$ relations for the $\tilde H$-like particles, while the allowed lower mass bound of $\tilde W$-like DM candidates is about $1100$ GeV, which is independent of the $\tan\beta$ relation
in the neutralino-like III and IV cases (see Figs.~\ref{fig:allow neutralino-like III}-\ref{fig:allow neutralino-like IV}).
On the contrary to the $\tilde B$-like DM candidates, $\tilde H$- and $\tilde W$-like DM candidates can be accessible in the direct search of $\sigma^{SI}_N$ as well as the indirect search of $\la\sigma(\chi{\chi}\rightarrow W^+W^-) v\ra$ in the near future.
Therefore without considering the outlier samples, the allowed mass regions for $\tilde H$-like, $\tilde B$-like and $\tilde W$-like
in Figs.~\ref{fig:allow neutralino-like II}-\ref{fig:allow neutralino-like IV} can be understood.
On the other hand, we find that the allowed $\tilde H$-like particles are highly pure, as
$98\%$, $97\%$, $99\%$ and $99.9\%$ of them are in the states of $\eta_1$ or $\eta_2$ with the composition fraction greater than $90\%$ in the cases of neutralino-like I - IV respectively. However, only $39\%$, $5\%$, $55\%$ and $99\%$ of the allowed $\tilde B$-like particles are in the state of $\eta_3$ with the composition fraction greater than $90\%$ in the cases of neutralino-like I - IV respectively. That is because either the GUT relation or the $\tan\beta$ relation is imposed in the cases of neutralino-like I - III. As for the allowed $\tilde W$-like particles, $99.9\%$ and $99.5\%$ of them are in the state of $\eta_5$ with the composition fraction greater than $90\%$ in neutralino-like III - IV, respectively.

\subsection{Case B: Reduced case}

\begin{figure}[h!]
\centering
\captionsetup{justification=raggedright}
 \subfigure[\ Constraint on $\Omega^{\rm{obs}}_\chi$]{
  \includegraphics[width=0.45\textwidth,height=0.13\textheight]{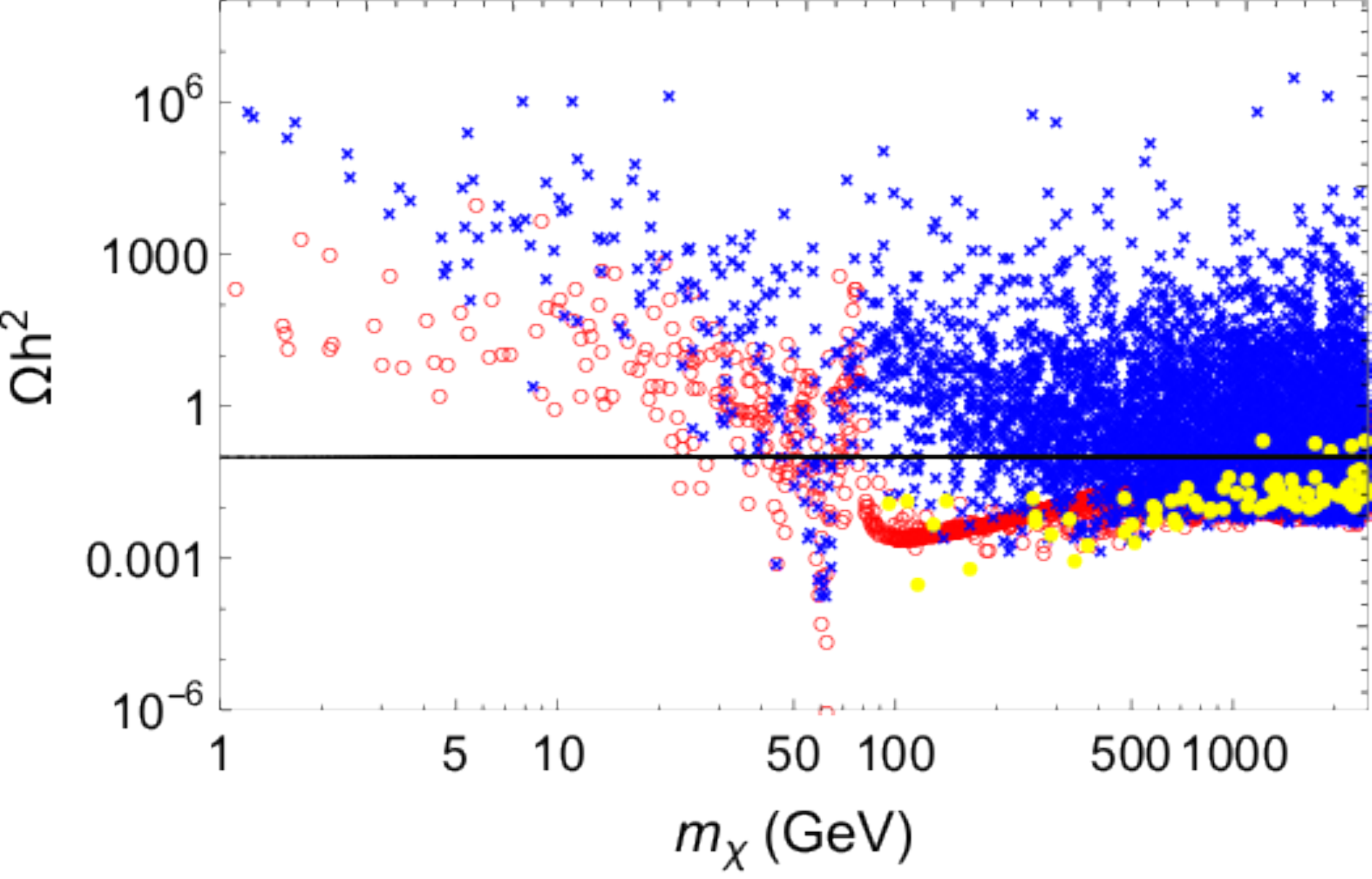}
}\subfigure[\ LUX constraint on $\sigma^{SI}$ with NB limit]{
  \includegraphics[width=0.45\textwidth,height=0.13\textheight]{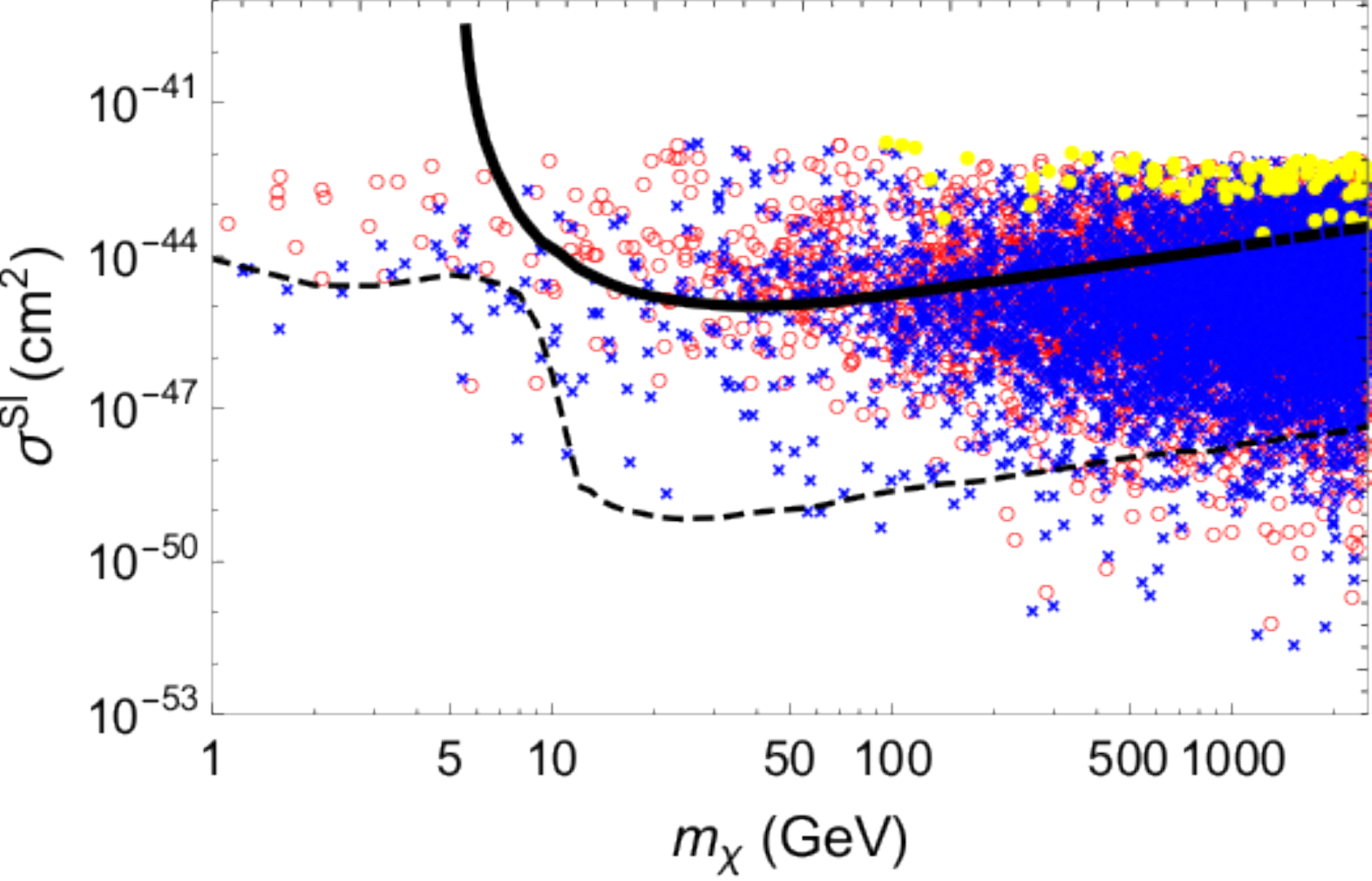}
}\\\subfigure[\ XENON100 constraint on $\sigma^{SD}_n$]{
  \includegraphics[width=0.45\textwidth,height=0.13\textheight]{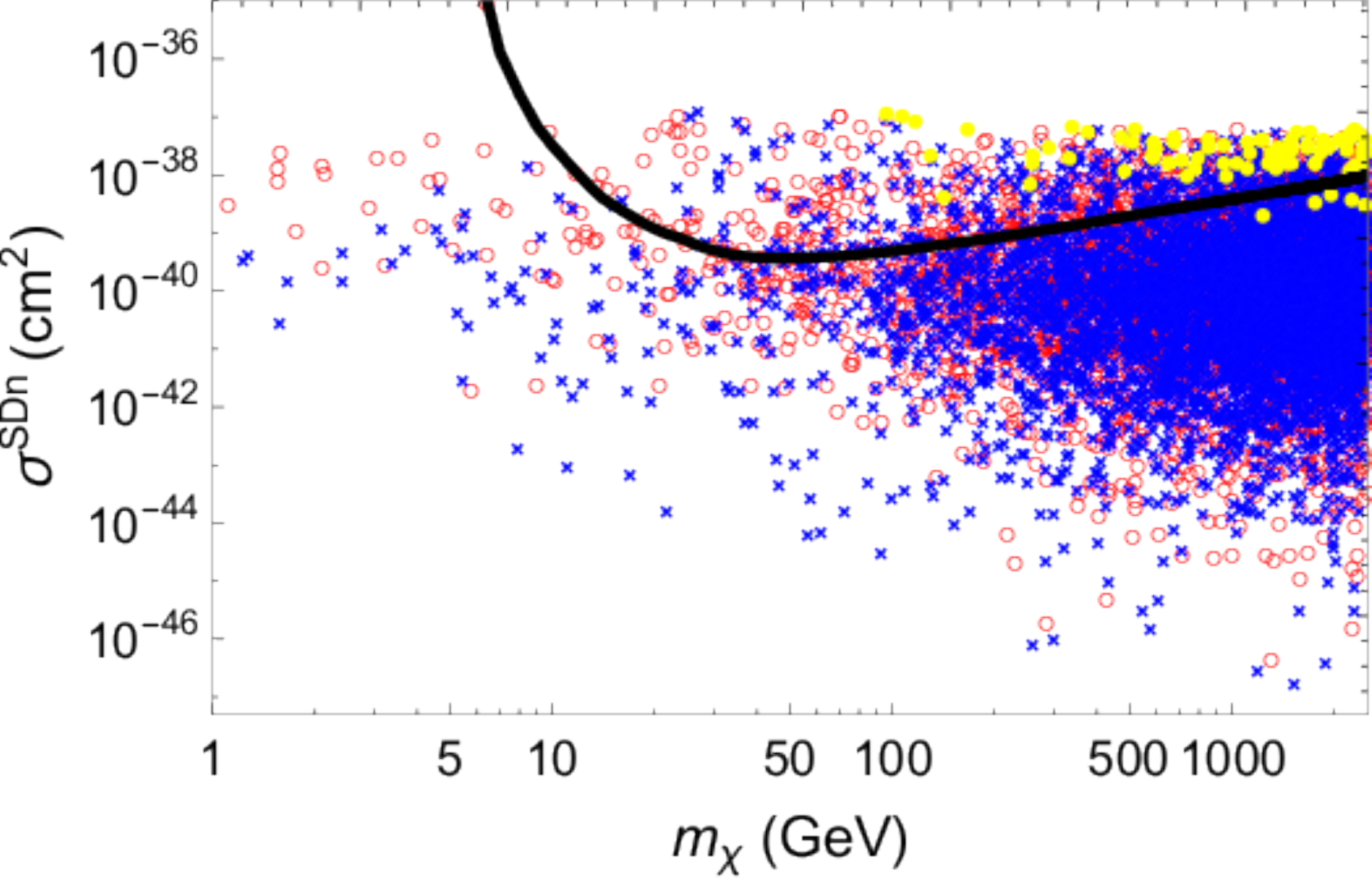}
}\subfigure[\ XENON100 constraint on $\sigma^{SD}_p$]{
  \includegraphics[width=0.45\textwidth,height=0.13\textheight]{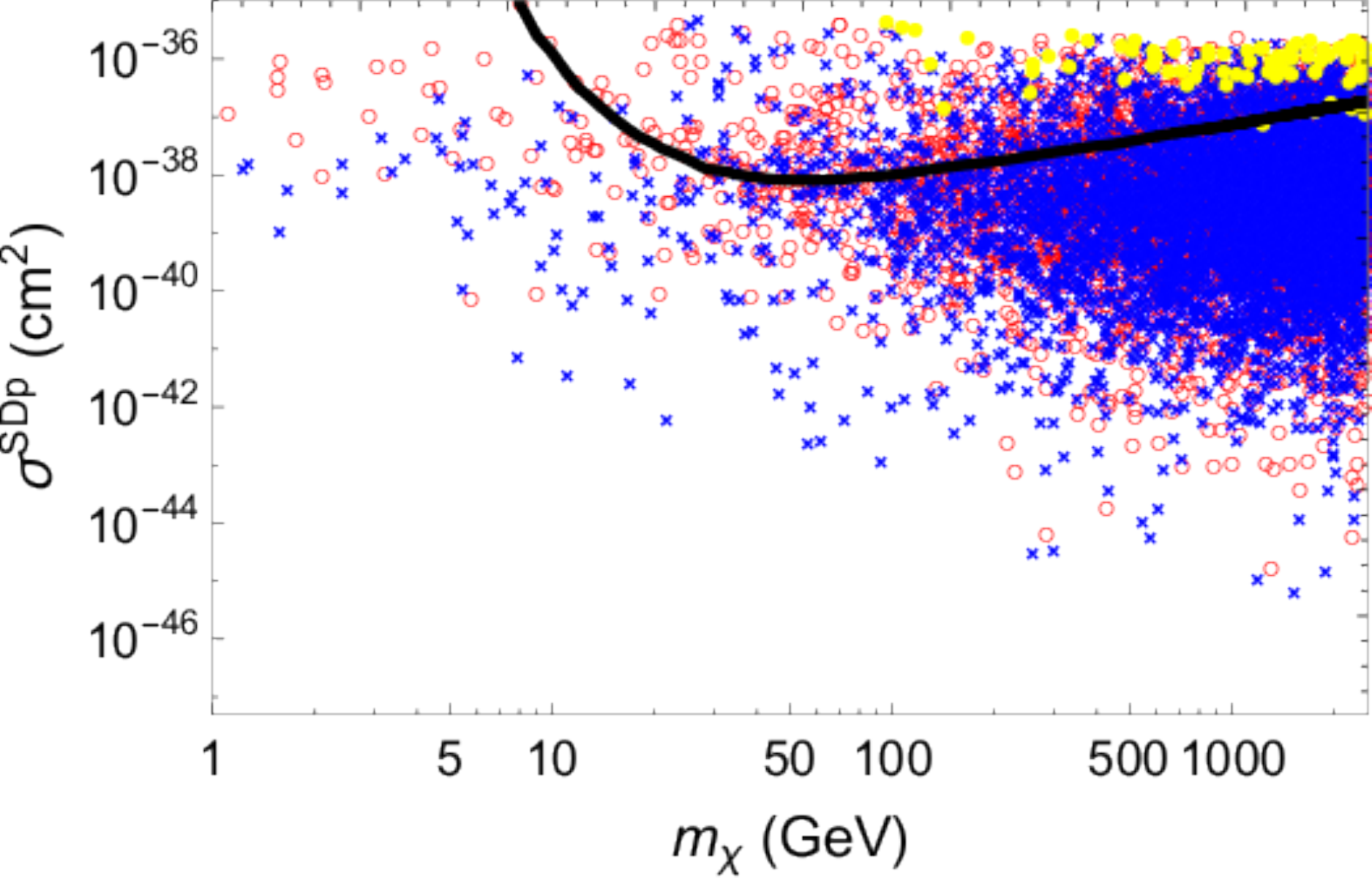}
}\\\subfigure[\ PICO-60 constraint on $\sigma^{SD}_p$]{
  \includegraphics[width=0.45\textwidth,height=0.13\textheight]{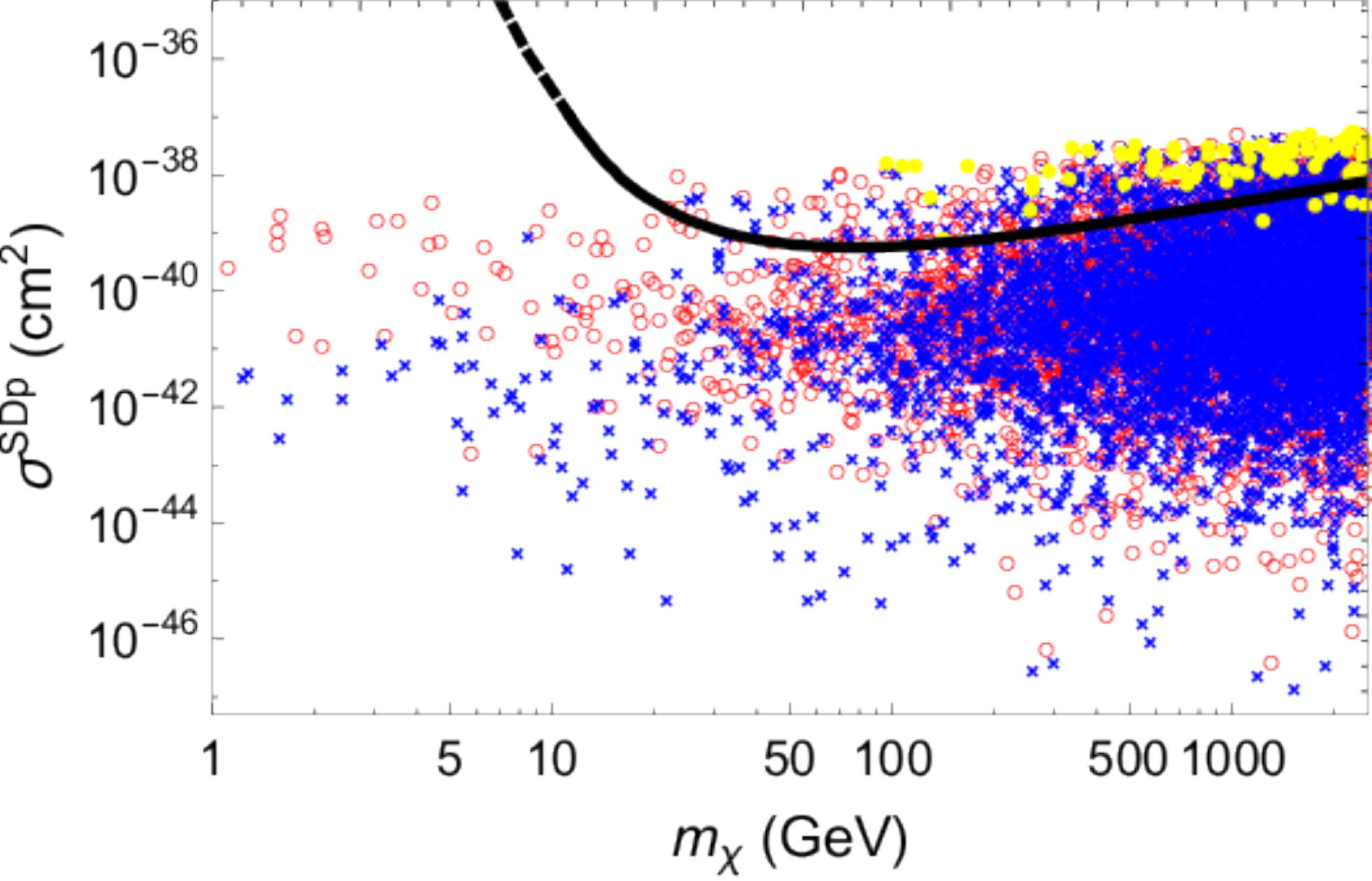}
}\subfigure[\ Fermi-LAT conststraint on $\chi^0 {\chi}^0\rightarrow W^+W^-$]{
  \includegraphics[width=0.45\textwidth,height=0.13\textheight]{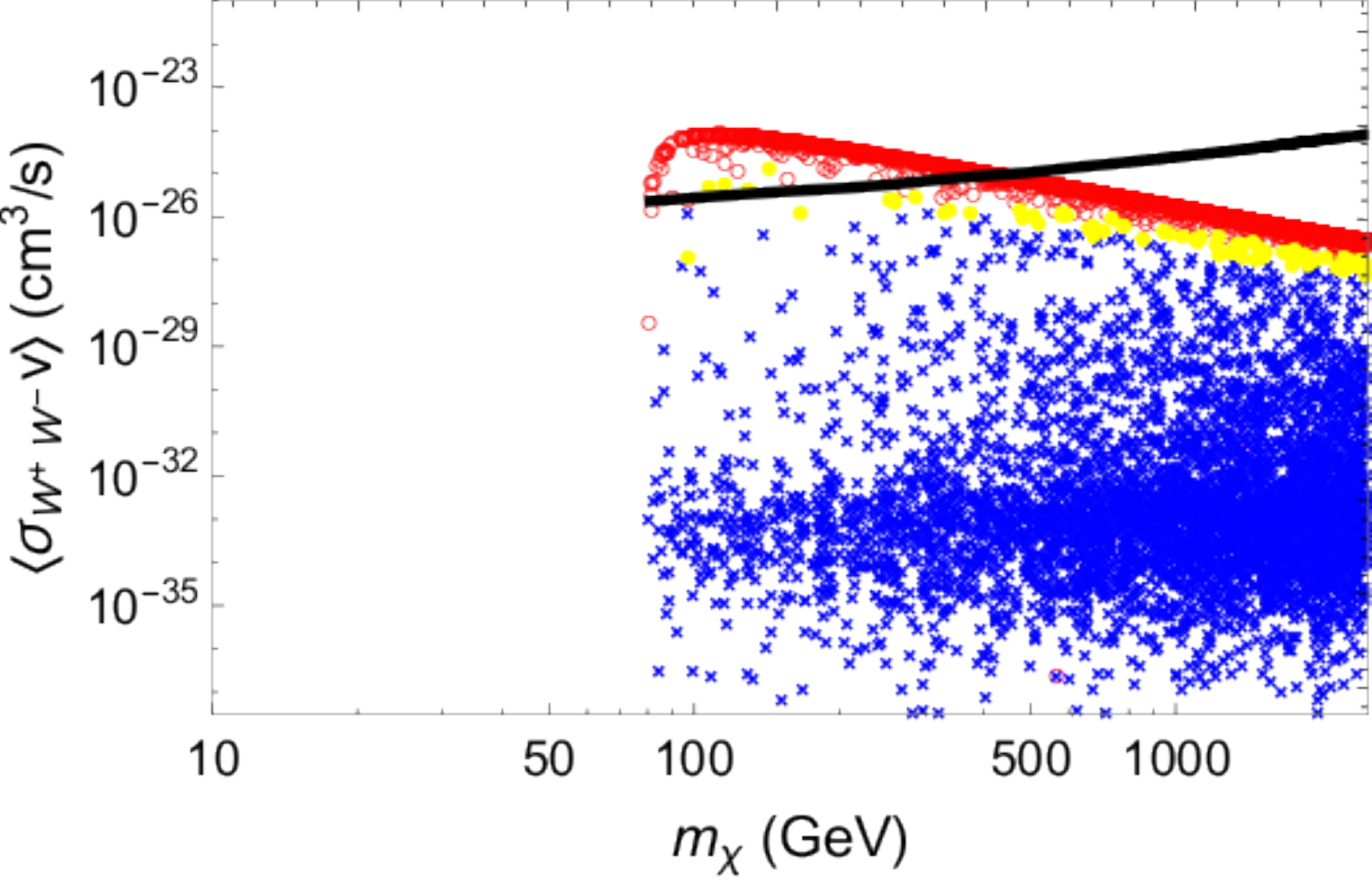}
}\\\subfigure[\ Fermi-LAT conststraint on $\chi^0 {\chi}^0\rightarrow b\bar{b}$]{
  \includegraphics[width=0.45\textwidth,height=0.13\textheight]{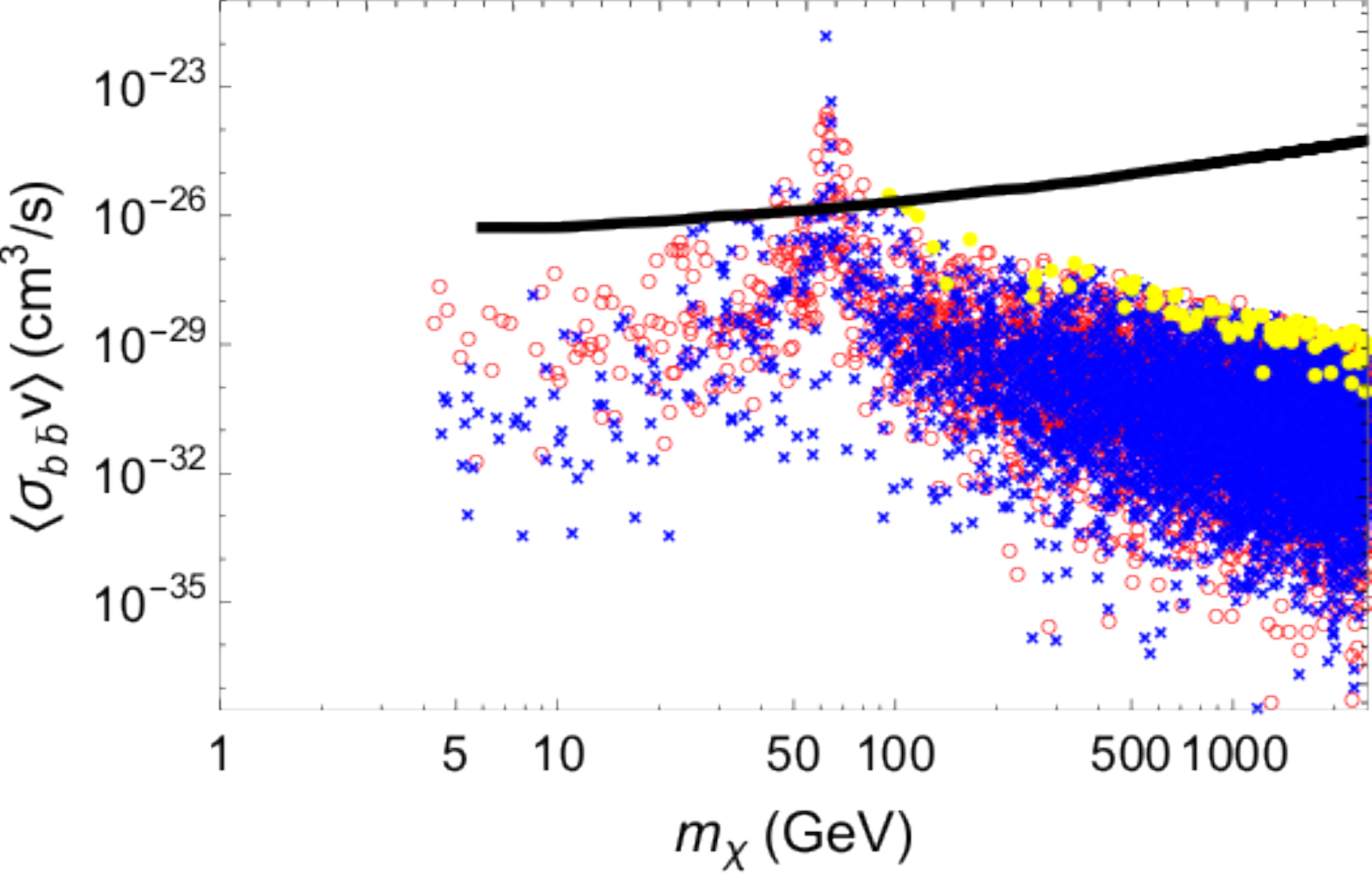}
}\
%\subfigure[\ Fermi-LAT conststraint on $\chi^0 \bar{\chi}^0\rightarrow u\bar{u}$]{
  %\includegraphics[width=0.45\textwidth,height=0.13\textheight]{12gCinduu2.pdf}
%}
\subfigure[\ Fermi-LAT conststraint on $\chi^0 {\chi}^0\rightarrow \tau^+\tau^-$]{
  \includegraphics[width=0.45\textwidth,height=0.13\textheight]{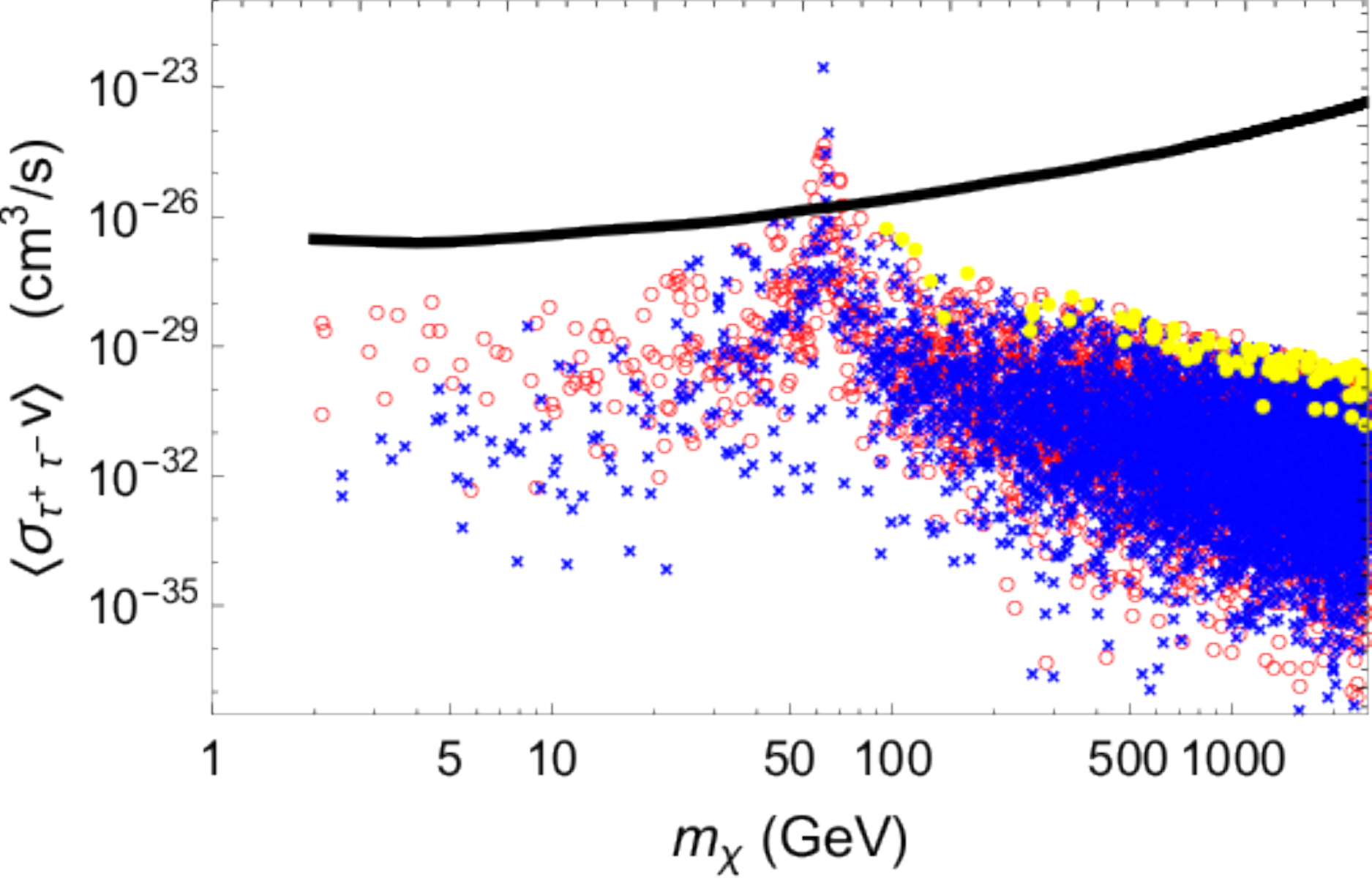}
}%\\\subfigure[\ Fermi-LAT conststraint on $\chi^0 \bar{\chi}^0\rightarrow \mu^+\mu^-$]{
 %\includegraphics[width=0.45\textwidth,height=0.13\textheight]{12iCindmumu2.pdf}
%}\subfigure[\ Fermi-LAT conststraint on $\chi^0 \bar{\chi}^0\rightarrow e^+e^-$]{
 %\includegraphics[width=0.45\textwidth,height=0.13\textheight]{12jCindee2.pdf}
%}
\caption{Results for all samples with constraints in the reduced  case
[{\color{red} $\circ$}:~higgsino-like,~
{\color{blue} $\times$}:~bino-like,
{\color{yellow} $\bullet$}:~mixed].}
\label{fig:Reduced}
\end{figure}

For the reduced case, it contains a minimal particle content $\eta_{1,2,3}$ ($\tilde H$- and $\tilde B$-like) with 4 free parameters $\mu_{1,2}$ and $g_{3,4}$. Since  $\eta_5$ ($\tilde W$-like) particles are absent, it is natural that the $\tilde W$-like particles do not appear in this case. We show the results in Fig.~\ref{fig:Reduced} with all samples.
As in the neutralino-like cases, we show that all values of $\la\sigma_{W^+W^-} v\ra$ for the $ \tilde B$-like particles should be less than those values for the $\tilde H$-like and the mixed particles in Fig.~\ref{fig:Reduced}(f) which is consistent with the fact that a $\tilde B$-like DM pair does not contribute to $s$-wave scattering amplitude.

As in the neutralino-like cases, 
we do not show the highly helicity suppressed
plots of $\la\sigma_{u{\bar u}} v\ra$, $\la\sigma_{\mu^+\mu^-} v\ra$ and $\la\sigma_{e^+e^-} v\ra$.
The reduced case contains more free parameters than the cases of neutralino-like I, II and III, so that it can have a wider spread in each scatter plot than the cases of neutralino-like I, II and III as the $\tan\beta$ relations are not imposed.
Therefore, although most $\tilde B$-like samples are ruled out by the $\Omega_{\chi} h^2$ constraint,
we can still have plenty of $\tilde B$-like particles being allowed.
As in the neutralino-like IV case, more $\tilde B$-like particles have lower values in $\Omega_{\chi} h^2$ and more $\tilde H$-like particles have larger values in $\sigma^{SI}_N$
% and more $\tilde H$-like particles have larger values in $\sigma^{SI}$ 
[see Fig.~\ref{fig:Reduced}(a,b)].
% and (b)].
%However, some $\tilde H$-like particles can also have smaller values in $\sigma^{SI}$ than those in the neutralino-like IV case.
Consequently, more $\tilde B$-like particles (relative to neutralino-like I, II and III) and less $\tilde H$-like particles (relative to neutralino-like I) are allowed.
% and more $\tilde H$-like particles (relative to neutralino-like IV) 
%are allowed and less $\tilde H$-like particles are allowed .
We find that about $48\%$ of $\tilde H$-like particles and $23\%$ of $\tilde B$-like particles could be DM candidates.

\begin{figure}[h!]
\centering
\captionsetup{justification=raggedright}
 \subfigure[\ Constraint on $\Omega^{\rm{obs}}_\chi$]{
  \includegraphics[width=0.45\textwidth,height=0.13\textheight]{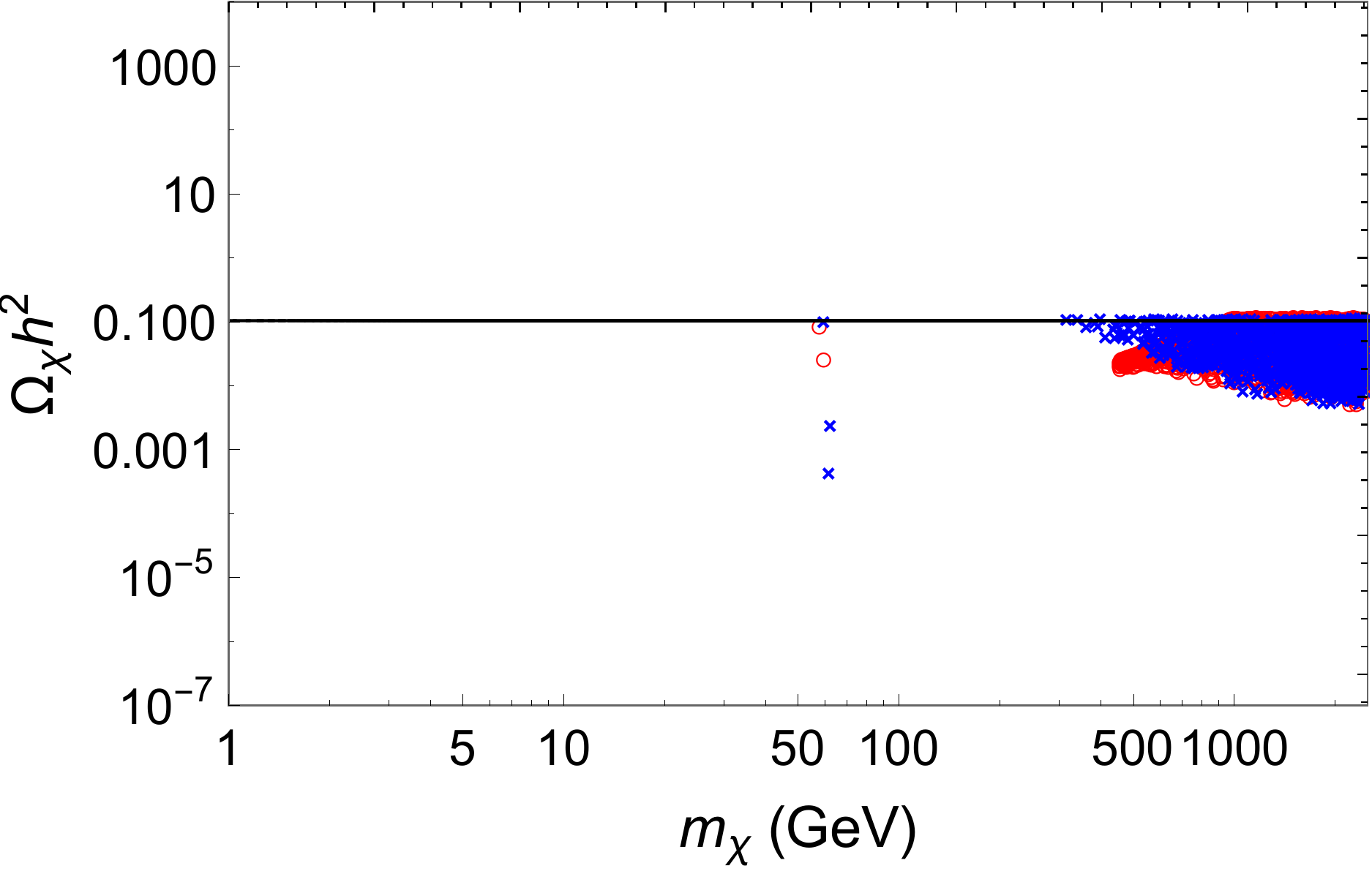}
}\subfigure[\ LUX constraint on $\sigma^{SI}$ with NB limit]{
  \includegraphics[width=0.45\textwidth,height=0.13\textheight]{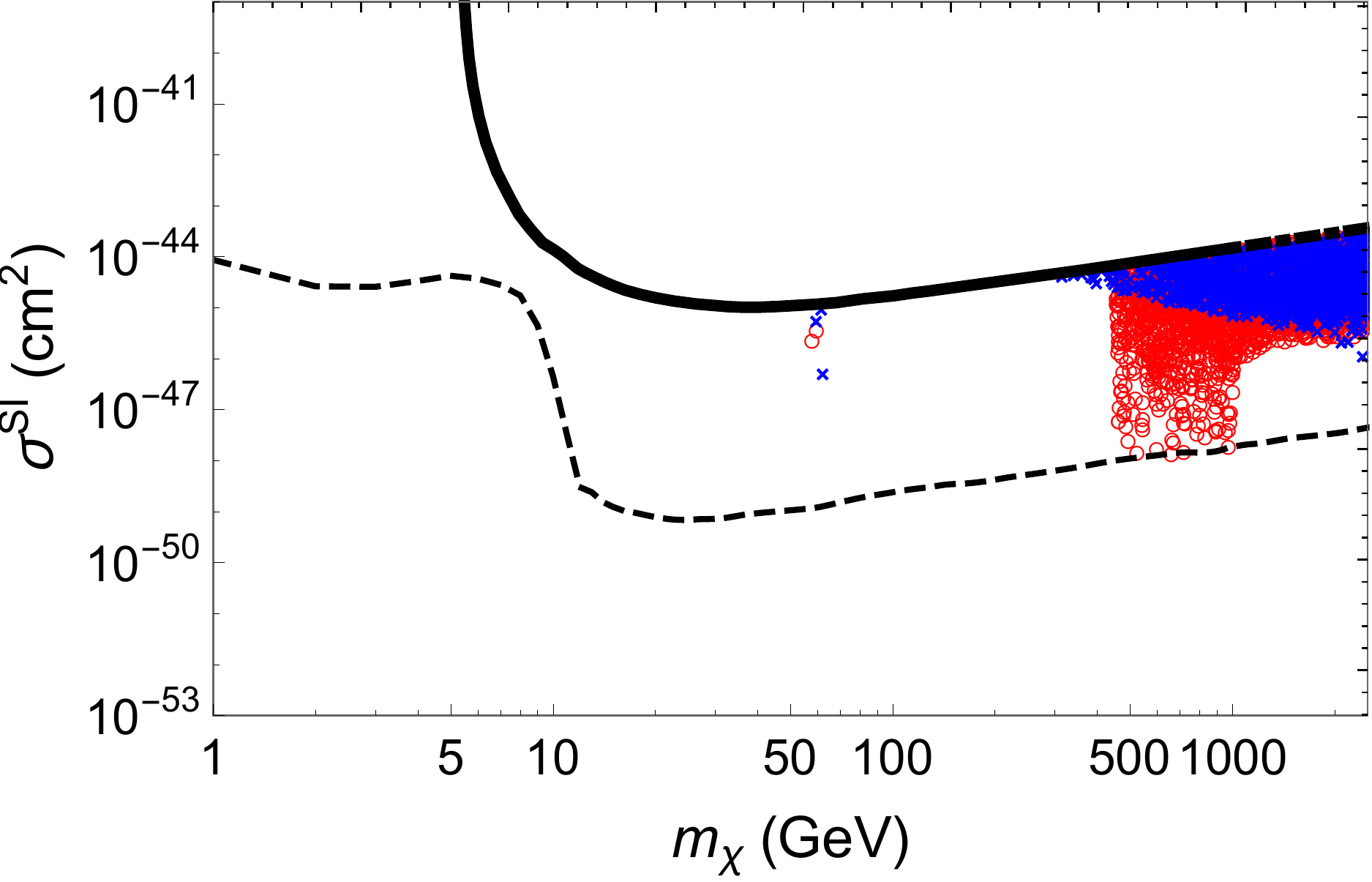}
}\\\subfigure[\ XENON100 constraint on $\sigma^{SD}_n$]{
  \includegraphics[width=0.45\textwidth,height=0.13\textheight]{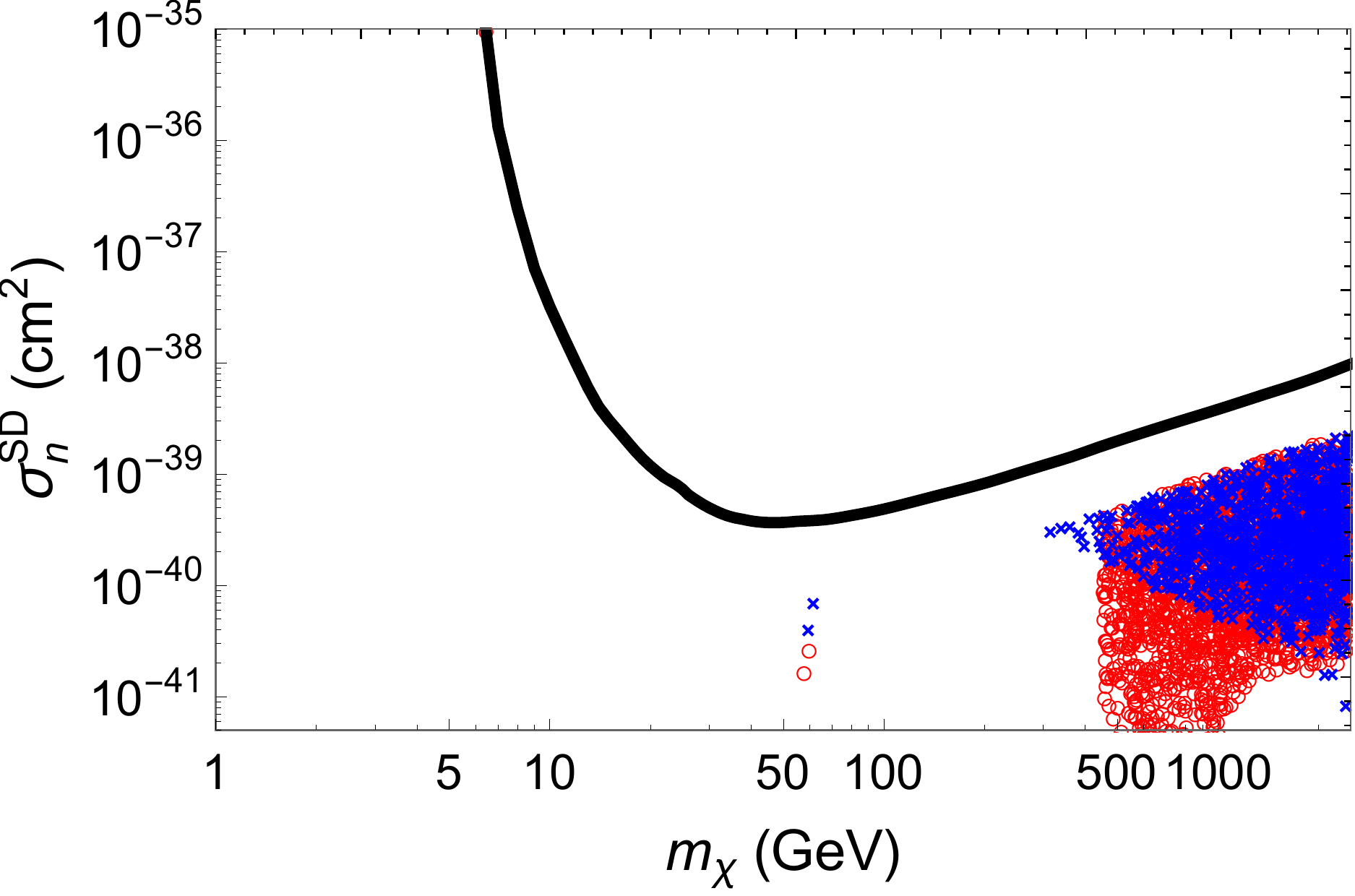}
}\subfigure[\ XENON100 constraint on $\sigma^{SD}_p$]{
  \includegraphics[width=0.45\textwidth,height=0.13\textheight]{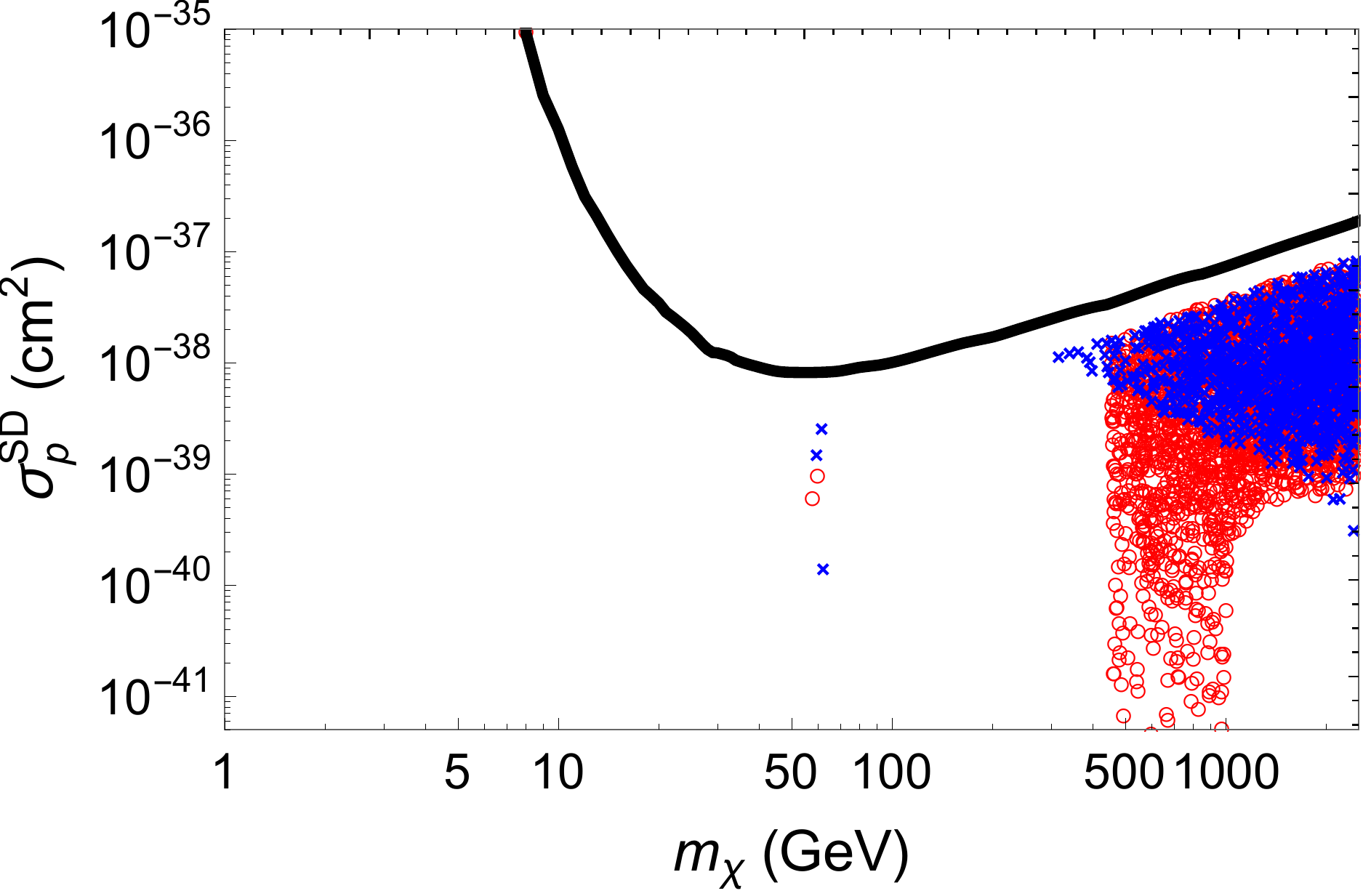}
}\\\subfigure[\ PICO-60 constraint on $\sigma^{SD}_p$]{
  \includegraphics[width=0.45\textwidth,height=0.13\textheight]{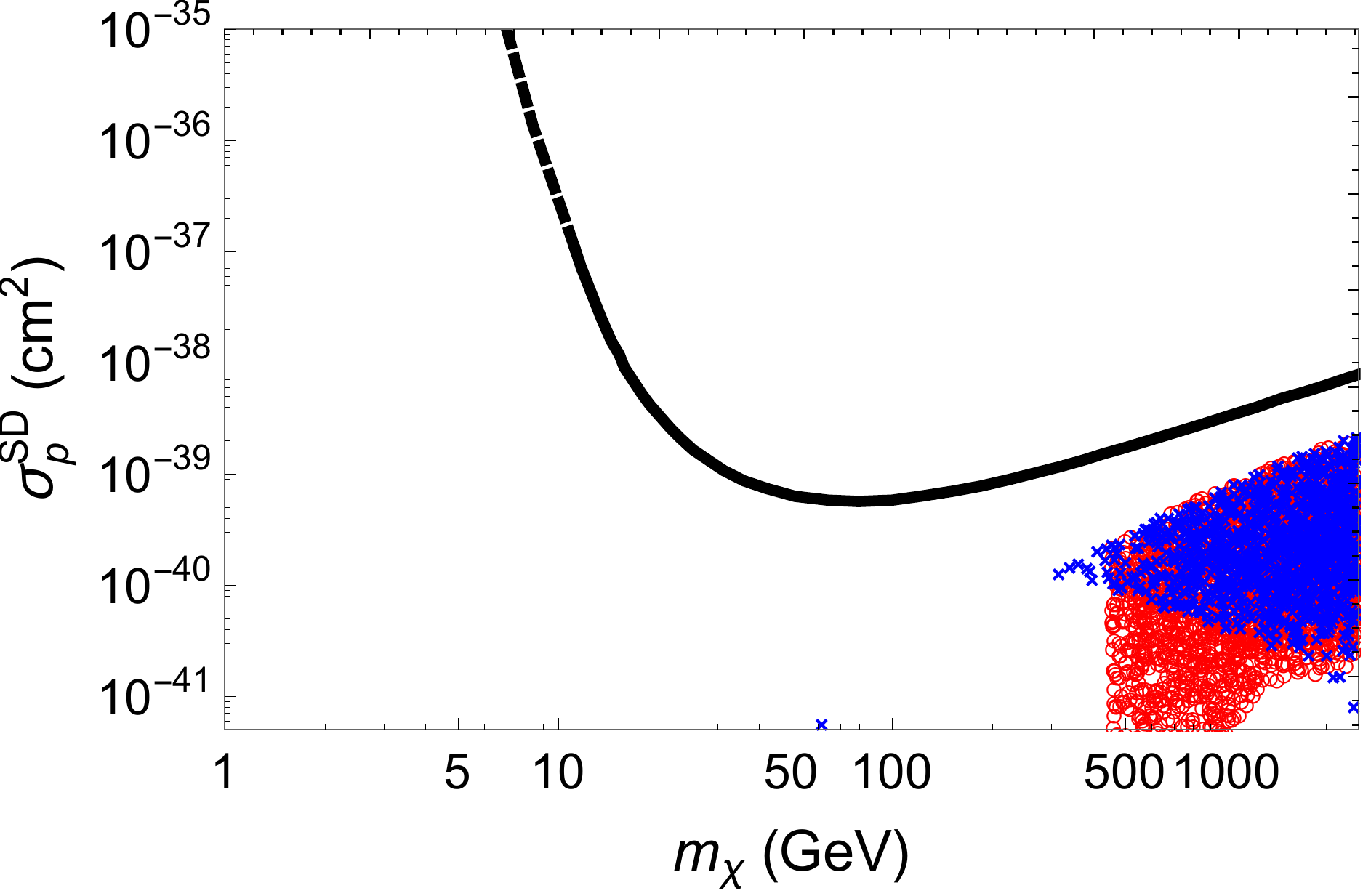}
}\subfigure[\ Fermi-LAT conststraint on $\chi^0 {\chi}^0\rightarrow W^+W^-$]{
  \includegraphics[width=0.45\textwidth,height=0.13\textheight]{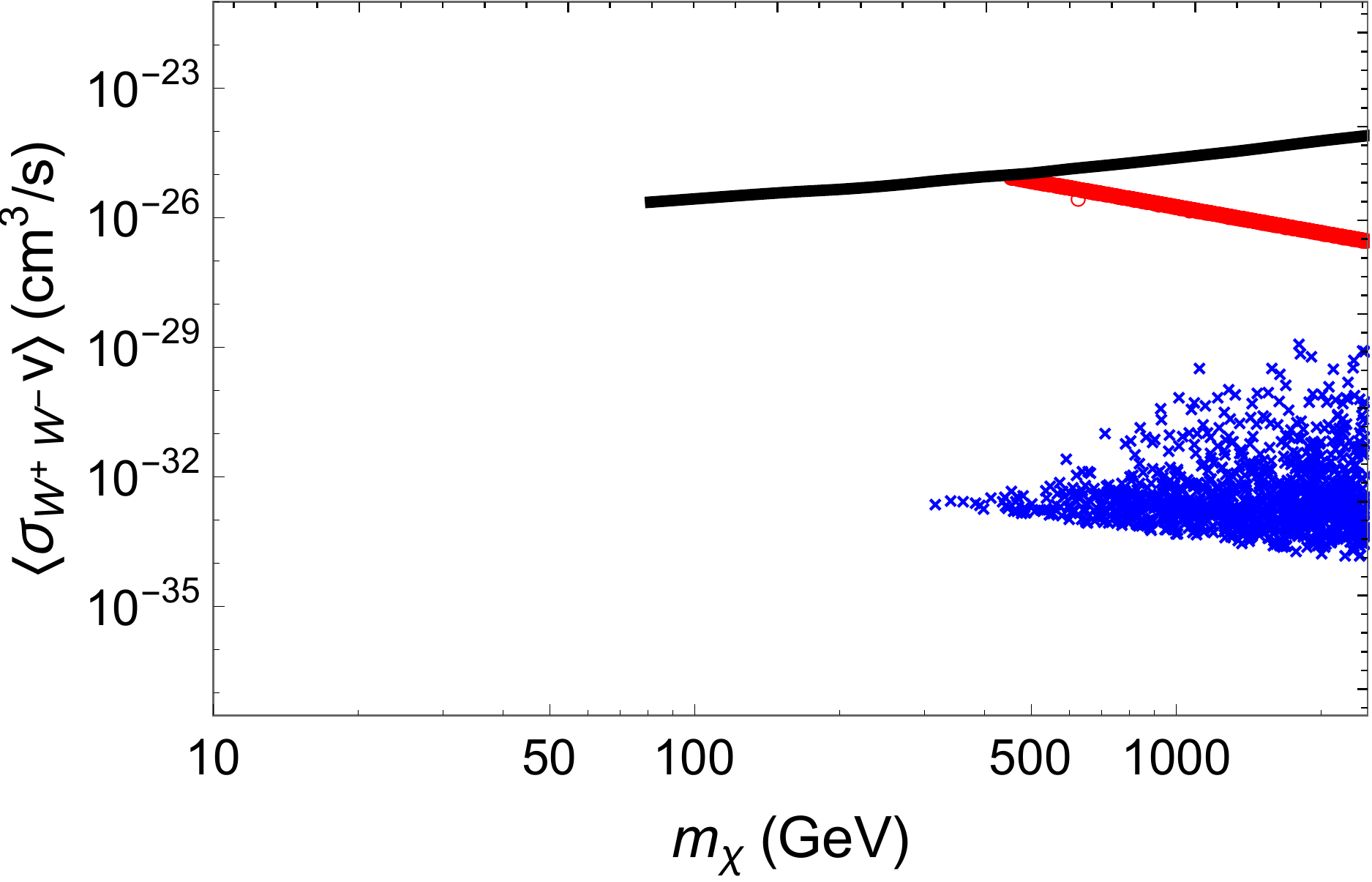}
}\\\subfigure[\ Fermi-LAT conststraint on $\chi^0 {\chi}^0\rightarrow b\bar{b}$]{
  \includegraphics[width=0.45\textwidth,height=0.13\textheight]{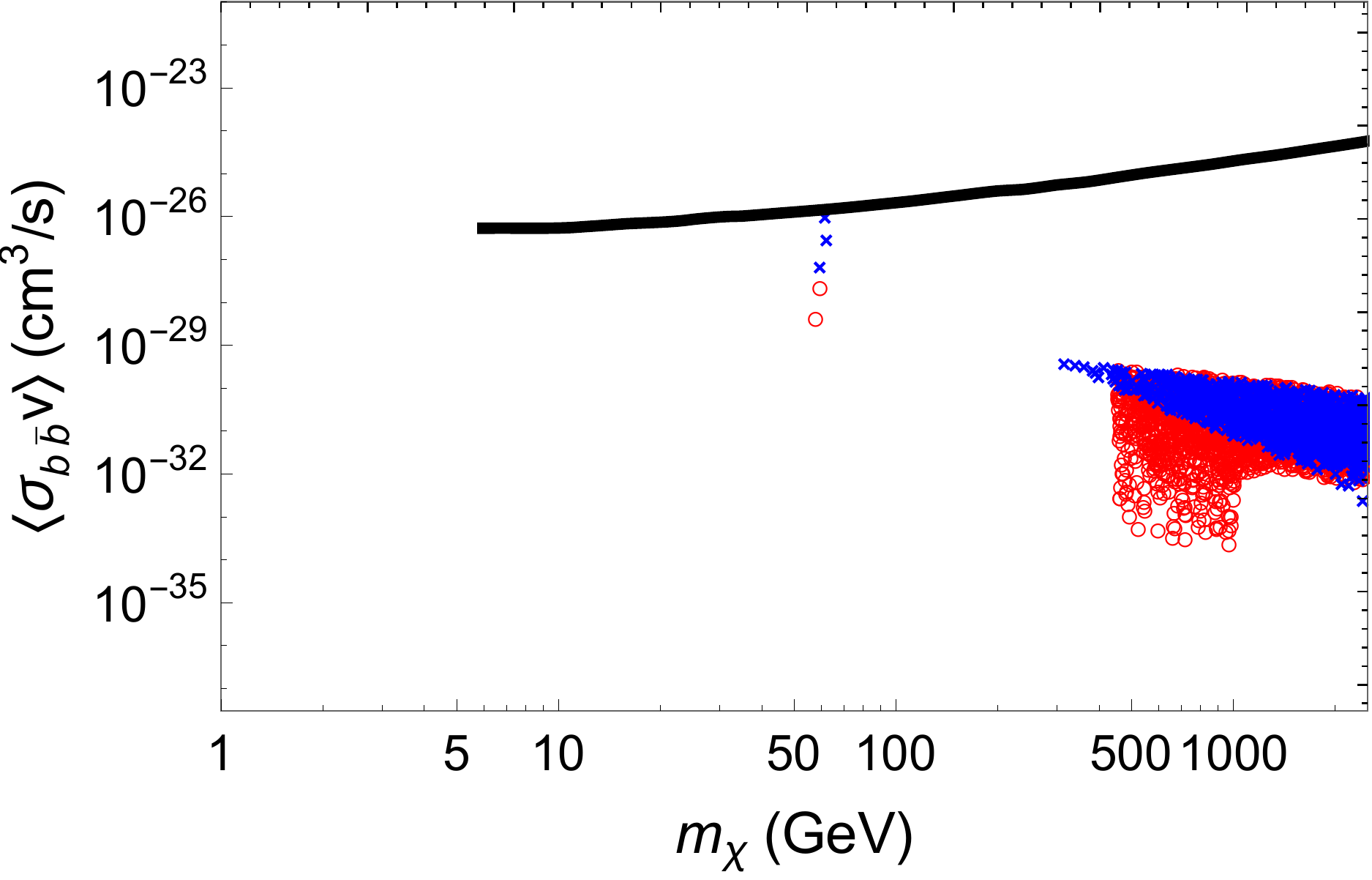}
}\
%\subfigure[\ Fermi-LAT conststraint on $\chi^0 \bar{\chi}^0\rightarrow u\bar{u}$]{
  %\includegraphics[width=0.45\textwidth,height=0.13\textheight]{13gCallowuu.pdf}
%}
\subfigure[\ Fermi-LAT conststraint on $\chi^0 {\chi}^0\rightarrow \tau^+\tau^-$]{
  \includegraphics[width=0.45\textwidth,height=0.13\textheight]{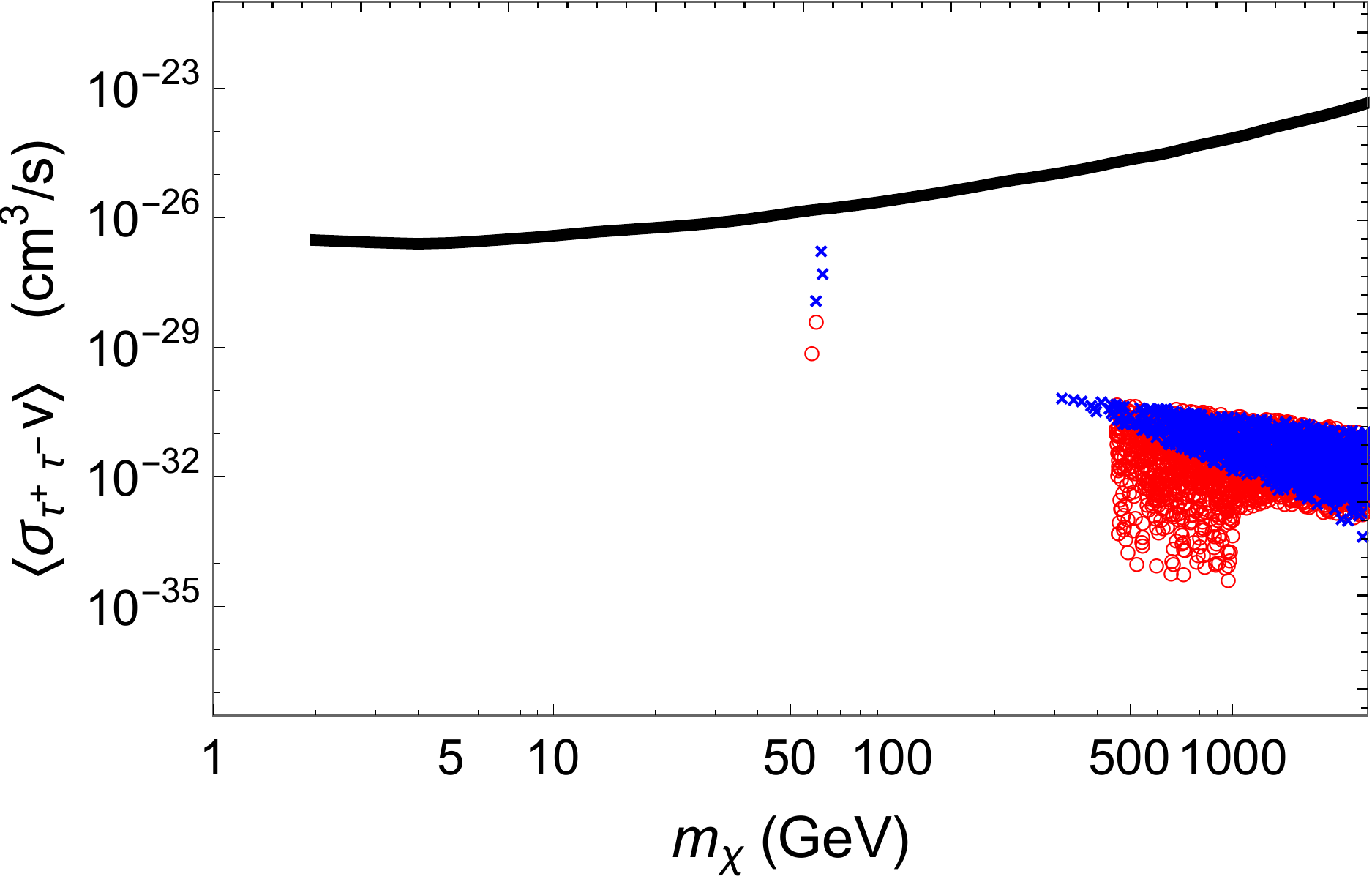}
}%\\\subfigure[\ Fermi-LAT conststraint on $\chi^0 \bar{\chi}^0\rightarrow \mu^+\mu^-$]{
  %\includegraphics[width=0.45\textwidth,height=0.13\textheight]{13iCallowmumu.pdf}
%}\subfigure[\ Fermi-LAT conststraint on $\chi^0 \bar{\chi}^0\rightarrow e^+e^-$]{
  %\includegraphics[width=0.45\textwidth,height=0.13\textheight]{13jCallowee.pdf}
%}
\caption{Results for allowed samples satisfying all constraints in the reduced case      
[{\color{red} $\circ$}:~higgsino-like,
{\color{blue} $\times$}:~bino-like,   
{\color{yellow} $\bullet$}:~mixed].}
\label{fig:allow Reduced}
\end{figure}

We redraw Fig.~\ref{fig:Reduced} in Fg.~\ref{fig:allow Reduced} but with the allowed samples only.
As in the neutralino-like cases, the direct detection of $\sigma^{SI}_N$ and the indirect detection of $\la\sigma(\chi{\chi}\rightarrow W^+W^-) v\ra$ are more accessible for DM searches in the near future. Similarly, the $\tilde B$-like particles can be sensitively detected only through the experiments of SI DM-nucleus scattering, while the $\tilde H$-like particles can be sensitively detected through both the direct search in the SI experiments of DM-nucleus scattering and the indirect search in the observation of DM annihilation to $W^+W^-$ channel in the near future.
Comparing Figs.~\ref{fig:allow neutralino-like I}, \ref{fig:allow neutralino-like II}-\ref{fig:allow neutralino-like IV} and \ref{fig:allow Reduced},
we see that this case is closer to the neutralino-like IV case, but without $\tilde W$-like particles.
%As noted 
Despite of the fact that most of  $\tilde B$-like particles are ruled out by the $\Omega_{\chi} h^2$ constraint, and further by LUX $\sigma^{SI}$ constraint, 
more allowed $\tilde B$-like particles can lower down the allowed mass range of $\tilde B$-like particles from $m_{\chi} \gtrsim 1$ TeV (as in the cases of neutralino-like I and II without the GUT relation) to $m_{\chi} \geq 317$ GeV.
On the other hand, the $\tilde H$-like particles with $m_{\chi} \lesssim M_W$ are ruled out by the relic density and the Fermi-LAT $\la\sigma (\chi{\chi}\rightarrow b\bar b) v\ra$ constraints, while the $\tilde H$-like particles with $m_{\chi} > M_W$ are subjected to the Fermi-LAT $\la\sigma (\chi{\chi}\rightarrow W^+W^-) v\ra$ and the LUX $\sigma^{SI}_N$ constraints, so that only the $\tilde H$-like particles with $m_{\chi} \gtrsim 454$ GeV could be the DM candidates.
We also find that the allowed $\tilde H$- and ${\tilde B}$-like particles are highly pure, as
$99.9\%$ of both $\tilde H$- and ${\tilde B}$-like particles are in the states of $\eta_{1,2}$ and $\eta_3$, respectively, with their composition fractions greater than $90\%$.

\vfill
\eject

\vfill
\eject

\subsection{Case C: Extended case}

\begin{figure}[h!]
\centering
\captionsetup{justification=raggedright}
 \subfigure[\ Constraint on $\Omega^{\rm{obs}}_\chi$]{
  \includegraphics[width=0.45\textwidth,height=0.13\textheight]{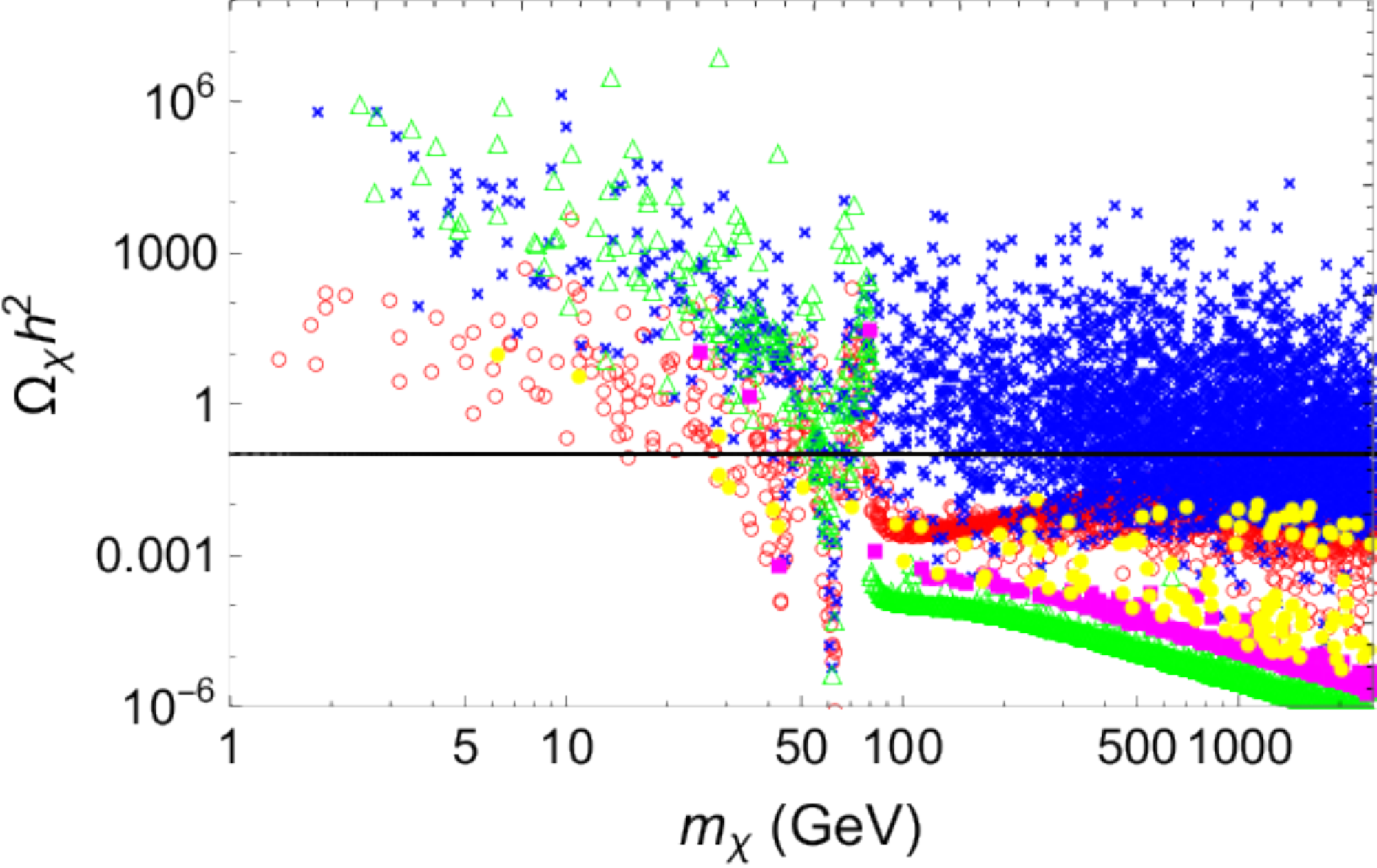}
}\subfigure[\ LUX constraint on $\sigma^{SI}$ with NB limit]{
  \includegraphics[width=0.45\textwidth,height=0.13\textheight]{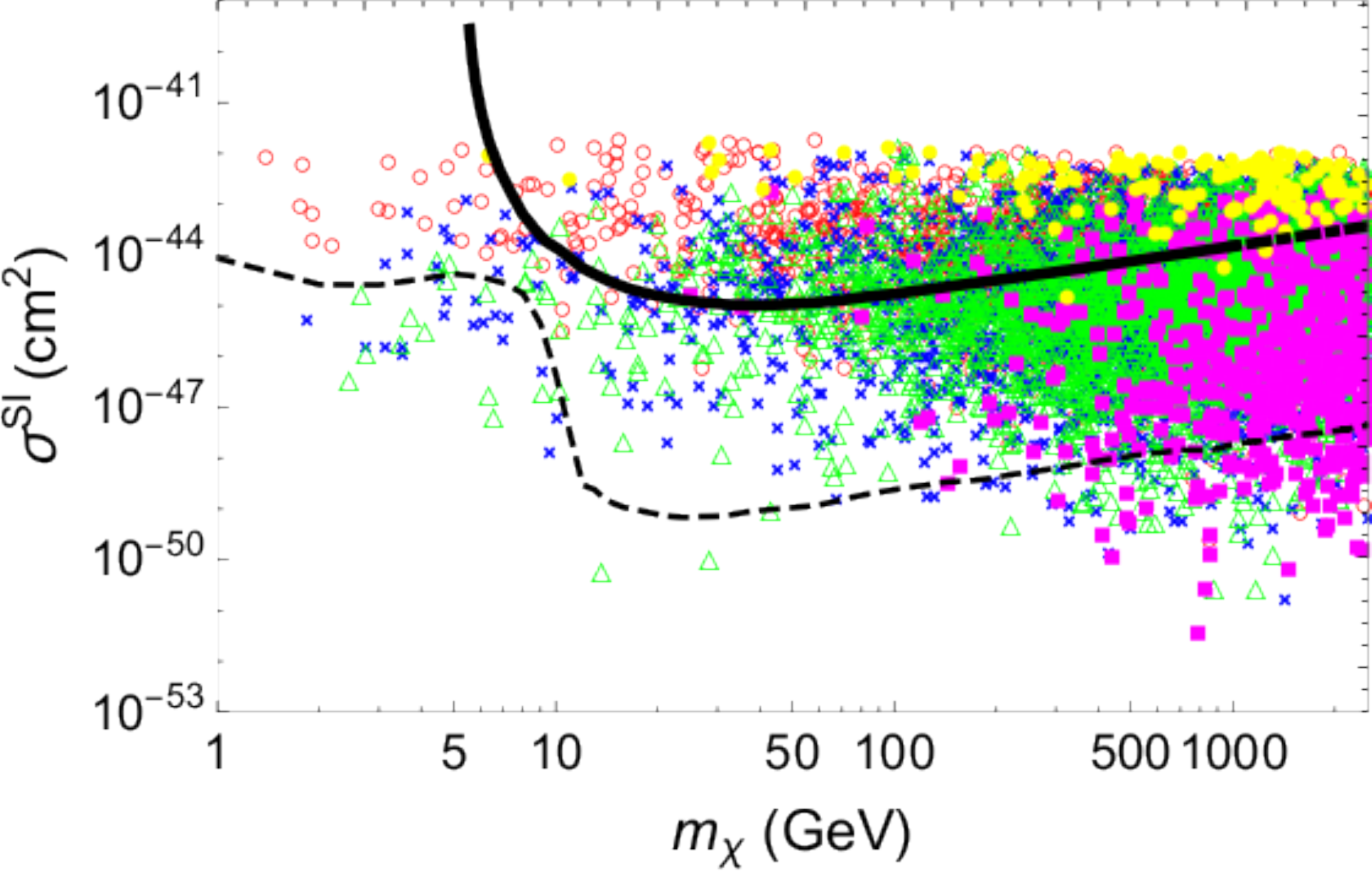}
}\\\subfigure[\ XENON100 constraint on $\sigma^{SD}_n$]{
  \includegraphics[width=0.45\textwidth,height=0.13\textheight]{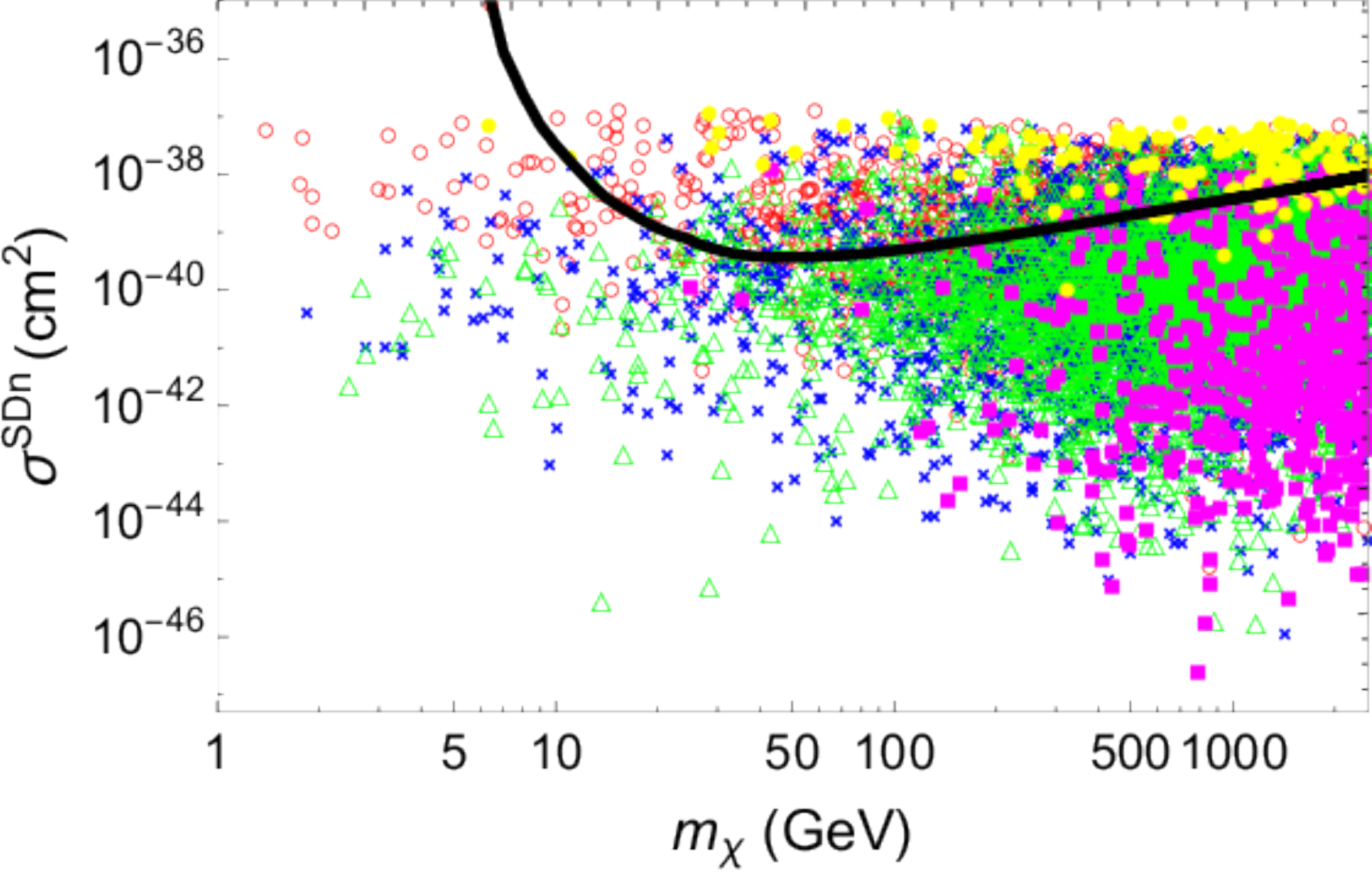}
}\subfigure[\ XENON100 constraint on $\sigma^{SD}_p$]{
  \includegraphics[width=0.45\textwidth,height=0.13\textheight]{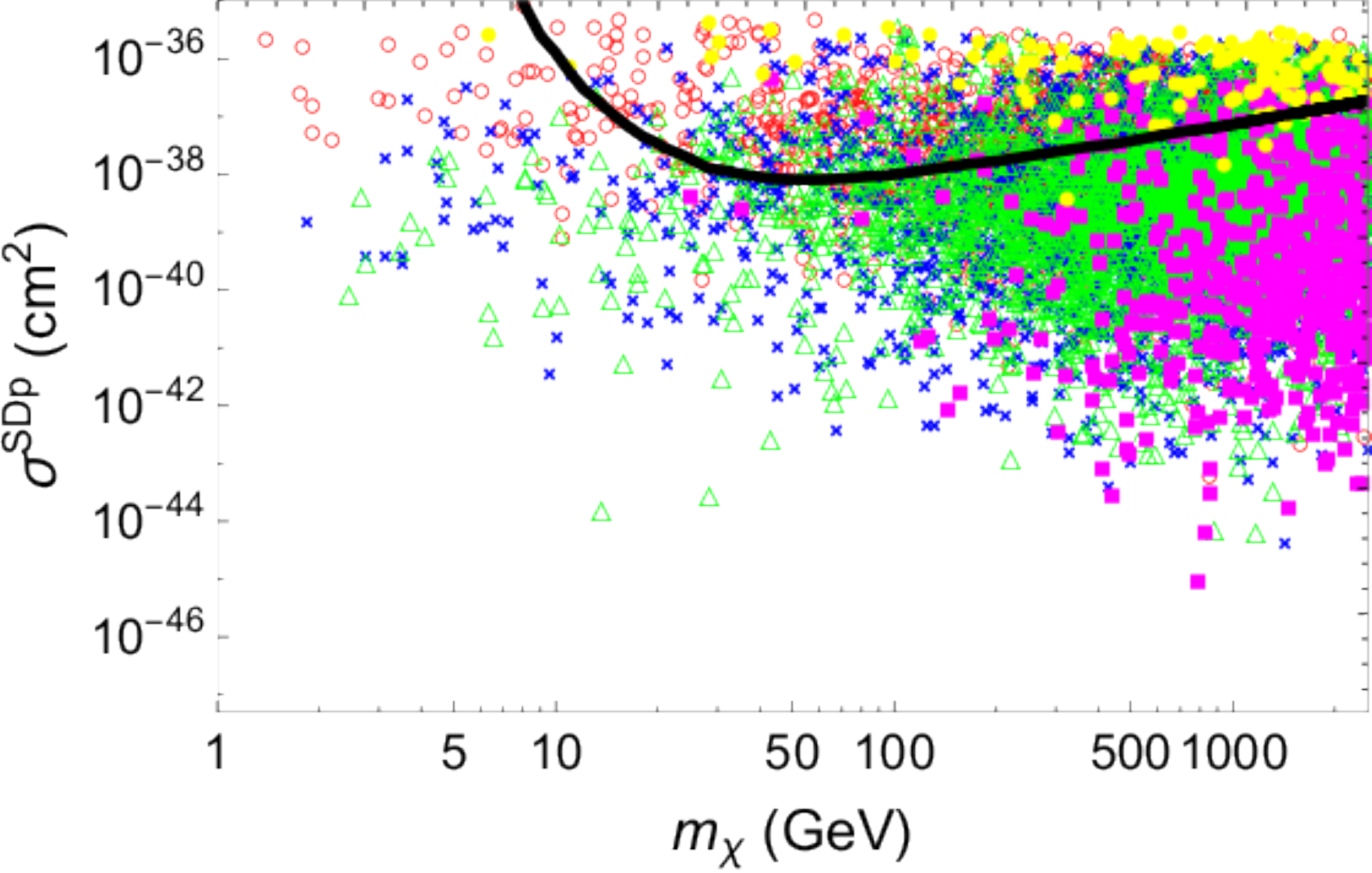}
}\\\subfigure[\ PICO-60 constraint on $\sigma^{SD}_p$]{
  \includegraphics[width=0.45\textwidth,height=0.13\textheight]{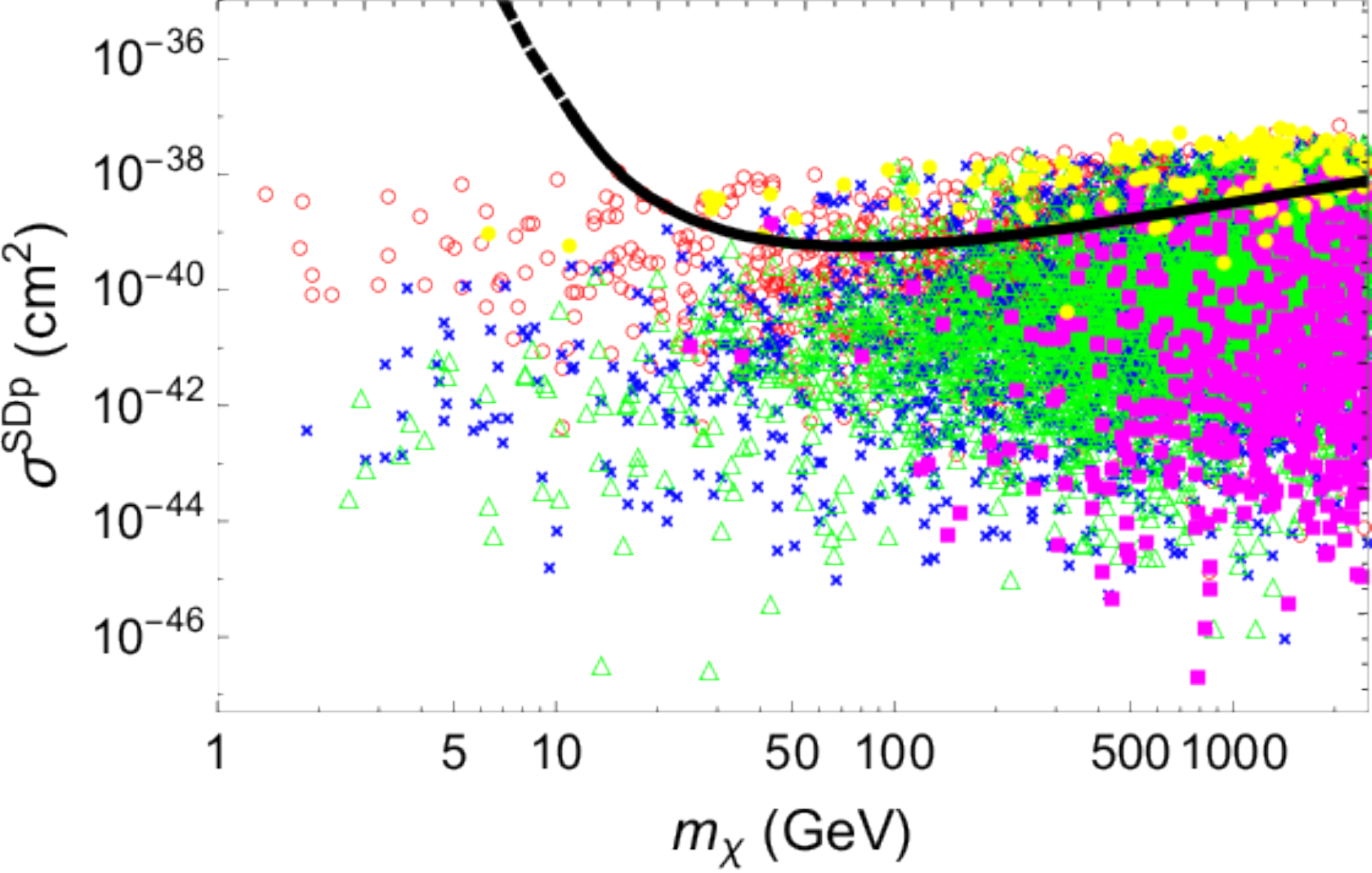}
}\subfigure[\ Fermi-LAT constraint on $\chi^0 {\chi}^0\rightarrow W^+W^-$]{
  \includegraphics[width=0.45\textwidth,height=0.13\textheight]{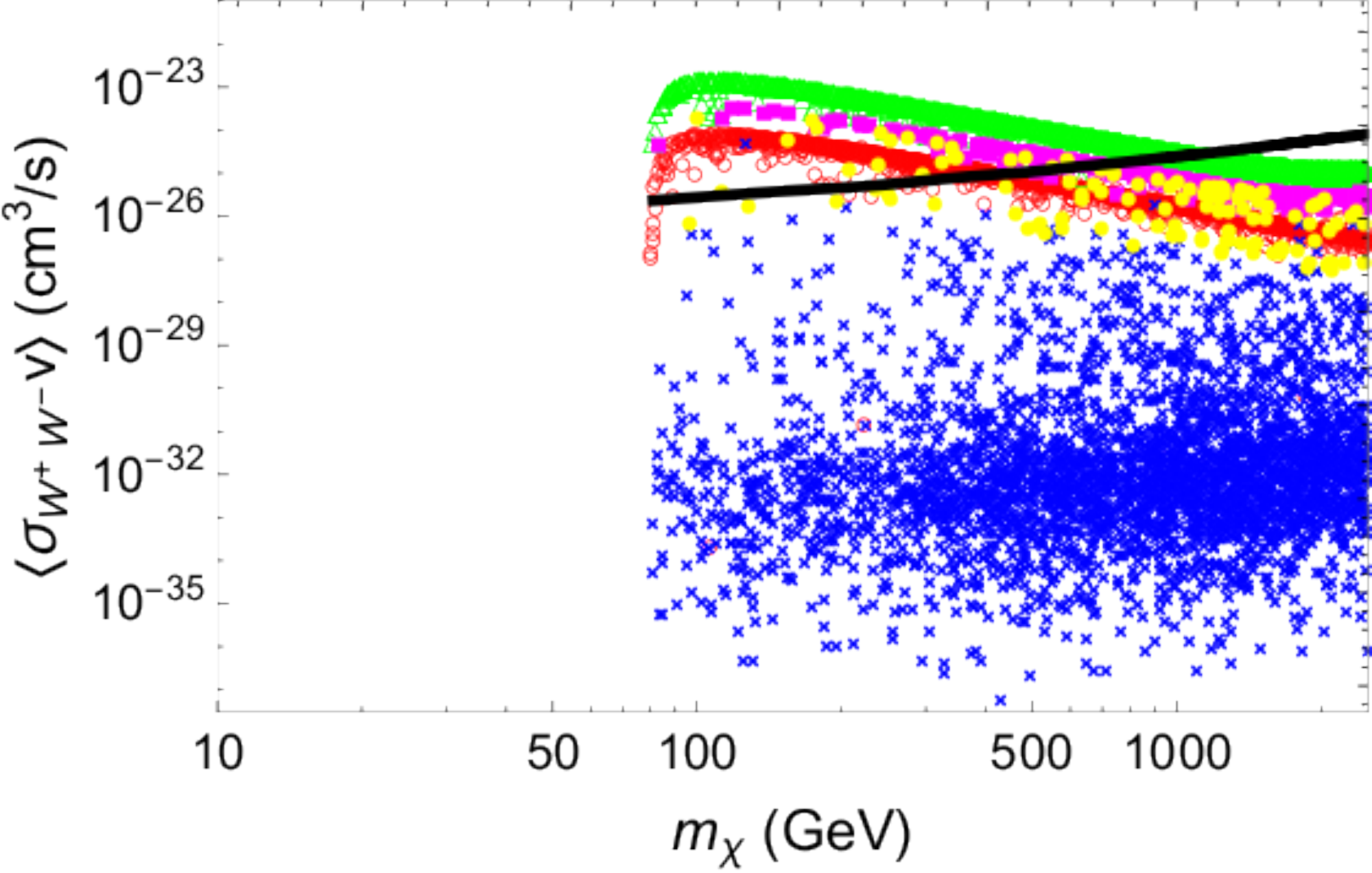}
}\\\subfigure[\ Fermi-LAT constraint on $\chi^0 {\chi}^0\rightarrow b\bar{b}$]{
  \includegraphics[width=0.45\textwidth,height=0.13\textheight]{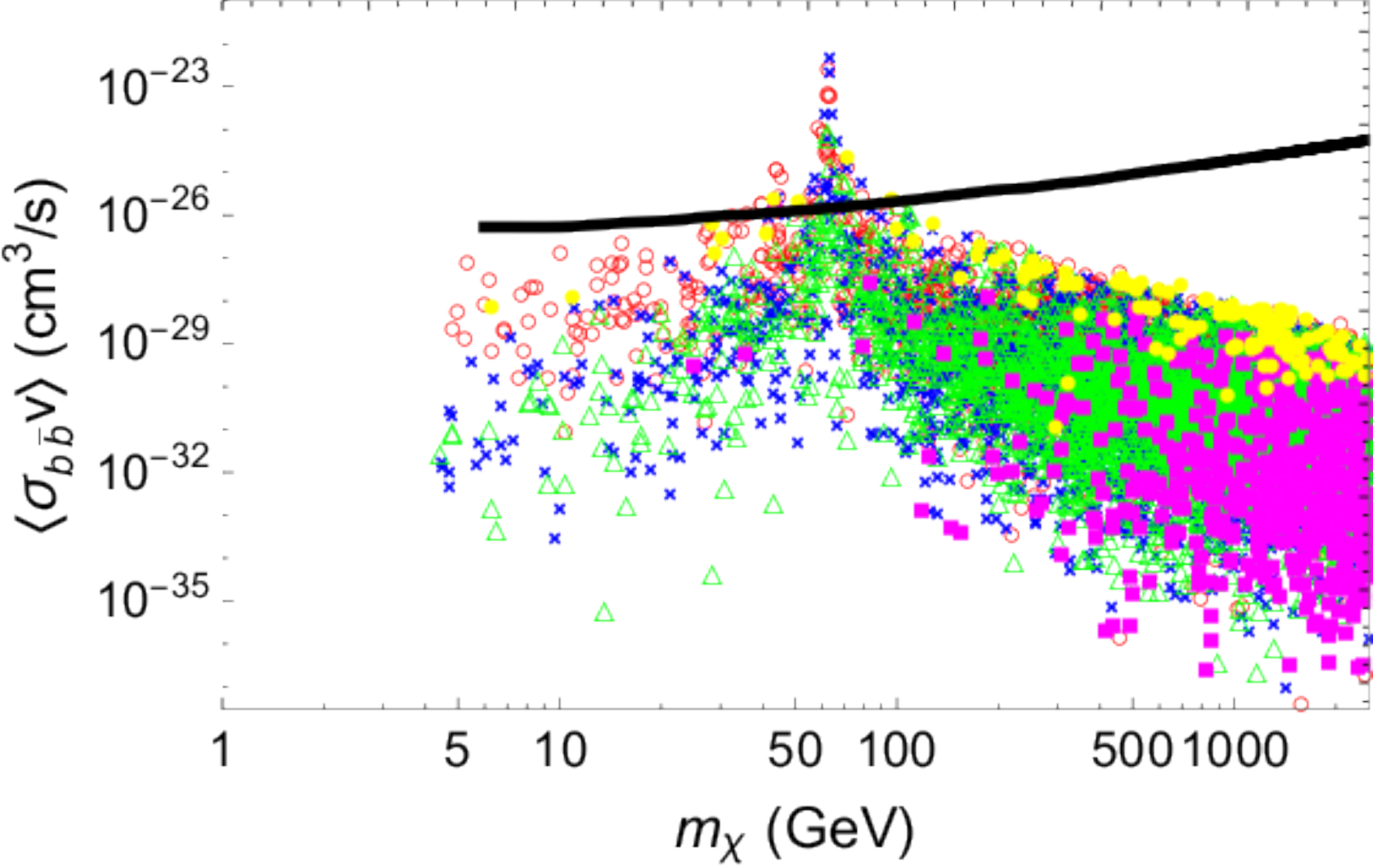}
}\
%\subfigure[\ Fermi-LAT constraint on $\chi^0 \bar{\chi}^0\rightarrow u\bar{u}$]{
  %\includegraphics[width=0.5\textwidth,height=0.13\textheight]{14gHinduu2.pdf}
%}
\subfigure[\ Fermi-LAT constraint on $\chi^0 {\chi}^0\rightarrow \tau^+\tau^-$]{
  \includegraphics[width=0.45\textwidth,height=0.13\textheight]{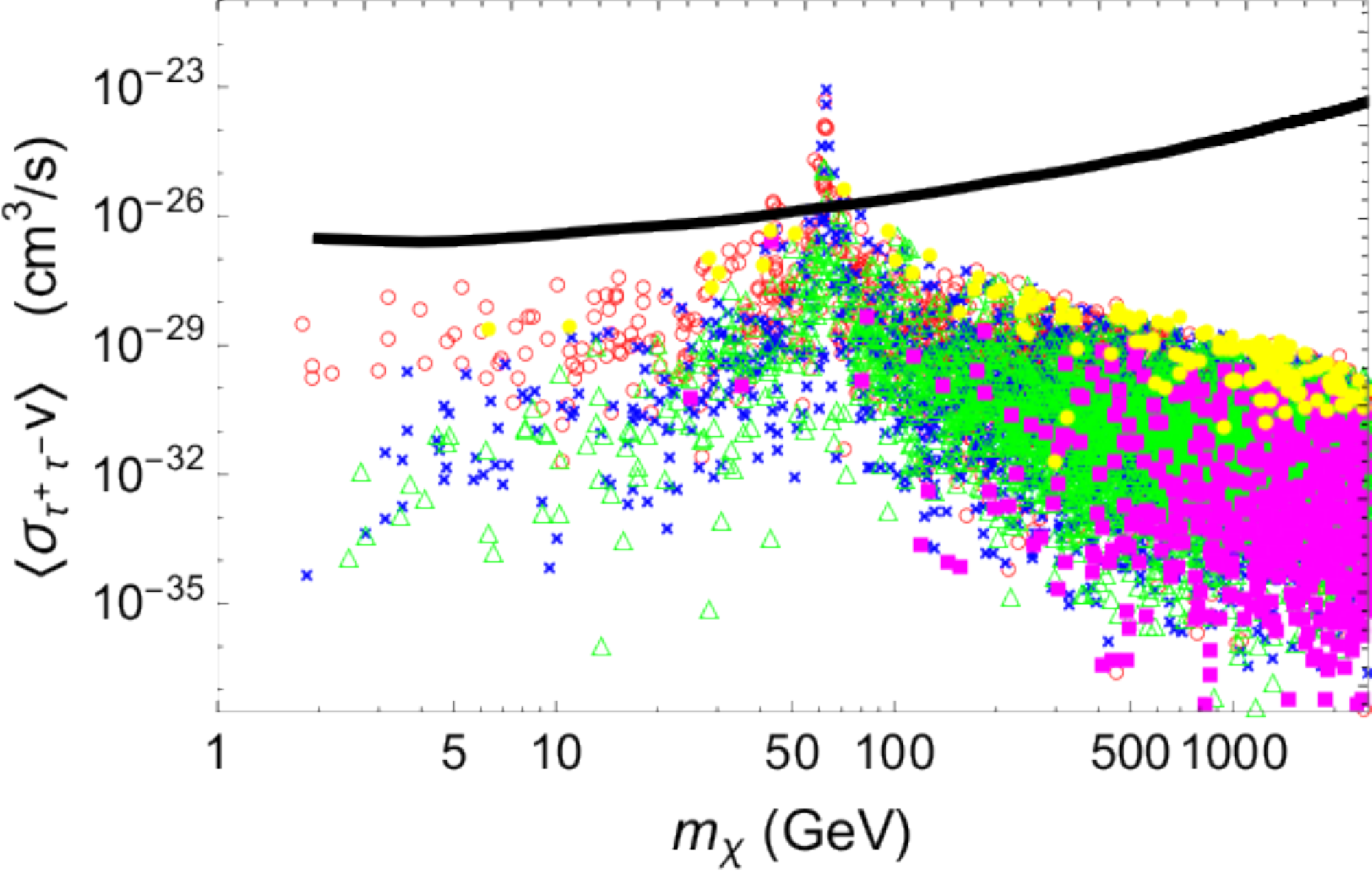}
}%\\\subfigure[\ Fermi-LAT constraint on $\chi^0 \bar{\chi}^0\rightarrow \mu^+\mu^-$]{
 %\includegraphics[width=0.5\textwidth,height=0.13\textheight]{14iHindmumu2.pdf}
%}\subfigure[\ Fermi-LAT constraint on $\chi^0 \bar{\chi}^0\rightarrow e^+e^-$]{
  %\includegraphics[width=0.5\textwidth,height=0.13\textheight]{14jHindee2.pdf}
%}
\caption{Results for all samples with constraints in the extended case
[{\color{red} $\circ$}:~higgsino-like,
{\color{blue} $\times$}:~bino-like,
{\color{green} $\triangle$}:~wino-like,
{\color{magenta} $\blacksquare$}:~non neutralino-like,
{\color{yellow} $\bullet$}:~mixed].}
\label{fig:Extended}
\end{figure}

For the extended case, it has a maximal particle content with $\eta_{1\sim 3,5,7\sim 10}$. In addition to the $\tilde W$-like particles ($\sim\eta_5$), the non neutralino-like $\tilde X$ particles ($\sim\eta_{9,10}$)\footnote{Note that $\eta_{7,8}$ do not have neutral particles and hence they do not contribute to the dark matter compositions.} also appear in this case and the latter contain about $5\%$ of the samples.
We show the results in Fig.~\ref{fig:Extended} with all samples.
As in other cases, we do not present the highly helicity suppressed plots of $\la\sigma_{u{\bar u}} v\ra$, $\la\sigma_{\mu^+\mu^-} v\ra$ and $\la\sigma_{e^+e^-} v\ra$, but we show that all values of $\la\sigma_{W^+W^-} v\ra$ for the $ \tilde B$-like particles should be less than those values for the $\tilde H$-like and the mixed particles in Fig.~\ref{fig:Extended}(f) which is consistent with the fact that a $\tilde B$-like DM pair does not contribute to $s$-wave scattering amplitude.
In this case, all model parameters, $\mu_{1\sim 5}$ and $g_{3\sim 6}$ are free (without the GUT and the $\tan\beta$ relations) so that it has the widest spread in each scatter plot among all cases.
%As in the cases 
Without the GUT and the $\tan\beta$ relations,
more $\tilde B$-like particles have lower values in $\Omega_{\chi} h^2$ and more $\tilde H$-like particles spread toward larger values in $\sigma^{SI}_N$. 
%and more $\tilde H$-like particles have larger values in $\sigma^{SI}$ 
%than those in the cases of neutralino-like I, II and III.
Consequently, more $\tilde B$-like particles (relative to neutralino-like I, II, III) and less $\tilde H$-like particles (relative to neutralino-like I) are allowed.
%, while less $\tilde H$-like particles can survive
[see Fig.~\ref{fig:Extended}(a,b) ].
%and (b)], 
We find that $43\%$ of $\tilde H$-like particles and up to $22\%$ of $\tilde B$-like particles could be DM candidates.

\begin{figure}[h!]
\centering
\captionsetup{justification=raggedright}
 \subfigure[\ Constraint on $\Omega^{\rm{obs}}_\chi$]{
  \includegraphics[width=0.45\textwidth,height=0.13\textheight]{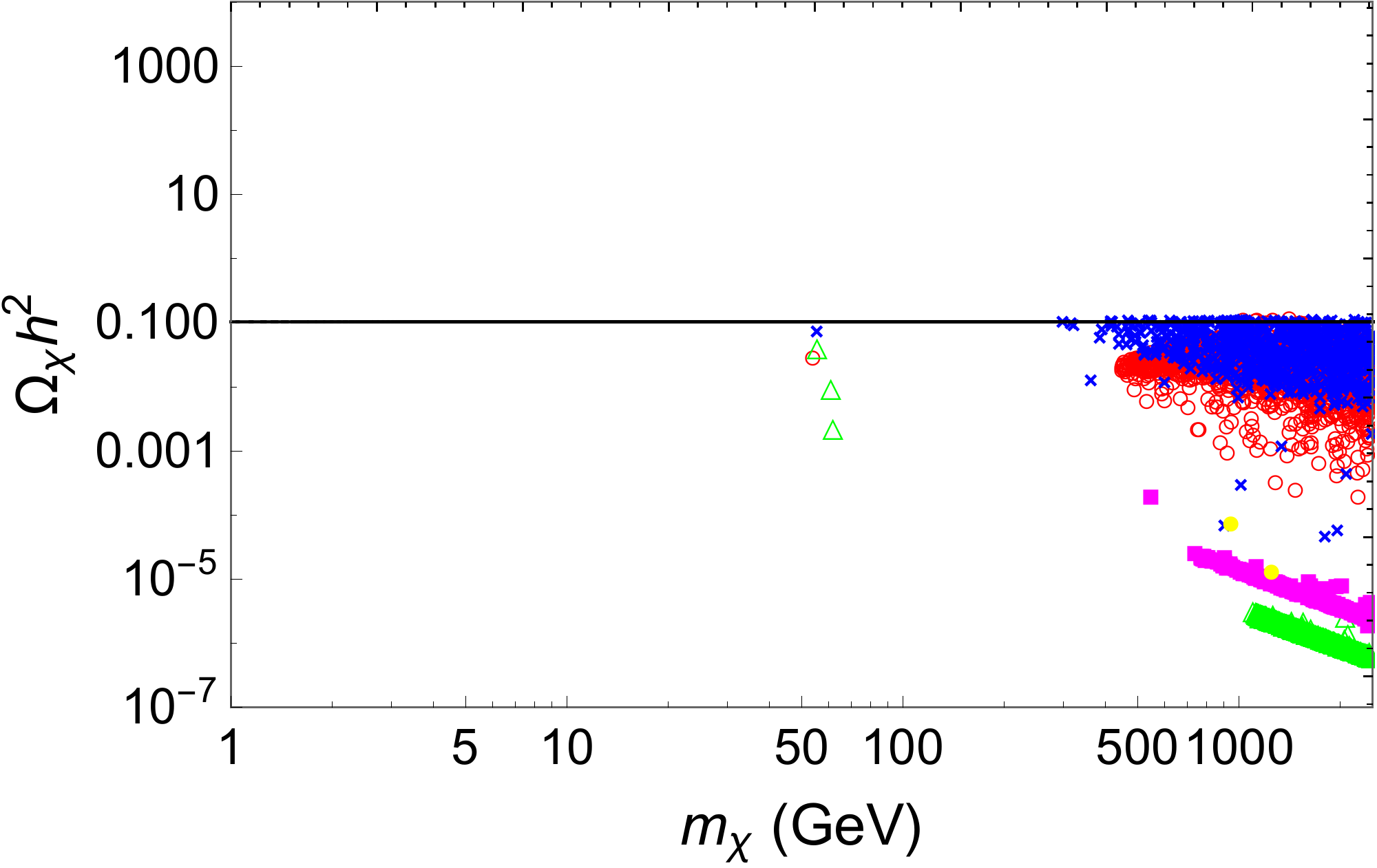}
}\subfigure[\ LUX constraint on $\sigma^{SI}$ with NB limit]{
  \includegraphics[width=0.45\textwidth,height=0.13\textheight]{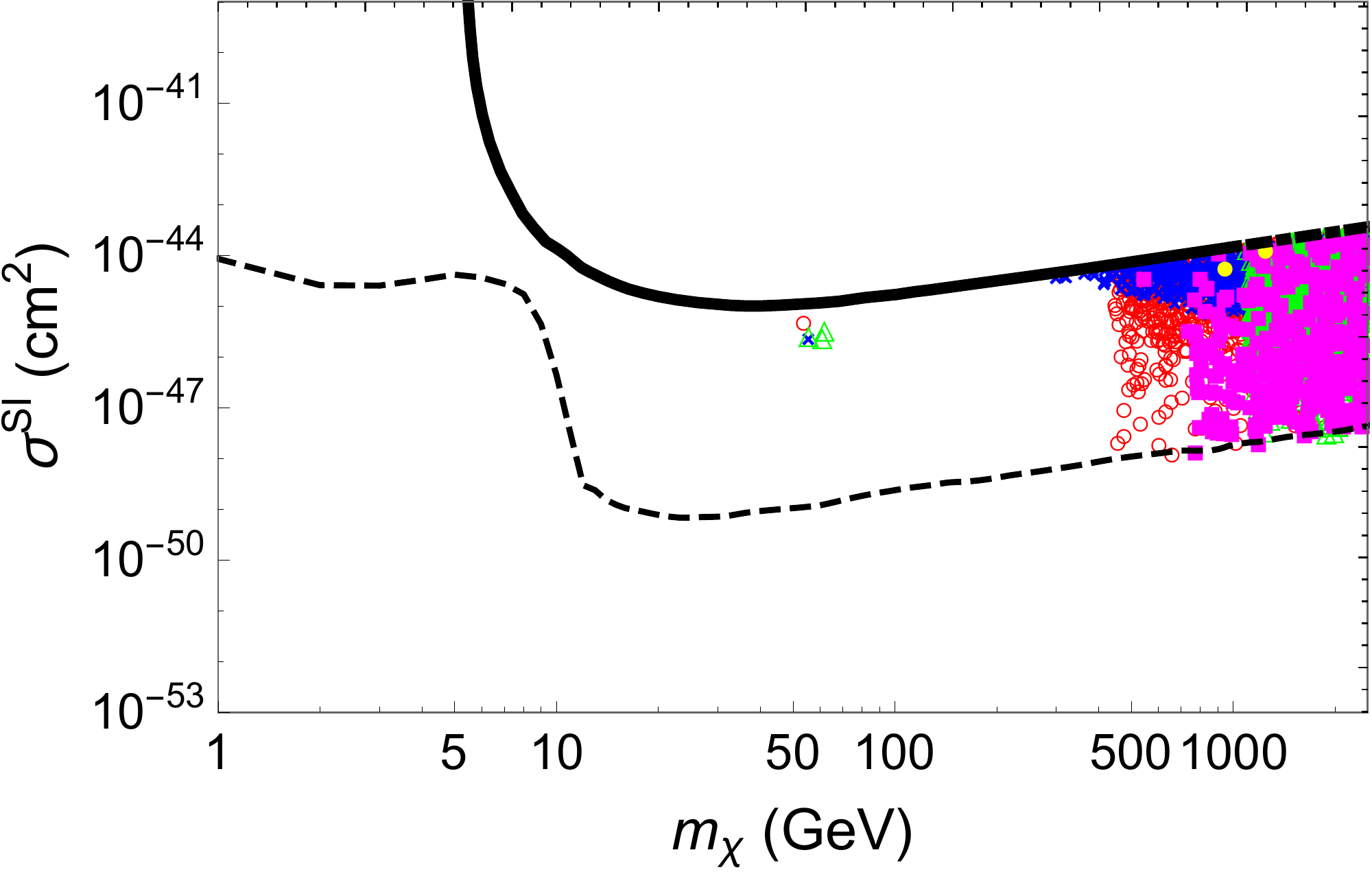}
}\\\subfigure[\ XENON100 constraint on $\sigma^{SD}_n$]{
  \includegraphics[width=0.45\textwidth,height=0.13\textheight]{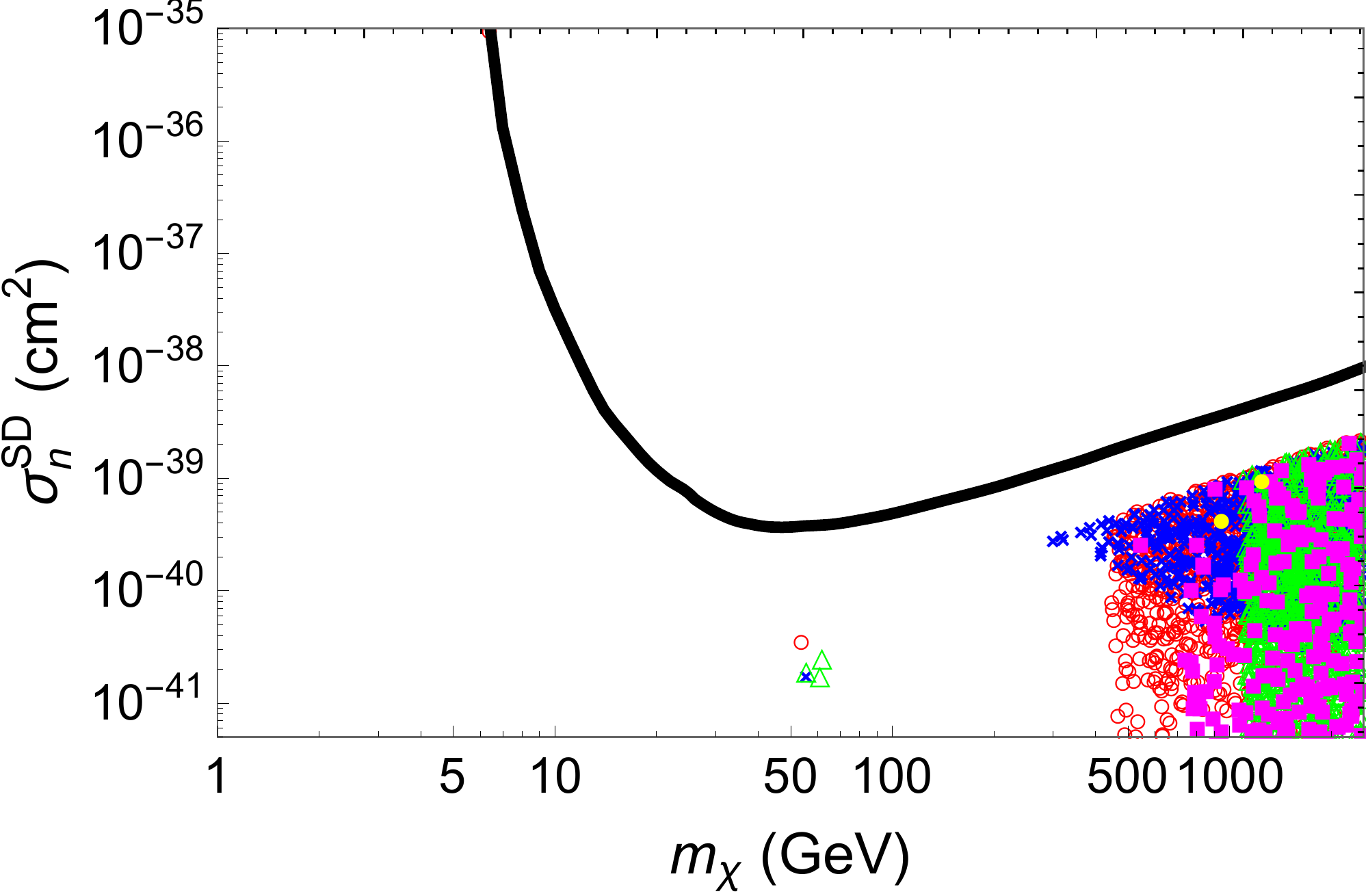}
}\subfigure[\ XENON100 constraint on $\sigma^{SD}_p$]{
  \includegraphics[width=0.45\textwidth,height=0.13\textheight]{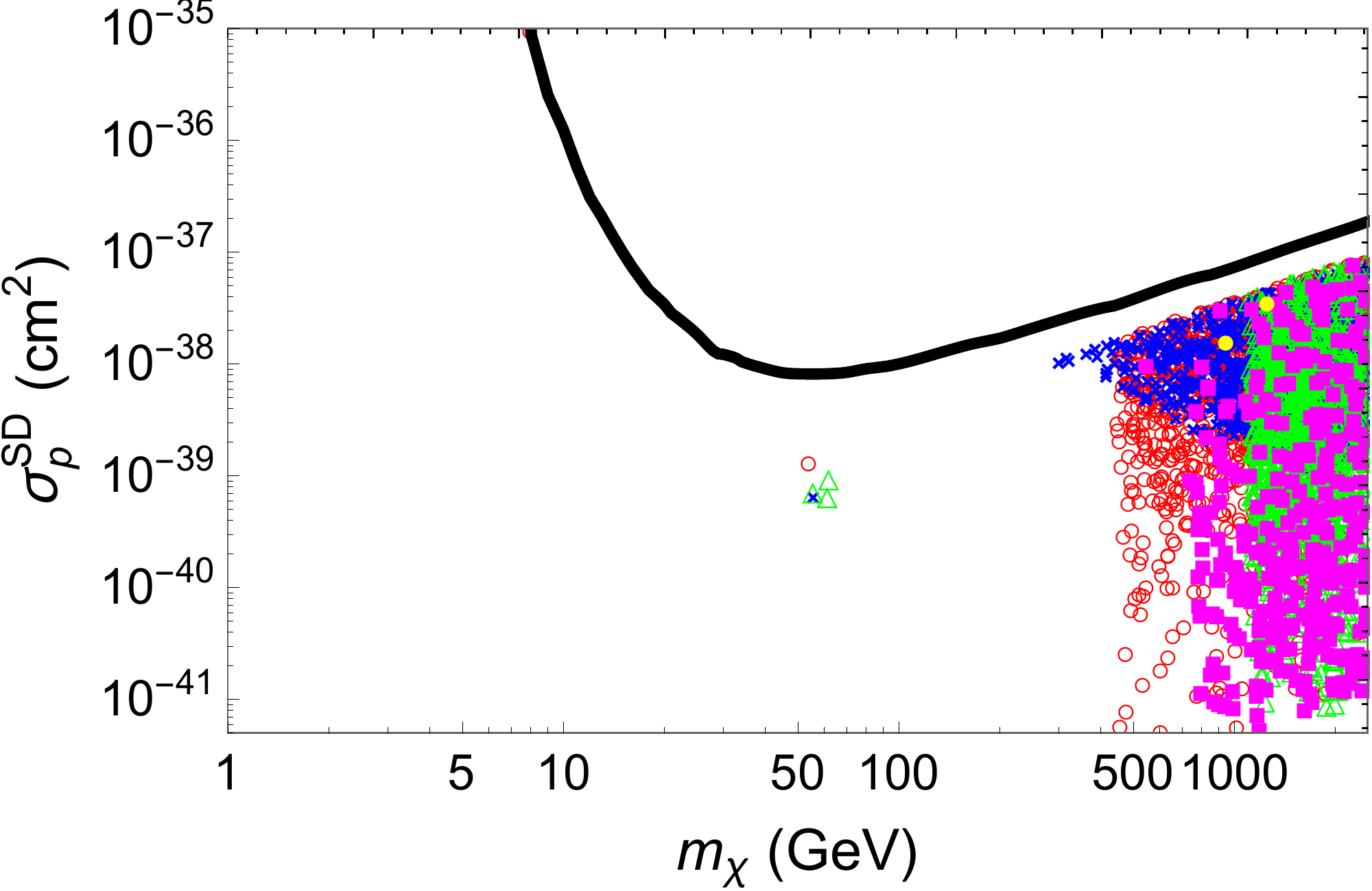}
}\\\subfigure[\ PICO-60 constraint on $\sigma^{SD}_p$]{
  \includegraphics[width=0.45\textwidth,height=0.13\textheight]{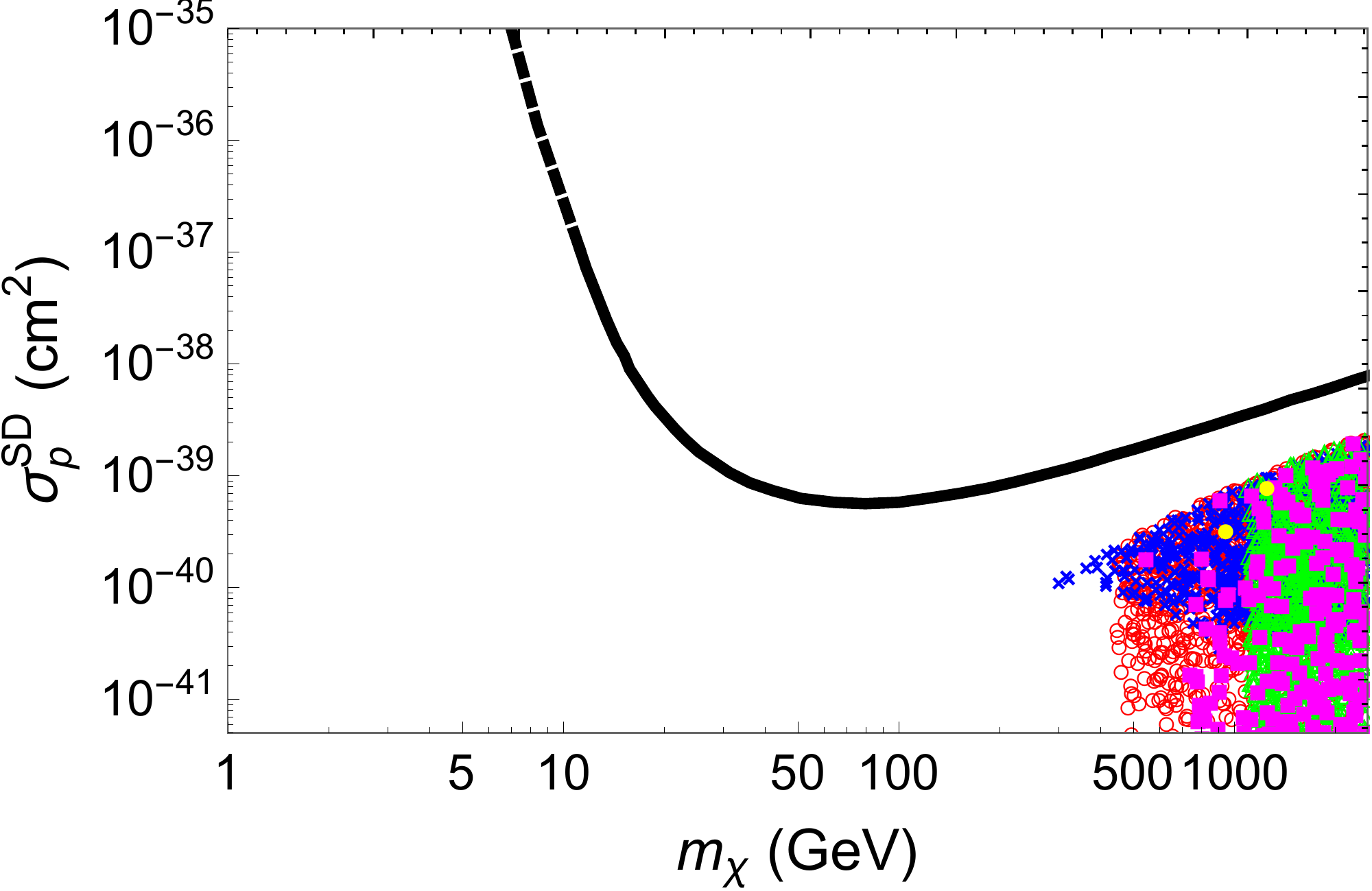}
}\subfigure[\ Fermi-LAT constraint on $\chi^0 {\chi}^0\rightarrow W^+W^-$]{
  \includegraphics[width=0.45\textwidth,height=0.13\textheight]{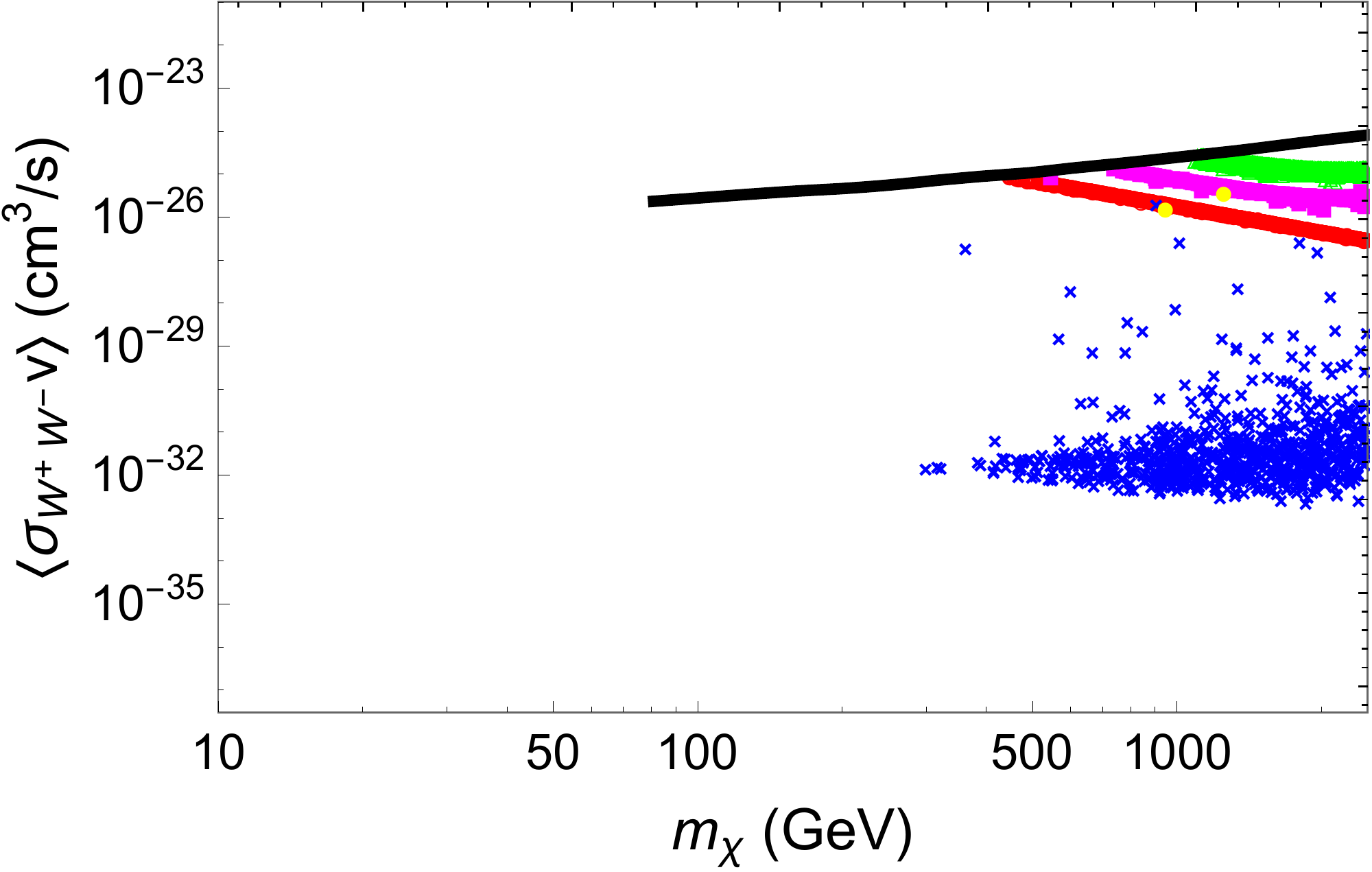}
}\\\subfigure[\ Fermi-LAT constraint on $\chi^0 {\chi}^0\rightarrow b\bar{b}$]{
  \includegraphics[width=0.45\textwidth,height=0.13\textheight]{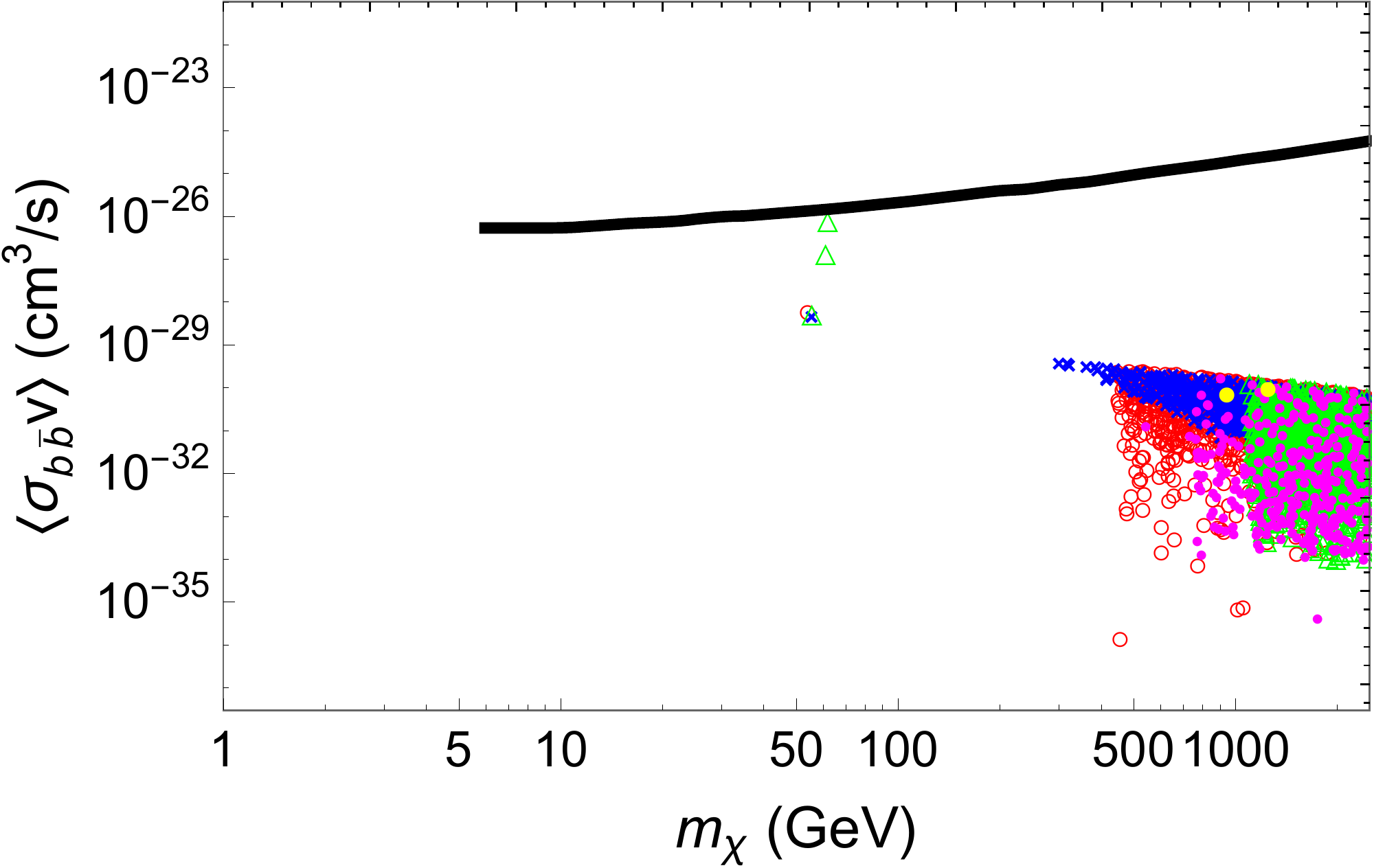}
}\
%\subfigure[\ Fermi-LAT constraint on $\chi^0 \bar{\chi}^0\rightarrow u\bar{u}$]{
  %\includegraphics[width=0.45\textwidth,height=0.13\textheight]{15gHallowuu.pdf}
%}
\subfigure[\ Fermi-LAT constraint on $\chi^0 {\chi}^0\rightarrow \tau^+\tau^-$]{
  \includegraphics[width=0.45\textwidth,height=0.13\textheight]{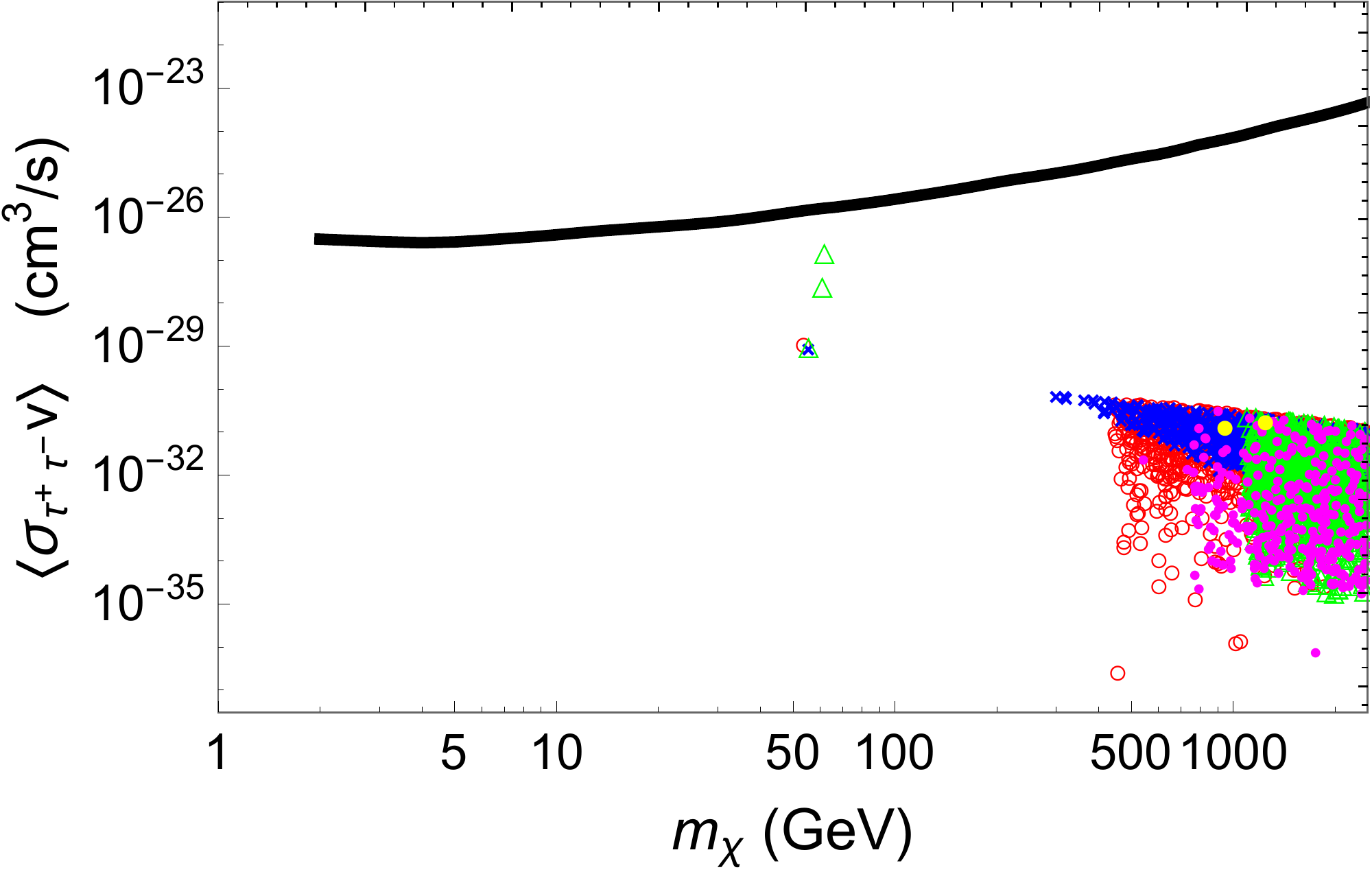}
}%\\\subfigure[\ Fermi-LAT constraint on $\chi^0 \bar{\chi}^0\rightarrow \mu^+\mu^-$]{
  %\includegraphics[width=0.45\textwidth,height=0.13\textheight]{15iHallowmumu.pdf}
%}\subfigure[\ Fermi-LAT constraint on $\chi^0 \bar{\chi}^0\rightarrow e^+e^-$]{
  %\includegraphics[width=0.45\textwidth,height=0.13\textheight]{15jHallowee.pdf}
%}
\caption{Results for allowed samples satisfying all constraints in the extended case
[{\color{red} $\circ$}:~higgsino-like,
{\color{blue} $\times$}:~bino-like,
{\color{green} $\triangle$}:~wino-like,
{\color{magenta} $\blacksquare$}:~non neutralino-like,
{\color{yellow} $\bullet$}:~mixed].}
\label{fig:allow Extended}
\end{figure}

We redraw the Fig.~\ref{fig:Extended} in Fg.~\ref{fig:allow Extended}, but with the allowed samples only. Similarly, we find that $\tilde B$-like DM candidates are accessible only in the SI experiments of DM-nucleus scattering, while all other types of DM candidates can be sensitively detected from both the direct search in the SI experiments of DM-nucleus scattering
and the indirect search in the observation of DM annihilation to $W^+W^-$ channel in the near future. Despite of the fact that most of  $\tilde B$-like particles are ruled out by the $\Omega_{\chi} h^2$ constraint, and further by LUX $\sigma^{SI}_N$ constraint, 
more allowed $\tilde B$-like DM candidates can lower down the allowed mass range of $\tilde B$-like particles from $m_{\chi} \gtrsim 1$ TeV (as in the cases with GUT relation) to $m_{\chi} \gtrsim 300$ GeV.
The $\tilde H$-like particles with $m_{\chi} \lesssim M_W$ are ruled out by the relic density and the Fermi-LAT $\la\sigma (\chi{\chi}\rightarrow b\bar b) v\ra$ constraints, while the $\tilde H$-like particles with $m_{\chi} > M_W$ are subjected to the Fermi-LAT $\la\sigma (\chi{\chi}\rightarrow W^+W^-) v\ra$ and the LUX $\sigma^{SI}_N$ constraints, so that only the $\tilde H$-like particles with $m_{\chi} \gtrsim 450$ GeV could be the DM candidates.
Similarly, the $\tilde W$-like particles and the non neutralino-like $\tilde X$ particles with $m_{\chi} \lesssim M_W$ are ruled out by the relic density and the Fermi-LAT $\la\sigma (\chi{\chi}\rightarrow b\bar b) v\ra$ constraints, while the $\tilde W$-like particles and the non neutralino-like $\tilde X$ particles with $m_{\chi} > M_W$ are subjected to the Fermi-LAT $\la\sigma (\chi{\chi}\rightarrow W^+W^-) v\ra$ and the LUX $\sigma^{SI}_N$ constraints, so that only the $\tilde W$-like particles and the non neutralino-like $\tilde X$ particles with $m_{\chi} \gtrsim 1107, 738$ GeV, respectively, could be the DM candidates. We also find that  about $31\%$ of $\tilde W$-like particles and $62\%$ of non neutralino-like $\tilde X$ particles are allowed to be DM candidates. Furthermore, we find that
the allowed $\tilde H$-, $\tilde B$-, $\tilde W$-like particles and the non neutralino-like $\tilde X$ particles are highly pure, as
$99.5\%$, $99.2\%$, $99.5\%$ and $95\%$ of them are in the states of $\eta_{1,2}$ ,$\eta_3$, $\eta_5$ and $\eta_{9,10}$, respectively,  with their composition fractions greater than $90\%$.

\vfill
\eject

\vfill
\eject

\subsection{Summary and Predictions}

\captionsetup{
justification=raggedright,
}
\begin{table}[b!]
\footnotesize{
\begin{tabular}{|c|cccc|c|c|}
  \hline
  &
  &
  & {\hspace{-2.7cm}Case A}
  &
  & Case B
  & Case C
  \\
  & neutralino-like I & neutralino-like II & neutralino-like III & neutralino-like IV & Reduced & Extended \\
  \hline
  $\tilde H$-like &
  \begin{tabular}[c]{@{}l@{}} \quad\ \ 456 \\ (456,\ \ 940)\end{tabular} &
  \begin{tabular}[c]{@{}l@{}} \quad\ \ 457 \\ (457,\ \ 937)\end{tabular} &
  \begin{tabular}[c]{@{}l@{}} \quad\ \ 457 \\ (457,\ \ 947)\end{tabular} &
  \begin{tabular}[c]{@{}l@{}} \quad\ \ 454 \\ (454,\ \ 947)\end{tabular} &
  \begin{tabular}[c]{@{}l@{}} \quad\ \ 454 \\ (454,\ \ 949)\end{tabular} &
  \begin{tabular}[c]{@{}l@{}} \quad\ \ 450 \\ (450,\ \ 927)\end{tabular} \\
  \hline
  $\tilde B$-like &
  \begin{tabular}[c]{@{}l@{}} 1411 \\ \ \ X \end{tabular} &
  \begin{tabular}[c]{@{}l@{}}  1258 \\ \ X \end{tabular} &
  \begin{tabular}[c]{@{}l@{}}  \ 341 \\ \ \ X \end{tabular} &
  \begin{tabular}[c]{@{}l@{}}  \ 288 \\ \ \ X \end{tabular} &
  \begin{tabular}[c]{@{}l@{}}  \ 317 \\ \ \  X \end{tabular} &
  \begin{tabular}[c]{@{}l@{}}  \ 299 \\ \ \  X \end{tabular} \\
  \hline
  $\tilde W$-like &
  \begin{tabular}[c]{@{}l@{}}   X \\   X \end{tabular} &
  \begin{tabular}[c]{@{}l@{}}   X \\   X \end{tabular} &
  \begin{tabular}[c]{@{}l@{}} \quad\ \ 1120 \\ (1120 2500\footnotemark[1])\end{tabular} &
  \begin{tabular}[c]{@{}l@{}} \quad\ \ 1090 \\ (1090, 2374)\end{tabular} &
  \begin{tabular}[c]{@{}l@{}}   X \\   X \end{tabular} &
  \begin{tabular}[c]{@{}l@{}} \quad\ \ 1107 \\ (1107, 2080)\end{tabular} \\
  \hline
  $\tilde X$-like & 
  \begin{tabular}[c]{@{}l@{}}   X \\   X \end{tabular} &
  \begin{tabular}[c]{@{}l@{}}   X \\   X \end{tabular} &
  \begin{tabular}[c]{@{}l@{}}   X \\   X \end{tabular} &
  \begin{tabular}[c]{@{}l@{}}   X \\   X \end{tabular} &
  \begin{tabular}[c]{@{}l@{}}   X \\   X \end{tabular} &
  \begin{tabular}[c]{@{}l@{}} \quad\ \ 738 \\ (738,\ \ 1563)\end{tabular} \\
  \hline
  \end{tabular}
  \footnotetext[1]{This value is originated from the limitation of our numerical analysis.}
 \caption{Allowed mass ranges according to particle attribute to detect DM in the near future. The upper values denote the lower mass bounds (in unit of GeV) to detect DM in the direct search of SI DM-nucleus scattering experiments and the lower intervals denote the mass interval (in unit of GeV) suitable to detect DM in the indirect search of DM annihilation process via $W^+W^-$ channel between the present limit and the projected limit which is taken to be one order of magnitude lower than the present one.}
\label{tab:testible}}
\end{table}

In this subsection, we will summarize the previous discussion and give some predictions.
The allowed samples must satisfy all the constraints simultaneously, namely, the observed relic density $\Omega^{\rm obs}_{\chi}h^2$ constraint (below $+3\sigma$), the LUX constraint on $\sigma^{SI}_N$, the XENON100 constraints on $\sigma^{SD}_{n,p}$, PICO-60 constraint on $\sigma^{SD}_{p}$, and the Fermi-LAT constraints on $\la\sigma(\chi {\chi} \rightarrow W^+W^-, b {\bar b}, u {\bar u}, \tau^+\tau^-, \mu^+\mu^-, e^+e^-) v \ra$.
For all cases, we find that
most of $\tilde B$-like particles are ruled out by the $\Omega_{\chi} h^2$ constraint, and further by the LUX $\sigma^{SI}_N$ constraint;
the $\tilde H$-like particles with $m_{\chi} \lesssim M_W$ are ruled out by the relic density and the Fermi-LAT $\la\sigma (\chi{\chi}\rightarrow b\bar b) v\ra$ constraints, while the $\tilde H$-like particles with $m_{\chi} > M_W$ are subjected to the Fermi-LAT $\la\sigma (\chi{\chi}\rightarrow W^+W^-) v\ra$ and the LUX $\sigma^{SI}_N$ constraints.
For all cases, all values in $\la\sigma_{W^+W^-} v\ra$ for the $\tilde B$-like particles are smaller than those values for the $\tilde H$-like particles due to the fact that a $\tilde B$-like DM pair does not contribute to $s$-wave scattering amplitude. Besides, the process of $\chi\chi\rightarrow f{\bar f}$ favors heavy fermions since the $s$-wave contribution is helicity suppressed.   
We see that the direct search of SI DM-nucleus elastic scattering and the indirect search of DM annihilation to $W^+W^-$ channel are more important. In other words, they are sensitive to the DM searches in the near future.

Without considering the outlier samples, we show the allowed mass range of different particle attribute to detect DM in direct as well as indirect searches in Table \ref{tab:testible}.
The upper values denote the lower mass bounds to detect DM in the direct search of SI DM-nucleus scattering experiments and the lower intervals denote the mass interval suitable to detect DM in the indirect search of DM annihilation process via $W^+W^-$ channel using the present limit and the projected limit, which is taken to be one order of magnitude lower than the present one. We see that the DM mass should be greater than $450, 288, 1090, 738$ GeV to detect the $\tilde H$-, $\tilde B$-, $\tilde W$-like DM particles, and the non neutralino-like $\tilde X$ DM particles, respectively.
Note that unlike the indirect case, we do not see the upper mass bound to detect DM in the direct search in this analysis. In other words, future direct searches can explore larger DM mass range than the indirect one.

\begin{figure}[t!]
\centering
\captionsetup{justification=raggedright}
  \subfigure{
  \includegraphics[width=0.3\textwidth,height=0.133\textheight]{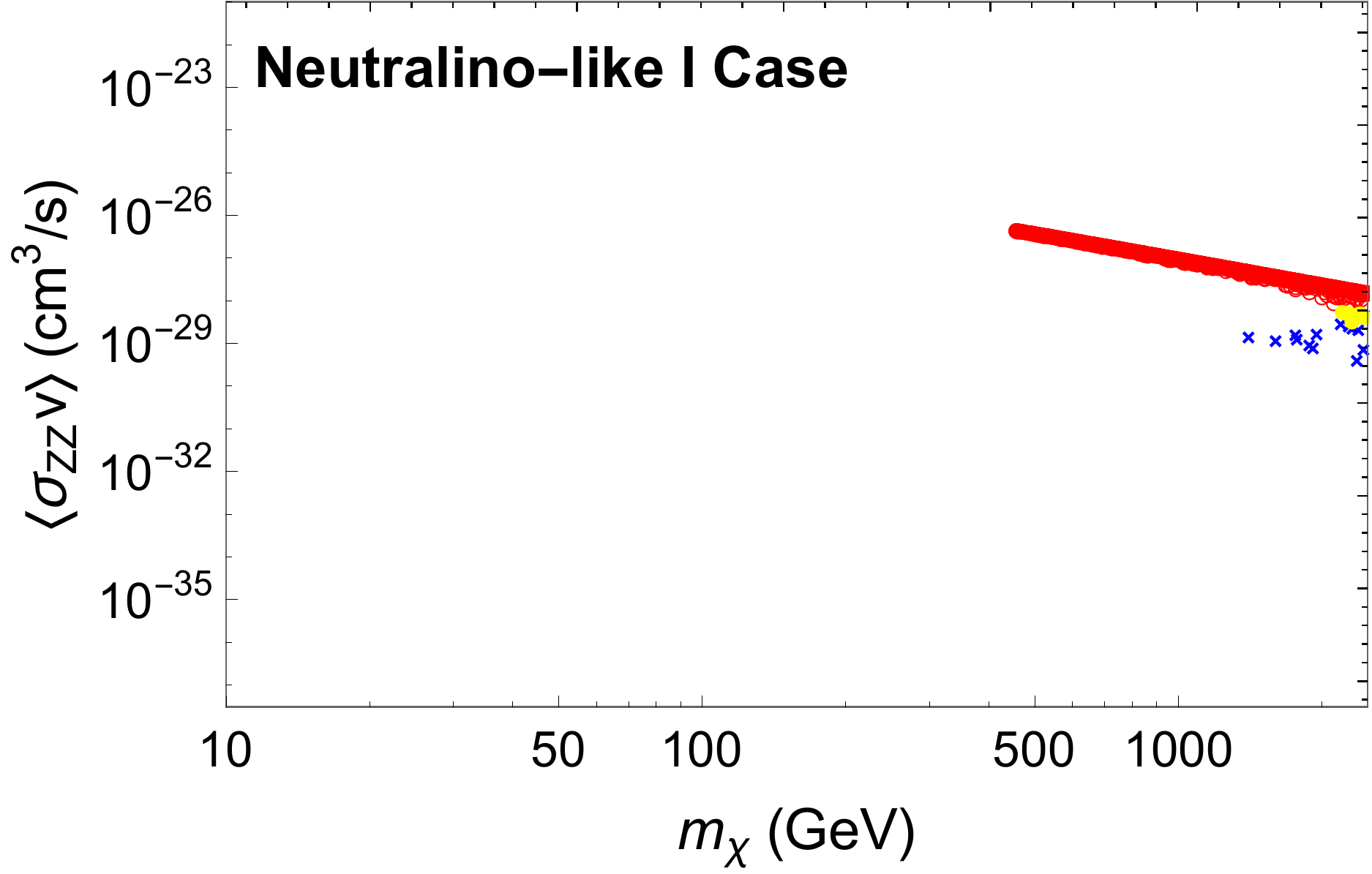}
}\subfigure{
  \includegraphics[width=0.3\textwidth,height=0.133\textheight]{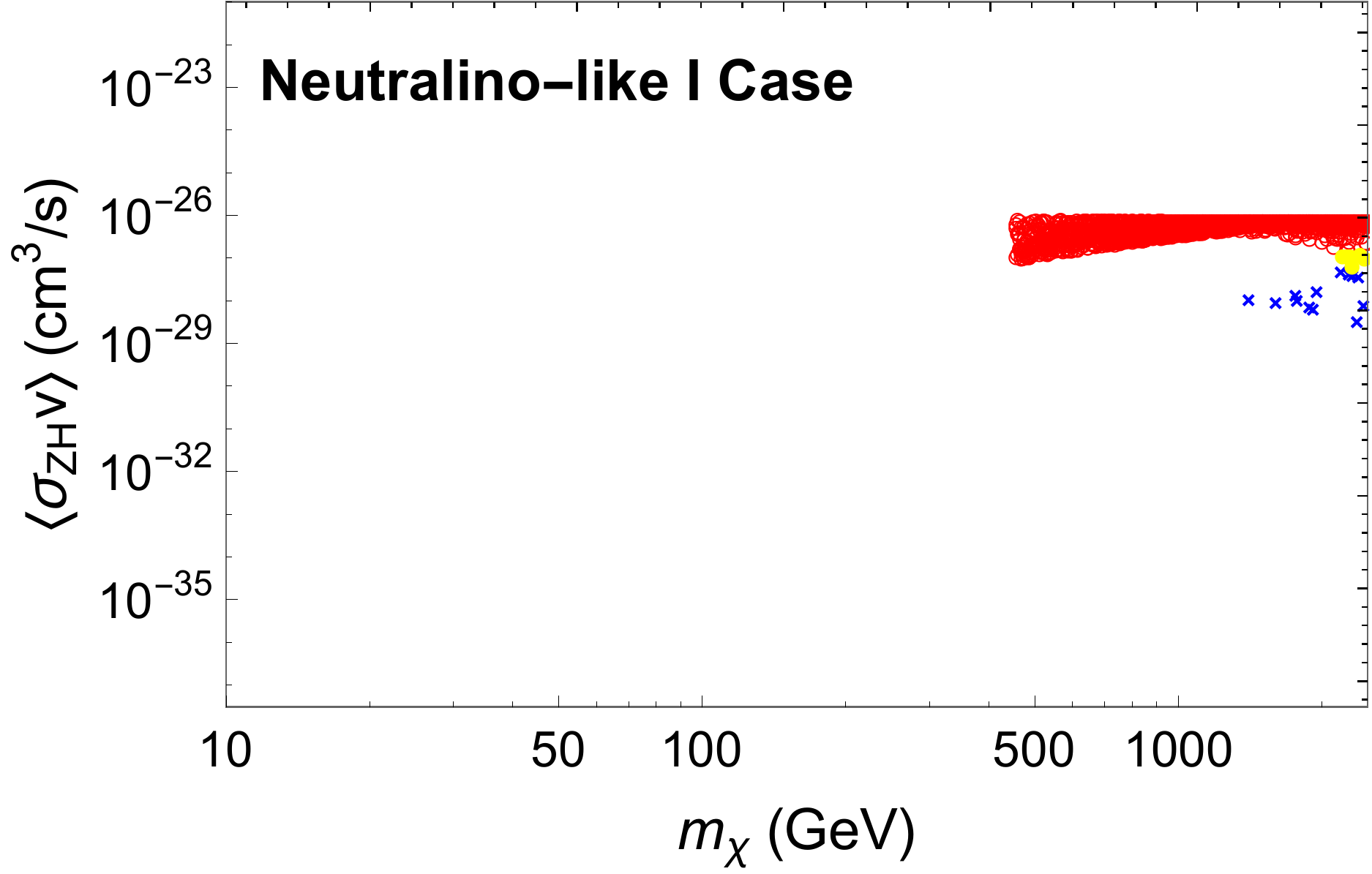}
}\subfigure{
  \includegraphics[width=0.3\textwidth,height=0.133\textheight]{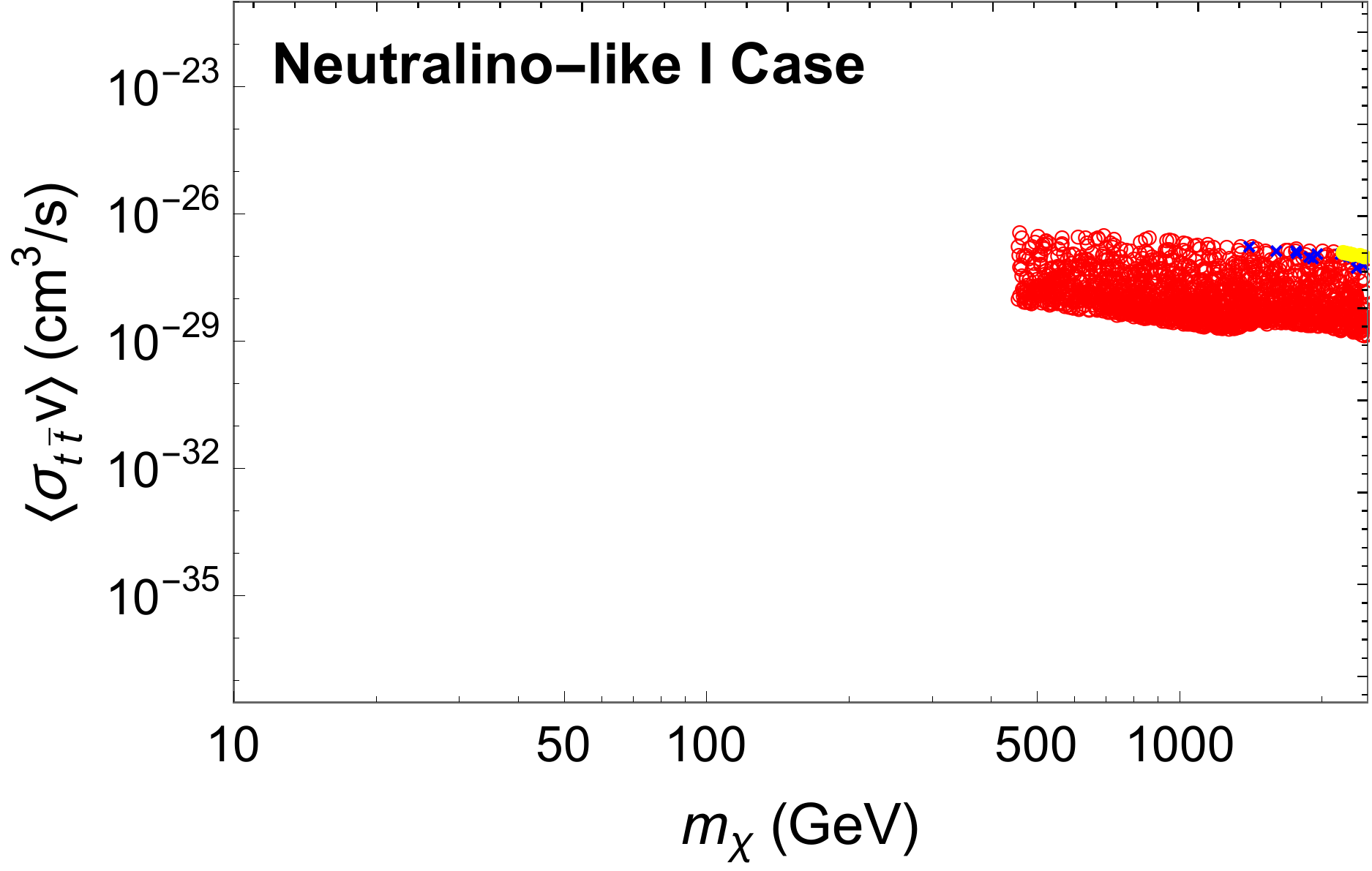}
}\\
  \subfigure{
  \includegraphics[width=0.3\textwidth,height=0.133\textheight]{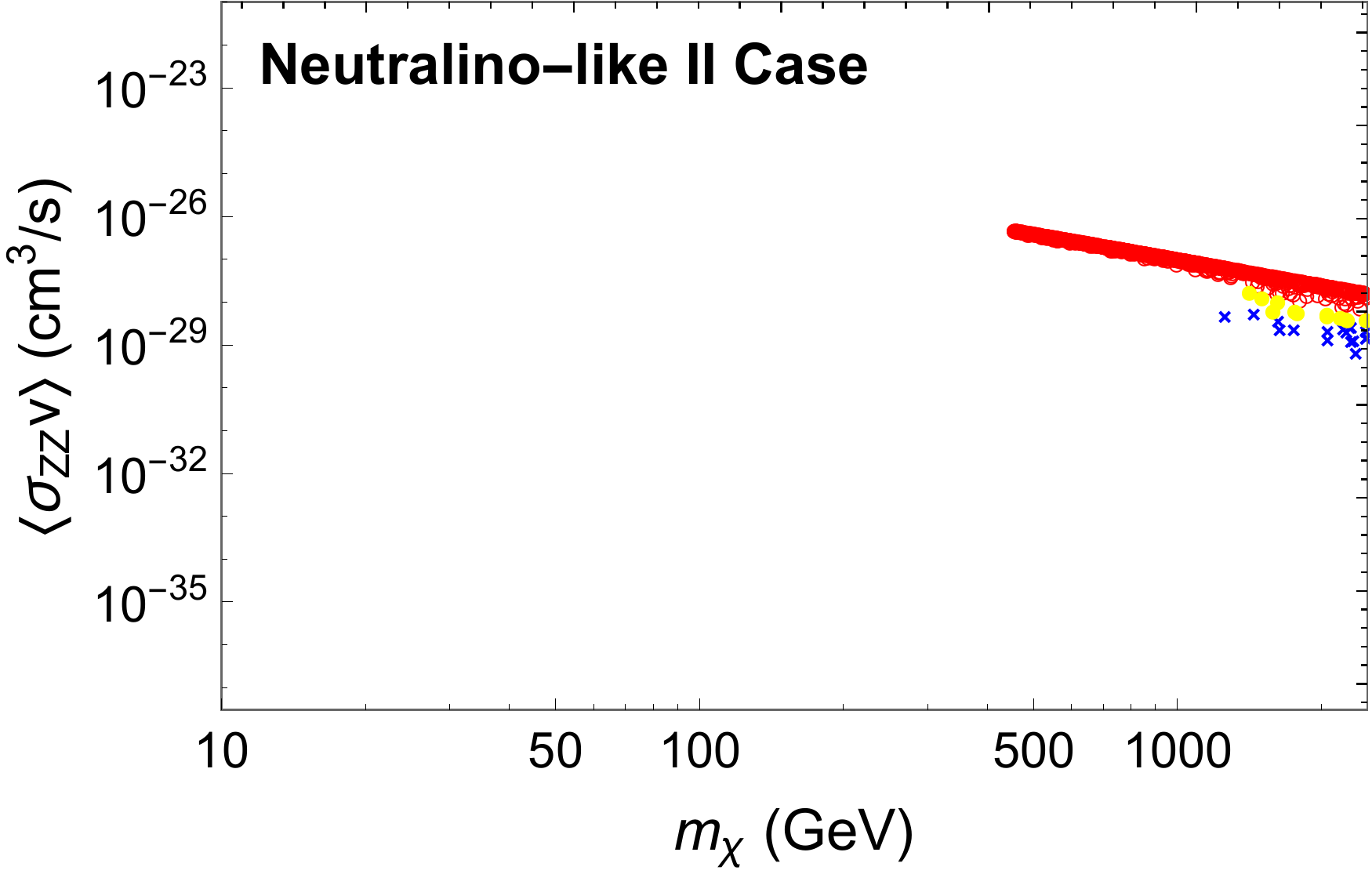}
}\subfigure{
  \includegraphics[width=0.3\textwidth,height=0.133\textheight]{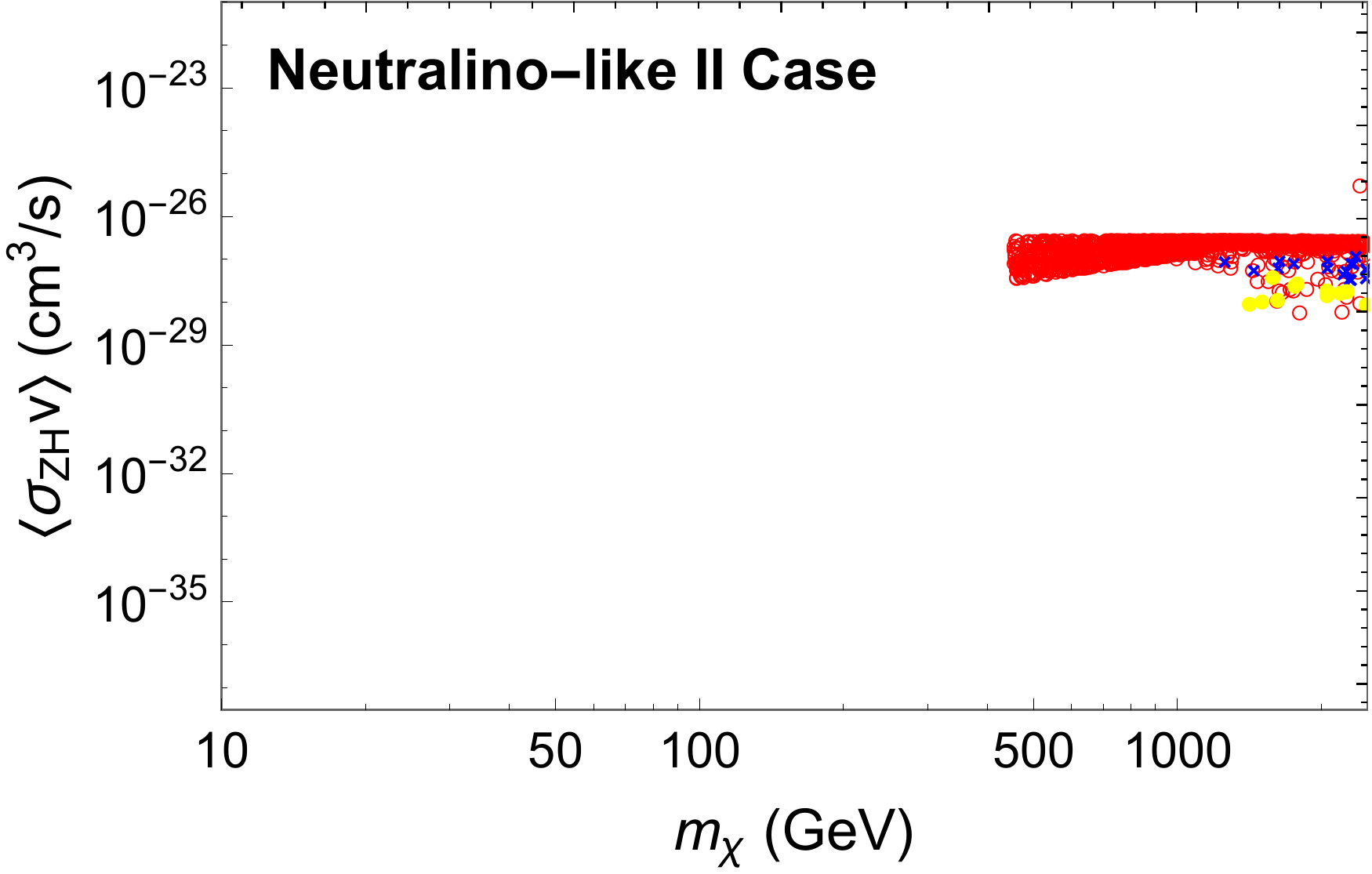}
}\subfigure{
  \includegraphics[width=0.3\textwidth,height=0.133\textheight]{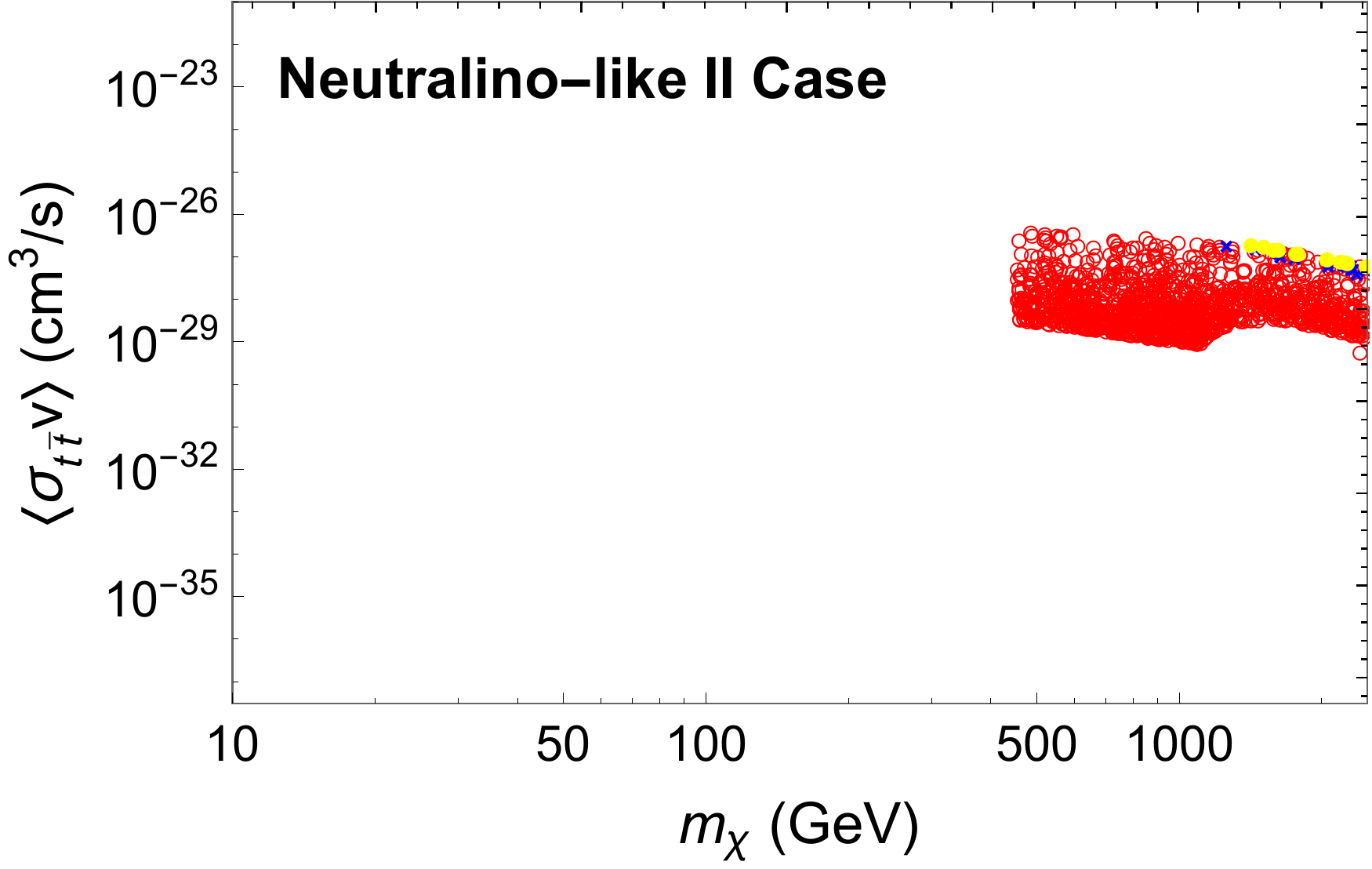}
}\\
  \subfigure{
  \includegraphics[width=0.3\textwidth,height=0.133\textheight]{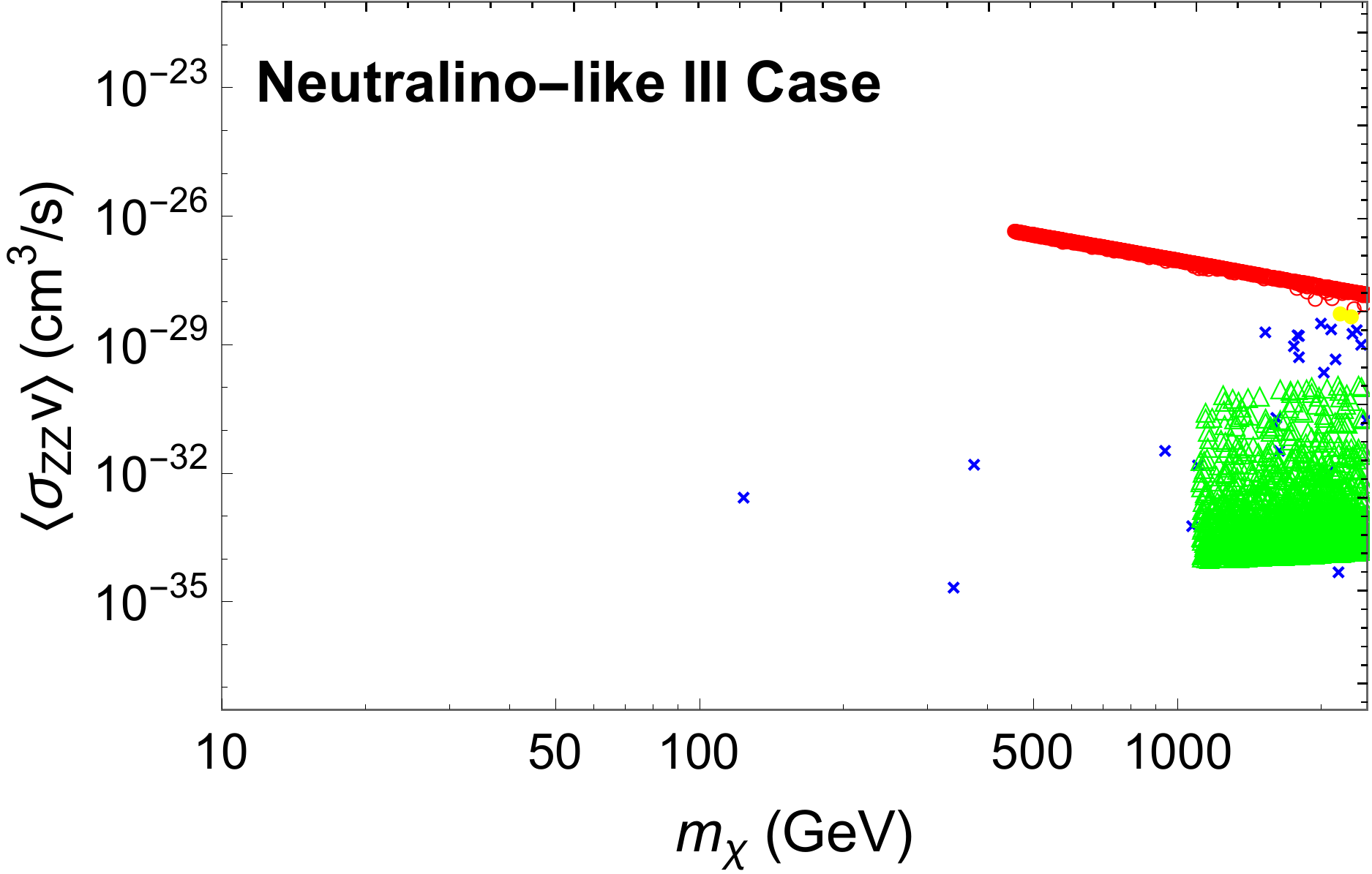}
}\subfigure{
  \includegraphics[width=0.3\textwidth,height=0.133\textheight]{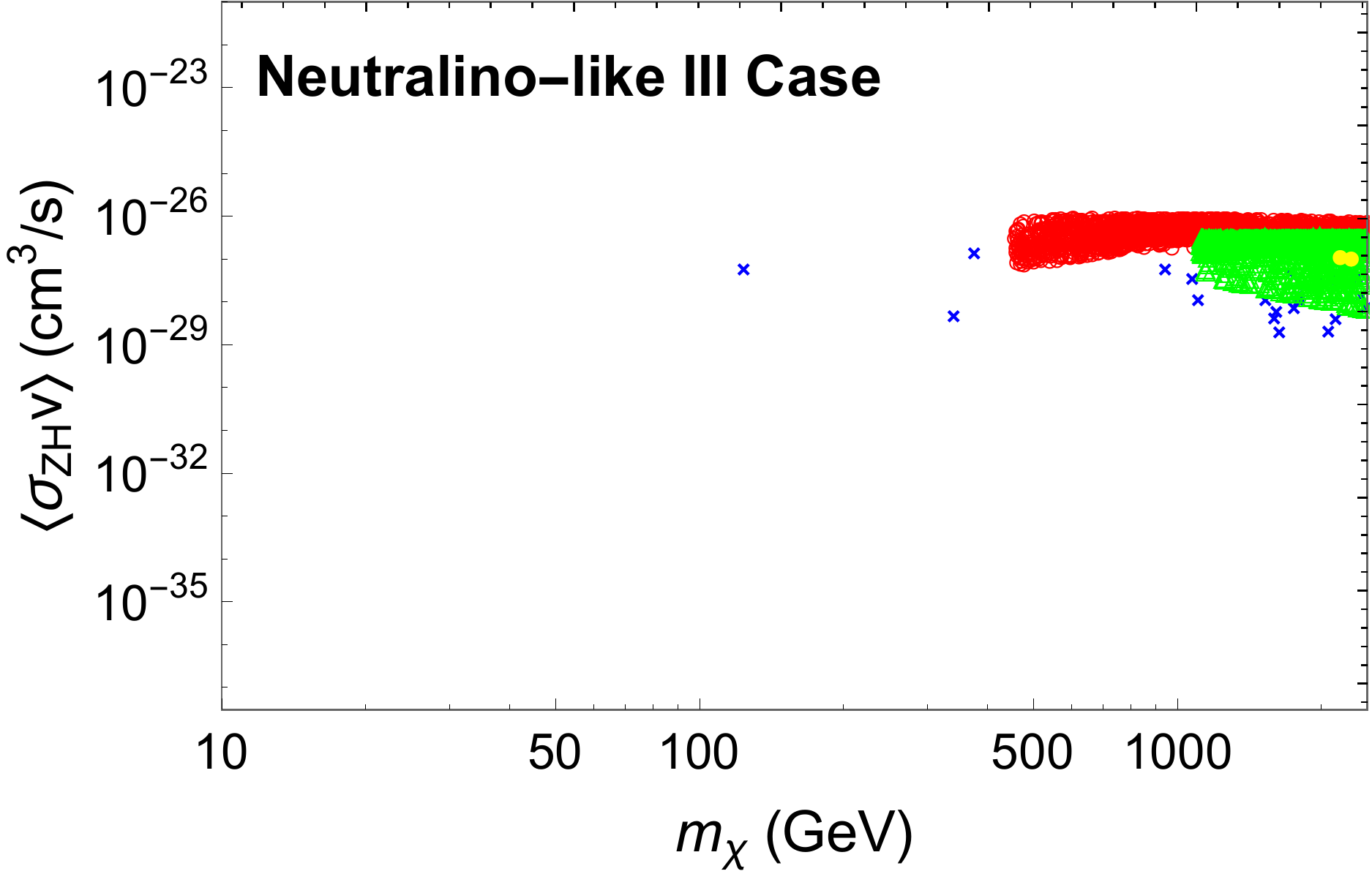}
}\subfigure{
  \includegraphics[width=0.3\textwidth,height=0.133\textheight]{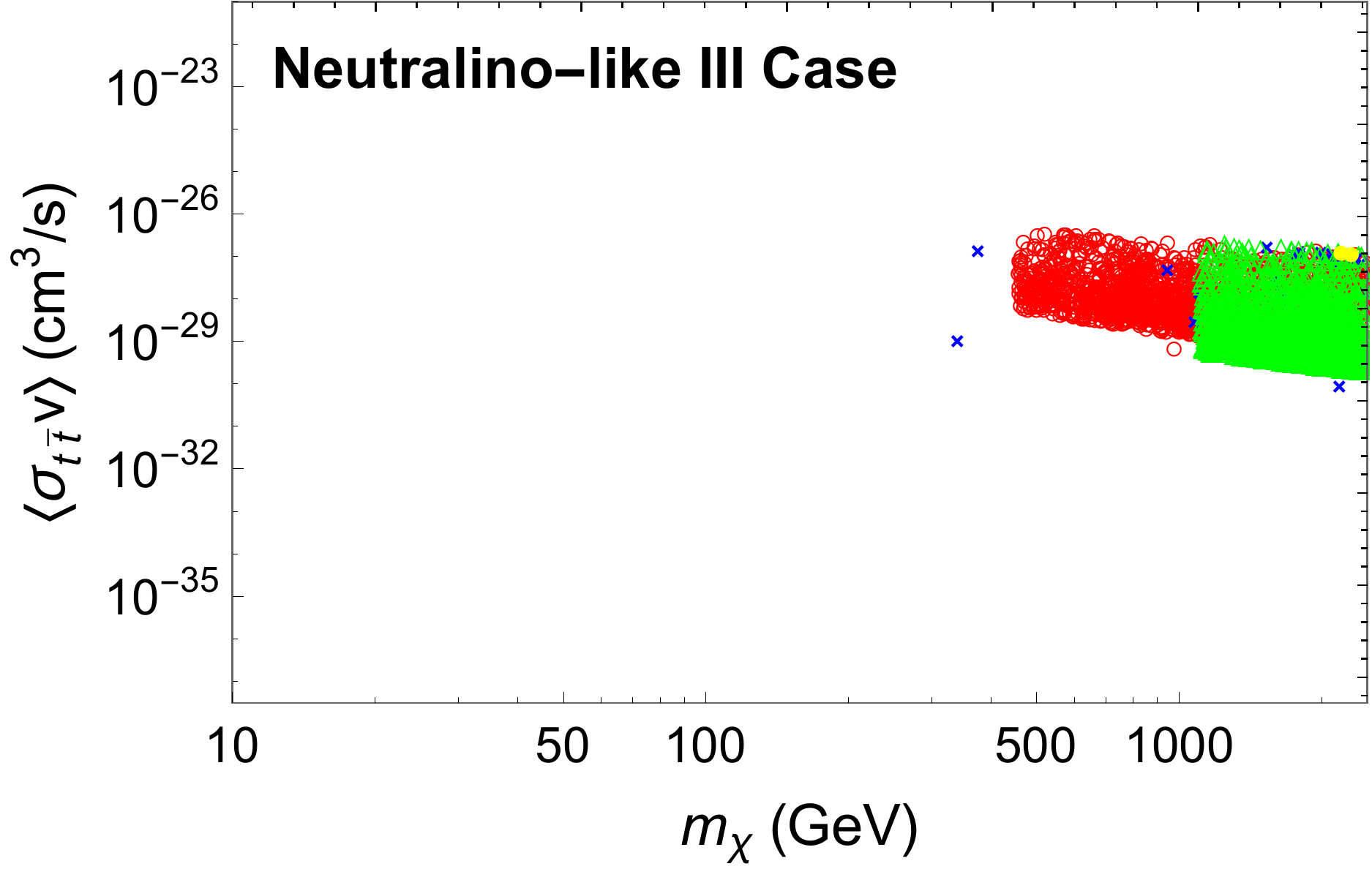}
}\\
  \subfigure{
  \includegraphics[width=0.3\textwidth,height=0.133\textheight]{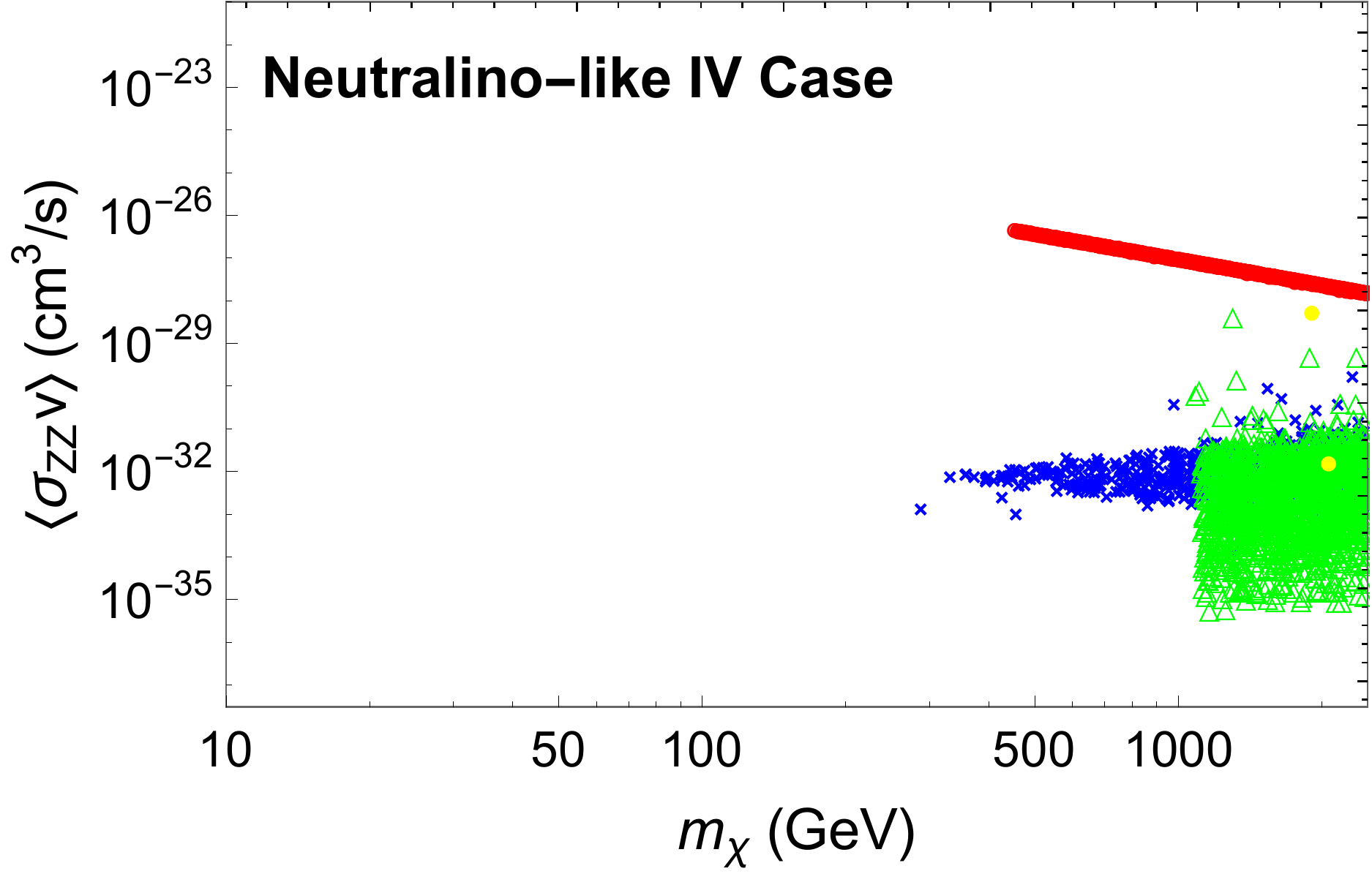}
}\subfigure{
  \includegraphics[width=0.3\textwidth,height=0.133\textheight]{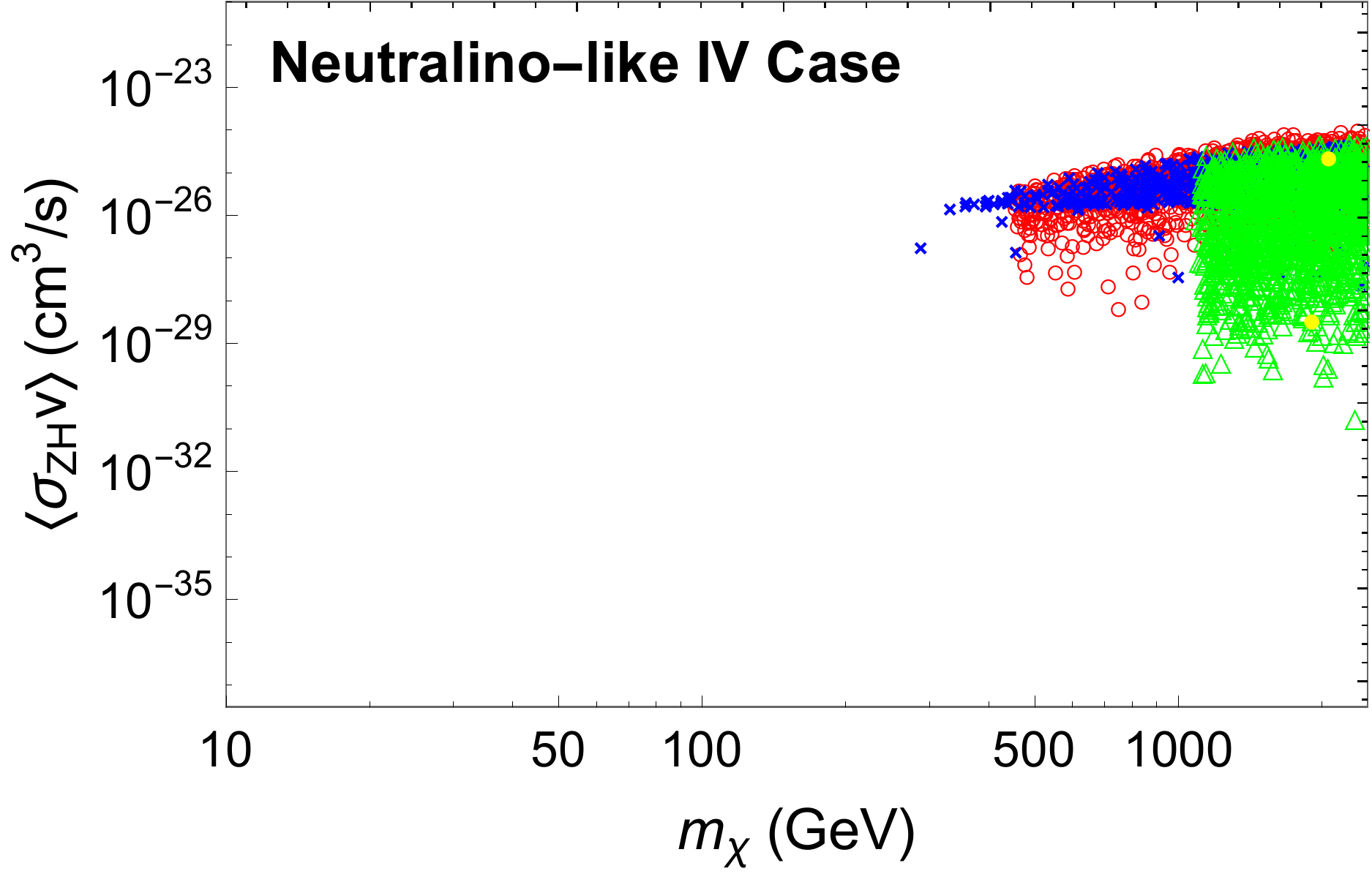}
}\subfigure{
  \includegraphics[width=0.3\textwidth,height=0.133\textheight]{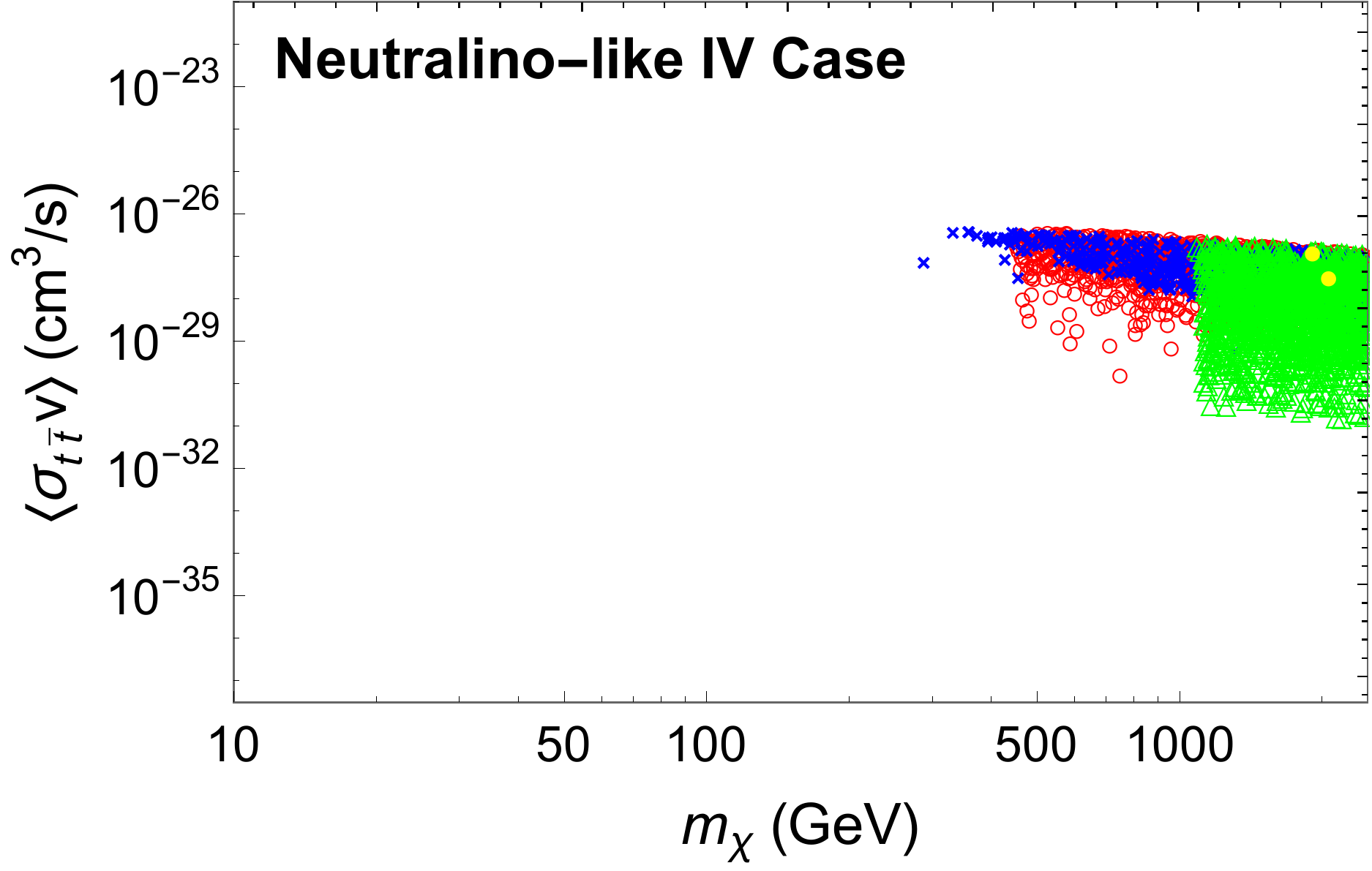}
}\\
  \subfigure{
  \includegraphics[width=0.3\textwidth,height=0.133\textheight]{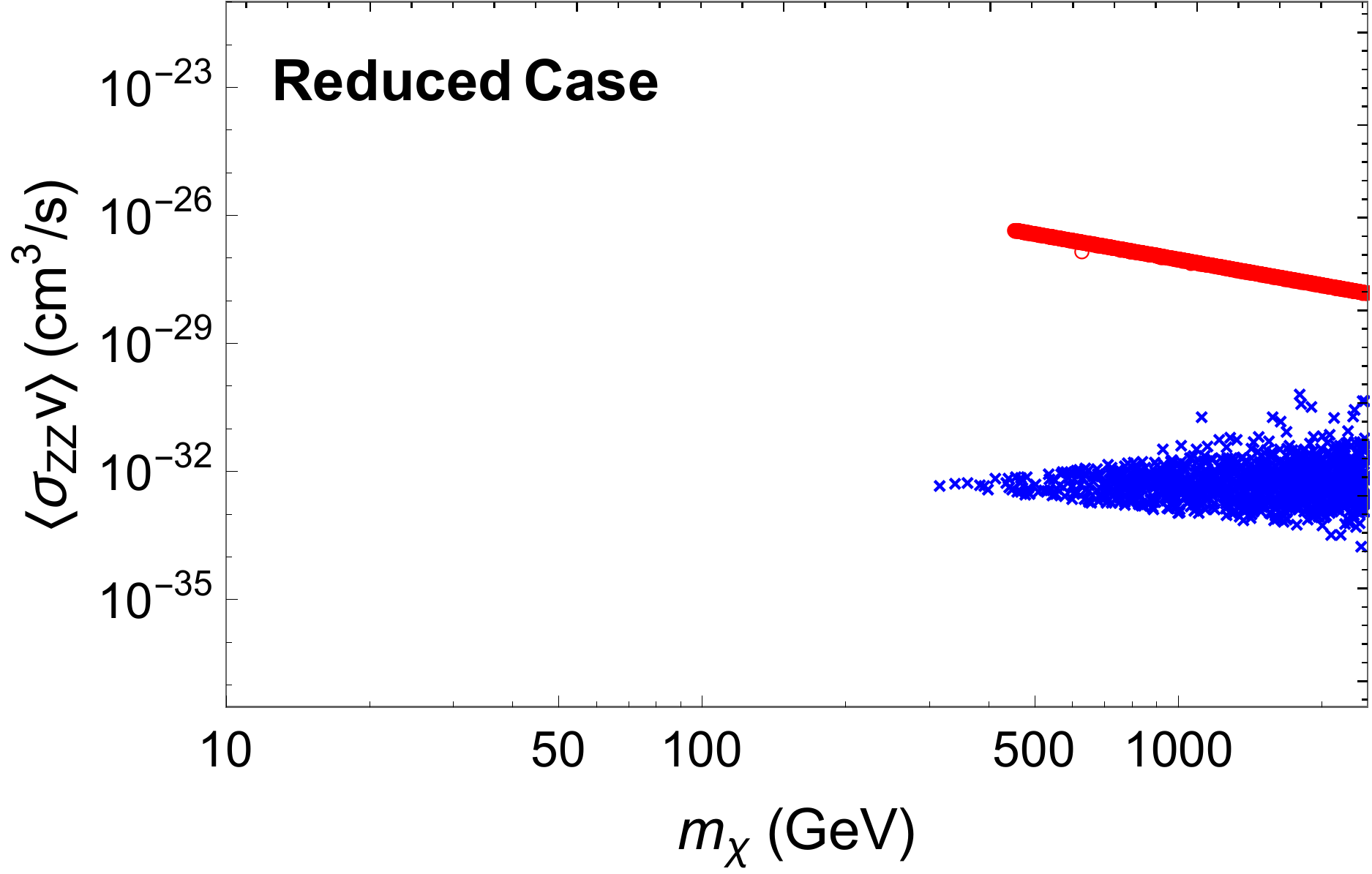}
}\subfigure{
  \includegraphics[width=0.3\textwidth,height=0.133\textheight]{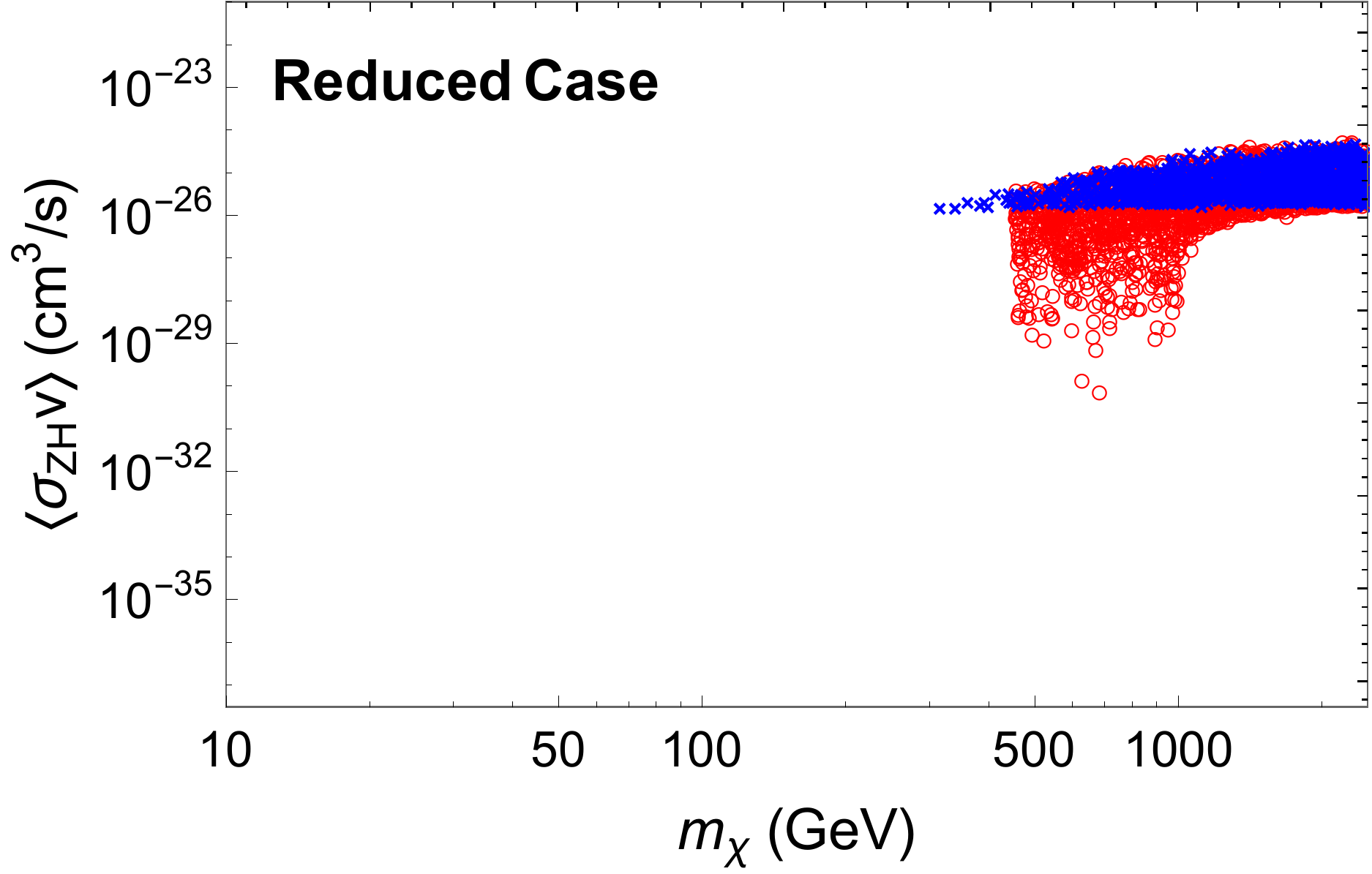}
}\subfigure{
  \includegraphics[width=0.3\textwidth,height=0.133\textheight]{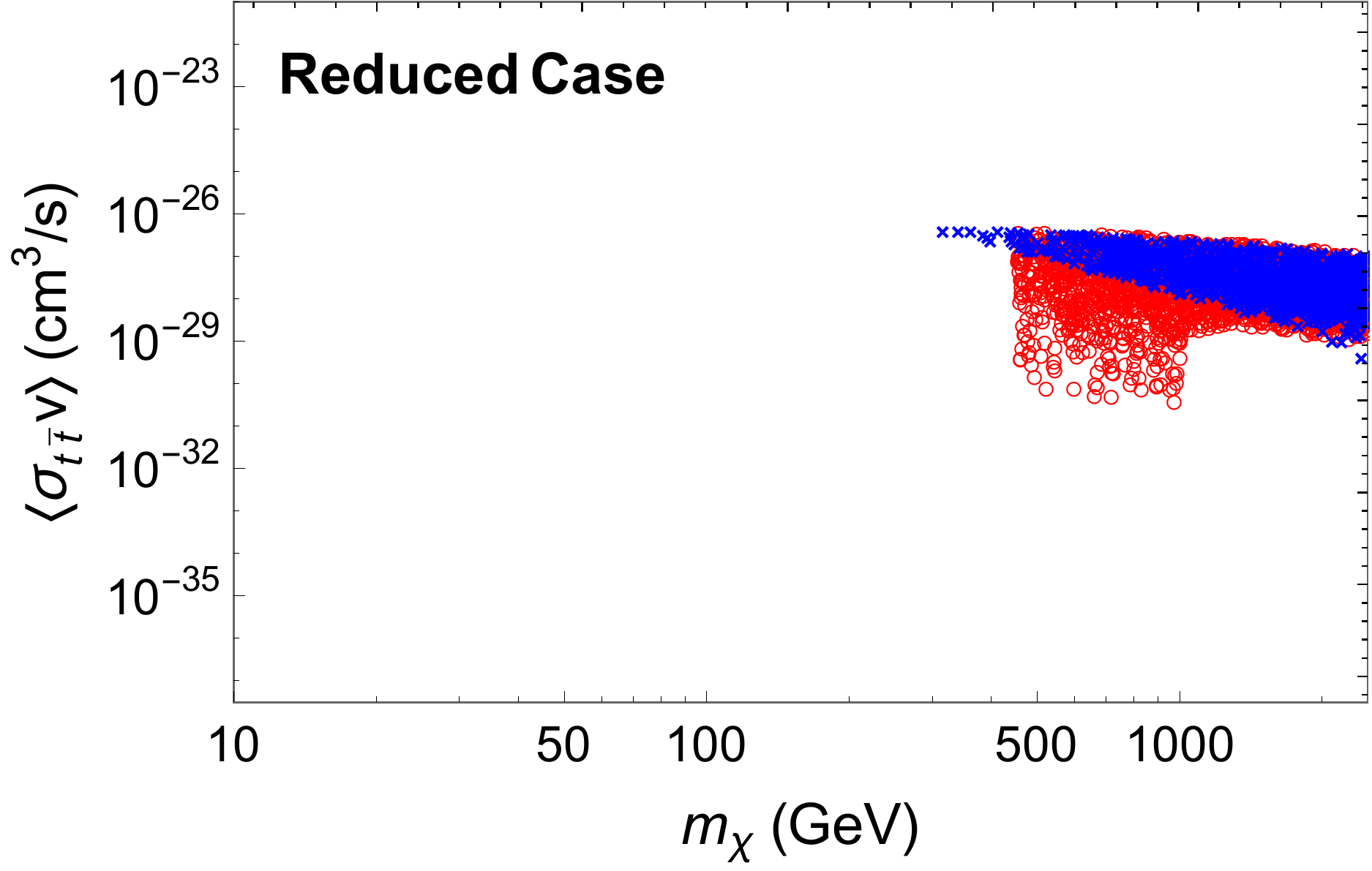}
}\\
  \subfigure{
  \includegraphics[width=0.3\textwidth,height=0.133\textheight]{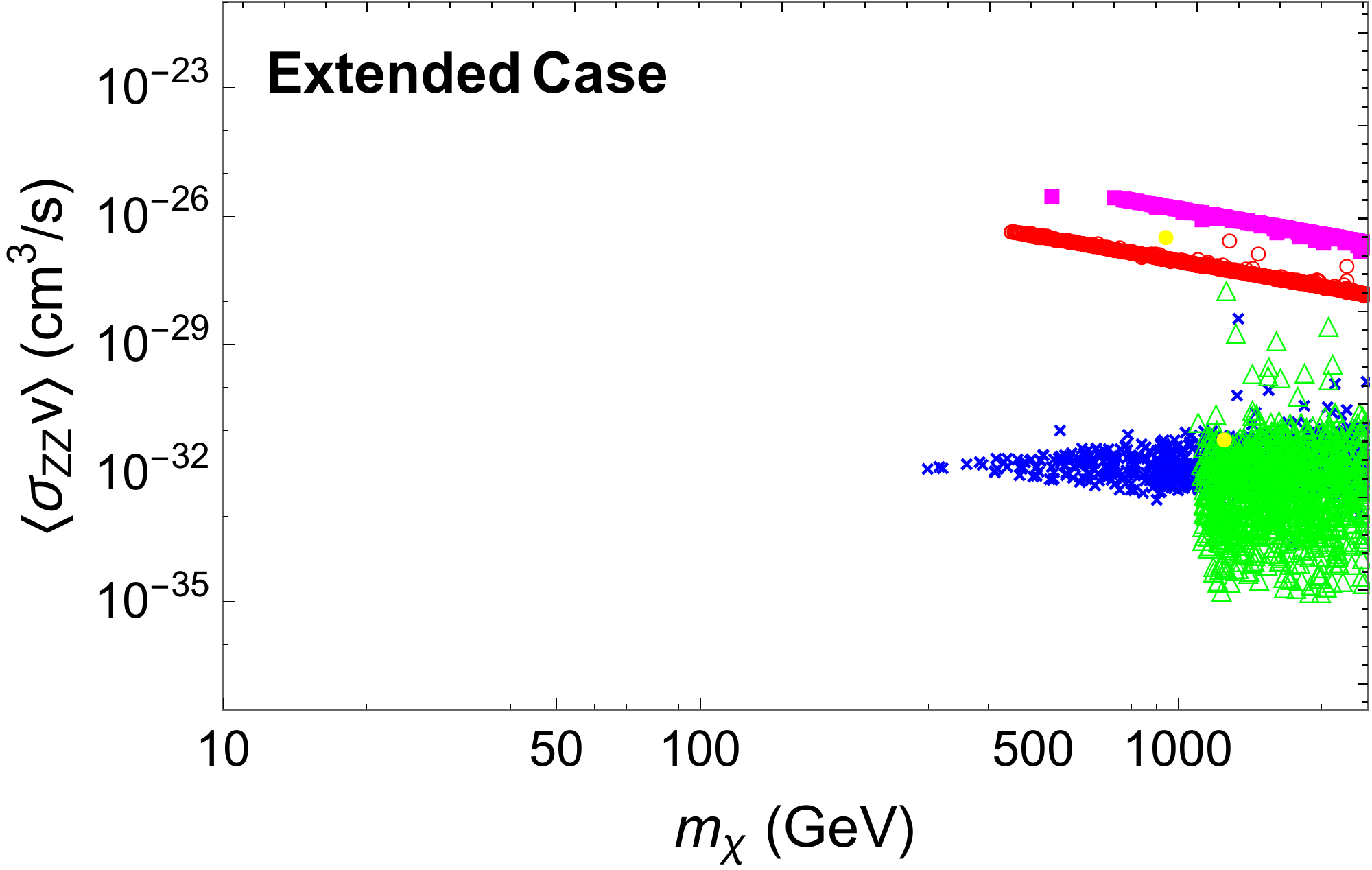}
}\subfigure{
  \includegraphics[width=0.3\textwidth,height=0.133\textheight]{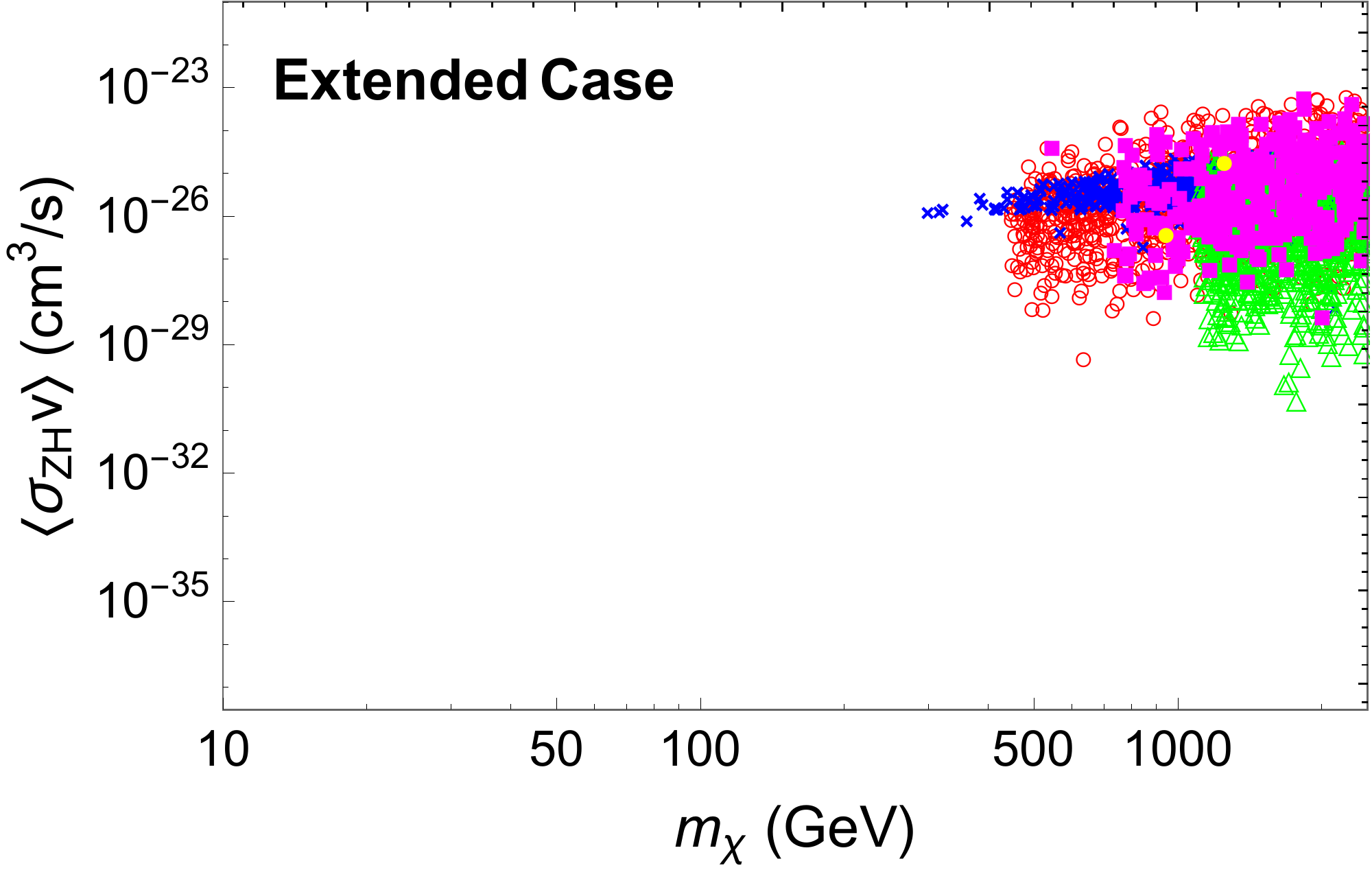}
}\subfigure{
  \includegraphics[width=0.3\textwidth,height=0.133\textheight]{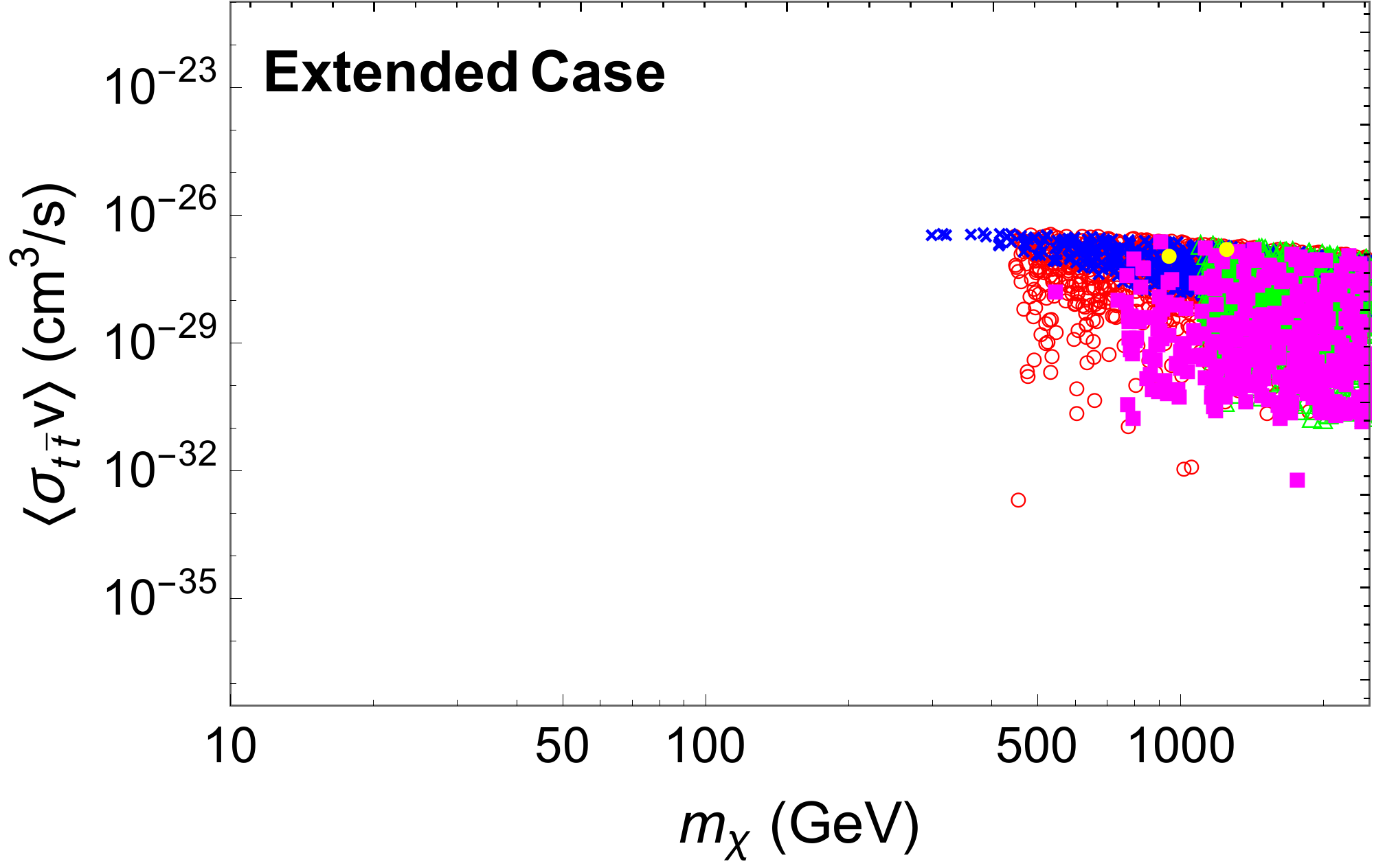}
}
\caption{Predictions of  $\la \sigma_{ZZ, ZH, t{\bar t}}\ v \ra$ versus $m_\chi$ for allowed DM candidates
[{\color{red} $\circ$}:~higgsino-like,~
{\color{blue} $\times$}:~bino-like,~
{\color{green} $\triangle$}:~wino-like,~
{\color{magenta} $\blacksquare$}:~non neutralino-like,
{\color{yellow} $\bullet$}:~mixed].}
\label{fig:allow ZZZHtt}
\end{figure}

The Fermi-LAT constraint on $\la\sigma (\chi{\chi}\rightarrow W^+W^-) v\ra$ is more useful than other Fermi-LAT constraints with light $f\bar f$ in the final states. 
On the other hand, from the discussion of the properties of DM annihilation processes $\chi\chi\rightarrow W^+W^-, ZZ, ZH, HH, f\bar f$ in Sec. II-B, we know that only the process of $\chi\chi\rightarrow HH$ has no $s$-wave contribution and the process $\chi\chi\rightarrow f\bar f$ favors heavy fermion pairs.
Hence it is also important to study DM annihilation to gauge boson and heavy quark processes. In Fig.~\ref{fig:allow ZZZHtt}, we show our predictions on
 $\la\sigma (\chi{\chi}\rightarrow ZZ, ZH, t{\bar t}) v\ra$ with the allowed samples. Their values of $\la\sigma v\ra$ can be as large as $10^{-26}$ cm$^3/$s. It will be useful to search DM with these processes.

\begin{table}[t!]
\footnotesize{
\begin{tabular}{|c|cccc|c|c|}
  \hline
  & 
  &
  & {\hspace{-2.7cm}Case A}
  &
  & Case B
  & Case C
  \\
  \%
  & neutralino-like I & neutralino-like II & neutralino-like III & neutralino-like IV & Reduced & Extended \\
  \hline
  $\tilde H$-like &
  \begin{tabular}[c]{@{}l@{}} (29, 18) \\ \quad \ 63 \end{tabular} &
  \begin{tabular}[c]{@{}l@{}} (28, 14) \\ \quad \ 49 \end{tabular} &
  \begin{tabular}[c]{@{}l@{}} (33, 15) \\ \quad \ 45 \end{tabular} &
  \begin{tabular}[c]{@{}l@{}} (31, 15) \\ \quad \ 46 \end{tabular} &
  \begin{tabular}[c]{@{}l@{}} (50, 24) \\ \quad \ 48 \end{tabular} &
  \begin{tabular}[c]{@{}l@{}} (29, 12) \\ \quad \ 43 \end{tabular} \\
 % \begin{tabular}[c]{@{}l@{}} (28.51, 17.95) \\ \quad\quad 62.96 \end{tabular} &
 % \begin{tabular}[c]{@{}l@{}} (27.59, 13.53) \\ \quad\quad 49.04 \end{tabular} &
  %\begin{tabular}[c]{@{}l@{}} (32.76, 14.72) \\ \quad\quad 44.93 \end{tabular} &
  %\begin{tabular}[c]{@{}l@{}} (31.26, 14.52) \\ \quad\quad 46.45 \end{tabular} &
  %\begin{tabular}[c]{@{}l@{}} (49.73, 23.70) \\ \quad\quad 47.66 \end{tabular} &
  %\begin{tabular}[c]{@{}l@{}} (28.62, 12.32) \\ \quad\quad 43.05 \end{tabular} \\
  \hline
  $\tilde B$-like &
  \begin{tabular}[c]{@{}l@{}} (71,  0.2) \\ \quad\ 0.3 \end{tabular} &
  \begin{tabular}[c]{@{}l@{}} (72,  0.2) \\ \quad\ 0.3 \end{tabular} &
  \begin{tabular}[c]{@{}l@{}} (33,  0.3) \\ \quad\ 0.9 \end{tabular} &
  \begin{tabular}[c]{@{}l@{}} (34,  8) \\ \quad\ 23 \end{tabular} &
  \begin{tabular}[c]{@{}l@{}} (49,11) \\ \quad\ 23 \end{tabular} &
  \begin{tabular}[c]{@{}l@{}} (34,  7) \\ \quad 22 \end{tabular} \\
  %\begin{tabular}[c]{@{}l@{}} (71.07,  0.22) \\ \quad\quad 0.31 \end{tabular} &
  %\begin{tabular}[c]{@{}l@{}} (72.07,  0.18) \\ \quad\quad 0.25 \end{tabular} &
  %\begin{tabular}[c]{@{}l@{}} (33.38,  0.30) \\ \quad\quad 0.90 \end{tabular} &
  %\begin{tabular}[c]{@{}l@{}} (33.53,  7.85) \\ \quad\quad 23.41 \end{tabular} &
  %\begin{tabular}[c]{@{}l@{}} (49.43,11.29) \\ \quad\quad 22.84 \end{tabular} &
 % \begin{tabular}[c]{@{}l@{}} (33.94,  7.42) \\ \quad\quad 21.86 \end{tabular} \\
  \hline
  $\tilde W$-like &
  \begin{tabular}[c]{@{}l@{}} X \\   X \end{tabular} &
  \begin{tabular}[c]{@{}l@{}} X \\   X \end{tabular} &
  \begin{tabular}[c]{@{}l@{}} (33, 15) \\ \quad\ 45 \end{tabular} &
  \begin{tabular}[c]{@{}l@{}} (34, 13) \\ \quad\ 39 \end{tabular} &
  %\begin{tabular}[c]{@{}l@{}} (33.31, 15.03) \\\ \quad\quad 45.12 \end{tabular} &
  %\begin{tabular}[c]{@{}l@{}} (33.97, 13.19) \\ \quad\quad 38.83 \end{tabular} &
  \begin{tabular}[c]{@{}l@{}} X \\   X \end{tabular} &
  \begin{tabular}[c]{@{}l@{}} (31, 10) \\ \quad\ 31 \end{tabular} \\
  %\begin{tabular}[c]{@{}l@{}} (30.93, 9.56) \\ \quad\quad 30.91 \end{tabular} \\
  \hline
  $\tilde X$-like &
  \begin{tabular}[c]{@{}l@{}}  X \\   X \end{tabular} &
  \begin{tabular}[c]{@{}l@{}}  X \\   X \end{tabular} &
  \begin{tabular}[c]{@{}l@{}}  X \\   X \end{tabular} &
  \begin{tabular}[c]{@{}l@{}}  X \\   X \end{tabular} &
  \begin{tabular}[c]{@{}l@{}}  X \\   X \end{tabular} &
  \begin{tabular}[c]{@{}l@{}} (5, 3) \\ \ \ 62 \end{tabular} \\
 % \begin{tabular}[c]{@{}l@{}} ( 5.25,  3.24) \\ \quad\quad 61.71 \end{tabular} \\
  \hline
  \end{tabular}
\caption{Particle attribute distribution of the allowed DM candidates. The values in the first row ``$\tilde H$-like" and the first column ``neutralino-like I" of the table mean that $29\%$ of the whole sample in neutralino-like I case are $\tilde H$-like and only $18\%$ of the whole sample are the allowed $\tilde H$-like particles, or equivalently, among the $\tilde H$-like particles, only $63\%$ of them are allowed.}
\label{tab:90main}}
\end{table}

In Table \ref{tab:90main}, we summarize the distribution of allowed samples satisfying all  constraints.
The two values in the parentheses of the table show the percentages (with regard to the whole sample) of a specified particle attribute before and after being subjected to the constraints respectively. For example,
in the first row ``$\tilde H$" and the first column ``neutralino-like I case" of the table, we see that there are $29\%$ of the whole sample in neutralino-like I case being $\tilde H$-like particles and only $18\%$ of the whole sample being allowed $\tilde H$-like particles.
Among the $\tilde H$-like particles, only $63\%$ of them survive under the constraints and this surviving rate is shown below the parenthesis. From this table, we see that less $\tilde H$-like particles are allowed (relative to neutralino-like I) and less $\tilde B$-like particles can survive in the cases with the $\tan\beta$ relation (neutralino-like I - III). As mentioned before, it is due to the fact that ``a higher $\tan\beta$ value" or ``without the GUT relation" can give us wider spreads in the scatter plots. It results in that more $\tilde H$-like particles spread into the prohibited region in the $\sigma^{SI}_N$ scatter plot.
On the other hand, with the $\tan\beta$ relation, less $\tilde B$-like particles can spread into the allowed region in the $\Omega_{\chi} h^2$ scatter plot. 

As shown in the table, in the neutralino-like III, IV and the extended cases, we have plenty of $\tilde W$-like particles.
The $\tilde W$-particles with $m_{\chi} \lesssim M_W$ are ruled out by the relic density and the Fermi-LAT $\la\sigma (\chi{\chi}\rightarrow b\bar b) v\ra$ constraints, while the $\tilde W$-particles with $m_{\chi} > M_W$ are subjected to the Fermi-LAT $\la\sigma (\chi{\chi}\rightarrow W^+W^-) v\ra$ constraint and the LUX $\sigma^{SI}_N$ constraint.
The fewer relations on model parameters give wider spread in the scatter plots of $\Omega_{\chi} h^2$, $\sigma^{SI}_N$ and $\la\sigma (\chi {\chi}\rightarrow W^+W^-)$, resulting in lower surviving rates of
$\tilde W$-like DM candidates, namely, $45\%$, $39\%$ and $31\%$ in the neutralino-like III, IV and the extended cases respectively.
As for the non neutralino-like $\tilde X$ particles, $62\%$ of them could be DM candidates.

\begin{table}[t!]
\scriptsize{
\begin{tabular}{|c|cccc|c|c|}
  \hline
  &
  &
  & {\hspace{-2.7cm}Case A}
  &
  & Case B
  & Case C
  \\
  & neutralino-like I & neutralino-like II & neutralino-like III & neutralino-like IV & Reduced & Extended \\
  \hline
  $m_{\chi}$ &
  \begin{tabular}[c]{@{}l@{}} (\ \ \ 51.1, 2495.1) \\ (1116.6, 2496.7)\end{tabular} &
  \begin{tabular}[c]{@{}l@{}} (\ \ \ 57.1, 2498.9) \\ (1018.2, 2471.4)\end{tabular} &
  \begin{tabular}[c]{@{}l@{}} (\ \ 58.9, 2498.3) \\ ( 947.7, 2454.6)\end{tabular} &
  \begin{tabular}[c]{@{}l@{}} (\ \ \ 54.8, 2499.9) \\ (  \ 332.6, 2464.6)\end{tabular} &
  \begin{tabular}[c]{@{}l@{}} (\ \ \ 57.8, 2498.8) \\ (  \ 316.6, 2481.7)\end{tabular} &
  \begin{tabular}[c]{@{}l@{}} (\ \ \ 54.0, 2499.0) \\ (  \ 299.0, 2383.5)\end{tabular} \\
  \hline
  $\mu_{1}$ &
  \begin{tabular}[c]{@{}l@{}} (\ \ \ 52.3, 6933.9) \\ (1118.2, 2499.3)\end{tabular} &
  \begin{tabular}[c]{@{}l@{}} (\ \ \ 58.1, 3655.4) \\ (1019.1, 2473.8)\end{tabular} &
  \begin{tabular}[c]{@{}l@{}} (\ \ \ 62.7, 7982.6) \\ (  980.4, 5452.8)\end{tabular} &
  \begin{tabular}[c]{@{}l@{}} ( 59.34, 7999.4) \\ ( 901.6, 7896.2)\end{tabular} &
  \begin{tabular}[c]{@{}l@{}} ( 58.56, 7997.0) \\ ( 953,9, 7953.3)\end{tabular} &
  \begin{tabular}[c]{@{}l@{}} ( 52.38, 7998.4) \\ (1040.5, 7858.5)\end{tabular} \\
  \hline
  $\mu_{2}$ &
  \begin{tabular}[c]{@{}l@{}} (\ \ \ 58.4, 3814.3) \\ (1430.8, 3811.4)\end{tabular} &
  \begin{tabular}[c]{@{}l@{}} ( 0.214, 3823.5) \\ (2027.4, 3802.9)\end{tabular} &
  \begin{tabular}[c]{@{}l@{}} ( 59.80, 7999.9) \\ (1781.6, 7979.6)\end{tabular} &
  \begin{tabular}[c]{@{}l@{}} ( \ 56.99, 7999.4) \\ ( \ 339.5, 7970.8)\end{tabular} &
  \begin{tabular}[c]{@{}l@{}} ( \ 61.67, 7998.2) \\ ( \ 323.4, 7977.4)\end{tabular} &
  \begin{tabular}[c]{@{}l@{}} ( 56.7, 7996.9) \\ (305.4, 7134.6)\end{tabular} \\
  \hline
  $\mu_{3}$ &
  \begin{tabular}[c]{@{}l@{}} (  122.2, 7978.8) \\ (2993.0, 7972.8)\end{tabular} &
  \begin{tabular}[c]{@{}l@{}} ( 0.447, 7998.2) \\ (4240.9, 7955.1)\end{tabular} &
  \begin{tabular}[c]{@{}l@{}} ( 5.328, 7999.5) \\ (1816.7, 7977.2)\end{tabular} &
  \begin{tabular}[c]{@{}l@{}} ( \ 63.0, 7998.6) \\ (  848.5, 7992.6)\end{tabular} &
  \begin{tabular}[c]{@{}l@{}} X \\ X \end{tabular} &
  \begin{tabular}[c]{@{}l@{}} ( 1.095, 7994.6) \\ (1305.3, 7951.2)\end{tabular} \\
  \hline
  $\mu_{4}$ &
  \begin{tabular}[c]{@{}l@{}} X \\ X \end{tabular} &
  \begin{tabular}[c]{@{}l@{}} X \\ X \end{tabular} &
  \begin{tabular}[c]{@{}l@{}} X \\ X \end{tabular} &
  \begin{tabular}[c]{@{}l@{}} X \\ X \end{tabular} &
  \begin{tabular}[c]{@{}l@{}} X \\ X \end{tabular} &
  \begin{tabular}[c]{@{}l@{}} (\ 681.80, 7999.4) \\ ( 681.80, 7633.5)\end{tabular} \\
  \hline
  $\mu_{5}$ &
  \begin{tabular}[c]{@{}l@{}} X \\ X \end{tabular} &
  \begin{tabular}[c]{@{}l@{}} X \\ X \end{tabular} &
  \begin{tabular}[c]{@{}l@{}} X \\ X \end{tabular} &
  \begin{tabular}[c]{@{}l@{}} X \\ X \end{tabular} &
  \begin{tabular}[c]{@{}l@{}} X \\ X \end{tabular} &
  \begin{tabular}[c]{@{}l@{}} (551.7, 7998.7) \\ (551.7, 7771.8)\end{tabular} \\
  \hline
  $g_{3}$ &
  \begin{tabular}[c]{@{}l@{}} 0.111\\ 0.111\end{tabular} &
  \begin{tabular}[c]{@{}l@{}} 0.012\\ 0.012\end{tabular} &
  \begin{tabular}[c]{@{}l@{}} 0.111\\ 0.111\end{tabular} &
  \begin{tabular}[c]{@{}l@{}} (1.40e-4, 0.999) \\ (2.61e-2, 0.998)\end{tabular} &
  \begin{tabular}[c]{@{}l@{}} (2.86e-4, 0.999) \\ (5.20e-4, 0.976)\end{tabular} &
  \begin{tabular}[c]{@{}l@{}} (2.57e-4, 0.999) \\ (7.76e-3, 0.975)\end{tabular} \\
  \hline
  $g_{4}$ &
  \begin{tabular}[c]{@{}l@{}} 0.221\\ 0.221\end{tabular} &
  \begin{tabular}[c]{@{}l@{}} 0.247\\ 0.247\end{tabular} &
  \begin{tabular}[c]{@{}l@{}} 0.221\\ 0.221\end{tabular} &
  \begin{tabular}[c]{@{}l@{}} (1.90e-4, 0.999) \\ (5.20e-2, 0.994)\end{tabular} &
  \begin{tabular}[c]{@{}l@{}} (5.12e-4, 0.999) \\ (8.50e-3, 0.996)\end{tabular} &
  \begin{tabular}[c]{@{}l@{}} (3.00e-4, 0.999) \\ (8.38e-3, 0.988)\end{tabular} \\
  \hline
  $g_{5}$ &
  \begin{tabular}[c]{@{}l@{}} 0.207\\ 0.207\end{tabular} &
  \begin{tabular}[c]{@{}l@{}} 0.023\\ 0.023\end{tabular} &
  \begin{tabular}[c]{@{}l@{}} 0.207\\ 0.207\end{tabular} &
  \begin{tabular}[c]{@{}l@{}} (1.26e-3, 0.999) \\ (5.51e-3, 0.979)\end{tabular} &
  \begin{tabular}[c]{@{}l@{}} X \\ X \end{tabular} &
  \begin{tabular}[c]{@{}l@{}} (6.00e-4, 0.999) \\ (1.03e-2, 0.993)\end{tabular} \\
  \hline
  $g_{6}$ &
  \begin{tabular}[c]{@{}l@{}} 0.413\\ 0.413\end{tabular} &
  \begin{tabular}[c]{@{}l@{}} 0.462\\ 0.462\end{tabular} &
  \begin{tabular}[c]{@{}l@{}} 0.413\\ 0.413\end{tabular} &
  \begin{tabular}[c]{@{}l@{}} (1.37e-5, 0.999) \\ (1.31e-2, 0.994)\end{tabular} &
  \begin{tabular}[c]{@{}l@{}} X \\ X \end{tabular} &
  \begin{tabular}[c]{@{}l@{}} (1.46e-4, 0.999) \\ (1.91e-2, 0.995)\end{tabular} \\
  \hline
  $g_{7}$ &
  \begin{tabular}[c]{@{}l@{}} X \\ X \end{tabular} &
  \begin{tabular}[c]{@{}l@{}} X \\ X \end{tabular} &
  \begin{tabular}[c]{@{}l@{}} X \\ X \end{tabular} &
  \begin{tabular}[c]{@{}l@{}} X \\ X \end{tabular} &
  \begin{tabular}[c]{@{}l@{}} X \\ X \end{tabular} &
  \begin{tabular}[c]{@{}l@{}} (6.47e-4, 0.999) \\ (8.79e-3, 0.985)\end{tabular} \\
  \hline
  $g_{8}$ &
  \begin{tabular}[c]{@{}l@{}} X \\ X \end{tabular} &
  \begin{tabular}[c]{@{}l@{}} X \\ X \end{tabular} &
  \begin{tabular}[c]{@{}l@{}} X \\ X \end{tabular} &
  \begin{tabular}[c]{@{}l@{}} X \\ X \end{tabular} &
  \begin{tabular}[c]{@{}l@{}} X \\ X \end{tabular} &
  \begin{tabular}[c]{@{}l@{}} (6.70e-4, 0.999) \\ (6.70e-4, 0.998)\end{tabular} \\
  \hline
  $g_{9}$ &
  \begin{tabular}[c]{@{}l@{}} X \\ X \end{tabular} &
  \begin{tabular}[c]{@{}l@{}} X \\ X \end{tabular} &
  \begin{tabular}[c]{@{}l@{}} X \\ X \end{tabular} &
  \begin{tabular}[c]{@{}l@{}} X \\ X \end{tabular} &
  \begin{tabular}[c]{@{}l@{}} X \\ X \end{tabular} &
  \begin{tabular}[c]{@{}l@{}} (1.02e-4, 0.999) \\ (1.45e-2, 0.981)\end{tabular} \\
  \hline
  $g_{10}$ &
  \begin{tabular}[c]{@{}l@{}} X \\ X \end{tabular} &
  \begin{tabular}[c]{@{}l@{}} X \\ X \end{tabular} &
  \begin{tabular}[c]{@{}l@{}} X \\ X \end{tabular} &
  \begin{tabular}[c]{@{}l@{}} X \\ X \end{tabular} &
  \begin{tabular}[c]{@{}l@{}} X \\ X \end{tabular} &
  \begin{tabular}[c]{@{}l@{}} (3.04e-5, 0.999) \\ (1.98e-2, 0.994)\end{tabular} \\
  \hline
  $|a_q/m_q|$ &
  \begin{tabular}[c]{@{}l@{}} (9.3e-10, 9.02e-8) \\ (6.15e-9, 6.96e-8)\end{tabular} &
  \begin{tabular}[c]{@{}l@{}} (2.5e-10, 7.05e-8) \\ (3.69e-9, 1.28e-8)\end{tabular} &
  \begin{tabular}[c]{@{}l@{}} (1.52e-9, 9.08e-8) \\ (2.40e-9, 6.86e-8)\end{tabular} &
  \begin{tabular}[c]{@{}l@{}} (7.5e-10, 9.52e-8) \\ (4.87e-9, 6.95e-8)\end{tabular} &
  \begin{tabular}[c]{@{}l@{}} (5.6e-10, 9.03e-8) \\ (6.6e-10, 9.31e-8)\end{tabular} &
  \begin{tabular}[c]{@{}l@{}} (1.5e-11, 9.18e-8) \\ (1.5e-9, 7.28e-8)\end{tabular} \\
  \hline
  $|d_q|$ &
  \begin{tabular}[c]{@{}l@{}} (9.8e-11, 4.31e-8) \\ (1.38e-9, 1.57e-8)\end{tabular} &
  \begin{tabular}[c]{@{}l@{}} (5.7e-10, 8.62e-8) \\ (3.04e-9, 7.66e-9)\end{tabular} &
  \begin{tabular}[c]{@{}l@{}} (1.59e-10, 3.20e-8) \\ (2.17e-10, 1.26e-8)\end{tabular} &
  \begin{tabular}[c]{@{}l@{}} (9.1e-12,  5.70e-8) \\ (2.05e-11, 2.46e-8)\end{tabular} &
  \begin{tabular}[c]{@{}l@{}} (3.8e-13, 5.02e-8)\\ (3.4e-11, 3.19e-8)\end{tabular} &
  \begin{tabular}[c]{@{}l@{}} (1.7e-12, 3.62e-7)\\ (2.9e-12, 3.52e-8)\end{tabular} \\
  \hline
\end{tabular}}
\newline
\caption{Allowed range for DM mass, model parameters and effective couplings.\qquad\qquad
The upper and lower intervals represent the allowed range for samples satisfying all the constraints with $\Omega_{\chi}h^2$ in the criteria C1 ( $\leq +3\sigma$) and C2 (within $\pm 3\sigma$) respectively. }
\label{tab:range-1}
\end{table}
\vfill
\eject
%%%%%%%%%%%%%%%%%%%%%%%%%%%%%%%%%%%%%%%%%%
\begin{table}[t!]
\scriptsize{
\begin{tabular}{|c|cccc|c|c|}
  \hline
  &
  &
  & {\hspace{-2.7cm}Case A}
  &
  & Case B
  & Case C
  \\
  & neutralino-like I & neutralino-like II & neutralino-like III & neutralino-like IV & Reduced & Extended \\
  \hline
  $|g^{L_{H}}_{11}|$ &
  \begin{tabular}[c]{@{}l@{}} (1.80e-3,1.75e-1) \\ (1.20e-2,1.35e-1)\end{tabular} &
  \begin{tabular}[c]{@{}l@{}} (4.84e-4,2.49e-1) \\ (7.17e-3,2.49e-1)\end{tabular} &
  \begin{tabular}[c]{@{}l@{}} (2.95e-3,1.77e-1) \\ (4.66e-3,1.33e-1)\end{tabular} &
  \begin{tabular}[c]{@{}l@{}} (1.45e-3,1.85e-1) \\ (9.46e-3,1.35e-1)\end{tabular} &
  \begin{tabular}[c]{@{}l@{}} (1.08e-3,1.81e-1) \\ (1.33e-3,1.81e-1)\end{tabular} &
  \begin{tabular}[c]{@{}l@{}} (2.85e-5,1.78e-1) \\ (2.94e-3,1.42e-1)\end{tabular} \\
  \hline
  $|g^{L_{Z}}_{11}|$ &
  \begin{tabular}[c]{@{}l@{}} (4.39e-6,1.93e-3) \\ (6.20e-5,7.03e-4)\end{tabular} &
  \begin{tabular}[c]{@{}l@{}} (2.57e-5,3.87e-3) \\ (1.37e-4,3.34e-4)\end{tabular} &
  \begin{tabular}[c]{@{}l@{}} (7.12e-6,1.43e-3) \\ (9.73e-6,5.62e-4)\end{tabular} &
  \begin{tabular}[c]{@{}l@{}} (4.07e-7,2.56e-3) \\ (9.20e-7,1.11e-3)\end{tabular} &
  \begin{tabular}[c]{@{}l@{}} (1.71e-8,2.25e-3) \\ (1.51e-6,1.43e-3)\end{tabular} &
  \begin{tabular}[c]{@{}l@{}} (7.81e-8,1.62e-2) \\ (1.30e-7,1.58e-3)\end{tabular} \\
  \hline
  $|g^{L_{W^-}}_{11}|$ &
  \begin{tabular}[c]{@{}l@{}} (3.95e-3,3.28e-1) \\ (1.42e-1,3.27e-1)\end{tabular} &
  \begin{tabular}[c]{@{}l@{}} (8.96e-4,3.30e-1) \\ (3.26e-1,3.27e-1)\end{tabular} &
  \begin{tabular}[c]{@{}l@{}} (1.22e-3,6.53e-1) \\ (3.02e-2,3.27e-1)\end{tabular} &
  \begin{tabular}[c]{@{}l@{}} (8.75e-6,6.54e-1) \\ (5.06e-5,3.27e-1)\end{tabular} &
  \begin{tabular}[c]{@{}l@{}} (1.35e-3,3.28e-1) \\ (5.88e-3,3.27e-1)\end{tabular} &
  \begin{tabular}[c]{@{}l@{}} (8.50e-7,6.53e-1) \\ (4.03e-6,3.26e-1)\end{tabular} \\
  \hline
  $|g^{R_{W^-}}_{11}|$ &
  \begin{tabular}[c]{@{}l@{}} (4.00e-3,3.27e-1) \\ (1.40e-1,3.27e-1)\end{tabular} &
  \begin{tabular}[c]{@{}l@{}} (1.07e-3,3.27e-1) \\ (3.26e-1,3,27e-1)\end{tabular} &
  \begin{tabular}[c]{@{}l@{}} (1.68e-3,6.53e-1) \\ (3.05e-2,3.32e-1)\end{tabular} &
  \begin{tabular}[c]{@{}l@{}} (2.04e-4,6.54e-1) \\ (2.45e-4,3.38e-1)\end{tabular} &
  \begin{tabular}[c]{@{}l@{}} (4.87e-3,3.28e-1) \\ (4.34e-3,3.27e-1)\end{tabular} &
  \begin{tabular}[c]{@{}l@{}} (1.72e-6,6.54e-1) \\ (1.72e-6,3.29e-1)\end{tabular} \\
  \hline
  $|g^{L_{H}}_{12}|$ &
  \begin{tabular}[c]{@{}l@{}} (3.20e-3,5.32e-2) \\ (3.32e-3,5.32e-2)\end{tabular} &
  \begin{tabular}[c]{@{}l@{}} (1.28e-6,1.08e-1) \\ (5.49e-3,1.43e-2)\end{tabular} &
  \begin{tabular}[c]{@{}l@{}} (1.20e-3,3.10e-1) \\ (2.65e-3,5.43e-2)\end{tabular} &
  \begin{tabular}[c]{@{}l@{}} (2.17e-5,9.35e-1) \\ (4.99e-4,6.38e-1)\end{tabular} &
  \begin{tabular}[c]{@{}l@{}} (7.49e-7,4.85e-1) \\ (6.10e-5,4.08e-1)\end{tabular} &
  \begin{tabular}[c]{@{}l@{}} (6.46e-7,8.89e-1) \\ (1.42e-5,6.62e-1)\end{tabular} \\
  \hline
  $|g^{L_{Z}}_{12}|$ &
  \begin{tabular}[c]{@{}l@{}} (8.67e-6,1.85e-1) \\ (7.99e-2,1.85e-1)\end{tabular} &
  \begin{tabular}[c]{@{}l@{}} (4.79e-5,1.85e-1) \\ (1.85e-1,1.86e-1)\end{tabular} &
  \begin{tabular}[c]{@{}l@{}} (6.36e-6,1.85e-1) \\ (1.63e-5,1.85e-1)\end{tabular} &
  \begin{tabular}[c]{@{}l@{}} (5.36e-8,1.85e-1) \\ (6.23e-6,1.85e-1)\end{tabular} &
  \begin{tabular}[c]{@{}l@{}} (2.20e-3,1.85e-1) \\ (3.68e-3,1.85e-1)\end{tabular} &
  \begin{tabular}[c]{@{}l@{}} (4.66e-7,3.71e-1) \\ (5.53e-7,1.86e-1)\end{tabular} \\
  \hline
  $|g^{L_{W^-}}_{12}|$ &
  \begin{tabular}[c]{@{}l@{}} (2.67e-3,5.49e-2) \\ (1.00e-2,2.00e-2)\end{tabular} &
  \begin{tabular}[c]{@{}l@{}} (5.49e-3,5.24e-2) \\ (6.49e-3,1.70e-2)\end{tabular} &
  \begin{tabular}[c]{@{}l@{}} (1.72e-4,2.84e-1) \\ (5.88e-3,1.27e-1)\end{tabular} &
  \begin{tabular}[c]{@{}l@{}} (1.27e-4,4.87e-1) \\ (4.91e-4,1.78e-1)\end{tabular} &
  \begin{tabular}[c]{@{}l@{}} X \\ X \end{tabular} &
  \begin{tabular}[c]{@{}l@{}} (5.47e-7,3.80e-1) \\ (1.07e-5,7.41e-2)\end{tabular} \\
  \hline
  $|g^{R_{W^-}}_{12}|$ &
  \begin{tabular}[c]{@{}l@{}} (1.29e-3,1.61e-2) \\ (2.72e-3,5.63e-3)\end{tabular} &
  \begin{tabular}[c]{@{}l@{}} (3.20e-4,1.54e-2) \\ (8.50e-4,3.78e-3)\end{tabular} &
  \begin{tabular}[c]{@{}l@{}} (1.00e-6,9.15e-2) \\ (2.71e-3,4.07e-2)\end{tabular} &
  \begin{tabular}[c]{@{}l@{}} (1.27e-6,1.63e-1) \\ (2.87e-5,5.58e-2)\end{tabular} &
  \begin{tabular}[c]{@{}l@{}} X \\ X \end{tabular} &
  \begin{tabular}[c]{@{}l@{}} (3.27e-7,1.24e-1) \\ (2.33e-5,4.08e-2)\end{tabular} \\
  \hline
  $|g^{L_{H}}_{13}|$ &
  \begin{tabular}[c]{@{}l@{}} (1.65e-3,1.66e-1) \\ (1.13e-1,1.66e-1)\end{tabular} &
  \begin{tabular}[c]{@{}l@{}} (1.46e-3,1.30e-1) \\ (1.28e-1,1.30e-1)\end{tabular} &
  \begin{tabular}[c]{@{}l@{}} (3.65e-3,3.10e-1) \\ (6.01e-2,3.10e-1)\end{tabular} &
  \begin{tabular}[c]{@{}l@{}} (4.55e-4,9.59e-1) \\ (3.00e-2,6.94e-1)\end{tabular} &
  \begin{tabular}[c]{@{}l@{}} (2.26e-2,9.81e-1) \\ (7.71e-2,8.07e-1)\end{tabular} &
  \begin{tabular}[c]{@{}l@{}} (1.61e-5,9.62e-1) \\ (6.51e-4,6.24e-1)\end{tabular} \\
  \hline
  $|g^{L_{Z}}_{13}|$ &
  \begin{tabular}[c]{@{}l@{}} (1.12e-6,1.06e-2) \\ (4.08e-4,6.10e-4)\end{tabular} &
  \begin{tabular}[c]{@{}l@{}} (1.63e-5,4.14e-3) \\ (9.45e-4,1.60e-3)\end{tabular} &
  \begin{tabular}[c]{@{}l@{}} (8.73e-6,7.07e-2) \\ (2.02e-4,2.51e-3)\end{tabular} &
  \begin{tabular}[c]{@{}l@{}} (7.90e-7,4.00e-2) \\ (5.62e-6,1.64e-2)\end{tabular} &
  \begin{tabular}[c]{@{}l@{}} (9.52e-8,6.86e-3) \\ (6.84e-6,5.36e-3)\end{tabular} &
  \begin{tabular}[c]{@{}l@{}} (2.19e-7,1.11e-1) \\ (1.18e-5,2.85e-2)\end{tabular} \\
  \hline
  $|g^{L_{W^-}}_{13}|$ &
  \begin{tabular}[c]{@{}l@{}} X \\ X \end{tabular} &
  \begin{tabular}[c]{@{}l@{}} X \\ X \end{tabular} &
  \begin{tabular}[c]{@{}l@{}} X \\ X \end{tabular} &
  \begin{tabular}[c]{@{}l@{}} X \\ X \end{tabular} &
  \begin{tabular}[c]{@{}l@{}} X \\ X \end{tabular} &
  \begin{tabular}[c]{@{}l@{}} (8.77e-8,1.72e-1) \\ (2.28e-5,5.64e-2)\end{tabular} \\
  \hline
  $|g^{R_{W^-}}_{13}|$ &
  \begin{tabular}[c]{@{}l@{}} X \\ X \end{tabular} &
  \begin{tabular}[c]{@{}l@{}} X \\ X \end{tabular} &
  \begin{tabular}[c]{@{}l@{}} X \\ X \end{tabular} &
  \begin{tabular}[c]{@{}l@{}} X \\ X \end{tabular} &
  \begin{tabular}[c]{@{}l@{}} X \\ X \end{tabular} &
  \begin{tabular}[c]{@{}l@{}} (5.66e-7,6.79e-2) \\ (2.46e-5,2.32e-2)\end{tabular} \\
  \hline
  $|g^{L_{H}}_{14}|$ &
  \begin{tabular}[c]{@{}l@{}} (1.20e-1,3.10e-1) \\ (1.24e-1,3.10e-1)\end{tabular} &
  \begin{tabular}[c]{@{}l@{}} (1.03e-1,2.42e-1) \\ (2.41e-1,2.42e-1)\end{tabular} &
  \begin{tabular}[c]{@{}l@{}} (3.15e-5,3.45e-1) \\ (1.11e-1,3.40e-1)\end{tabular} &
  \begin{tabular}[c]{@{}l@{}} (1.88e-5,9.84e-1) \\ (3.49e-4,7.46e-1)\end{tabular} &
  \begin{tabular}[c]{@{}l@{}} X \\ X \end{tabular} &
  \begin{tabular}[c]{@{}l@{}} (5.15e-5,9.84e-1) \\ (3.39e-4,7.25e-1)\end{tabular} \\
  \hline
  $|g^{L_{Z}}_{14}|$ &
  \begin{tabular}[c]{@{}l@{}} (2.39e-4,3.05e-3) \\ (3.72e-4,6.72e-4)\end{tabular} &
  \begin{tabular}[c]{@{}l@{}} (5.74e-4,5.01e-3) \\ (1.10e-3,1.78e-3)\end{tabular} &
  \begin{tabular}[c]{@{}l@{}} (3.70e-8,1.44e-2) \\ (2.74e-4,1.18e-3)\end{tabular} &
  \begin{tabular}[c]{@{}l@{}} (2.05e-7,5.46e-3) \\ (5.59e-7,4.34e-3)\end{tabular} &
  \begin{tabular}[c]{@{}l@{}} X \\ X \end{tabular} &
  \begin{tabular}[c]{@{}l@{}} (2.61e-10,8.05e-2) \\ (1.16e-6,1.99e-2)\end{tabular} \\
  \hline
  $|g^{L_{W^-}}_{14}|$ &
  \begin{tabular}[c]{@{}l@{}} X \\ X \end{tabular} &
  \begin{tabular}[c]{@{}l@{}} X \\ X \end{tabular} &
  \begin{tabular}[c]{@{}l@{}} X \\ X \end{tabular} &
  \begin{tabular}[c]{@{}l@{}} X \\ X \end{tabular} &
  \begin{tabular}[c]{@{}l@{}} X \\ X \end{tabular} &
  \begin{tabular}[c]{@{}l@{}} (2.79e-7,1.66e-1) \\ (8.19e-6,2.31e-2)\end{tabular} \\
  \hline
  $|g^{R_{W^-}}_{14}|$ &
  \begin{tabular}[c]{@{}l@{}} X \\ X \end{tabular} &
  \begin{tabular}[c]{@{}l@{}} X \\ X \end{tabular} &
  \begin{tabular}[c]{@{}l@{}} X \\ X \end{tabular} &
  \begin{tabular}[c]{@{}l@{}} X \\ X \end{tabular} &
  \begin{tabular}[c]{@{}l@{}} X \\ X \end{tabular} &
  \begin{tabular}[c]{@{}l@{}} (3.33e-7,4.35e-2) \\ (3.40e-6,2.30e-2)\end{tabular} \\
  \hline
  $|g^{L_{H}}_{15}|$ &
  \begin{tabular}[c]{@{}l@{}} X \\ X \end{tabular} &
  \begin{tabular}[c]{@{}l@{}} X \\ X \end{tabular} &
  \begin{tabular}[c]{@{}l@{}} X \\ X \end{tabular} &
  \begin{tabular}[c]{@{}l@{}} X \\ X \end{tabular} &
  \begin{tabular}[c]{@{}l@{}} X \\ X \end{tabular} &
  \begin{tabular}[c]{@{}l@{}} (5.86e-5,9.49e-1) \\ (5.62e-4,6.25e-1)\end{tabular} \\
  \hline
  $|g^{L_{Z}}_{15}|$ &
  \begin{tabular}[c]{@{}l@{}} X \\ X \end{tabular} &
  \begin{tabular}[c]{@{}l@{}} X \\ X \end{tabular} &
  \begin{tabular}[c]{@{}l@{}} X \\ X \end{tabular} &
  \begin{tabular}[c]{@{}l@{}} X \\ X \end{tabular} &
  \begin{tabular}[c]{@{}l@{}} X \\ X \end{tabular} &
  \begin{tabular}[c]{@{}l@{}} (3.24e-7,2.63e-2) \\ (1.20e-6,1.46e-2)\end{tabular} \\
  \hline
  $|g^{L_{H}}_{16}|$ &
  \begin{tabular}[c]{@{}l@{}} X \\ X \end{tabular} &
  \begin{tabular}[c]{@{}l@{}} X \\ X \end{tabular} &
  \begin{tabular}[c]{@{}l@{}} X \\ X \end{tabular} &
  \begin{tabular}[c]{@{}l@{}} X \\ X \end{tabular} &
  \begin{tabular}[c]{@{}l@{}} X \\ X \end{tabular} &
  \begin{tabular}[c]{@{}l@{}} (2.97e-5,9.84e-1) \\ (1.91e-4,8.18e-1)\end{tabular} \\
  \hline
  $|g^{L_{Z}}_{16}|$ &
  \begin{tabular}[c]{@{}l@{}} X \\ X \end{tabular} &
  \begin{tabular}[c]{@{}l@{}} X \\ X \end{tabular} &
  \begin{tabular}[c]{@{}l@{}} X \\ X \end{tabular} &
  \begin{tabular}[c]{@{}l@{}} X \\ X \end{tabular} &
  \begin{tabular}[c]{@{}l@{}} X \\ X \end{tabular} &
  \begin{tabular}[c]{@{}l@{}} (1.85e-8,1.20e-2) \\ (2.23e-6,3.24e-3)\end{tabular} \\
  \hline
\end{tabular}
\caption{Allowed range for the coupling strengths. The upper and lower intervals represent the allowed range for samples satisfying all the constraints with $\Omega_{\chi}h^2$ in the criteria C1 ( $\leq +3\sigma$) and C2 (within $\pm 3\sigma$) respectively.}
\label{tab:range-2}}
\end{table}
%%%%%%%%%%%%%%%%%%%%%%%%%%%%%%%%%%%%%%%%%
Including the allowed outlier samples, we show the allowed ranges of DM mass, mass parameters ($\mu_i$), Yukawa couplings ($g_i$) and the effective couplings ($|a_q/m_q|$ and $|d_q|$) used in the calculation of DM scattering off $^{129, 131}{\rm Xe}$ nuclei and ${\rm CF_3I}$ nuclei in Table \ref{tab:range-1} , and the allowed ranges for the coupling strengths used in the calculation of DM annihilation processes in Table \ref{tab:range-2}. In Table \ref{tab:range-2}, we have used the following definitions:
$g^{L_H}_{1j}=O^{L_H}_{1j}$,
$g^{L_Z}_{1j}=\frac{g}{2\cos\theta_W}O^{L_Z}_{1j}$
and
$g^{L,R_{W^-}}_{1j}=\frac{g}{\sqrt 2}O^{L,R_{W^-}}_{1j}$.
The allowed DM relic density should satisfy the condition: $\Omega_{\chi}h^2\le 0.1198+3\times 0.0026$.
We consider two criterions: C1 having a less stringent constraint of the relic density with its value less than $+3  \sigma$, and C2 having a more stringent constraint of the relic density with its value within $\pm 3\sigma$, respectively, from the observed mean value.
In Table \ref{tab:range-1} and \ref{tab:range-2}, the upper and lower intervals represent the allowed range for samples satisfying all the constraints with $\Omega_{\chi}h^2$ falling into the criteria C1 and C2 respectively. 

\section{DIscussions and Conclusions}

\subsection{Coannihilation}

In addition to the annihilation, the coannihilation, namely, the annihilation from the other WIMPs, 
%since all WIMPs would decay into DM particles, 
may affect the DM relic density in some parameter region. The coannihilation becomes significantly important when the WIMPs are nearly mass degenerate with DM~\cite{GS}. In this subsection, we preliminarily explore the variation on the calculation of DM relic density when including the coannihilation. To see the leading effect of coannihilation, we consider two lightest neutral as well as two single charged WIMPs annihilating to the SM fermions through the s-channel in the neutralino-like I case. The corresponding Feynman diagrams and Lagrangian are shown in Fig.~\ref{fey2} and Appendix C, respectively. The matrix elements for coannihilation are shown in Appendix H.
The formulation for coannihilation is presented in Appendix G. To simplify the calculation of coannihilation, we have set the freeze-out temperature parameter $x_f=25$.
\begin{figure}[t]
  \centering
  \includegraphics[width=14cm]{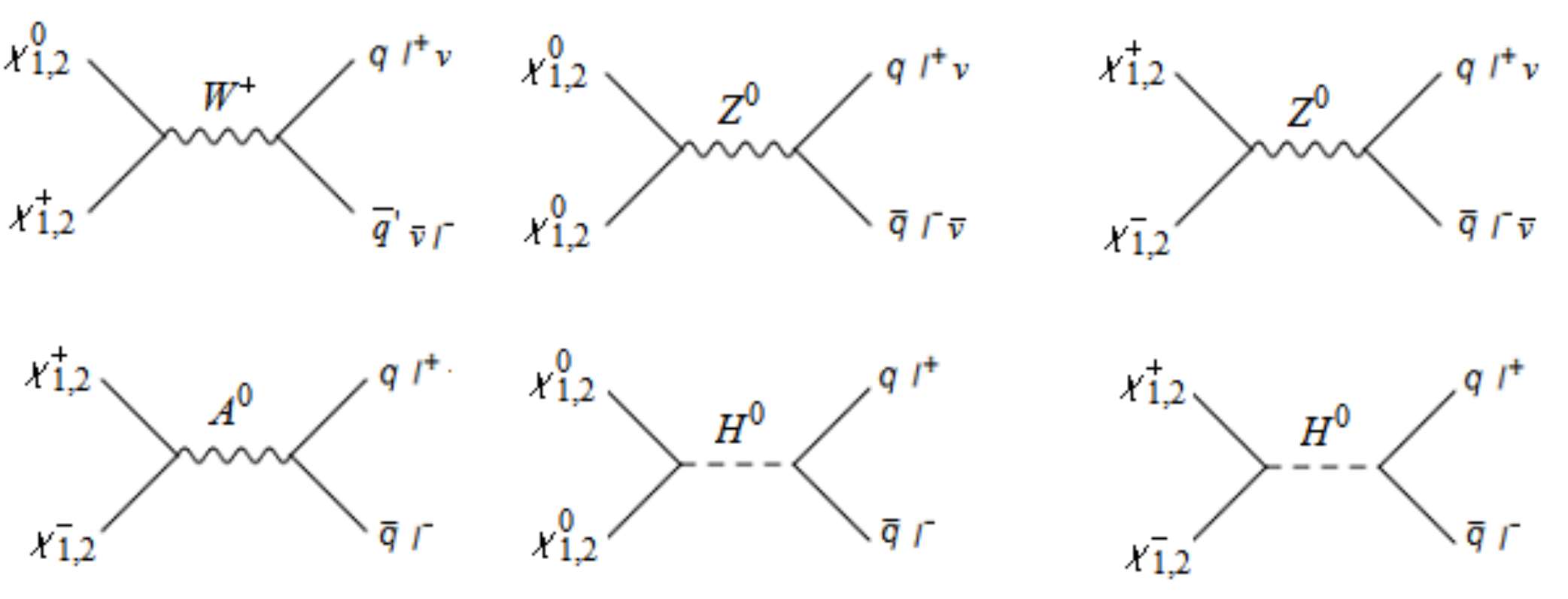}\\
  \caption{The coannihilation processes $(\chi{\chi}\rightarrow {\rm SM \ fermions})$ through s-channel}\label{17fey2}
\end{figure}

\begin{figure}[h!]
\centering
\captionsetup{justification=raggedright}
 \subfigure[\ Without coannihilation]{
  \includegraphics[width=0.45\textwidth,height=0.15\textheight]{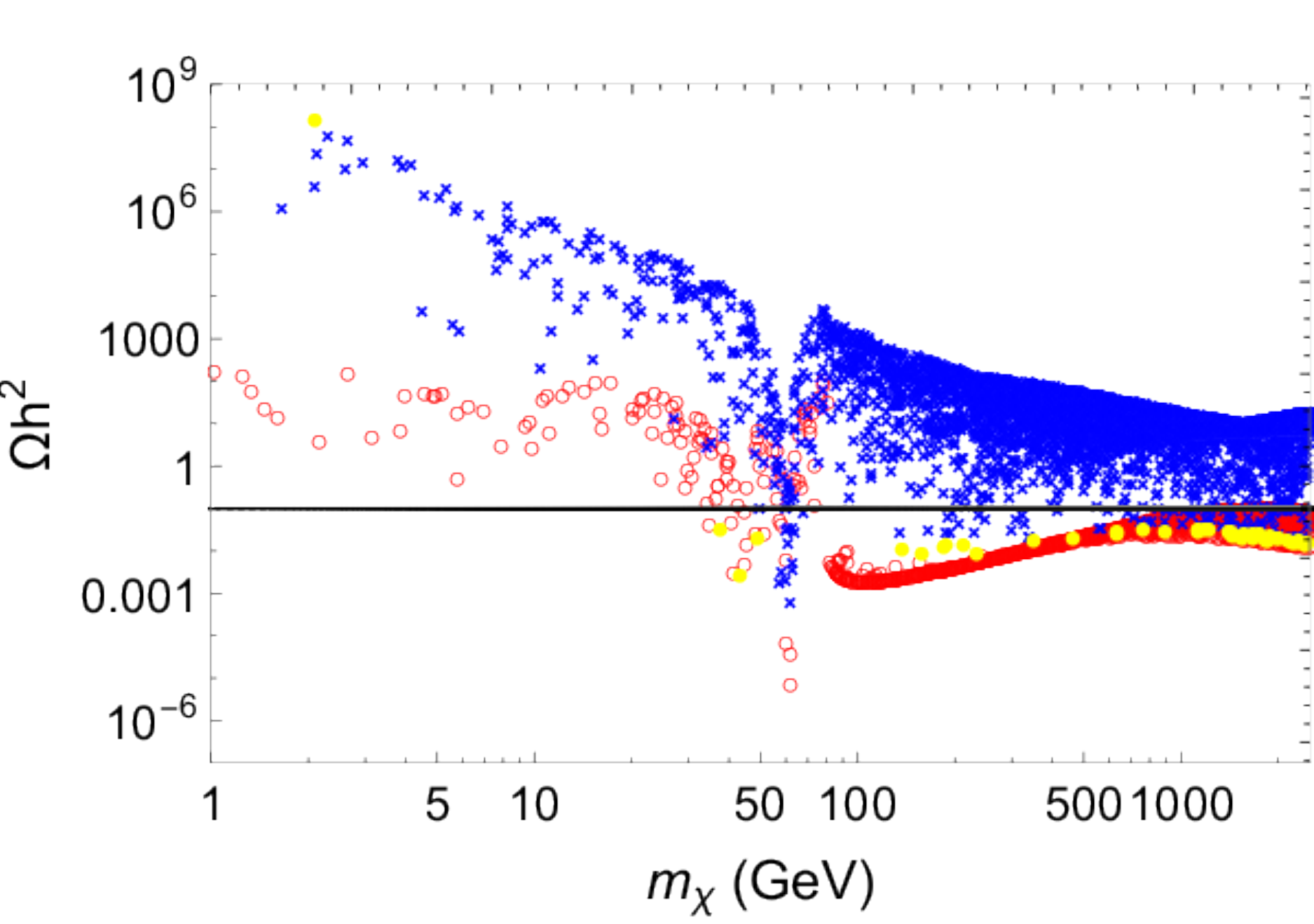}
}\subfigure[\ With coannihilation]{
  \includegraphics[width=0.45\textwidth,height=0.15\textheight]{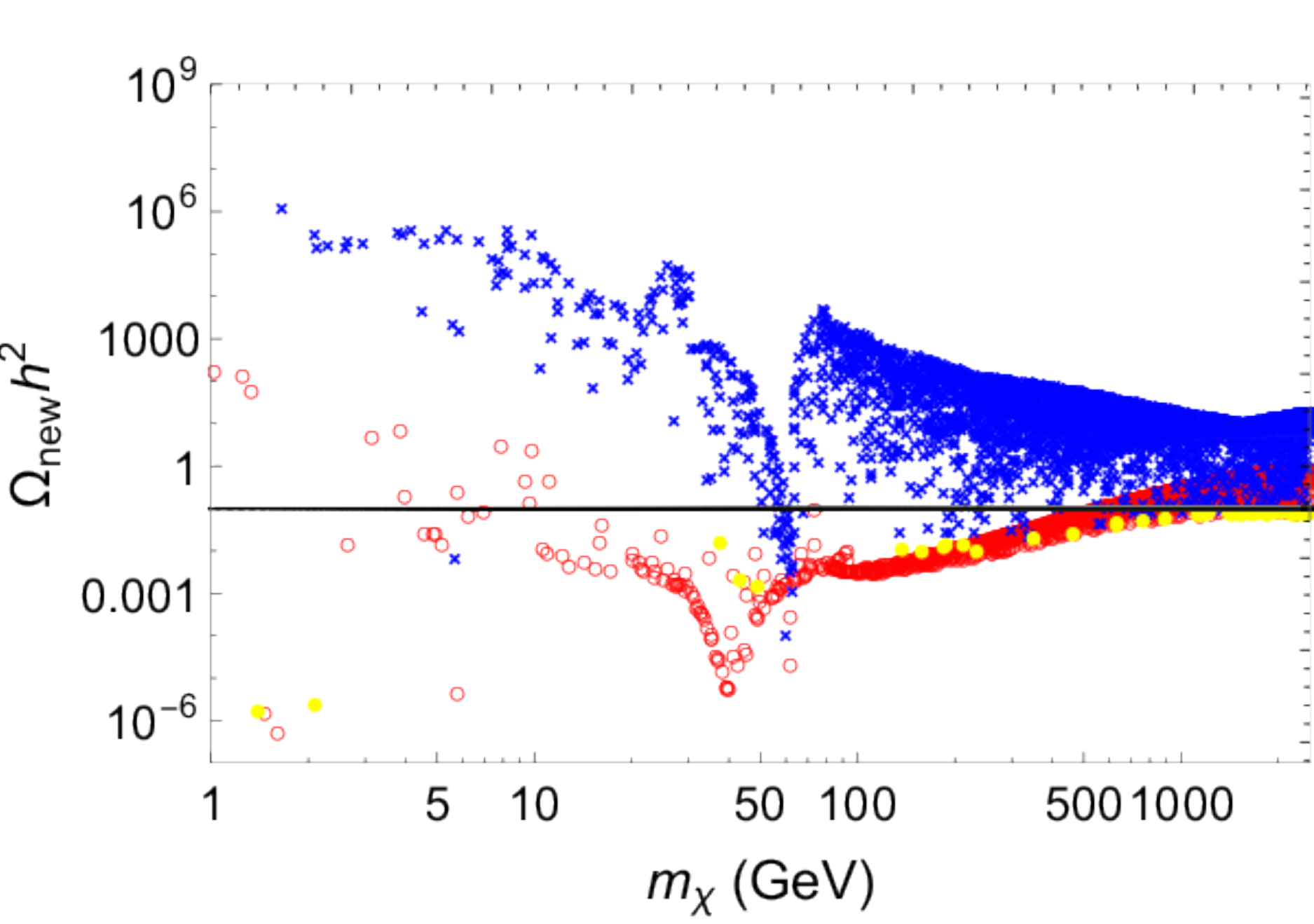}
}
\caption{Scatter plots of DM relic abundance before and after considering coannihilation in the neutralino-like I case
[{\color{red} $\circ$}:~higgsino-like,~
{\color{blue} $\times$}:~bino-like,
{\color{yellow} $\bullet$}:~mixed].}
\label{fig:coannihilation-1}
\end{figure}

Figs.~\ref{fig:coannihilation-1}(a), \ref{fig:coannihilation-1}(b) show the scatter plots of relic density without and with coannihilation respectively. 
We see that the $\Omega h^2$ constraint affects a little on the selection of the $\tilde B$-like particles, but a lot on the selection of the $\tilde H$-like particles. Most $\tilde H$-like particles with mass less than $M_W$ ruled out originally become allowed now, while part of $\tilde H$-like particles with mass greater than $M_W$ allowed originally become ruled out now when including the leading effect of coannihilation.

%\vfill
%\eject

\begin{figure}[t!]
\centering
\captionsetup{justification=raggedright}
\subfigure[{\ \color{red} $\circ$} / {\color{blue} $\times$}:~with/without coannihilation]{
 \includegraphics[width=0.45\textwidth,height=0.15\textheight]{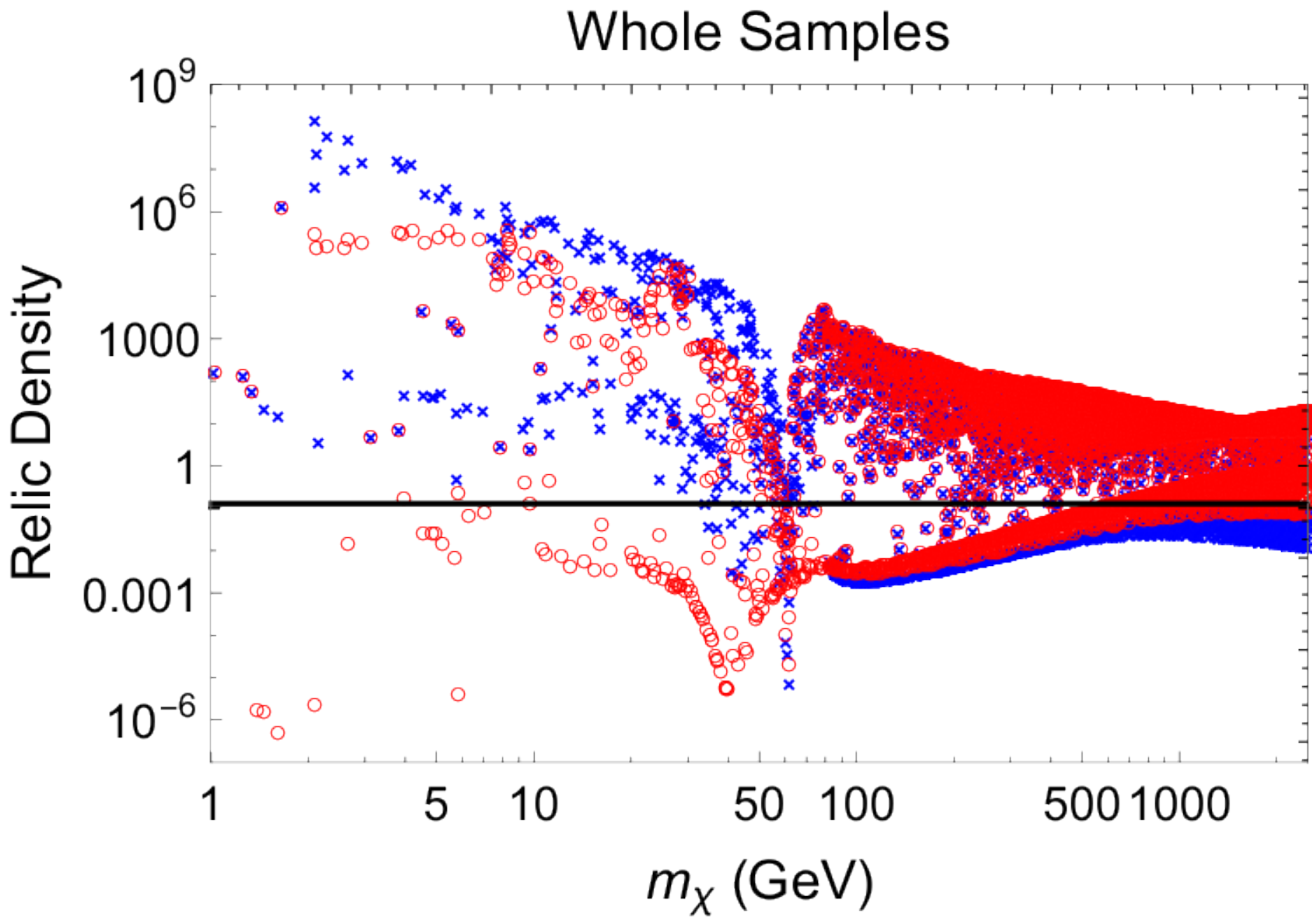}
}\subfigure[{\ \color{red} $\circ$} / {\color{blue} $\times$}:~with/without coannihilation]{
  \includegraphics[width=0.45\textwidth,height=0.15\textheight]{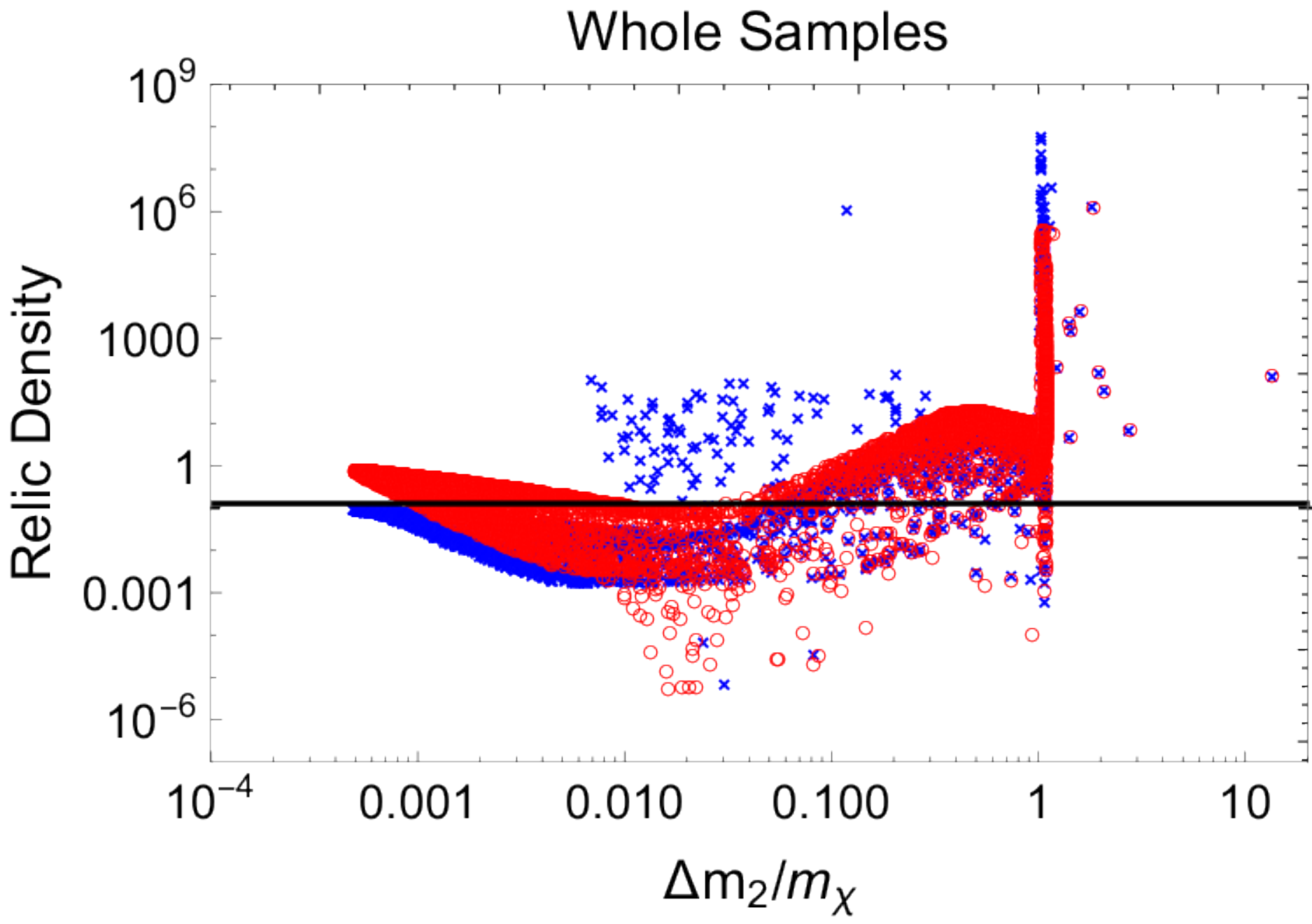}
}\\\subfigure[]{
  \includegraphics[width=0.45\textwidth,height=0.15\textheight]{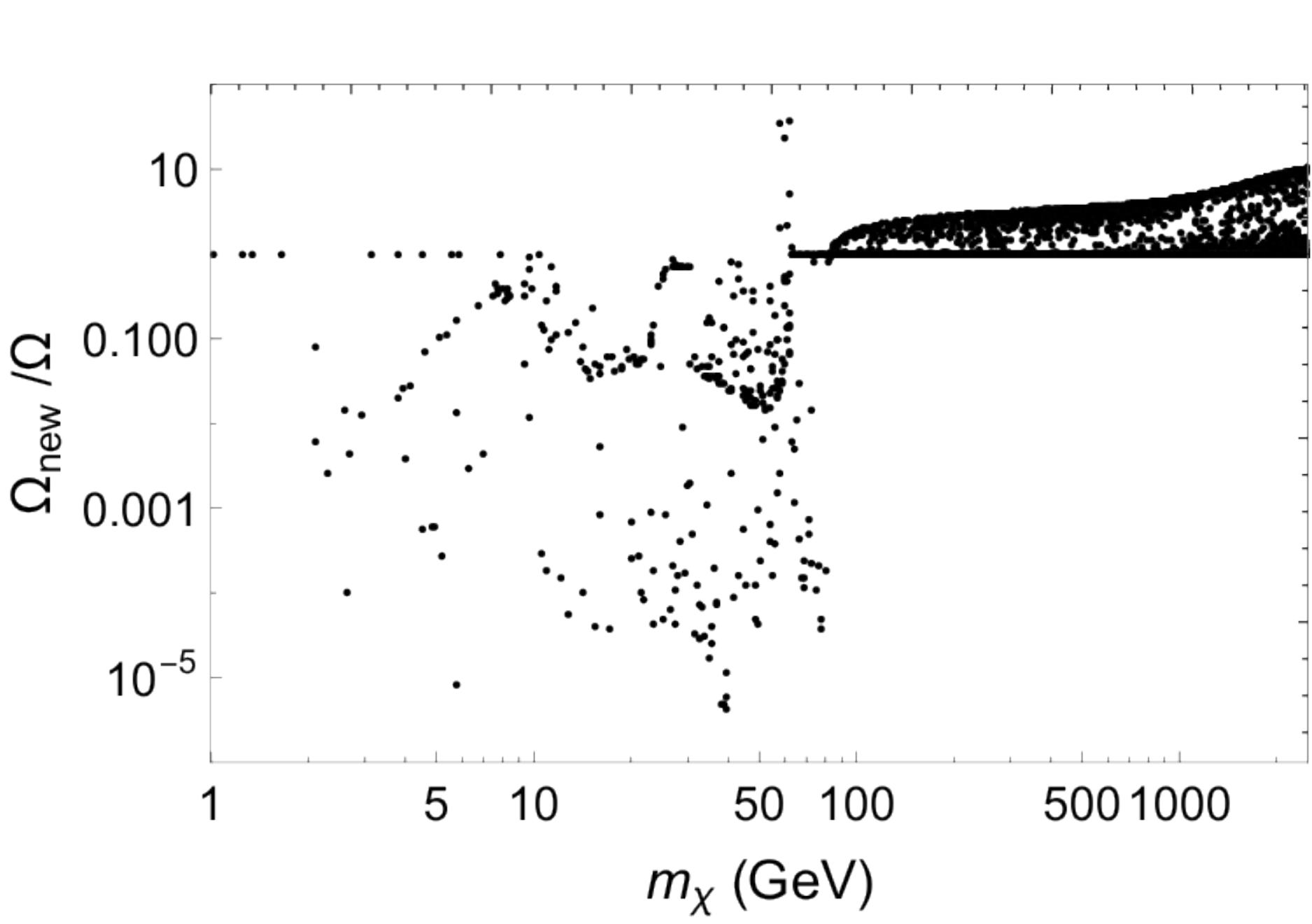}
}\subfigure[]{
  \includegraphics[width=0.45\textwidth,height=0.15\textheight]{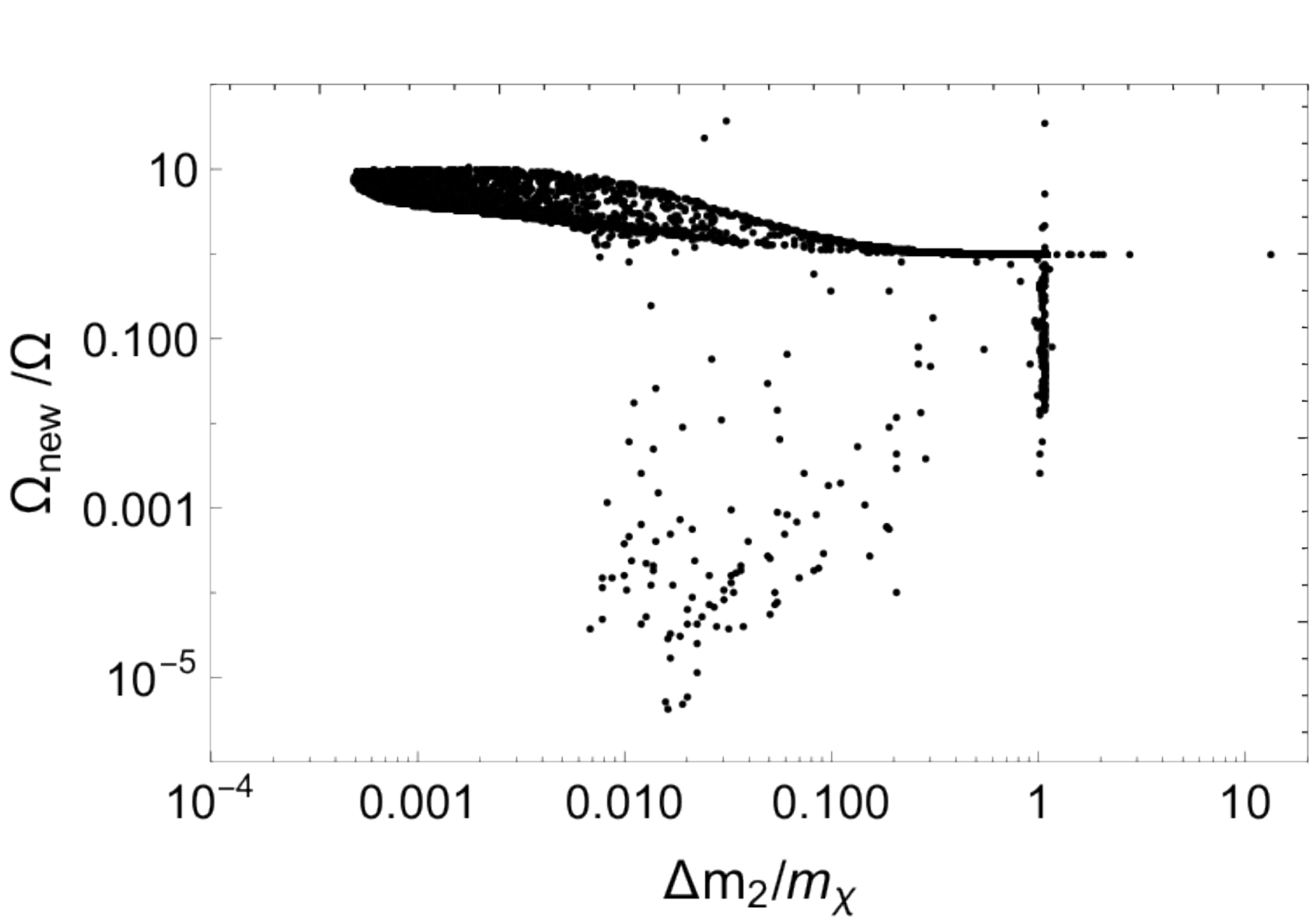}
}\\\subfigure[{\ \color{red} $\circ$} / {\color{blue} $\times$}:~with/without coannihilation]{
  \includegraphics[width=0.45\textwidth,height=0.15\textheight]{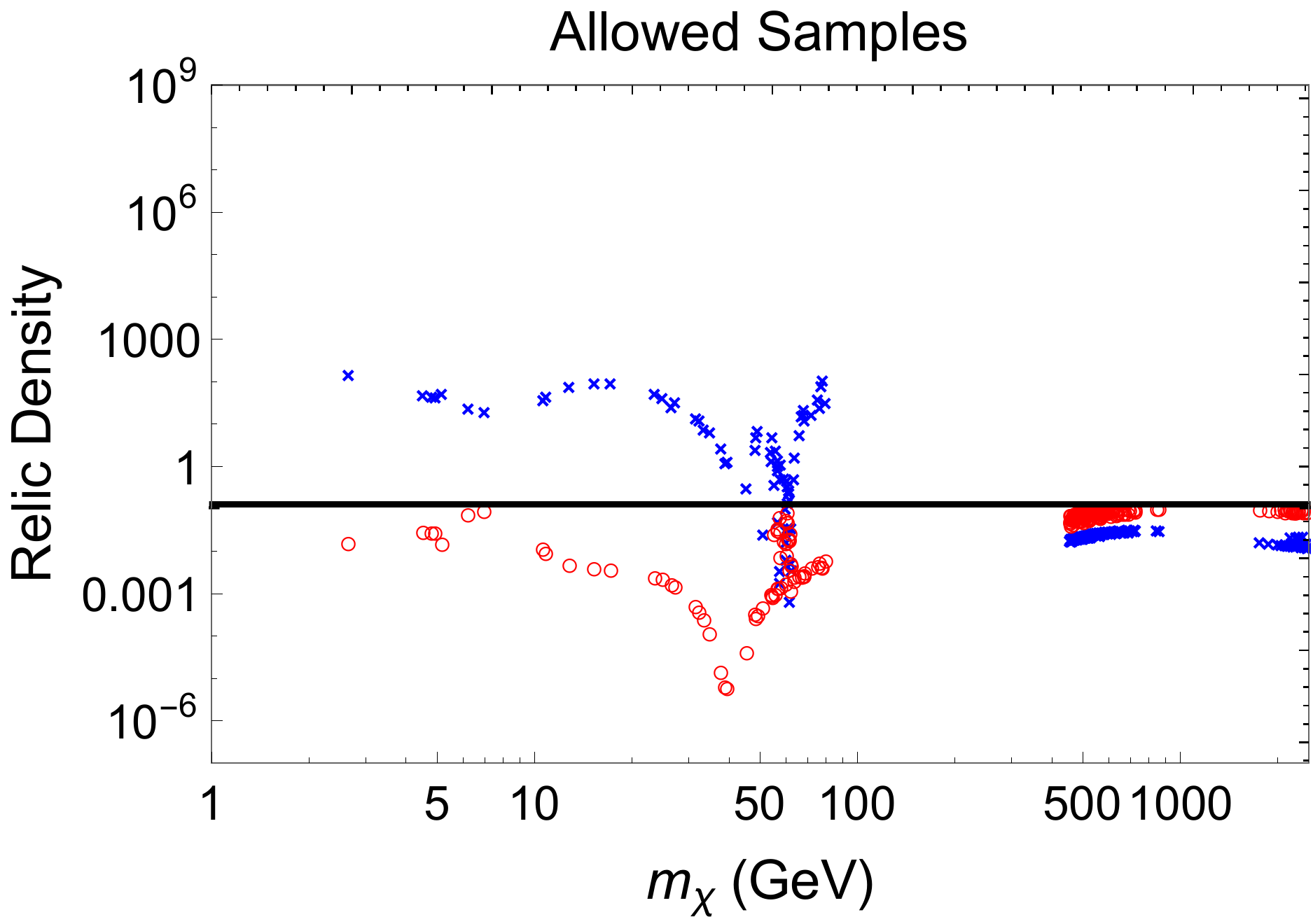}
}\subfigure[{\ \color{red} $\circ$} / {\color{blue} $\times$}:~with/without coannihilation]{
  \includegraphics[width=0.45\textwidth,height=0.15\textheight]{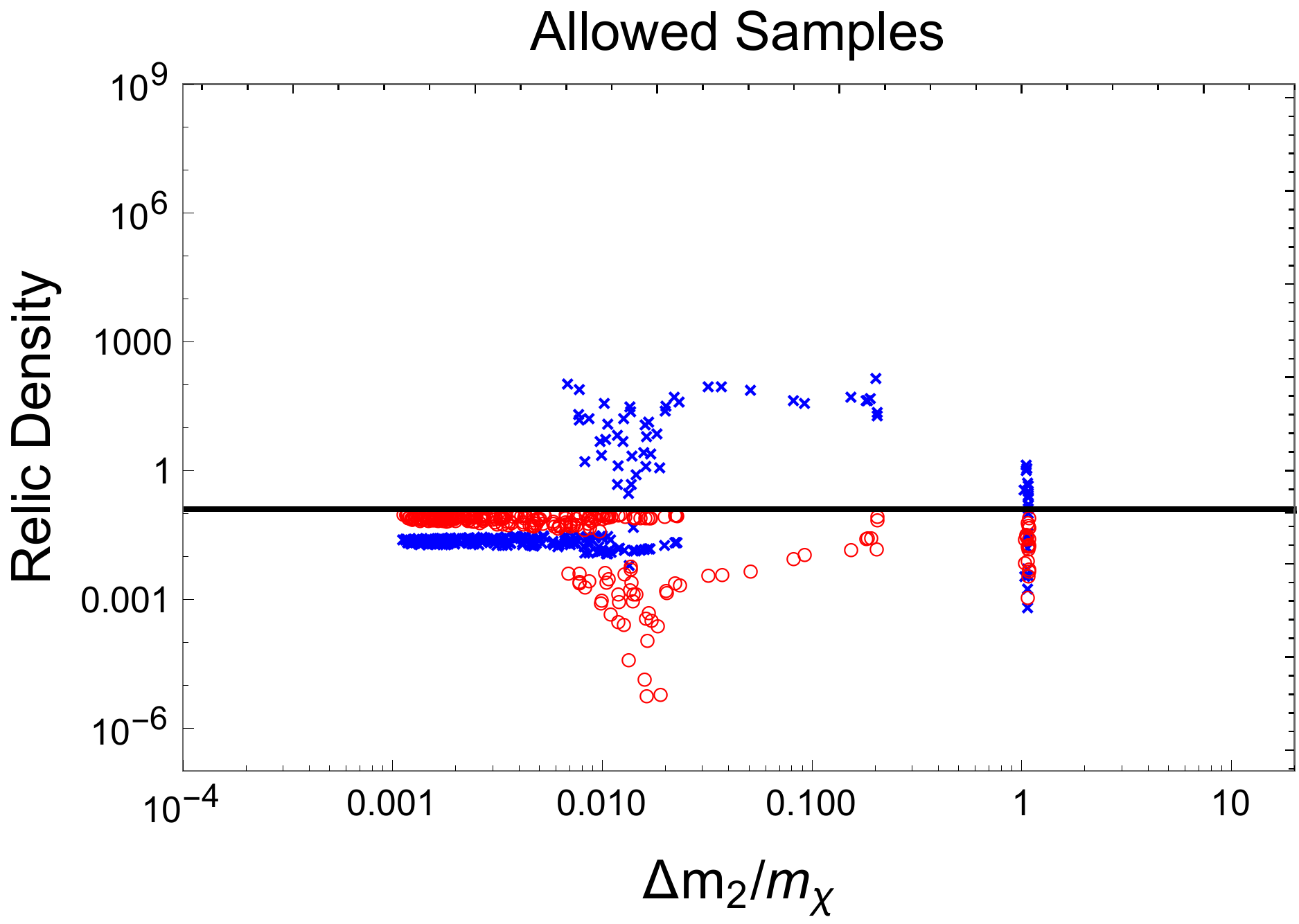}
}\\\subfigure[]{
  \includegraphics[width=0.45\textwidth,height=0.15\textheight]{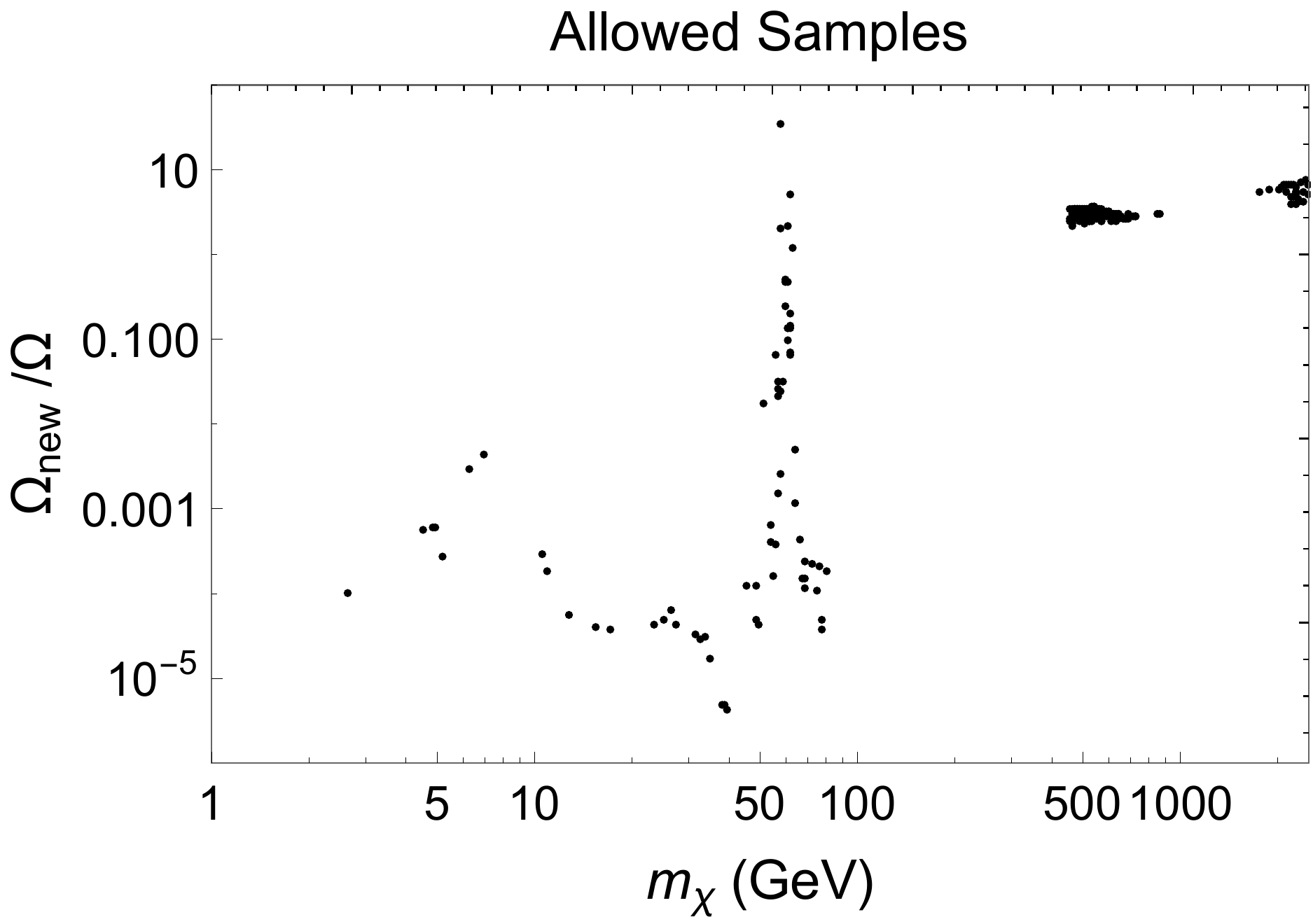}
}\subfigure[]{
  \includegraphics[width=0.45\textwidth,height=0.15\textheight]{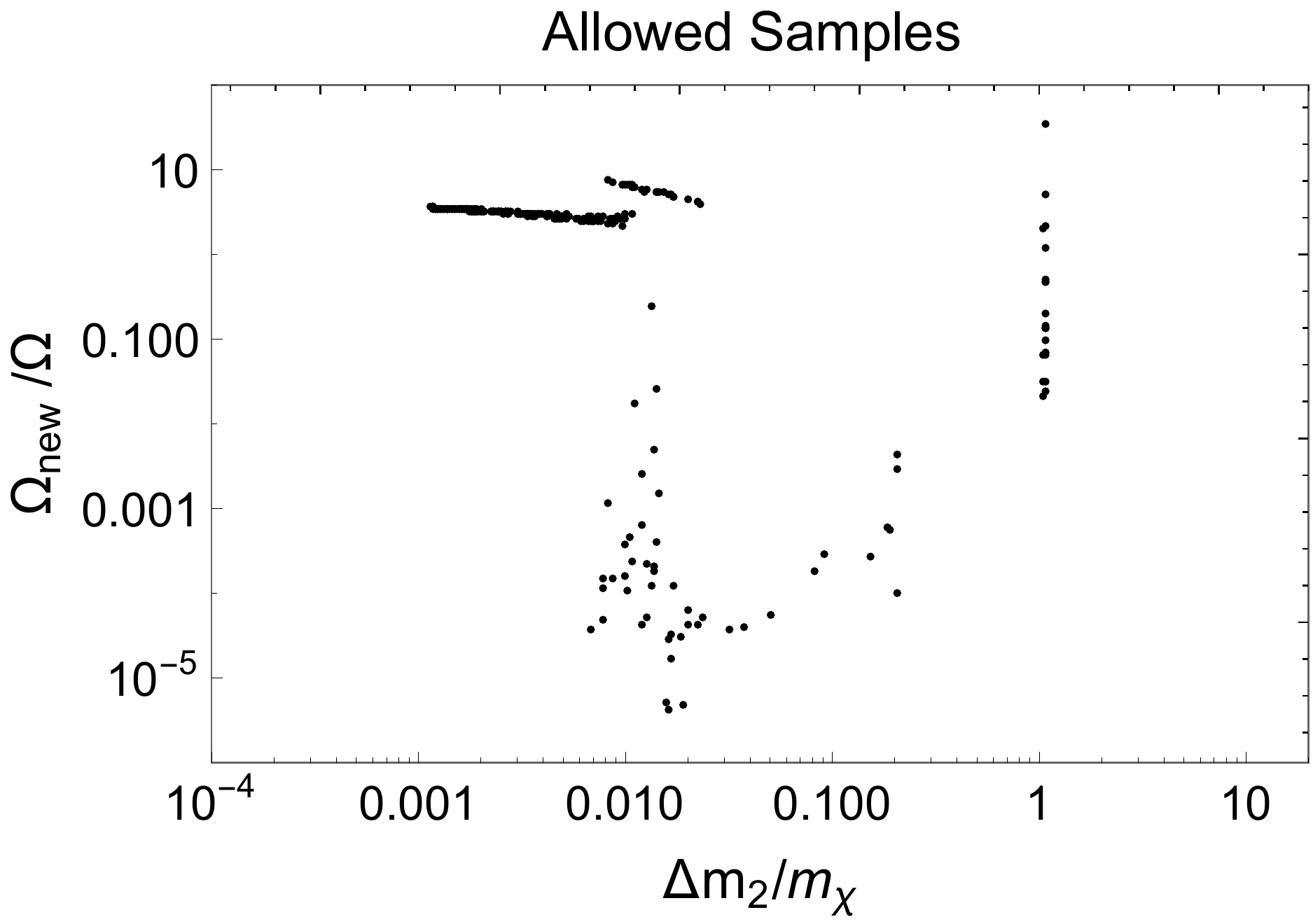}
}
\caption{Leading effect of coannihilation on DM relic abundance at $x_f=25$ in the neutralino-like I case}
%[{\color{red} $\circ$},~
%{\color{blue} $\times$}:~without caannihilation
%,{\color{yellow} $\bullet$}:~mixed
%].}
\label{fig:coannihilation-2}
\end{figure}

To see the variation of DM relic density, we overlap the Figs.~\ref{fig:coannihilation-1}(a) (in {\color{blue} $\times$}) and \ref{fig:coannihilation-1}(b) (in {\color{red} $\circ$}) in Fig.~\ref{fig:coannihilation-2}(a). 
We also show the variation of DM relic density versus the mass fraction $\Delta m_2/m_\chi\equiv ({\rm Min}[m_{\chi^0_2},m_{\chi^\pm_{1,2}}]-m_{\chi^0_1})/m_{\chi^0_1}$ in Fig.~\ref{fig:coannihilation-2}(b). Let $\Omega_{\rm new}$ and $\Omega$ denote the relic density with and without considering the coannihilation respectively. 
Apart from a few samples around the poles, we find that $\Omega_{\rm new}\geq \Omega$ with $m_\chi \gtrsim m_W$, while $\Omega_{\rm new}\leq \Omega$ with $m_\chi \lesssim m_W$ in Fig.~\ref{fig:coannihilation-2}(c). 
We also find that the smaller mass fraction usually gives the greater value in $\Omega_{\rm new}/\Omega$ as shown in Fig.~\ref{fig:coannihilation-2}(d). 
We show the relic density versus DM mass $m_\chi$ and mass fraction $\Delta m_2/m_\chi$ with allowed samples which satisfy all constraints in Figs.~\ref{fig:coannihilation-2}(e) and (f) respectively, and Figs.~\ref{fig:coannihilation-2}(g) and (h) for $\Omega_{\rm new}/\Omega$.
In Figs.~\ref{fig:coannihilation-2}(e) and (f), the sample marked with ``{\color{red} $\circ$}" are allowed when including the coannihilation, and the sample marked with ``{\color{blue} $\times$}" correspond to the sample marked with ``{\color{red} $\circ$}" but only considering the annihilation. 

%\vfil
%\eject

\begin{figure}[t!]
\centering
\captionsetup{justification=raggedright}
 \subfigure[\ Constraint on $\Omega^{\rm{obs}}_\chi$]{
  \includegraphics[width=0.45\textwidth,height=0.15\textheight]{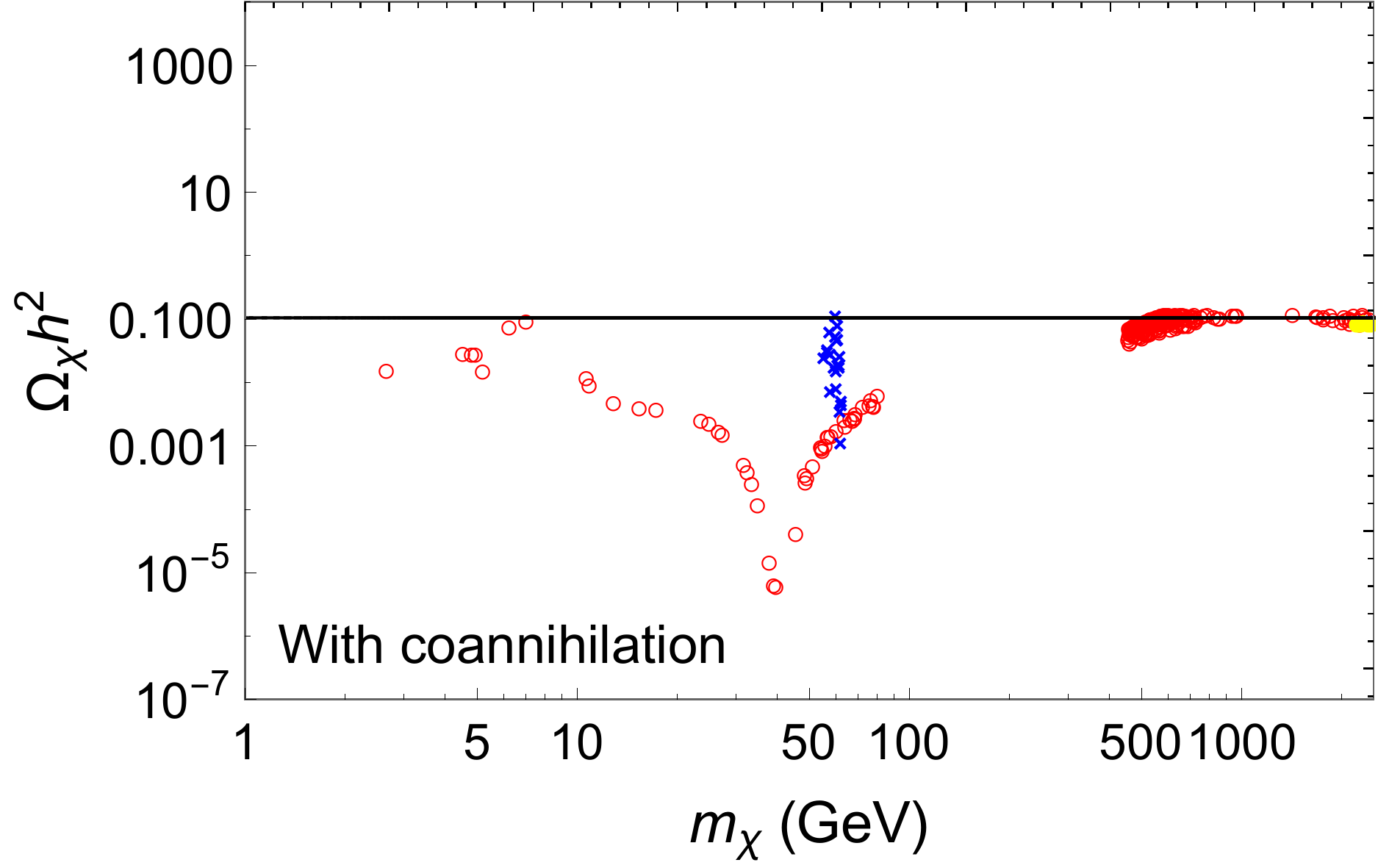}
}\subfigure[\ LUX constraint on $\sigma^{SI}$ with NB limit]{
  \includegraphics[width=0.45\textwidth,height=0.15\textheight]{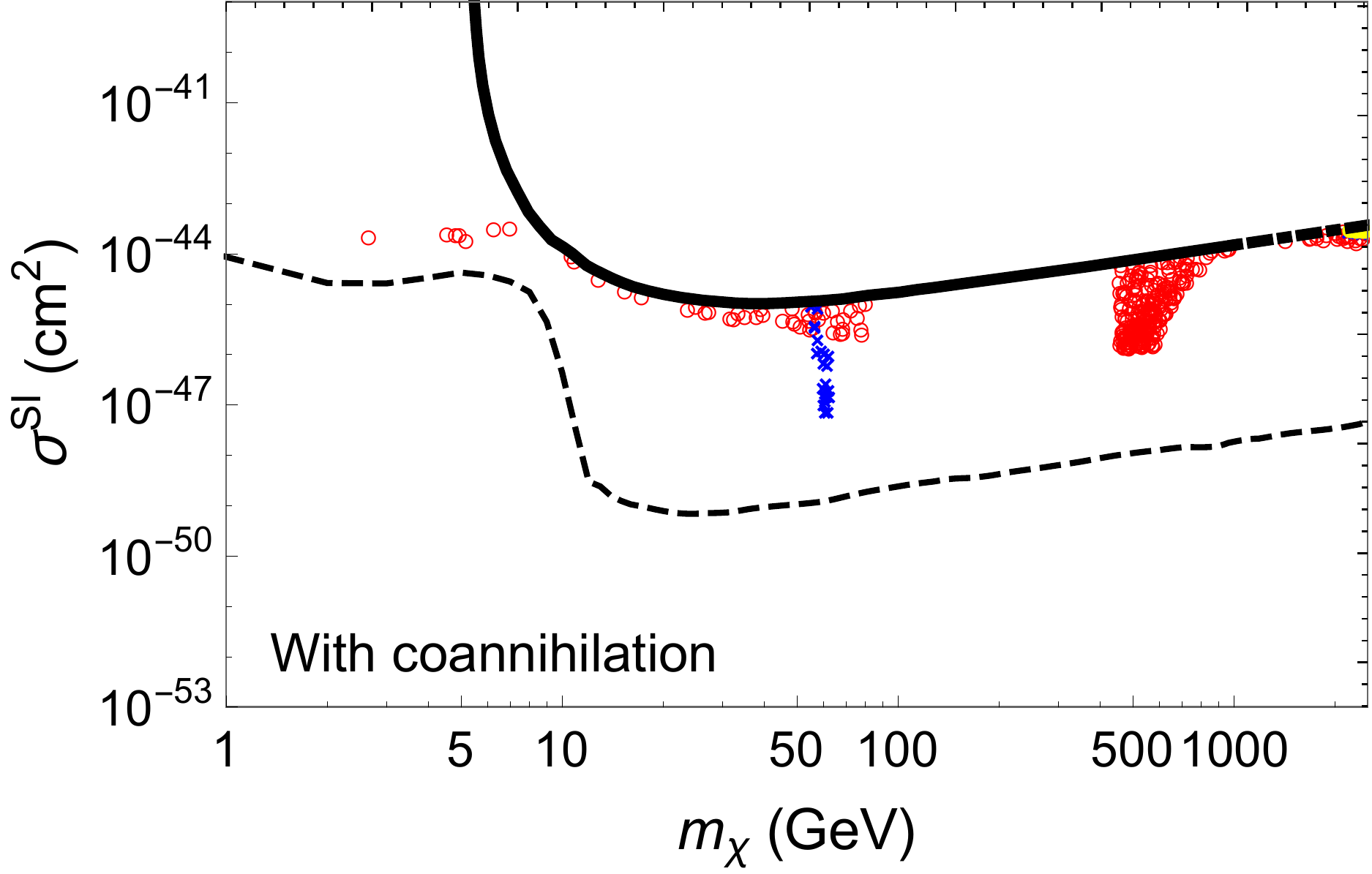}
}\\\subfigure[\ Fermi-LAT constraint on $\chi^0 {\chi}^0\rightarrow W^+W^-$]{
  \includegraphics[width=0.45\textwidth,height=0.15\textheight]{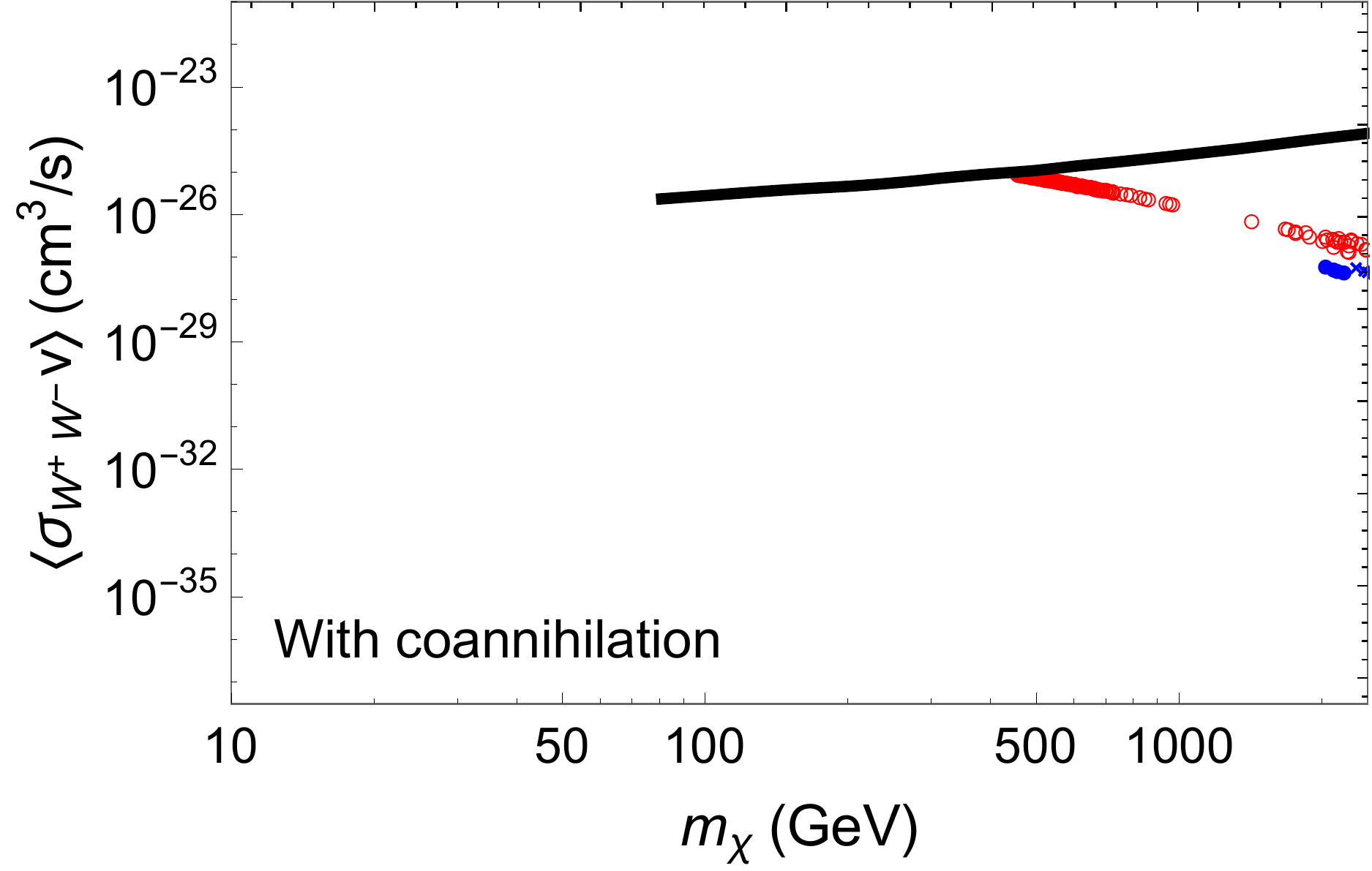}
}\subfigure[\ Fermi-LAT constraint on $\chi^0 {\chi}^0\rightarrow b\bar{b}$]{
  \includegraphics[width=0.45\textwidth,height=0.15\textheight]{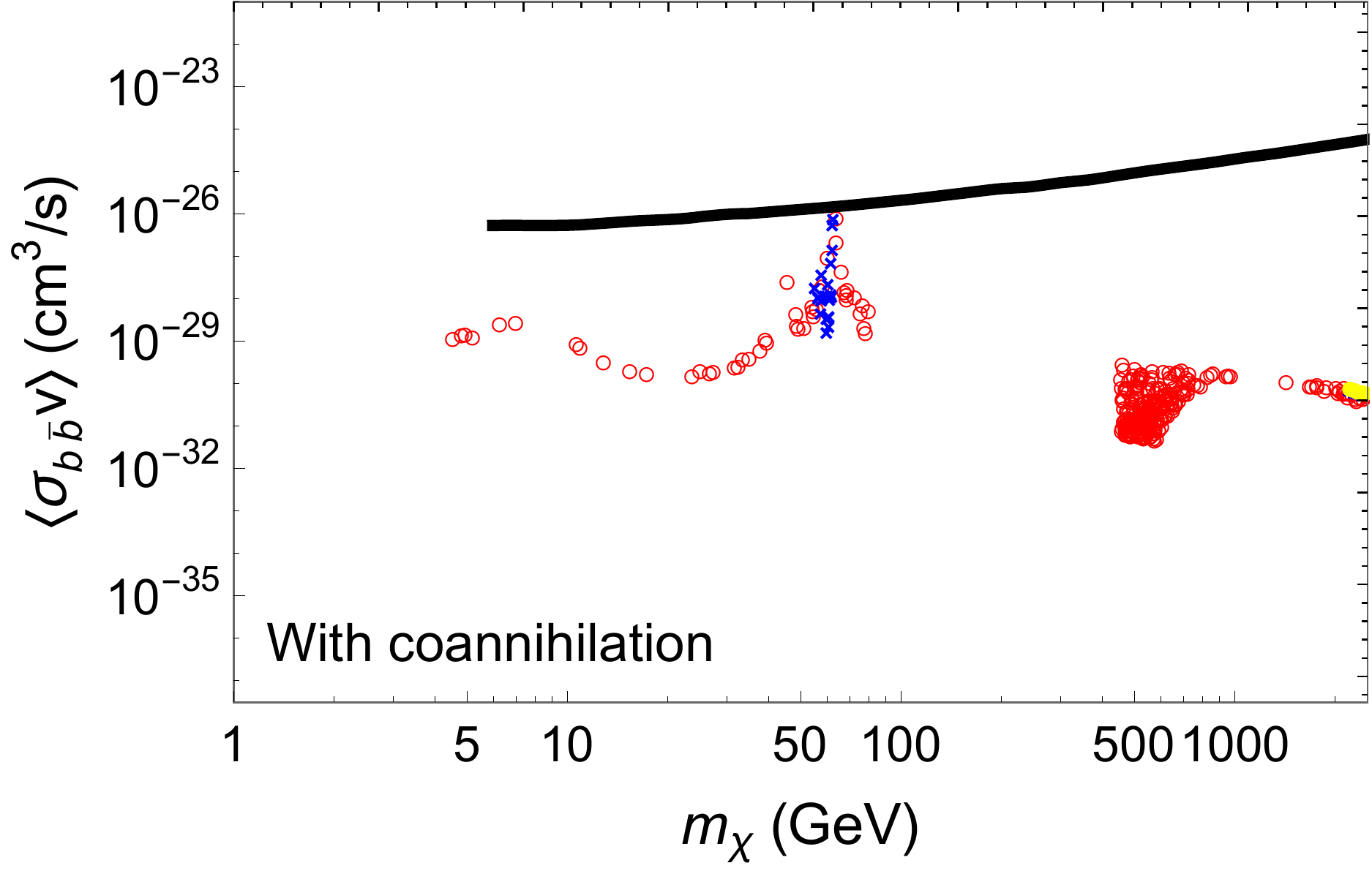}
}
\caption{Results for allowed samples satisfying all constraints in the neutralino-like I case
[{\color{red} $\circ$}:~higgsino-like,
{\color{blue} $\times$}:~bino-like,   {\color{yellow} $\bullet$}:~mixed].}
\label{fig:allowco neutralino-like I}
\end{figure}

In Figs.~\ref{fig:allowco neutralino-like I}(a-d), we only show the allowed samples which the allowed regions touch the experimental upper limits, namely, in the plots of $\Omega_\chi h^2$, $\sigma^{SI}$, $\la\sigma_{W^+W^-} {v}\ra$ and  $\la\sigma_{b\bar b} {v}\ra$ versus DM mass 
$m_\chi$, respectively. By comparing with plots in Fig.~\ref{fig:allow neutralino-like I}, we see that the $\tilde H$-like particles with $10\  {\rm GeV} \lesssim m_\chi \lesssim m_W$ could be detected now  through the direct search experiment of SI DM-nucleus elastic scattering in the near future, while originally detectable $\tilde H$-like particles with mass $950 \lesssim m_\chi \lesssim 1680$ GeV in the SI DM-nucleon scattering experiment can not be detected now when considering the leading effect of coannihilation.

\subsection{Conclusions}

In this work,
we construct a generic model of Majorana fermionic dark matter. Starting with two Weyl spinor multiplets $\eta_{1,2}\sim (I,\mp Y)$ coupled to the standard model Higgs, six additional Weyl spinor multiplets with $(I\pm 1/2, \pm(Y\pm 1/2))$ are needed in general. It has 13 parameters in total, five mass parameters and eight Yukawa couplings.
The DM sector of the minimal supersymmetric standard model is a special case of the model with $(I,Y)=(1/2,1/2)$. Therefore, this model can be viewed as an extension of the neutralino DM sector .
Nevertheless, this model does not have sfermions and the second Higgs as in the MSSM, but have more $Z_2$-odd fermions.
We consider three typical cases: the neutralino-like, the reduced and the extended cases.
For the neutralino-like case, we study four different scenarios (neutralino-like I-IV) according to whether the GUT relation on mass parameters or the $\tan\beta$ relation on the Yukawa couplings is imposed or not.
For the reduced case, it has the minimal particle content, while the extended case has the maximal particle content. For each case, we generate 10000 samples from the parameter space and survey the DM mass in the range of $(1,2500)$ GeV. For each sample, we calculate the DM relic density $\Omega_{\chi}h^2$, the SI, SD DM-nucleon elastic scattering cross sections for direct search and the velocity averaged cross section of DM annihilation processes $\langle \sigma (\chi{\chi}\rightarrow W^+W^-, ZZ, ZH, HH, f{\bar f}) v\rangle$ for indirect search.
We compare our results with eleven constraints from the observed DM relic density, the direct search of LUX, XENON100 and PICO-60 experiments, and the indirect search of Fermi-LAT data, respectively.
We investigate the interplay of these three complementary searching strategies and tell the differences among the cases. For each case, we find the allowed DM candidates satisfying all the constraints, and obtain the lower mass bounds of finding the $\tilde H$-, $\tilde B$-, $\tilde W$- and non neutralino-like 
DM particles.
We discuss the properties of DM annihilation processes $\chi\chi\rightarrow W^+W^-, ZZ, ZH, HH, f{\bar f}$. We see that the processes of $\tilde B$-like particles annihilating to $W^+W^-$ and $ZZ$ do not have $s$-wave contribution. The process $\chi{\chi}\rightarrow ZH$ is allowed to have $s$-wave contribution, while the process $\chi{\chi}\rightarrow HH$ does not have $s$-wave contribution. We also see that the process of $\chi\chi\rightarrow f{\bar f}$ have a helicity suppressed $s$-wave contribution.    
We find that the $\tilde H$- and $\tilde B$-like particles appear in all cases, plenty of $\tilde W$-like particles can appear in the neutralino-like III, IV cases with the GUT relation relaxed and in the extended case. The non-MSSM like $\tilde X$ particle can only appear in the extended case. We find that most of $\tilde B$-like particles are ruled out by the $\Omega_{\chi} h^2$ constraint, and further by the LUX constraint;
the $\tilde H$-, $\tilde W$-like particles and the non neutralino-like $\tilde X$ particles with $m_{\chi} \lesssim M_W$ are ruled out by the $\Omega_{\chi} h^2$ and the Fermi-LAT $\la\sigma (\chi{\chi}\rightarrow b\bar b) v\ra$ constraints, while the $\tilde H$, $\tilde W$-like particles and the non neutralino-like $\tilde X$ particles with $m_{\chi} > M_W$ are constrained by the Fermi-LAT $\la\sigma (\chi{\chi}\rightarrow W^+W^-) v\ra$ and the LUX $\sigma^{SI}$ bounds.
We note that
in general the allowed $\tilde H$-, $\tilde W$-like particles and the non neutralino-like $\tilde X$ particles are highly pure with composition fraction $\geq 90\%$. It is also true for $\tilde B$-like particles in the cases without GUT and $\tan\beta$ relations.

When without considering the coannihilation, we find the lower mass bounds to detect DM in the SI DM-nucleus scattering experiments, and the suitable mass ranges to detect DM in the DM annihilation to $W^+W^-$ channel using the present limit and the projected limit (taken to be one order of magnitude lower than the present one).
Apart from the outlier samples, the masses for finding the $\tilde H$-, $\tilde B$-, $\tilde W$-like DM particles and the non neutralino-like $\tilde X$ DM particles are given.
The $\tilde H$-like particles can be detected
with DM mass $\gtrsim 450$ GeV in all cases. The $\tilde B$-like particles can be detected with mass $\gtrsim 1258$ GeV in the cases of neutralino-like I and II. On the other hand, the lower mass bound of $\tilde B$ can be lower down to 341 GeV to detect them for other cases. The $\tilde W$-like particles can be detected with DM mass $\gtrsim 1120$ GeV in the neutralino-like III, IV and the extended cases. Of course, the non neutralino-like particles $\tilde X$ can only be detected with DM mass $\gtrsim 738$ GeV in the extended case. 
We also give the predictions on
 $\la\sigma (\chi{\chi}\rightarrow ZZ, ZH, t{\bar t}) v\ra$ in the indirect search.
The most rewarding way to find the DM particles in this model in the near future is from the direct search of SI DM-nucleus scattering experiments and/or from the indirect search of DM annihilation processes via $W^+W^-, ZZ, ZH, t{\bar t}$ channels.
We also investigate the leading effect of coannihilation in the neutralino-like I case. The change is that the $\tilde H$-like particles with $10\  {\rm GeV} \lesssim m_\chi \lesssim m_W$ can also be detected through the  direct search of SI DM-nucleus scattering experiment in the near future, while $\tilde H$-like particles with mass $950 \lesssim m_\chi \lesssim 1680$ GeV now become undetectable. 
The study of the generic Majorana fermion DM model can be further extended. The whole calculation of coannihilation is worthy of being probed further.
The non-perturbative Sommerfeld effect also has not been implemented. These studies will be presented elsewhere.
This work concentrates on $(I,Y)=(1/2,1/2)$, but the formalism is generic and can be used to study with arbitrary quantum numbers.

\section*{Acknowledgments}

We thanks Yi-Chin Yeh and Chung Kao for discussions.
This research was supported by the Ministry of Science and Technology of R.O.C. under
Grant Nos. 104-2811-M-033-005 and in part by 103-2112-M-033-002-MY3.

\appendix

\section{Neutral and Charged WIMP Masses with $I=Y=1/2$}

For $I=Y=\frac{1}{2}$, $\eta_7$ and $\eta_8$ are singlets with charge $\mp 1$, in other words, $\eta_7^1$ and $\eta_8^{-1}$ are absent. The Lagrangian for neutral WIMP mass term is modified as
\be
-{\cal L}^0_m&=&
\mu_1\lambda^{1}_{-\frac{1}{2},\frac{1}{2}}\eta^{-\frac{1}{2}}_2\eta^{\frac{1}{2}}_1
+\frac{1}{2}\mu_2\lambda^{2}_{0,0}\eta^{0}_4\eta^{0}_3
+\frac{1}{2}\mu_3\lambda^{3}_{0,0}\eta^{0}_6\eta^{0}_5
+\mu_5\lambda^{5}_{-1,1}\eta^{-1}_{10}\eta^{1}_{9}
\non\\
&&
+g_3\lambda^2_{\frac{1}{2},-{\frac{1}{2}},0}\la\tilde\phi^{\frac{1}{2}}\ra \eta^{-{\frac{1}{2}}}_2\eta^{0}_3
+g_4\lambda^2_{-\frac{1}{2},{\frac{1}{2}},0}\la\phi^{-\frac{1}{2}}\ra\eta^{\frac{1}{2}}_1\eta^{0}_4
\non\\
&&
+g_5\lambda^3_{\frac{1}{2},-{\frac{1}{2}},0}\la\tilde\phi^{\frac{1}{2}}\ra \eta^{-{\frac{1}{2}}}_2\eta^{0}_5
+g_6\lambda^3_{-\frac{1}{2},{\frac{1}{2}},0}\la\phi^{-\frac{1}{2}}\ra\eta^{\frac{1}{2}}_1\eta^{0}_6
\non\\
&&
+g_{9}\lambda^5_{-\frac{1}{2},-{\frac{1}{2}},1}\la\phi^{-\frac{1}{2}}\ra\eta^{-{\frac{1}{2}}}_2\eta^{1}_{9}
+g_{10}\lambda^5_{\frac{1}{2},{\frac{1}{2}},-1}\la\tilde\phi^{\frac{1}{2}}\ra \eta^{\frac{1}{2}}_1\eta^{-1}_{10}
+h.c..
\en
It can be simplified as
\be
-{\cal L}^0_m&=&
-\mu_1\eta^{-\frac{1}{2}}_2\eta^{\frac{1}{2}}_1
+\frac{1}{2}\mu_2\eta^{0}_4\eta^{0}_3
-\frac{1}{2}\mu_3\eta^{0}_6\eta^{0}_5
+\mu_5\eta^{-1}_{10}\eta^{1}_{9}
\non\\
&&
+g_3\la\tilde\phi^{\frac{1}{2}}\ra \eta^{-{\frac{1}{2}}}_2\eta^{0}_3
-g_4\la\phi^{-\frac{1}{2}}\ra\eta^{\frac{1}{2}}_1\eta^{0}_4
\non\\
&&
-g_5\la\tilde\phi^{\frac{1}{2}}\ra \eta^{-{\frac{1}{2}}}_2\eta^{0}_5
-g_6\la\phi^{-\frac{1}{2}}\ra\eta^{\frac{1}{2}}_1\eta^{0}_6
\non\\
&&
+g_{9}\sqrt{2}\la\phi^{-\frac{1}{2}}
\ra\eta^{-{\frac{1}{2}}}_2\eta^{1}_{9}
+g_{10}\sqrt{2}\la\tilde\phi^{\frac{1}{2}}\ra \eta^{\frac{1}{2}}_1\eta^{-1}_{10}
+h.c..
\label{eq: L0m2}
\en
With the basis $\Psi^{0T}_i=(\eta^{1/2}_1,\eta^{-1/2}_2,\eta^0_3,\eta^0_5,
\eta^1_9,\eta^{-1}_{10})$, the above Eq. (\ref{eq: L0m2}) can be written as
\be
{\cal L}^0_m=-\frac{1}{2}\Psi^{0T}Y\Psi^0+h.c.,
\label{eq: L1m2}
\en
where the corresponding mass matrix $Y$ takes the form
\be
\footnotesize
\left(
\begin{array}{cccccc}
0
 &-\mu_1
 &-\frac{g_4 v}{\sqrt2}
 &\frac{g_6 v}{\sqrt2}
 &0
 &g_{10} v
 \\
-\mu_1
 &0
 &\frac{ g_3 v}{\sqrt2}
 &\frac{-g_5 v}{\sqrt2}
 &g_9 v
 &0
 \\
-\frac{g_4 v}{\sqrt2}
 &\frac{g_3 v}{\sqrt2}
 &\mu_2
 &0
 &0
 &0
\\
\frac{g_6 v}{\sqrt2}
 &\frac{-g_5 v}{\sqrt2}
 &0
 &\mu_3
 &0
 &0
 \\
 0
 &g_9 v
 &0
 &0
 &0
 &\mu_5
 \\
 g_{10} v
 &0
 &0
 &0
 &\mu_5
 &0
\end{array}
\right).
\non \\
\label{eq: MY2}
\en

For $I=Y=\frac{1}{2}$,
the Lagrangian for single charged WIMP mass term is modified as
\be
-{\cal L}^{\pm}_m&=&
\mu_1\eta^{\frac{1}{2}}_2\eta^{-\frac{1}{2}}_1
+\frac{1}{2}\mu_3(\eta^{1}_6\eta^{-1}_5+\eta^{-1}_6\eta^{1}_5)
+\mu_4\eta^{0}_8\eta^{0}_7
-\mu_5\eta^{0}_{10}\eta^{0}_{9}
\non\\
&&
+g_5
  \sqrt{2}\la\tilde\phi^{\frac{1}{2}}\ra \eta^{\frac{1}{2}}_2\eta^{-1}_5
+g_6
\sqrt{2}\la\phi^{-\frac{1}{2}}\ra\eta^{-\frac{1}{2}}_1\eta^{1}_6
-g_7
  \la\phi^{-\frac{1}{2}}\ra\eta^{-\frac{1}{2}}_2\eta^{0}_7
\non\\
&&
+g_8
\la\tilde\phi^{\frac{1}{2}}\ra \eta^{\frac{-1}{2}}_1\eta^{0}_8
-g_{9}
\la\phi^{-\frac{1}{2}}\ra\eta^{\frac{1}{2}}_2\eta^{0}_{9}
-g_{10}
\la\tilde\phi^{\frac{1}{2}}\ra \eta^{-\frac{1}{2}}_1\eta^{0}_{10}
+h.c..
\label{eq: L2m2}
\en
With the basis
$\Psi^{+T}_i=(\eta'^{1/2}_2,\eta'^1_5,\eta'^0_8,\eta'^0_{10})$ and
$\Psi^{-T}_i=(\eta^{-1/2}_1,\eta^{-1}_5,\eta^0_7,
\eta^0_9)$,
the above 
Lagrangian becomes
\be
-{\cal L}^{\pm}_m&=&
\mu_1\eta^+_2\eta^-_1
+\frac{1}{2}\mu_3(\eta'^+_5\eta^-_5+\eta^-_5\eta'^+_5)
-\mu_4\eta'^{+}_8\eta^-_7
+\mu_5\eta'^{+}_{10}\eta^{-}_{9}
\non\\
&&
+g_5
  \sqrt{2}\la\tilde\phi^{0}\ra \eta^{+}_2\eta^{-}_5
+g_6
\sqrt{2}\la\phi^{0}\ra\eta^{-}_1\eta'^{+}_5
-g_7
  \la\phi^{0}\ra\eta^{+}_2\eta^{-}_7
\non\\
&&
-g_8
\la\tilde\phi^{0}\ra \eta^-_1\eta'^{+}_8
-g_{9}
\la\phi^{0}\ra\eta^{+}_2\eta^{-}_{9}
+g_{10}
\la\tilde\phi^{0}\ra \eta^{-}_1\eta'^{+}_{10}
+h.c..
\label{eq: L3m}
\en
Hence it can be written as the compact form as follows
\be
{\cal L}^{\pm}_m=-\frac{1}{2}(\Psi^+, \Psi^-)
\left(
\begin{array}{cc}
0&X^T
\\
X&0
\end{array}
\right)
\left(
\begin{array}{c}
\Psi^+
\\
\Psi^-
\end{array}
\right)+h.c.,
\label{eq: L4m}
\en
where $X$ takes the form
\be
\footnotesize
\left(
\begin{array}{cccc}
 \mu_1
 &g_6 v
  &\frac{-g_8 v}{\sqrt{2}}
  &\frac{g_{10} v}{\sqrt{2}}
 \\
 g_5 v
 &\mu_3
 &0
 &0
 \\
 \frac{-g_7 v}{\sqrt{2}}
  &0
 &-\mu_4
 &0
 \\
 \frac{-g_9 v}{\sqrt{2}}
 &0
 &0
 &\mu_5
\end{array}
\right).
\non \\
\label{eq: MX2}
\en

\section{Mass Eigenstates for the Neutral and Charged WIMPs}

For the neutral WIMPs, their 4-component representations are of the form
\be
%\tilde
\psi^0_k=
\left(
\begin{array}{c}
\eta^{q_k}_k
\\
\bar\eta^{q_k}_k
\end{array}
\right).
\en
In the above, $q_k$ is defined as the third component of isospin of $\eta_k$ as mentioned in the text. For $I=Y=1/2$ with the basis $\Psi^{0T}_i=(\eta^{1/2}_1,\eta^{-1/2}_2,\eta^0_3,\eta^0_5,
\eta^1_9,\eta^{-1}_{10})$, $q_i=(1/2,-1/2,0,0,1,-1)$.
The 4-component mass eigenstates can be obtained by doing a transformation with a unitary matrix $N$ as follows:
\be
%\tilde
\chi^0_i
\equiv 
\left(
\begin{array}{c}
\zeta^0_i%\chi^+_i
\\
\bar\zeta^0_j%\bar\chi^-_i
\end{array}
\right)
=(N_{ij}P_L+N^*_{ij}P_R)
%\tilde
\psi^0_j,
\en
so that $M^0_D\equiv N^*YN^\dagger$ is a diagonal matrix with nonnegative entries $m_{\chi^0_k}$.
Hence the mass term in Eq. (\ref{eq: L1m}) becomes
\be
{\cal L}^0_m&=&-\frac{1}{2}\Psi^{0T}Y\Psi^0+h.c.
\non\\
&=&-\frac{1}{2}\sum_k m_{\chi^0_k}\bar{
%\tilde
\chi}^0_k{
%\tilde 
\chi}^0_k.
\en
For the single charged WIMPs, their 4-component representations are of the form
\be
%\tilde
\psi^+_k=
\left(
\begin{array}{c}
\eta'^{q_k+1}_k
\\
\bar\eta^{q_{k-1}}_k
\end{array}
\right)
{\rm and}\ \
%\tilde
\psi^-_k=
\left(
\begin{array}{c}
\eta^{q_k-1}_k
\\
\bar\eta'^{q_{k+1}}_k
\end{array}
\right).
%\non\\
\en
For $I=Y=1/2$, with the basis
$\Psi^{+T}_i=(\eta'^{1/2}_2,\eta'^1_5,\eta'^0_8,\eta'^0_{10})$
, $q_i=(-1/2,0,-1,-1)$
and
$\Psi^{-T}_i=(\eta^{-1/2}_1,\eta^{-1}_5,\eta^0_7,\eta^0_9)$,
$q_i=(1/2,0,1,1)$,
the 4-component mass eigenstates can be obtained by doing the transformation with two unitary matrices
$U$ and $V$ as follows:
\be
%\tilde
\chi_i\equiv
\left(
\begin{array}{c}
\zeta^+_i%\chi^+_i
\\
\bar\zeta^-_j%\bar\chi^-_i
\end{array}
\right)
=(V_{ij}P_L+U^*_{ij}P_R)
%\tilde
\psi^+_j\ \
{\rm and}\ \
%\tilde
\chi^c_i=
\left(
\begin{array}{c}
\zeta^-_j%\chi^-_i
\\
\bar\zeta^+%\chi^+_i
\end{array}
\right)=(U_{ij}P_L+V^*_{ij}P_R)
%\tilde
\psi^-_j,
\en
so that $M^{\pm}_D\equiv U^*XV^\dagger$ is a diagonal matrix with nonnegative entries $m_{\chi^{+}_k}$.
Hence the mass term in Eq. (\ref{eq: L4m}) becomes
\be
{\cal L}^{\pm}_m&=&-\frac{1}{2}(\Psi^+, \Psi^-)
\left(
\begin{array}{cc}
0&X^T
\\
X&0
\end{array}
\right)
\left(
\begin{array}{c}
\Psi^+
\\
\Psi^-
\end{array}
\right)
+h.c.
\non\\
&=&-\sum_k m_{\chi^{+}_k}\bar{
%\tilde
\chi}_k{
%\tilde
\chi}_k.
\en

%\section{Indirect searches}
\section{Lagrangian for WIMPs interacting with SM particles}

The Lagrangian for WIMPs interacting with the SM gauge bosons in 4-component notation can be derived from the following gauge invariance terms with 2-component notation\cite{Haber} using the generic Lagrangian in Eq.(\ref{eq: generic}) and Appendices A and B:
\be
-(gT^a_{ij}V^a_\mu +g'y_i\delta_{ij}V'_\mu ){\bar\psi}^i{\bar\sigma^\mu}\psi^j.
\en
In the following, we just write down the results.

%\subsection{Lagrangian for WIMPs with $I=Y=1/2$}

For $I=Y=1/2$, the Lagrangian of the W-boson interaction with the neutral and single charged WIMPs can be written as

\be
{\cal L}_{\chi^0\chi^{\mp}W^\pm}=-\frac{g}{\sqrt2}\{
W^-_{\mu}[{\bar{
%\tilde
{\chi}}^0_i}\gamma^{\mu}
(O^{L_{W^-}}_{ij}P_L+O^{R_{W^-}}_{ij}P_R)
%\tilde
{\chi}^+_j]+
W^+_{\mu}[{\bar{
%\tilde
{\chi}}^+_i}\gamma^{\mu}
(O^{L_{W^+}}_{ij}P_L+O^{R_{W^+}}_{ij}P_R)
%\tilde
{\chi}^0_j]\},
\non\\
\en
where
\be
 \left\{
 \begin{array}{lr}
 O^{L_{W^-}}_{ij}=\sum^6_{k=1}\sum^4_{l=1}
 (-1)^{{\rm mod}(2I_j,2)+1}N_{ik}T^{+0T}_{ki}V^\dagger_{lj}
 \ \ {\rm with}\ \ O^{L_{W^+}}_{ij}=(O^{L_{W^-}})^\dagger_{ij}\\
 \\
 O^{R_{W^-}}_{ij}=-\sum^6_{k=1}\sum^4_{l=1}
 N^{*}_{ik}T^{0-}_{kj}U^T_{lj}
 \ \ {\rm with}\ \ O^{R_{W^+}}_{ij}=(O^{R_{W^-}})^\dagger_{ij},
\end{array}
\right.
\en
and
\be
T^{+0T}_{kl}=
\left(
\begin{array}{cccc}
0&0&0&0
\\
1&0&0&0
\\
0&0&0&0
\\
0&\sqrt2&0&0
\\
0&0&0&0
\\
0&0&0&\sqrt2
\end{array}
\right),\ \ \
T^{0-}_{kl}=
\left(
\begin{array}{cccc}
1&0&0&0
\\
0&0&0&0
\\
0&0&0&0
\\
0&\sqrt2&0&0
\\
0&0&0&0
\\
0&0&0&\sqrt2
\end{array}
\right).
\en

The Lagrangian of the $Z$-boson interaction with the neutral WIMPs is
\be
{\cal L}_{\chi^0_i\chi^0_jZ}=\frac{g}{2\cos\theta_W}Z_{\mu}\bar{
%\tilde
{\chi}}^0_i\gamma^{\mu}
(O^{L_Z}_{ij}P_L+O^{R_Z}_{ij}P_R)
%\tilde
{\chi}^0_j,
\label{eq: chi0chi0Z}
\en
 where
\be
O^{L_Z}_{ij}=\sum^6_{k=1}
q_kN_{ik}N^\dagger_{kj}\ \ \ {\rm with} \ \ \ O^{R_{Z}}_{ij}=-O^{L_{Z^*}}_{ij}.
\en
On the other hand,  $O^{L_Z}_{11}=O^{L_Z*}_{11}$. Hence the Lagrangian for the stable dark matter annihilation via the Z boson can be further simplified as
\be
{\cal L}_{\chi^0_1\chi^0_1Z}=
-\frac{g}{2\cos\theta_W}O^{L_Z}_{11}Z_{\mu}\bar{
%\tilde
{\chi}}^0_1\gamma^{\mu}
\gamma^5
%\tilde
{\chi}^0_1.
\en
The Lagrangian of the Higgs-boson interactions with the neutral WIMPs is
\be
{\cal L}_{\chi^0_i\chi^0_jH^0}=
-H^0{\bar{
%\tilde
{\chi}}}^0_i(O^{L_H}_{ij}P_L+O^{R_H}_{ij}P_R)
%\tilde
{\chi}^0_j,
\en
where
\be
O^{L_H}_{ij}=N^*_{ik}f_{kl}N^{\dagger}_{lj}\ \ \ {\rm with} \ \ \
O^{R_H}_{ij}=(O^{L_H}_{ij})^*,
\en
and
\be
\footnotesize
f_{kl}=
\left(
\begin{array}{cccccccc}
0
 &0
 &-\frac{g_4}{\sqrt2}
 &\frac{g_6}{\sqrt2}
 &0
 &g_{10}
 \\
0
 &0
 &\frac{g_3}{\sqrt2}
 &-\frac{g_5}{\sqrt2}
 &g_9
 &0
 \\
-\frac{g_4}{\sqrt2}
 &\frac{g_3}{\sqrt2}
 &0
 &0
 &0
 &0
\\
\frac{g_6}{\sqrt2}
 &\frac{g_5}{\sqrt2}
 &0
 &0
 &0
 &0
 \\
0
 &g_9
 &0
 &0
 &0
 &0
 \\
g_{10}
 &0
 &0
 &0
 &0
 &0
\end{array}
\right).
%\non \\
\en

For coannihilation, we need the Lagrangian of the $Z$-boson interaction with the single charged WIMPs 
\be
{\cal L}_{\chi^-_i\chi_j^+Z}=\frac{g}{\cos\theta_W}Z_{\mu}\bar{
%\tilde
{\chi}}^+_i\gamma^{\mu}
(O^{L^+_Z}_{ij}P_L+O^{R^+_Z}_{ij}P_R)
%\tilde
{\chi}^+_j-eA_\mu\bar{\tilde{\chi}}^+_i\gamma^{\mu}
%\tilde
{\chi}^+_i,
\en
 where
\be
O^{L^+_Z}_{ij}&=&-\frac{1}{2}V_{i1}V^*_{j1}-V_{i2}V^*_{j2}+\delta_{ij}\sin^2\theta_W,
\non\\
O^{R^+_Z}_{ij}&=&-\frac{1}{2}U^*_{i1}U_{j1}-U^*_{i2}U_{j2}+\delta_{ij}\sin^2\theta_W.
\en
We also need the Lagrangian of the Higgs-boson interaction with the single charged WIMPs
\be
{\cal L}_{\chi^-_i\chi^+_jH^0}=
H^0{\bar{
%\tilde
{\chi}}}^+_i(O^{L^+_H}_{ij}P_L+O^{R^+_H}_{ij}P_R)
%\tilde
{\chi}^+_j,
\en
where
\be
O^{L^+_H}_{ij}=U^*_{ik}h^L_{kl}V^{\dagger}_{lj}\ \ \ {\rm with} \ \ \
O^{R^+_H}_{ij}=V_{ik}h^R_{kl}U^T_{lj}\ \ \
\en
and
\be
\footnotesize
h^L_{lk}=
\left(
\begin{array}{cccccccc}
0
 &-g_6
 &\frac{g_8}{\sqrt2}
 &-\frac{g_10}{\sqrt2}
 \\
-g_5
 &0
 &0
 &0
 \\
\frac{g_7}{\sqrt2}
 &0
 &0
 &0
\\
\frac{g_9}{\sqrt2}
 &0
 &0
 &0
\end{array}
\right)
{\rm and \ \ \ }
h^R_{lk}=h^L_{kl}.
%\non \\
\en

\section{Matrix elements for dark matter annihilation}

\subsubsection{$\chi^0_1\chi^0_1\rightarrow W^+W^-$}
The dark matter can annihilate into $W^+W^-$ via the t-channel exchange of  a single charged WIMP and the s-channel exchange of a $Z^0$ boson or $H^0$ scalar corresponding to the following matrix element:
\be
M(\chi^0_1\chi^0_1\rightarrow W^+W^-)=M_{1a}+M_{1b}+2M_{2a}+2M_{3a},
\en
where
\be
M_{1a}&=&-i\sum_k\frac{g^2}{2}\frac{1}{t-m^2_{\chi^+_k}}
[{\bar v}(p_1)\gamma^{\mu}
(O^{L_{W^-}}_{1k}P_L+O^{R_{W^-}}_{1k}P_R)
(\pslash_3-\pslash_1+m_{\chi^+_k})
\non\\
&&\qquad\qquad\qquad\qquad
\times
\gamma^{\nu}
(O^{L_{W^+}}_{k1}P_L+O^{R_{W+}}_{k1}P_R)u(p_2)\epsilon^*_{\mu}(p_3)\epsilon^*_{\nu}(p_4)],
\non\\
M_{1b}&=&-i\sum_k\frac{g^2}{2}\frac{1}{u-m^2_{\chi^+_k}}[{\bar v}(p_1)\gamma^{\nu}
(O^{L_{W^+}}_{k1}P_L+O^{R_{W^+}}_{k1}P_R)
(\pslash_4-\pslash_1+m_{\chi^+_k})
\non\\
&&\qquad\qquad\qquad\qquad
\times
\gamma^{\mu}
(O^{L_{W^-}}_{1k}P_L+O^{R_{W-}}_{1k}P_R)u(p_2)\epsilon^*_{\mu}(p_3)\epsilon^*_{\nu}(p_4)],
\non\\
M_{2a}&=&-i\frac{g^2}{2}O^{L_Z}_{11}\frac{1}{s-M^2_Z+iM_Z\Gamma_Z}{\bar v}(p_1)\gamma^5
[(\pslash_3-\pslash_4)(\epsilon^*(p_3)\cdot\epsilon^*(p_4))
\non\\
&&\qquad\qquad\qquad\qquad
-\eslash^*(p_4)(p_4\cdot\epsilon^*(p_3))+\eslash^*(p_3)(p_3\cdot\epsilon^*(p_4))],
\non\\
M_{3a}&=&-igM_W\frac{1}{s-M^2_H+iM_H\Gamma_H}
{\bar v}(p_1)(O^{L_H}_{11}P_L+O^{R_H}_{11}P_R)u(p_2)\epsilon^*_{\mu}(p_3)\cdot\epsilon^*_{\nu}(p_4).
\en

\subsubsection{$\chi^0_1\chi^0_1\rightarrow H^0H^0$}
The dark matter can annihilate into $H^0H^0$ via the s-channel exchange of a $H^0$ scalar and the 
t-change exchange of a neutral WIMP corresponding to the following matrix element:
\be
M(\chi^0_1\chi^0_1\rightarrow H^0H^0)=2M_{1a}+M_{2a}+M_{2b}+M_{2c}+M_{2d},
\en
where
\be
M_{1a}&=&-ig\frac{3m^2_H}{2M_W}\frac{1}{s-m^2_H+im_H\Gamma_H}
{\bar v}(p_1)(O^{L_H}_{11}P_L+O^{R_H}_{11}P_R)u(p_2),
\non\\
M_{2a}&=&-i\sum_k\frac{1}{t-m^2_{\chi^0_k}}{\bar v}(p_1)(O^{L_H}_{1k}P_L+O^{R_H}_{1k}P_R)
(\pslash_3-\pslash_1+m_{\chi^+_0})(O^{L_H}_{k1}P_L+O^{R_H}_{k1}P_R)u(p_2),
\non\\
M_{2b}&=&-i\sum_k\frac{1}{u-m^2_{\chi^0_k}}{\bar v}(p_1)(O^{L_H}_{k1}P_L+O^{R_H}_{k1}P_R)
(\pslash_4-\pslash_1+m_{\chi^+_0})(O^{L_H}_{1k}P_L+O^{R_H}_{1k}P_R)u(p_2),
\non\\
M_{2c}&=&-i\sum_k\frac{1}{u-m^2_{\chi^0_k}}{\bar v}(p_1)(O^{L_H}_{1k}P_L+O^{R_H}_{1k}P_R)
(\pslash_4-\pslash_1+m_{\chi^+_0})(O^{L_H}_{k1}P_L+O^{R_H}_{k1}P_R)u(p_2),
\non\\
M_{2d}&=&-i\sum_k\frac{1}{t-m^2_{\chi^0_k}}{\bar v}(p_1)(O^{L_H}_{k1}P_L+O^{R_H}_{k1}P_R)
(\pslash_3-\pslash_1+m_{\chi^+_0})(O^{L_H}_{1k}P_L+O^{R_H}_{1k}P_R)u(p_2).
\non\\
\en

\subsubsection{$\chi^0_1\chi^0_1\rightarrow Z^0Z^0$}

The dark matter can annihilate into $Z^0Z^0$ via the t-channel exchange of a neutral WIMP and the s-channel exchange of a $H^0$ scalar corresponding to the following matrix element:
\be
M(\chi^0_1\chi^0_1\rightarrow Z^0Z^0)=M_{1a}+M_{1b}+M_{1c}+M_{1d}+4M_{2a},
\en
where
\be
M_{1a}&=&-i(\frac{g}{2\cos\theta_W})^2\frac{1}{t-m^2_{\chi^0_k}}[{\bar v}(p_1)\gamma^{\mu}
(O^{L_Z}_{1k}P_L+O^{R_Z}_{1k}P_R)(\pslash_3-\pslash_1+m_{\chi^0_k})
\non\\
&&\qquad\qquad\qquad
\times
\gamma^{\nu}(O^{L_Z}_{k1}P_L+O^{R_Z}_{k1}P_R)u(p_2)]\epsilon^*_{\mu}(p_3)\epsilon^*_{\nu}(p_4),
\non\\
M_{1b}&=&-i(\frac{g}{2\cos\theta_W})^2\frac{1}{u-m^2_{\chi^0_k}}[{\bar v}(p_1)\gamma^{\nu}
(O^{L_Z}_{k1}P_L+O^{R_Z}_{k1}P_R)(\pslash_4-\pslash_1+m_{\chi^0_k})
\non\\
&&\qquad\qquad\qquad
\times
\gamma^{\mu}(O^{L_Z}_{1k}P_L+O^{R_Z}_{1k}P_R)u(p_2)]\epsilon^*_{\mu}(p_3)\epsilon^*_{\nu}(p_4),
\non\\
M_{1c}&=&-i(\frac{g}{2\cos\theta_W})^2\frac{1}{u-m^2_{\chi^0_k}}[{\bar v}(p_1)\gamma^{\nu}
(O^{L_Z}_{1k}P_L+O^{R_Z}_{1k}P_R)(\pslash_4-\pslash_1+m_{\chi^0_k})
\non\\
&&\qquad\qquad\qquad
\times
\gamma^{\mu}(O^{L_Z}_{k1}P_L+O^{R_Z}_{k1}P_R)u(p_2)]\epsilon^*_{\mu}(p_3)\epsilon^*_{\nu}(p_4),
\non\\
M_{1d}&=&-i(\frac{g}{2\cos\theta_W})^2\frac{1}{t-m^2_{\chi^0_k}}[{\bar v}(p_1)\gamma^{\mu}
(O^{L_Z}_{k1}P_L+O^{R_Z}_{k1}P_R)(\pslash_3-\pslash_1+m_{\chi^0_k})
\non\\
&&\qquad\qquad\qquad
\times
\gamma^{\nu}(O^{L_Z}_{1k}P_L+O^{R_Z}_{1k}P_R)u(p_2)]\epsilon^*_{\mu}(p_3)\epsilon^*_{\nu}(p_4),
\non\\
M_{2a}&=&i(\frac{g}{2\cos\theta_W})M_Z\frac{1}{s-M^2_H+iM_H\Gamma_H}
[{\bar v}(p_1)(O^{L_H}_{11}P_L+O^{R_H}_{11}P_R)u(p_2)].
\en

\subsubsection{$\chi^0_1\chi^0_1\rightarrow H^0Z^0$}

The dark matter can annihilate into $H^0Z^0$ via the t-channel exchange of a neutral WIMP and s-channel exchange of a $Z^0$ boson corresponding to the following matrix element:
\be
M(\chi^0_1\chi^0_1\rightarrow HZ)=M_{1a}+M_{1b}+4M_{2a},
\en
where
\be
M_{1a}&=&i\frac{g}{2\cos\theta_W}\frac{1}{t-m^2_{\chi^0_k}}[{\bar v}(p_1)\gamma^{\mu}
(O^{L_Z}_{1k}P_L+O^{R_Z}_{1k}P_R)(\pslash_3-\pslash_1+m_{\chi^0_k})
\non\\
&&\qquad\qquad\qquad
\times
(O^{L_H}_{k1}P_L+O^{R_H}_{k1}P_R)u(p_2)]\epsilon^*_{\mu}(p_3),
\non\\
M_{1b}&=&i\frac{g}{2\cos\theta_W}\frac{1}{u-m^2_{\chi^0_k}}[{\bar v}(p_1)
(O^{L_H}_{k1}P_L+O^{R_H}_{k1}P_R)(\pslash_3-\pslash_1+m_{\chi^0_k})
\non\\
&&\qquad\qquad\qquad
\times
(O^{L_Z}_{1k}P_L+O^{R_Z}_{1k}P_R)\gamma^{\mu}u(p_2)]\epsilon^*_{\mu}(p_3),
\non\\
M_{2a}&=&-i(\frac{g}{2\cos\theta_W})^2O^{L_Z}_{11}M_Z\frac{1}{s-M^2_Z+iM_Z\Gamma_Z}
[{\bar v}(p_1)\gamma^{\alpha}{\gamma^5}u(p_2)]\epsilon^*_{\alpha}(p_3).
\en

\subsubsection{$\chi^0_1\chi^0_1\rightarrow f\bar f$}

The dark matter can annihilate into $f\bar f$ via the s-channel exchange of a $Z^0$ boson or a $H^0$ scalar corresponding to the following matrix element:
\be
M(\chi^0_1\chi^0_1\rightarrow f\bar f)=2M_{1a}+2M_{2a},
\en
where
\be
M_{1a}&=&i(\frac{g}{2\cos\theta_W})^2O^{L_Z}_{11}M_Z\frac{1}{s-M^2_Z+iM_Z\Gamma_Z}
g_{\alpha\mu}[{\bar v}(p_1)\gamma^{\alpha}{\gamma^5}u(p_2)]
[\bar u(p_3)\gamma^{\mu}](g^f_V+g^f_A\gamma^5)v(p_4)],
\non\\
M_{2a}&=&-i\frac{gm_f}{2M_W}\frac{1}{s-M^2_H+iM_H\Gamma_H}
[{\bar v}(p_1)(O^{L_H}_{11}P_L+O^{R_H}_{11}P_R)u(p_2)][\bar u(p_3)v(p_4)]
\en
with $g^f_V=\frac{1}{2}T^f_{3L}-Q^f\sin^2\theta_W$, and $g^f_A=-\frac{1}{2}T^f_{3L}$.

\section{CP symmetry}

Before transforming the gauge eigenstates to mass eigenstates, all parameters in the Lagrangian are assumed to be real in this model. The Lagrangian is CP conserved. After field redefinition, some parameters become purely imaginary. The Lagrangian should still be CP conserved. We explicitly show this and a useful application in below.

The CP transformation of a four component field is given by
\be
{\cal CP}
%\tilde 
\chi_i(x){\cal P}^\dagger{\cal C}^\dagger
=\rho_{CP,\chi_i} \gamma_0%\tilde 
\chi_i^c(\tilde x)
=\rho_{CP,\chi_i} \gamma_0 C\bar{
%\tilde 
\chi}_i^T(\tilde x),
\label{eq: CP2}
\en
with the phase $\rho_{CP,\chi_i}$ for $
%\tilde
\chi_i$, $\tilde x^\mu\equiv x_\mu$ and $C=i\gamma_2\gamma_0$.
For a Majorana field we have $
%\tilde 
\chi_i^c=\rho_{M,\chi_i}{
%\tilde 
\chi_i}$, where $\rho_{M,\chi_i}$ is a phase.
Eq. (\ref{eq: CP2}) implies that $\rho_{CP,\chi_i}\rho_{M,\chi_i}$ is purely imaginary~\cite{Kayser:1984ge}.
This can be seen by using
%one still has
$v(\vec p,s)=C\bar u^T(\vec p,s)$,
$u(\vec p,s)=C\bar v^T(\vec p,s)$,
$\gamma_0 u(\vec p,s)=u(-\vec p,-s)$,
$\gamma_0 v(\vec p,s)=-v(\vec p,-s)$,
\be
{\cal CP}
%\tilde
\chi_i(x){\cal P}^\dagger{\cal C}^\dagger
&=&\rho_{CP,\chi_i} \gamma_0
%\tilde
\chi_i^c(\tilde x)=
\rho_{CP,\chi_i} \rho_{M,\chi_i}\gamma_0
%\tilde
\chi_i(\tilde x)
\en
and
\be
%\tilde
\chi_i(x)&=&\int \frac{d^3 p }{(2\pi)^3 2E} \left(b_i({\vec p,s}) u(\vec p,s)e^{-ip\cdot x}
                     + b^\dagger_i({\vec p,s}) \rho^*_{M\chi_i} v(\vec p,s)e^{ip\cdot x}\right),
\en
which imply
%\be
%{\cal CP} b(\vec p, s_z)({\cal CP})^\dagger=\eta_{CP}\eta_M b(\vec p,s_z),
%\quad
%\eta_M^*{\cal CP} b^\dagger(\vec p, s_z)({\cal CP})^\dagger=-\eta_{CP} b^\dagger(\vec p,s_z).
%\en
%Implies
\be
{\cal CP} b^\dagger_i(\vec p, s)({\cal CP})^\dagger=\rho^*_{CP,\chi_i}\rho^*_{M,\chi_i} b_i^\dagger(-\vec p,-s)
=-\rho_{CP,\chi_i}\rho_{M,\chi_i} b_i^\dagger(-\vec p,-s).
\en
Hence the phase $\rho_{CP,\chi_i}\rho_{M,\chi_i}$ is purely imaginary.

As shown in Appendix B, 
the neutral WIMP mass eigenstates are defined (in two-component notation) by
\be
\zeta^0_i%\chi^0_i
=N_{ij}\eta_j^0.
\en
%where $N$ is a unitary matrix satisfying
In the above, the superscript in $\eta_j$ here denotes the charge in stead of the third component of isospin, and $N$ is a unitary matrix satisfying
\be
N^* Y N^{*T}=M^0_D,
\en
where $M^0_D$ is a diagonal matrix with nonnegative entries.
Note in the case one obtains a negative mass in the first place, the negative sign in front of the mass $m_i$ can be absorbed in $N_{ij}$ with $N_{ij}$ being purely imaginary in the corresponding $i$-row.

The four component neutral Majorana states are defined as %(in the chiral representation)
%\be
%\tilde\chi^0_i
%=
%\left(
%\begin{array}{c}
%\chi^0_i\\
%\bar\chi^0_i
%\end{array}
%\right)
%=
%\left(
%\begin{array}{c}
%N_{ij}\psi^0_i\\
%N^*_{ij}\bar\psi^0_i
%\end{array}
%\right)=N_{ij} P_L\Psi^0_i+N^*_{ij} P_R \Psi^0_i,
%\en
%where
%\be
%\Psi^0_i
%\equiv
%\left(
%\begin{array}{c}
%\psi^0_i\\
%\bar\psi^0_i
%\end{array}
%\right).
%\en
%
\be
%\tilde
\chi^0_i
=
\left(
\begin{array}{c}
\zeta^0_i\\%\chi^0_i\\
\bar\zeta^0_i%\chi^0_i
\end{array}
\right)
=
\left(
\begin{array}{c}
N_{ij}\eta^0_i\\
N^*_{ij}\bar\eta^0_i
\end{array}
\right)=N_{ij} P_L
%\tilde
\psi^0_i+N^*_{ij} P_R %\tilde
\psi^0_i,
\en
where
\be
%\tilde
\psi^0_i
\equiv
\left(
\begin{array}{c}
\eta^0_i\\
\bar\eta^0_i
\end{array}
\right).
\en
From above definition, we have $
%\tilde
\psi_i^{0c}=
%\tilde
\psi^0_i$ and $
%\tilde
\chi^{0c}_i=
%\tilde
\chi^0_i$  so that 
%one obtains
$
\rho_{M,\chi_i}=\rho_{M,
%\tilde
\psi_i}=1.
$

For some given $i$, $N_{ij}$ are real, which gives
\be
%\tilde
\chi^0_i
=
\left(
\begin{array}{c}
N_{ij}\eta^0_i\\
N_{ij}\bar\eta^0_i
\end{array}
\right)
=N_{ij}
%\tilde
\psi^0_i
=N_{ij} (P_L
%\tilde
\psi^0_i+P_R 
%\tilde
\psi^0_i),
\en 
We now assume that $
%\tilde
\psi^0_i$ has common $\rho_{CP,
%\tilde
\psi}$ for all $i$.
Therefore, we have
\be
{\cal CP}
%\tilde
\psi^0_i(x){\cal P}^\dagger{\cal C}^\dagger
&=&\rho_{CP,
%\tilde
\psi}\gamma_0 
%\tilde
\psi^0_i(\tilde x).
\label{eq: CP psi Majorana 1}
\en
For the case of real $N_{ij}$ for some $i$, we now have
\be
{\cal CP}
%\tilde 
\chi^0_i(x){\cal P}^\dagger{\cal C}^\dagger
&=&N_{ij}{\cal CP}
%\tilde
\psi^0_i(x){\cal P}^\dagger{\cal C}^\dagger
\non\\
&=&\rho_{CP,
%\tilde
\psi}\gamma_0 N_{ij} 
%\tilde
\psi^0_i(\tilde x).
%\non\\
%&=&\eta_{P,\tilde\psi}\eta_{C,\tilde\psi}\eta_{M,\tilde\psi}\gamma_0 \tilde \chi^0_i(\tilde x).
\en
We obtain for the real $N_{ij}$ case:
\be
\rho_{CP,\chi_i}=\rho_{CP,
%\tilde
\psi}.
\label{eq: zetachireal}
\en 

If for some $i$, $N_{ij}$ are imaginary, i.e. $N^*_{ij}=-N_{ij}$, and we now have 
\be
%\tilde
\chi^0_i
=
\left(
\begin{array}{c}
N_{ij}\eta^0_i\\
N^*_{ij}\bar\eta^0_i
\end{array}
\right)
=
\left(
\begin{array}{c}
N_{ij}\eta^0_i\\
-N_{ij}\bar\eta^0_i
\end{array}
\right)
=N_{ij}(-\gamma_5)
%\tilde
\psi^0_i
=N_{ij} (P_L
%\tilde
\psi^0_i- P_R 
%\tilde
\psi^0_i).
%\non\\
\en 
Note that the relative sign of $P_L\tilde\psi^0_i$ and $P_R \tilde\psi^0_i$ is the key to absorb the minus of the mass term, which consists of left-handed and right-handed field at the same time.
The CP transformation of $
%\tilde
\chi^0_i$ is
\be
{\cal CP}
%\tilde 
\chi^0_i(x){\cal P}^\dagger{\cal C}^\dagger
&=&N_{ij}(-\gamma_5){\cal CP}
%\tilde
\psi^0_i(x){\cal P}^\dagger{\cal C}^\dagger
\non\\
&=&\rho_{CP,
%\tilde
\psi}(-\gamma_5)\gamma_0 N_{ij} %\tilde
\psi^0_i(\tilde x)
\non\\
&=&-\rho_{CP,
%\tilde
\psi}\gamma_0 
%\tilde 
\chi^0_i(\tilde x).
\en
We obtain for the imaginary $N_{ij}$ case:
\be
\rho_{CP,\chi_i}=-\rho_{CP,
%\tilde
\psi}.
\label{eq: zetaimaginary}
\en

Consider a hermitian operator ${\cal O}(x)$:
\be
{\cal O}_{ij}(x)&=&v_{ij}\bar{
%\tilde
\chi}_i(x)\gamma_\mu T^a A^{a\mu}(x) {
%\tilde
\chi}_j(x)+a_{ij}\bar{
%\tilde
\chi}_i(x)\gamma_\mu \gamma_5T^a A^{a\mu}(x) {
%\tilde
\chi}_j(x)
\non\\
&&+[v_{ij}\bar{%\tilde
\chi}_i(x)\gamma_\mu T^a A^{a\mu}(x) {
%\tilde
\chi}_j(x)+a_{ij}\bar{
%\tilde
\chi}_i(x)\gamma_\mu \gamma_5T^a A^{a\mu}(x) {
%\tilde
\chi}_j(x)]^\dagger.
%\non\\
\en
For example, in Eq. (\ref{eq: chi0chi0Z}), we have
\be
v_{ij}=\frac{g}{4\cos\theta_W}(O^{L_Z}_{ij}+O^{R_Z}_{ij}),
%\non\\
\quad
a_{ij}=\frac{g}{4\cos\theta_W}(-O^{L_Z}_{ij}+O^{R_Z}_{ij})
\en
with $O^{L_Z}_{ij}=\sum^6_{k=1}
q_kN_{ik}N^\dagger_{kj}$ and  $O^{R_{Z}}_{ij}=-O^{L_{Z^*}}_{ij}$.

Under CP transformation the operator transforms as
\be
{\cal CP}{\cal O}_{ij}(x){\cal P}^\dagger{\cal C}^\dagger
&=&
v_{ij}\rho^*_{CP,\chi_i}\rho_{CP,\chi_j}[\bar{
%\tilde
\chi}_i(\tilde x)\gamma_\mu T^a A^{a\mu}(\tilde x) {
%\tilde
\chi}_j(\tilde x)]^\dagger
\non\\
&&
+a_{ij}\rho^*_{CP,\chi_i}\rho_{CP,\chi_j}[\bar{
%\tilde
\chi}_i(\tilde x)\gamma_\mu \gamma_5T^a A^{a\mu}(\tilde x) {
%\tilde
\chi}_j(\tilde x)]^\dagger
\non\\
&&+v^*_{ij}\rho_{CP,\chi_i}\rho^*_{CP,\chi_j}
\bar{
%\tilde
\chi}_i(\tilde x)\gamma_\mu T^a A^{a\mu}(\tilde x) {
%\tilde
\chi}_j(\tilde x)
\non\\
&&+a^*_{ij}\rho_{CP,\chi_i}\rho^*_{CP,\chi_j}
\bar{
%\tilde
\chi}_i(\tilde x)\gamma_\mu \gamma_5T^a A^{a\mu}(\tilde x) {
%\tilde
\chi}_j(\tilde x).
%\non\\
\en
To have CP symmetry, one requires 
\be
{\cal CP}\int d^4x{\cal O}_{ij}(x){\cal P}^\dagger{\cal C}^\dagger
=\int d^4x{\cal O}_{ij}(\tilde x)
=\int d^4x{\cal O}_{ij}(x),
\en
%which requires
so that
\be
v_{ij}\rho^*_{CP,\chi_i}\rho_{CP,\chi_j}=v^*_{ij},
\quad
a_{ij}\rho^*_{CP,\chi_i}\rho_{CP,\chi_j}=a^*_{ij}.
\label{eq: v a}
\en

%In SUSY model neutralino fields are mixed states of gauginos and higgsinos: 
%(in two component notation)

%In the model with $I=Y=1/2$, the neutral WIMPs are the mixture of $\tilde H$-, $\tilde B$-, $\tilde W$-like and non neutralino-like $\tilde X$ states,
%\be
%\frac{1}{2}(\Psi^0)^TY\Psi^0+h.c., 
%\en
%where $\Psi^{0T}_i=(\eta^{1/2}_1,\eta^{-1/2}_2,\eta^0_3,\eta^0_5,
%\eta^1_9,\eta^{-1}_{10})$ and $Y$ a $4\times 4$ matrix.

%We now return to the CP transformation of ${\cal O}$,
%\be
%{\cal O}_{ij}(x)&=&v_{ij}\bar{\tilde\chi}^0_i(x)\gamma_\mu T^a A^{a\mu}(x) \tilde\chi^0_j(x)%+a_{ij}\bar{\tilde\chi}^0_i(x)\gamma_\mu \gamma_5T^a A^{a\mu}(x) \chi^0_j(x)
%\non\\
%&&+[v_{ij}\bar{\tilde\chi}^0_i(x)\gamma_\mu T^a A^{a\mu}(x) \tilde\chi^0_j(x)+a_{ij}\bar{\tilde\chi}^0_i(x)\gamma_\mu \gamma_5T^a A^{a\mu}(x) \tilde\chi^0_j(x)]^\dagger.
%\non\\
%\en
In the case both $
%\tilde
\chi^0_i$ and $
%\tilde 
\chi^0_j$ contain only real (or imaginary) $N_{ik}$, $N_{jr}$, 
Eq. (\ref{eq: v a}) gives
\be
v^*_{ij}=v_{ij},
\quad
a^*_{ij}=a_{ij},
\en 
and with Eq. (\ref{eq: zetachireal}) 
% they imply
we have
\be
v_{ij}\rho^*_{CP,\chi_i}\rho_{CP,\chi_j}=v^*_{ij},
\quad
a_{ij}\rho^*_{CP,\chi_i}\rho_{CP,\chi_j}=a^*_{ij}.
\en
Hence ${\cal O}$ is CP conserved.

In the case that
$
%\tilde
\chi^0_i$ is with a real $N_{ik}$, but $
%\tilde
\chi^0_j$ is with a imaginary $N_{jr}$,
Eq. (\ref{eq: v a}) gives
\be
v^*_{ij}=-v_{ij},
\quad
a^*_{ij}=-a_{ij}.
\en
and Eqs.  (\ref{eq: zetachireal}) and (\ref{eq: zetaimaginary}) give
\be
\rho_{CP,\chi_i}=\rho_{CP,\tilde\psi}=-\rho_{CP,\chi_j}.
\en
It implies
\be
v_{ij}\rho^*_{CP,\chi_i}\rho_{CP,\chi_j}=v^*_{ij},
\quad
a_{ij}\rho^*_{CP,\chi_i}\rho_{CP,\chi_j}=a^*_{ij}.
\en
The operator ${\cal O}$ is CP conserved as expected.

In the center of mass frame of two Majorana particles, which are in a definite angular momentum configuration, the state is given by
\be
|^{2S+1}L_J,J_z>=\sum_{m,s_z,s_z'}
\int \frac{d^3p}{2E} f_{J_z m}(\vec p)
{\cal S}_m(s_z,s_z')
b^\dagger(\vec p,s_z)b^\dagger(-\vec p,s_z')|0\ra,
\en
where 
\be
f_{J_z m}(\vec p)\equiv \la L, M, S, m|L, S; J, J_z\ra Y_{LM}(\hat p) R(|\vec p|),
\en
with $\la L, M, S, m|L, S; J, J_z\ra$ is the Clebsch-Gordan coefficient and $R(|\vec p|)$ the radial wave function. 
Note that it is easier to use spin instead of helicity basis here.
For the spin wave function, we have 
\be
{\cal S}_m(s_z,s_z')=(-1)^{S+1}{\cal S}_m(s_z',s_z).
\en
The spherical harmonic wave function has the following property:
\be
Y_{LM}(-\hat p)=(-1)^L Y_{LM}(\hat p).
\en 
Note that
\be
b^\dagger(\vec p,s_z)b^\dagger(-\vec p,s_z')|0\ra
=-b^\dagger(-\vec p,s'_z)b^\dagger(\vec p,s_z)|0\ra,
\en
the above relations of $\chi_m$ and $Y_{LM}$ lead to
\be
(-)^{L+S}=1.
\label{eq: Fermi}
\en

Since
\be
{\cal CP}b^\dagger(\vec p,s_z)b^\dagger(-\vec p,s_z')|0\ra
%&=&(\rho^*_{CP}\rho^*_M)^2b^\dagger(-\vec p,s_z)b^\dagger(\vec p,s_z')|0\ra
%\non\\
&=&-b^\dagger({-\vec p,s_z})b^\dagger({\vec p,s_z'})|0\ra,
\en
where we use the fact that the phase $\rho_{CP}\rho_M$ is purely imaginary,
we have
\be
{\cal CP}|^{2S+1}L_J,J_z\ra&=&-\sum_{m,s_z,s_z'}\int \frac{d^3p}{2E} f_{J_z m}(\vec p){\cal S}_m(s_z,s_z')
b^\dagger(-\vec p,s_z)b^\dagger(\vec p,s_z')|0\ra
%\non\\
%&=&\sum_{m,s_z,s_z'}\int \frac{d^3p}{2E} f_{J_z m}(-\vec p)\chi_m(s_z,s_z')
%b^\dagger(\vec p,s_z)b^\dagger(-\vec p,s_z')|0\ra
\non\\
&=&(-)^{L+1}|^{2S+1}L_J,J_z\ra,
\label{eq: DM CP}
\en
where we have made use of $f_{J_z m}(-\vec p)=(-)^L f_{J_z m}(\vec p)$.
Note that for a $f\bar f$ pair, similar argument leads to  ${\cal CP}|^{2S+1}L_J,J_z\ra=(-)^{S+1}|^{2S+1}L_J,J_z\ra$.

As CP is a good quantum number, it can can be used as a selection rule in dark matter annihilation processes, when the initial state has a specific $L$ (and $S$) configuration.

\section{Formulae for DM-nucleus elastic scattering cross section}

The derivation of the DM-nucleus elastic scattering cross section in the literature are scattered and usually with different approximations, normalizations and notations. It will be useful to rederive the formulas here.

\subsection{Kinematics}

We consider the elastic scattering of
\be
\chi(p_\chi)+{\cal N}(p)\to \chi(p'_\chi)+{\cal N}(p').
\en
We define
\be
&q\equiv p'-p=p_\chi-p'_\chi,
\quad
P\equiv p+p',
\quad
P_\chi\equiv p_\chi+p'_\chi,&
\non\\
&S\equiv (p^{(\prime)}+p^{(\prime)}_\chi)^2=m_{\cal N}^2+m_\chi^2+2 p^{(\prime)}\cdot p^{(\prime)}_\chi.&
\en
In particular, we have
\be
q^2&=&2m_{\cal N}^2-2 p\cdot p'=2 m_\chi^2-2 p_\chi\cdot p'_\chi,
\en
and, in the center of mass frame,
\be
q^2=(E'-E)^2-(|\vec p'_{cm}|^2+|\vec p_{cm}|^2-2\vec p'_{cm}\cdot\vec p_{cm})=2|\vec p_{cm}|^2(\cos\theta-1).
\en
When $q^2=0$, we must have $|\vec p_{cm}|=0$ or $\cos\theta=1$. In either case, it gives $q=0$.
Therefore, in elastic scattering, $q^2=0$ implies $q=0$ in the center of mass frame and in all other frames.

In the lab frame $p=(m_{\cal N},\vec 0)$ and $p_\chi=(m_\chi+m_\chi v^2/2, m_\chi \vec v)$. We obtain
\be
S&=&(m_{\cal N}+m_\chi)^2(1+\frac{\mu_{\cal N}}{m_{\cal N}+m_\chi} v^2),
\label{eq: S NR}
\en
where $\mu_{\cal N}\equiv m_\chi m_{\cal N}/(m_{\cal N}+m_\chi)$ is the reduced mass.
The center of mass energy of the whole system is
\be
E_{cm}=\sqrt s=m_{\cal N}+m_\chi+\frac{1}{2}\mu_{\cal N} v^2,
\en
as expected.

The center of mass velocity in the lab frame is $m_\chi \vec v/(m_{\cal N}+m_\chi)$. Boost the frame by $-m_\chi \vec v/(m_{\cal N}+m_\chi)$, we obtain the velocity of $p$ and $p_\chi$ at the center of mass frame as $-m_\chi\vec v/(m_{\cal N}+m_\chi)$ and $\vec v m_{\cal N}/(m_{\cal N}+m_\chi)$, respectively. Hence, we have
\be
|\vec p_{cm}|=\mu_{\cal N} v,
\label{eq: pcm NR}
\en
and
$q^2=2 \mu_{\cal N}^2 v^2(\cos\theta-1)$.

\subsection{Effective Lagrangian for Direct Searches}

In this model, we have scalar-scalar, psudo scalar-scalar, axial-axial and axial-vector interactions for direct searches. The process of DM-nucleus scattering is non-relativistic so that we can use the effective Lagrangian which can be derived from the Lagrangian in Appendix C to calculate the related SI and SD cross sections. We just give the results as below. The effective Lagrangian for scalar-scalar and pseudo scalar-scalar interactions are
\be
{\cal L}^{SS}=\sum_qa^q\bar{
%\tilde
\chi}^0_1\chi^0_1\bar qq,
\quad
{\cal L}^{PS}=\sum_qa'^q\bar{
%\tilde
\chi}^0_1\gamma_5\chi^0_1\bar qq,
\en
where 
\be
a^q=i\frac{gm_q}{2M_Wm^2_H}{\rm Re}(O^{L_H}_{11}),
\quad
a'^q=\frac{gm_q}{2M_Wm^2_H}{\rm Im}(O^{L_H}_{11}),
\en
the effective Lagrangian for axial-axial and axial-vector interactions are
\be
{\cal L}^{AA}=\sum_qd^q\bar{
%\tilde
\chi}^0_1\gamma^{\mu}\gamma^5\chi^0_1
\bar q\gamma_{\mu}\gamma^5q,
\quad
{\cal L}^{AV}=\sum_qb^q \bar{
%\tilde
\chi}^0_1\gamma^{\mu}\gamma^5\chi^0_1
\bar q\gamma_{\mu}q,
\en
where 
\be
d^q=-\frac{i}{2}(\frac{g}{M_W})^2O^{L_Z}_{11}g_A,
\quad
b^q=-2i(\frac{g}{M_W})^2O^{L_Z}_{11}g_V
\en
with $g_A=-\frac{1}{2}T^q_{3L}$ and $g_V=\frac{1}{2}T^q_{3L}-\sin^2\theta_WQ^q$.

\subsection{Vector, axial vector current, scalar and pseudoscalar matrix elements in the $q=0$ limit}

Using parity transformation, one see that the matrix elements of vector ($j_{Vh}$), axial vector current ($j_{Ah}$), scalar ($s_h$) and pseudoscalar ($p_h$) matrix elements should satisfy the following relations,
\be
\la {\cal N}(p',s')|j_{V(A)h,\mu}(x)|{\cal N}(p,s)\ra
&=&\la {\cal N}(p',s')|P^\dagger P j_{V(A)h,\mu}(x)P^\dagger P |{\cal N}(p,s)\ra
\non\\
&=&\pm\eta_P^*\eta_P
\la {\cal N}(\tilde p',s')|j^\mu_{V(A)h}(\tilde x)|{\cal N}(\tilde p,s)\ra
\non\\
&=&\pm
\la {\cal N}(\tilde p',s')|j^\mu_{V(A)h}(\tilde x)|{\cal N}(\tilde p,s)\ra,
\non\\
\la {\cal N}(p',s')|s_h(p_h)(x)|{\cal N}(p,s)\ra
&=&\pm \la {\cal N}(\tilde p',s')|s_h(p_h)(\tilde x)|{\cal N}(\tilde p,s)\ra,
\label{eq: parity}
\en
where $\tilde p^\mu,\tilde x^\mu\equiv p_\mu,x_\mu$, $\eta$s are phases and $s,s'$ are spin ($S_z$) quantum numbers.

From Eq. (\ref{eq: parity}) it is clear that in the case of $p=p'$ and in the momentum rest frame, $p=(m_{\cal N},\vec 0)$, we have
\be
\la {\cal N}(m_{\cal N},s')|j_{V(A)h,\mu}(0)|{\cal N}(m_{\cal N},s)\ra
&=&\pm
\la {\cal N}(m_{\cal N},s')|j^\mu_{V(A)h}(0)|{\cal N}(m_{\cal N},s)\ra,
\en
which gives
\be
\la {\cal N}(m_{\cal N},s')|j_{Vh,i}(0)|{\cal N}(m_{\cal N},s)\ra=0,
\quad
\la {\cal N}(m_{\cal N},s')|j_{Ah,0}(0)|{\cal N}(m_{\cal N},s)\ra=0.
\label{eq: q=0}
\en
These imply that $\la {\cal N}(p',s')|j_{Vh,i}(x)|{\cal N}(p,s)\ra$ and $\la {\cal N}(p',s')|j_{Ah,0}(x)|{\cal N}(p,s)\ra$ are suppressed in the non relativistic limit: $p\simeq p'\simeq(m_{\cal N},\vec 0)$.

We consider the vector current case first.
From the first equation of Eq. (\ref{eq: q=0}), we obtain
\be
\la {\cal N}(m_{\cal N},s')|j_{Vh,\mu}(0)|{\cal N}(m_{\cal N},s)\ra=(2m_{\cal N}\delta_{s's} F_{\cal N}(0),\vec 0),
\label{eq: F}
\en
where $F_{\cal N}$ is the form factor and the $\delta_{ss'}$ factor is obtained as $j_{Vh,0}$ is a singlet under rotation.
We can write it in a covariant form:
\be
\la {\cal N}(p,s')|j_{Vh,\mu}(0)|{\cal N}(p,s)\ra=2p_\mu F_{\cal N}(0)\delta_{s's}.
\en
In the case of non-vanishing but small $q$, we have
\be
\la {\cal N}(p',s')|j_\mu(x)|{\cal N}(p,s)\ra\simeq (p_\mu+p'_\mu) F_{\cal N}(q^2) \delta_{s,s'} \exp[i(p'-p)\cdot x].
\en
Now we want to find $F_{\cal N}(0)$. From $Q\equiv\int d^3 x j_{Vh,0}(0,\vec x)$, we have
\be
\int d^3 x \la {\cal N}(p',s')|j_{Vh,0}(x)|{\cal N}(p,s)\ra=
 (p_0+p'_0) F_{Vh}(q^2) \delta_{s,s'}\int d^3 x \exp[i(p'-p)\cdot x]+\cdots,
\en
giving
\be
\la {\cal N}(p',s')|Q|{\cal N}(p,s)\ra=(E+E') \delta_{s,s'} F_{\cal N}(q^2) \exp[i(E'-E)t](2\pi)^3
\delta^3(\vec p-\vec p{}').
\en
Therefore, we have
\be
Q_{\cal N} \la {\cal N}(p',s')|{\cal N}(p,s)\ra
=F_{\cal N}(0) \delta_{s,s'} (2\pi)^3 2 E \delta^3(\vec p-\vec p{}'),
\en
which implies
\be
F_{\cal N}(0)=Q_{\cal N},
\label{eq: Q}
\en
and, hence, for vector current matrix elements in $q=0$ case and the $p^{(\prime)}$ rest frame is
\be
\la {\cal N}(m_{\cal N},s')|j_{Vh,0}(0)|{\cal N}(m_{\cal N},s)\ra
&=&2m_{\cal N}\delta_{ss'} F(0)
= 2m_{\cal N}\delta_{ss'}Q_{V\cal N},
\non\\
\la {\cal N}(m_{\cal N},s')|j_{Vh,i}(0)|{\cal N}(m_{\cal N},s)\ra&=&0.
\label{eq: jV at q=0}
\en
These results will be useful in later discussion.
For
\be
j_{hV,\mu}=b^q j_{qV,\mu}=b^u \bar u\gamma_\mu u+ b^d \bar d\gamma_\mu d+\cdots,
\en
it can be proved, by using isospin invariant, that
\be
Q_{Vp}=2 b^u+ b^d\equiv f_{Vp},
\quad
Q_{Vn}= b^u+2 b^d\equiv f_{Vn}.
\en
Hence, the corresponding charge is
\be
Q_{V\cal N}=Z Q_{Vp}+(A-Z) Q_{Vn}=Z(2b^u+b^d)+(A-Z)(b^u+2 b^d).
\en

We now turn to the axial vector case. We start from
\be
\la {\cal N}(p', s')|j_{Aq}^i(0)|{\cal N}(p,s)\ra
&=&\la {\cal N}(p', s')|\bar q\gamma^i\gamma_5 q(0)|{\cal N}(p,s)\ra
\non\\
&=&2\la {\cal N}(p', s')|\bar q\gamma_0\frac{\vec\Sigma}{2} q(0)|{\cal N}(p,s)\ra
\non\\
&\simeq&2\la {\cal N}(p', s')|\bar q\frac{\vec\Sigma}{2} q(0)|{\cal N}(p,s)\ra,
\en
where the non-relativistic approximation is used in the last line and note that the operator is spin density in quark degree of freedom.
Changing the degree of freedom from quark to nucleon, as one usually do in effective theory, we have
\be
\la {\cal N}(p', s')|\bar q\frac{\vec\Sigma}{2} q(0)|{\cal N}(p,s)\ra
&=&
\la {\cal N}(p', s')|(\Delta_q^p \bar p\frac{\vec\Sigma}{2} p(0)+
\Delta_q^n \bar n\frac{\vec\Sigma}{2} n(0))|{\cal N}(p,s)\ra
\non\\
&\equiv&
\la {\cal N}(p', s')|(\Delta_q^p \vec s_p(0)+
\Delta_q^n \vec s_n(0))|{\cal N}(p,s)\ra,
\en
where $\Delta_{p(n)}^q$ is the quark spin proportion in a proton (neutron).

Note that spin operators $S_{p,n,{\cal N}}$ are related to $\vec s_{p,n,\cal N}$ by
\be
\vec S_{p,n,{\cal N}}=\int d^3 x \vec s_{p,n,{\cal N}}(0,\vec x).
\en
We consider the non relativistic case,
$p\simeq(m_{\cal N},\vec 0)$, $q\simeq0$,
\be
\la {\cal N}(p,s')|\vec s_{p,n,\cal N}(x)|{\cal N}(p,s)\ra
&\simeq&2m_{\cal N} \la J_{\cal N}, s'|\vec S_{p,n,\cal N}| J_{\cal N}, s\ra \exp(i q\cdot x).
\en
From Wigner-Eckart theorem, as the rotational properties of the above matrix element is well understood and are identical to that of the matrix element of any vector operator. Explicitly, from the Wigner-Eckart theorem, we have
\be
\la J_{\cal N}, s'|(\vec S_{p,n})_m|J_{\cal N},s\ra
&=&\la J_{\cal N} 1; s m|J_{\cal N} 1; J_{\cal N} s'\ra \la J_{\cal N}||S_{p,n}|| J_{\cal N}\ra,
\non\\
\la J_{\cal N},s'|(\vec S_{\cal N})_m|J_{\cal N},s\ra
&=&\la J_{\cal N} 1; s m|J_{\cal N} 1; J_{\cal N} s'\ra \la J_{\cal N}||S_{\cal N}|| J_{\cal N}\ra,
\en
with $(\vec S_{p,n,\cal N})_{m=0,\pm 1}=(\vec S_{p,n,\cal N})_z, \mp[(\vec S_{p,n,\cal N})_x\pm i (\vec S_{p,n,\cal N})_y]/\sqrt2$.
Since the double line matrix elements are independent $s$, $s'$ (with $m=s'-s$), so does the ratio
\be
\frac{\la J_{\cal N}, s'|(\vec S_{p,n})_m|J_{\cal N},s\ra}
{\la J_{\cal N},s'|(\vec S_{\cal N})_m|J_{\cal N},s\ra}
=\frac{\la J_{\cal N}||S_{p,n}|| J_{\cal N}\ra}
{\la J_{\cal N}||S_{\cal N}|| J_{\cal N}\ra}
\equiv\lambda_{p,n}.
\en
Consequently, its value can be obtained by taking a convenient choice of $s,s'$ as $s=s'=J_{\cal N}$ and $m=0$.
In other words, we have
\be
\la J_{\cal N}, s'|\vec S_{p,n}|J_{\cal N},s\ra
&=&\lambda_{p,n}\la J_{\cal N},s'|\vec S_{\cal N}|J_{\cal N},s\ra,
\en
with
\be
\lambda_{p,n}=\frac{\la J_{\cal N}, s=J_{\cal N}|(S_{p,n})_z|J_{\cal N},s=J_{\cal N}\ra}
{\la J_{\cal N}, s=J_{\cal N}|(S_{\cal N})_z|J_{\cal N},s=J_{\cal N}\ra}
\equiv\frac{\la S_{p,n,z}\ra}{J_{\cal N}}.
\en
When the contributions of the two body current are included,
one needs to change $\la S_{p,n}\ra$ in $\lambda_{p,n}$ into effective $\la S_{p,n}\ra_{\rm eff}$,
where we have
\be
\la S_{p(n)}\ra_{\rm eff}\equiv  \la S_{p(n)}\ra\pm\delta a_1\frac{\la S_p\ra-\la S_n\ra}{2},
\en
and $\delta a_1$ is the fraction contributing to the isovector coupling~\cite{Menendez}.
We use the predicted spin expectation values in Ref.~\cite{Menendez,XENON100SD} for the calculation.
Putting everything together in the $q=0$ limit and the $p^{(\prime)}$ rest frame, we obtain
\be
\la {\cal N}(m_{\cal N}, s')|j_{Aq}^i(0)|{\cal N}(m_{\cal N},s)\ra
&=&
4m (\Delta_q^p \lambda_p+\Delta_q^n \lambda_n) \la J_{\cal N}, s'|(\vec S_{\cal N})_i|J_{\cal N},s\ra,
\non\\
\la {\cal N}(m_{\cal N}, s')|j_{Aq}^0(0)|{\cal N}(m_{\cal N},s)\ra&=&0,
\label{eq: jA at q=0}
\en
where Eq. (\ref{eq: q=0})
has been used. These results will be useful later.

Similarly, from Eq.~(\ref{eq: parity}), we have
\be
\la {\cal N}(m_{\cal N},s')|s_h(0)|{\cal N}(m_{\cal N},s)\ra
&=&2m_{\cal N} f_{s\cal N}\delta_{ss'},
\non\\
\la {\cal N}(m_{\cal N},s')|p_h(0)|{\cal N}(m_{\cal N},s)\ra&=&0,
\label{eq: s,p}
\en
where $s_h=a^q \bar q q$. Using (no sum on $q$)
\be
\la p(p,s')|m_q \bar q q(0)|p(p,s)\ra
&=&2 E\delta_{ss'} m_q f_{sp,q}
\non\\
&=&
2 E\delta_{ss'}
\left\{
\begin{array}{lr}
m_p f^{(p)}_{Tq},
& q=u,d,s,\\
\frac{2}{27} m_p \left(1-\sum_{q=u,d,s} f^{(p)}_{Tq}\right),
&q=c,b,t.
\end{array}
\right.
\en
In the above, the matrix elements of the light-quark currents in the proton or neutron are obtained in chiral perturbation theory from measurements of the pion-nucleon sigma term~\cite{Cheng1,Cheng2,GLS}. The heavy quark contribution to the mass of the nucleon through the triangle diagram~\cite{SVZ}.
Consequently, we have
\be
f_{s\cal N}&=&(Z f_{sp}+(A-Z) f_{sn}),
\non\\
f_{s p(n)}&=&a^q f_{sp,q}=\sum_{q=u,d,s} a^q \frac{m_{p(n)}}{m_q} f^{(p(n))}_{Tq}+
\sum_{q=c,b,t} a^q\frac{2}{27} \frac{m_{p(n)}}{m_q} \left(1-\sum_{q'=u,d,s} f^{(p(n))}_{Tq'}\right).
\en
These matrix elements at $q=0$ are used in Eq.~(\ref{eq:MAq=0}) in Sec.~\ref{Direct} to obtain the DM-nucleus scattering differential cross section at $q^2=0$.

\subsection{Total cross section $\sigma$ and $\sigma_0$}

Using the standard formula, we find that the differential cross
section in the center of mass frame is given by
\be
\frac{d\sigma(q^2=0)}{d\cos\theta}
=\frac{1}{32\pi S}\frac{p'_\chi}{p_\chi} \overline{\sum}|M_{fi}(q^2=0)|^2
\simeq\frac{\mu^2_{\cal N}}{32\pi m^2_Nm^2_\chi}\overline{\sum}|M_{fi}(q^2=0)|^2,
\en
where $\mu_{\cal N}$ is the reduced mass of $m_\chi$ and $m_N$.
The explicit expression of $M_{fi}$ is given in Eq.~(\ref{eq: iM^2}).
It is useful to define $\sigma_0$ as~\cite{JKG}
\be
\sigma_0\equiv\left|\frac{d\sigma (q^2=0)}{d|{\bf q}|^2}\right| \int^{4\mu_{\cal N}^2v^2}_0d|{\bf q}|^2.
\en
Recall that we have $|{\bf q}|^2=-q^2=2\mu_{\cal N} v^2(1-\cos\theta)$ and, consequently, the Jacobian
$d|{\bf q}|^2/d\cos\theta=-2\mu_{\cal N}^2 v^2$ is a constant. The quantity $\sigma_0$ can now be expressed as
\be
\sigma_0=\left|\frac{d\sigma (q^2=0)}{d\cos\theta}\right|
\int_{-1}^{1} d{\cos\theta}
\simeq\frac{\mu^2_{\cal N}}{16\pi m^2_Nm^2_\chi}\overline{\sum}|M_{fi}(q^2=0)|^2.
\en
The differential cross section $d\sigma/d |{\bf q}|^2$ with nonzero momentum transfer are parametrized as~\cite{JKG}
\be
\frac{d\sigma (q^2)}{d|{\bf q}|^2}
=\frac{d\sigma (q^2=0)}{d|{\bf q}|^2} F^2(|{\bf q}|^2)
\en
with $F^2(|{\bf q}|^2)$ a form factor,
giving
\be
\sigma=\int^{4\mu^2v^2}_0d|{\bf q}|^2
\frac{d\sigma (q^2)}{d|{\bf q}|^2}
=\int^{4\mu^2v^2}_0d|{\bf q}|^2 F^2(|{\bf q}|)
\frac{d\sigma (q=0)}{d|{\bf q}|^2}
=\frac{\sigma_0}{4\mu^2v^2}\int^{4\mu^2v^2}_0d|{\bf q}|^2 F^2(|{\bf q}|).
\en

\subsection{Normalizing $\sigma$}

The generic form of SI cross section $\sigma_0$ of DM scattering off the nucleus $A$ with the $i^{\rm th}$ isotope induced by spin independent interaction is
\be
\sigma^{SI}_{0,A_i}
&\simeq&
\frac{\mu^2_{\cal N}}{16\pi}\frac{\overline{\sum}|M^{SI}_{fi}(q^2=0)|^2}{m_{\cal N}^2 m_\chi^2}
\non\\
&=&\frac{\mu^2_{A_i}}{16\pi} (Q_{VA_i}^2+Q_{SA_i}^2)
\non\\
&\equiv&\frac{\mu^2_{A_i}}{16\pi}\sum_{X=V,S}C_X[f_{Xp} Z+ f_{Xn} (A_i-Z)]^2,
\en
where
\be
C_V=16\kappa^2_\chi \frac{v^2}{1-v^2}\quad {\rm and}\quad
C_S=16\kappa^2_\chi.
\en
For proton ($A=1, Z=1$) and neutron ($A=1, Z=0$), the above formulae give
\be
\sigma^{SI}_{0,p}&=&\sum_{X=V,S}\sigma^{SI(X)}_{0,p}
=\frac{\mu^2_p}{16\pi} (C_V f_{Vp}^2+C_S f_{Sp}^2),
\non\\
\sigma^{SI}_{0,n}&=&\sum_{X=V,S}\sigma^{SI(X)}_{0,n}
=\frac{\mu^2_n}{16\pi} (C_V f_{Vn}^2+C_S f_{Sn}^2).
\en
For the nucleus with atomic mass number $A_i$ and isotope abundance $\eta_i$,
we define a scaled cross section as the following
\be
\sigma^Z_N&\equiv&\frac{\sum_i \eta_i\sigma^{SI}_{A_i}}
{\sum_j \eta_j  A^2_j\frac{\mu^2_{A_j}}{\mu^2_p}},
\label{eq: sigma Z N}
\en
with the SI DM-nucleus cross section defined as
\be
\sigma^{SI}_{A_i}\equiv
\int \frac{d|{\bf q}|^2}{4\mu^2_{A_i}v^2}\sigma^{SI}_{0,A_i} F^2_{SI}(|{\bf q}|),
\en
so that
\be
\sigma^Z_{0,N}=
\frac{\sum_{X=V,S}\sigma^{SI(X)}_{0,p}\sum_i\eta_i\mu^2_{A_i} [Z+(A_i-Z)\frac{f_{Xn}}{f_{Xp}}]^2}{\sum\eta_j \mu_{A_j}^2 A_j^2}.
\en
In the isospin limit,
\be
\frac{f_{Xp}}{f_{Xn}}\to 1,
\en
we have
\be
\sigma^Z_{0,N}\to\sigma^{SI}_{0,p}=\sigma^{SI}_{0,n}.
\en
Data obtained from different experiments can be compared using $\sigma^Z_N$ defined in Eq. (\ref{eq: sigma Z N}).
Note that
even if the isospin limit is not satisfied, we can still normalize $\sigma^{SI}_A$ to $\sigma^Z_N$ as in Eq. (\ref{eq: sigma Z N}) and compare it to the experimental result by taking $\sigma^Z_N$ as some sort of scaled cross section, but losing the generality among different experiments.

For spin dependent interaction,
from Eq. (\ref{eq: iM^2}),
we obtain
\be
\sigma^{SD}_{0,A_i}\simeq
\frac{\mu^2_{A_i}}{16\pi}
64 \kappa^2_\chi
{d^q d^{q'}(\Delta_q^p \la S_p\ra_{\rm eff}+\Delta_q^n \la S_n \ra_{\rm eff})
(\Delta_{q'}^p \la S_p\ra_{\rm eff}+\Delta_{q'}^n \la S_n\ra_{\rm eff})}\frac{J_{A_i}+1}{J_{A_i}}.
\en
When DM scatters off a proton (neutron) target, we have
\be
\sigma^{SD}_{0,p(n)}&\simeq&
\frac{\mu^2_{p(n)}}{16\pi}
64  \kappa^2_\chi
{d^q d^{q'}(\Delta_q^{p(n)} \frac{1}{2})
(\Delta_{q'}^{p(n)} \frac{1}{2})}\frac{(1/2)+1}{1/2}
\non\\
&\simeq&
\frac{\mu^2_{p(n)}}{16\pi}
64  \kappa^2_\chi
{d^q d^{q'}(\Delta_q^{p(n)})
(\Delta_{q'}^{p(n)})}\frac{3}{4}.
\en
Now return to the generic case, but observe that in the case the
proton (neutron) contribution dominates the interaction ($|d_q \Delta^q_{p(n)}|\gg |d_q \Delta^q_{n(p)}|$),
we have
\be
\sigma^{SD}_{0,A_i} \to
\frac{4\mu_{A_j}^2 \la S_{p,n}\ra^2_{\rm eff} (J_{A_j}+1)}{3\mu_{p,n}^2J_{A_j}}\sigma^{SD}_{0,p(n)}.
\en
Given the above result,
it will be useful to define the normalized DM-nucleus cross section as~\cite{nSD1,nSD2,nSD3}
\be
\sigma^{SD}_{p,n}\equiv (\sum_i\eta_i\sigma^{SD}_{A_i})
\left(
\sum_j \eta_j\frac{4\mu_{A_j}^2 \la S_{p,n}\ra^2_{\rm eff}  (J_{A_j}+1)}{3\mu_{p,n}^2J_{A_j}}\right)^{-1},
\label{eq: SD p n}
\en
with the DM-nucleus SD cross section
\be
\sigma^{SD}_{A_i}\equiv
\int \frac{d|{\bf q}|^2}{4\mu^2_{A_i}v^2}\sigma^{SD}_{0,A_i} F^2_{SD}(|{\bf q}|).
\en

In the above, the form factor is related to the structure function by ~\cite{XENON100SD,Ressell}
\be
F^2_{SD}(|{\bf q}|)=\frac{S_A(|{\bf q}|)}{S_A(0)}\quad {\rm so\ that}\quad
F^2_{SD}(0)=1,
\en
where
\be
S_A(0)=\frac{(2J+1)(J+1)}{\pi J}[a_p\la S_p\ra_{\rm eff} +a_n\la S_n\ra_{\rm eff}].
\label{eq: SA0}
\en
The axial-vector structure function $S_A(|{\bf q}|)$ can be written in terms of its isoscalar/isovector (0/1) structure factors $S_{00}(|{\bf q}|), S_{01}(|{\bf q}|)$ and $S_{11}(|{\bf q}|)$ as follows~\cite{Menendez}
\be
S_A(|{\bf q}|)=a^2_0S_{00}(|{\bf q}|)+a_0a_1S_{01}(|{\bf q}|) +a^2_1S_{11}(|{\bf q}|),
\label{eq: SAq}
\en
where the isoscalar and isovector couplings in this model are given by
\be
a_{0,1}=a_p\pm a_n=\frac{d^q}{G_F/\sqrt{2}}(\Delta^p_q\pm\Delta^n_q).
\en
In fact, the form factor can be defined as
\be
 F^2_{SD}(|{\bf q}|)&\equiv&\frac{(d^q \Delta_{q}^p)^2 \la S_p\ra^2_{\rm eff} F^2_{pp}(|{\bf q}|)
+2d^q d^{q'}\Delta_q^p \Delta_{q'}^n \la S_p\ra_{\rm eff}\la S_n \ra_{\rm eff} F^2_{pn}(|{\bf q}|)+
(d^q \Delta_{q}^n)^2 \la S_n\ra^2_{\rm eff}F^2_{nn}(|{\bf q}|)}
{(d^q \Delta_{q}^p)^2 \la S_p\ra^2_{\rm eff} F^2_{pp}(0)
+2d^q d^{q'}\Delta_q^p\Delta_{q'}^n  \la S_p\ra_{\rm eff}\la S_n \ra_{\rm eff} F^2_{pn}(0)+
(d^q \Delta_{q}^n)^2 \la S_n\ra^2_{\rm eff}F^2_{nn}(0)}
\non\\
&=&\frac{(d^q \Delta_{q}^p)^2 \la S_p\ra^2_{\rm eff} F^2_{pp}(|{\bf q}|)
+2d^q d^{q'}\Delta_q^p \Delta_{q'}^n \la S_p\ra_{\rm eff}\la S_n \ra_{\rm eff} F^2_{pn}(|{\bf q}|)+
(d^q \Delta_{q}^n)^2 \la S_n\ra^2_{\rm eff}F^2_{nn}(|{\bf q}|)}
{(d^q \Delta_{q}^p \la S_p\ra_{\rm eff} +d^q \Delta_{q}^n \la S_n\ra_{\rm eff})^2}.
\non\\
\en
where
\be
F^2_{pp(nn)}(|{\bf q}|)&\equiv&\frac{S_{00}(|{\bf q}|)+S_{11}(|{\bf q}|)\pm S_{01}(|{\bf q}|)}
{S_{00}(0)+S_{11}(0)\pm S_{01}(0)},
\quad
F^2_{pn}(|{\bf q}|)\equiv\frac{S_{00}(|{\bf q}|)-S_{11}(|{\bf q}|)}{S_{00}(0)-S_{11}(0)}.
\en
Using the following relations
\be
S_{00}(0)+S_{11}(0)\pm S_{01}(0)&=&\frac{(2 J_{A_i}+1)(J_{A_i}+1)}{\pi J_{A_i}}\la S_{p,n}\ra_{\rm eff}^2,
\non\\
S_{00}(0)-S_{11}(0)&=&\frac{(2 J_{A_i}+1)(J_{A_i}+1)}{\pi J_{A_i}}\la S_p\ra_{\rm eff} \la S_n\ra_{\rm eff},
\en
which the former is derived from Eq. (\ref{eq: SA0}) and the latter is from (\ref{eq: SAq}),
we recover the usual expression,
\be
F^2_{SD}(|{\bf q}|)=\frac{a^2_0 S_{00}((|{\bf q}|)+a_1 a_0 S_{01}((|{\bf q}|)+a_1^2 S_{11}((|{\bf q}|)}
{a^2_0 S_{00}(0)+a_1 a_0 S_{01}(0)+a_1^2S_{11}(0)},
\quad
a_{0,1}=\frac{d^q}{G_F/\sqrt{2}}(\Delta^p_q\pm\Delta^n_q).
\en
One may define another normalized SD cross section $\sigma^{\prime SD}$ by attempting to remove the $q^2$ dependence,
\be
\sigma^{\prime SD}_{p,n}\equiv
\sum_i\eta_i \int \frac{d|{\bf q}|^2}{4\mu_{A_i}^2 v^2}\sigma^{SD}_{0,A_i} F^2_{SD}(|{\bf q}|)
\left(
\sum_j \eta_j\frac{4\mu_{A_j}^2 \la S_{p,n}\ra^2_{\rm eff}  F^2_{pp(nn)}(|{\bf q}|)(J_{A_j}+1)}{3\mu_{p,n}^2J_{A_j}}\right)^{-1},
\label{eq: SD p n}
\en
Although $F_{SD}(|{\bf q}|)$ gives a compact expression for the relation between $\sigma^{SD}$ and $\sigma^{SD}_0$, it is not universal as it depends on the coupling $d^q$;
nevertheless,
$F^2_{pp,nn,pn}(|{\bf q}|)$ do not depend on the coupling $d^q$.
We will give another expression in below.

In the case with both spin-independent and spin-dependent interactions,
we have
\be
\frac{d\sigma_{A_i}}{d |{\bf q}|^2}
=\frac{1}{4\mu_{A_i}^2 v^2} (\sigma^{SD}_0 F^2_{SI}(|{\bf q}|)
+\sigma^{SD}_{0,pp} F^2_{pp}(|{\bf q}|)+\sigma^{SD}_{0,nn} F^2_{nn}(|{\bf q}|)+\sigma^{SD}_{0,pn} F^2_{pn}(|{\bf q}|)),
\en
where
\be
\sigma_0^{SI}
&=&\frac{\mu_{A_i}^2}{\pi}\kappa^2_\chi
\left[
\frac{v^2}{1-v^2} Q_{V A_i}^2
+f^2_{s A_i}\right],
\non\\
\sigma^{SD}_{0,pp(nn)}&=&\frac{\mu_{A_i}^2}{\pi} \kappa^2_\chi
\left[\left(4+\frac{4v^2}{3(1-v^2)}\right)  (\sum d^q \Delta_q^{p(n)})^2 \lambda^2_{p(n)}
J_{A_i}(J_{A_i}+1)
\right],
\non\\
\sigma^{SD}_{0,pn}&=&\frac{\mu_{A_i}^2}{\pi} \kappa^2_\chi
\left[\left(4+\frac{4v^2}{3(1-v^2)}\right)  2(\sum d^q d^{q'} \Delta_q^p \Delta_{q'}^n) \lambda_p\lambda_n
J_{A_i}(J_{A_i}+1)
\right].
\label{eq: dsigma}
\en
Consequently, we have
\be
\sigma_{A_i}=\int d|{\bf q}|^2 \frac{d\sigma}{d|{\bf q}|^2}=
 (\sigma^{SI}_0 r_{SI} +\sigma^{SD}_{0,pp} r_{pp}
+\sigma^{SD}_{0,nn} r_{nn}
+\sigma^{SD}_{0,pn} r_{pn}) ,
\en
where
\be
r_j\equiv \int_0^{4\mu_{A_i}^2 v^2} \frac{d|{\bf q}|^2}{4\mu_{A_i}^2 v^2} F^2_{j}(|{\bf q}|),
\en
with $j=SI, pp, nn, pn$.

We defined scaled cross sections as following:
\be
\sigma^Z_N&\equiv&\frac{\sum_i \eta_i\sigma_{A_i}}
{\sum_j \eta_j  A^2_j\frac{\mu^2_{A_j}}{\mu^2_p}}.
\en
and
\be
\sigma^{SD}_{p,n}\equiv
(\sum_i\eta_i \sigma_{A_i})
\left(
\sum_j \eta_j\frac{4\mu_{A_j}^2 \la S_{p,n}\ra^2  (J_{A_j}+1)}{3\mu_{p,n}^2J_{A_j}}\right)^{-1},
\label{eq: SD p n}
\en
or
\be
\sigma^{\prime SD}_{p,n}\equiv
(\sum_i\eta_i \sigma'_{A_i, p(n)})
\left(
\sum_j \eta_j\frac{4\mu_{A_j}^2 \la S_{p,n}\ra^2  (J_{A_j}+1)}{3\mu_{p,n}^2J_{A_j}}\right)^{-1},
\label{eq: SD p n 1}
\en
with
\be
\sigma'_{A_i, p(n)}&=& (\sigma^{SI}_0 r'_{SI,p(n)}
+\sigma^{SD}_{0,pp} r'_{pp,p(n)}
+\sigma^{SD}_{0,nn} r'_{nn, p(n)}
+\sigma^{SD}_{0,pn} r'_{pn, p(n)}),
\non\\
r'_{j, p(n)}&=& \int _0^{4\mu_{A_i}^2 v^2}\frac{d|{\bf q}|^2}{4\mu_{A_i}^2 v^2} \frac{F^2_j(|{\bf q}|)}{F^2_{pp(nn)}(|{\bf q}|)}.
\en
Data obtained from different experiments can be compared using $\sigma^Z_N$ and $\sigma^{SD}_{p,n}$ or $\sigma^{\prime SD}_{p,n}$.

\section{Coannihilation formulation}

It has been mentioned~\cite{GS} that coannihilation becomes significantly important if the mass splitting $\delta m\simeq T_f$ between the dark matter particle $\chi^0_1$ and one of the other WIMPs in this generic model. Let $\chi_1$ be the dark matter and 
$\chi_i (i=1,2,...,N)$ be the WIMPs having the masses with $m_i<m_j$ for $i<j$ and the internal degree of freedom $g_i$. Let $n_i$ denote the number density of $\chi_i$. We only need to consider the total number density $n=\sum_{i=1}^Nn_i$ since all WIMPs $\chi_i$ will eventually decay to the dark matter $\chi_1$. With the assumption $n_i/n\approx n_i^{\rm eq}/n^{\rm eq}$ before and after freeze-out, we have the following Boltzmann equation~\cite{GS} :
\be
\frac{dn}{dt}+3Hn=-<\sigma_{\rm eff} v_{\rm M\phi l}>(n^2-n^2_{\rm eq}),
\label{eq: Boltzmann2}
\en
where $<\sigma_{\rm eff}v_{\rm M\phi l}>=\sum_{i,j=1}^N\sigma_{ij}(\chi_i\chi_j\rightarrow XX')r_ir_jv_{ij}$. 
The $X$ and $X'$ denote the SM particles and the $r_i$ is the ratio of $n_i^{\rm eq}/n^{\rm eq}$. 
Let the mass fraction be $\Delta_i\equiv (m_i-m_1)/m_1$ so that $r_i$ can be given by
\be
r_i\equiv n_i^{\rm eq}/n^{\rm eq}=\frac{g_i(1+\Delta_i)^{3/2}\exp(-x\Delta_i)}{\sum_{i=1}^Ng_i(1+\Delta_i)^{3/2}\exp(-x\Delta_i)}\equiv \frac{g_i(1+\Delta_i)^{3/2}\exp(-x\Delta_i)}{g_{\rm eff}},
\en
Here we only consider the leading effect, namely, we only consider the effect of the WIMPs, $\chi^0_2$ and $\chi^\pm_{1,2}$, with two SM particles in the final states through the s-channel interaction. Similarly,
we do not take the Taylor series expansion on $v^2$ in s-channel and put a step function for the allowed threshold energy for each interaction channel in the non-relativistic thermal averaged cross section as follows:
\be
<\sigma_{\rm eff}v_{\rm M\phi l}>_{\rm n.r.}&=&\sum_{i,j}\frac{x_{ij}^{3/2}}{2\sqrt{\pi}}\sum_{A,B}\int_0^{\infty}dv v^2 e^{-x_{ij} v^2/4}
\non\\
& &\times [\sigma (\chi_i \bar{\chi_j}\rightarrow A+B)v] \theta[m_i^2+m_j^2+\frac{2m_i m_j}{\sqrt{1-v^2}}-(m_A+m_B)^2].
\en 
In the above, $x_{ij}=\frac{\mu_{ij}}{\mu_{11}}x$ and $\mu_{ij}$ is the reduced mass of $\chi_{i}$ and $\chi_{j}$.
From the freeze-out condition, the new freeze-out temperature parameter $x_f$ can be solved numerically by the following equation
\be
x_f={\rm ln}\left [c(c+2)\sqrt{\frac{45}{8}}\frac{g_{\rm eff} m_\chi M_{\rm PL}<\sigma_{\rm eff} v>_{x=x_f}}{2\pi^3 \sqrt{g_*(m_\chi )}x_f^{1/2}}\right ].
\label{eq:cxf}
\en
The relic density now becomes 
\be
\Omega_{\rm DM}h^2\approx 1.04\times 10^9\frac{\rm GeV^{-1}}{M_{\rm PL}\sqrt{g_*(m_\chi)}J(x_f)},
\label{eq:Omegah2}
\en
where
\be
J(x_f)\equiv \int_{x_f}^\infty\frac{<\sigma_{\rm eff}v>}{x^2}dx.
\en
Note that the $\sigma_{\rm eff}$ is not only the function of $v$, but also a function of $x$ in coannilation.

\section{Matrix elements for WIMP coannihilation}

In this article, we only consider the leading effect of coannihilation with the first two lightest neutral as well as single charged WIMP annihilating to SM fermions through s-channel.  

\subsection{$\chi^0_j\chi^+_k\rightarrow {\bar q'}q, l^+\nu$}

The neutral WIMP $\chi^0_j$ and the single charged WIMP $\chi^+_k$ can annihilate into SM fermions through the s-channel exchange of a $W^+$ boson corresponding to the following matrix element:

\be
M(\chi^0_j\chi^+_k\rightarrow {\bar q'}q)&=&
-i(\frac{g}{\sqrt{2}})^2V_{qq'}\frac{1}{s-M^2_W+iM_W\Gamma_W}g_{\mu\nu}
\non\\
&&
\times 
[\bar v_j(p_1)\gamma^\mu (O^{L_{W-}}_{jk}P_L+O^{R_{W-}}_{jk}P_R)u_k(p_2)]
[\bar u(p_4)\gamma^\nu P_Lv(p_3)],
\en
and
\be
M(\chi^0_j\chi^+_k\rightarrow  l^+\nu)&=&
-i(\frac{g}{\sqrt{2}})^2\frac{1}{s-M^2_W+iM_W\Gamma_W}g_{\mu\nu}
\non\\
&&
\times
[\bar v_j(p_1)\gamma^\mu (O^{L_{W-}}_{jk}P_L+O^{R_{W-}}_{jk}P_R)u_k(p_2)]
[\bar u(p_4)\gamma^\nu P_Lv(p_3)].
\en

\subsection{$\chi^0_k\chi^-_j\rightarrow q'{\bar q}, l^-\bar{\nu}$}

Similarly,
the neutral WIMP $\chi^0_k$ and the single charged WIMP $\chi^-_j$ can annihilate into SM fermions via the s-channel exchange of a $W^-$ boson corresponding to the following matrix element:
\be
M(\chi^0_k\chi^-_j\rightarrow q'{\bar q})&=&
-i(\frac{g}{\sqrt{2}})^2V^*_{qq'}\frac{1}{s-M^2_W+iM_W\Gamma_W}g_{\mu\nu}
\non\\
&&
\times
[\bar v_j(p_2)\gamma^\mu (O^{L_{W+}}_{jk}P_L+O^{R_{W+}}_{jk}P_R)u_k(p_1)]
[\bar u(p_3)\gamma^\nu P_Lv(p_4)],
\en
and
\be
M(\chi^0_k\chi^-_j\rightarrow  l^-\bar{\nu})&=&
-i(\frac{g}{\sqrt{2}})^2\frac{1}{s-M^2_W+iM_W\Gamma_W}g_{\mu\nu}
\non\\
&&
\times
[\bar v_j(p_2)\gamma^\mu (O^{L_{W+}}_{jk}P_L+O^{R_{W+}}_{jk}P_R)u_k(p_1)]
[\bar u(p_3)\gamma^\nu P_Lv(p_4)].
\en

\subsection{$\chi^0_j\chi^0_k\rightarrow f\bar f$}

 The neutral WIMPs $\chi^0_j$ and $\chi^0_k$ can annihilate into $f\bar f$ through the s-channel exchange of a $Z^0$ boson or a $H^0$ scalar corresponding to the following matrix element:
 
\be
M(\chi^0_j\chi^0_k\rightarrow f\bar f)=M_{1a}+M_{1b}+2M_{2a},
\en
where
\be
M_{1a}&=&-\frac{i}{2}(\frac{g}{\cos\theta_W})^2g_{\alpha\nu}\frac{1}{s-M^2_Z+iM_Z\Gamma_Z}
\non\\
&&
\times
[v_j(p_1)\gamma^\alpha [(O^{L_Z}_{jk}P_L+O^{R_Z}_{jk})P_R)u_k(p_2)]
[\bar u(p_3)\gamma^\nu (g^f_V+g^f_A\gamma^5)v(p_4)],
\non\\
M_{1b}&=&-\frac{i}{2}(\frac{g}{\cos\theta_W})^2g_{\alpha\nu}\frac{1}{s-M^2_Z+iM_Z\Gamma_Z}
\non\\
&&
\times
[v_k(p_1)\gamma^\alpha [(O^{L_Z}_{kj}P_L+O^{R_Z}_{kj})P_R)u_j(p_2)]
[\bar u(p_3)\gamma^\nu (g^f_V+g^f_A\gamma^5)v(p_4)],
\non\\
M_{2a}&=&i\frac{gm_f}{2M_W}\frac{1}{s-M^2_H+iM_H\Gamma_H}
[\bar v_j(p_1)(O^{L_{H}}_{jk}P_L+O^{R_{H}}_{jk}P_R)u_k(p_2)][\bar u(p_3)v(p-4)].
\en

\subsection{$\chi^-_j\chi^+_k\rightarrow f\bar f$}

The single charged WIMPs $\chi^-_j$ and $\chi^+_k$ can annihilate into $f\bar f$ through the s-channel exchange of a $Z^0$ boson, a $A^0$ boson, or a $H^0$ scalar corresponding to the following matrix element:

\be
M(\chi^-_j\chi^+_k\rightarrow f\bar f)=M_1+M_2+M_3,
\en
where
\be
M_1&=&i(\frac{g}{\cos\theta_w})^2\frac{1}{s-M^2_Z+iM_Z\Gamma_Z}g_{\alpha\mu}
\non\\
&&
\times
[\bar u(p_3)\gamma^\alpha (g^f_V+g^f_A\gamma^5)v(p_4)][\bar v_j(p_2)\gamma^\mu (O^{L^+_Z}_{jk}P_L+O^{R^+_Z}_{jk}P_R)j_k(p_1),
\non\\
M_2&=&-ie^2g_{\alpha\mu}\frac{1}{s}\delta_{jk}
[\bar u(p_3)\gamma\alpha v(p_4)][\bar v_j(p_2)\gamma^\mu u_k(p_1),
\non\\
M_3&=&-i\frac{gm_f}{2M_W}\frac{1}{s-M^2_H+iM_H\Gamma_H}
[\bar u(p_3)v(p_4)][\bar v_j(p_2)(O^{L^+_H}_{jk}P_L+O^{R^+_H}_{jk}P_R)u_k(p_1)].
\en

\vfill
\eject
\end{document}